\documentclass
    [phd,
     smplmath,
     indenttoc,
     zdraft]
    {iitmdissertation}

\usepackage{multirow}
\usepackage{bigints}
\usepackage{aas_macros}
\usepackage{relsize}
\usepackage[figuresright]{rotating}
\usepackage{tablefootnote}
\usepackage{xurl}
\usepackage{physics}
\usepackage{bm}
\usepackage[hyperpageref]{backref}

\definecolor{lightgray}{gray}{0.9}
\definecolor{darkteal}{HTML}{076568}
\def\mystrut{\vrule height 15pt depth 6pt width 0pt}

\newenvironment{PE_table}{\renewcommand{\arraystretch}{1.5}\setlength{\tabcolsep}{6.5pt}}{}

\renewcommand*{\backrefalt}[4]{%
\ifcase #1 %
No citations.%
\or
(cited on pg. #2).%
\else
(cited on pgs. #2).%
\fi
}

\newcommand{\takenfrom}[1]{\textbf{[}{\sf{\textit{Taken from #1}}}\textbf{]}}

\DeclareSymbolFont{toneitalic}{T1}{\familydefault}{m}{it}
\DeclareMathSymbol{\cpartial}{\mathord}{toneitalic}{"F0}

\dissertationDepartment      {Department of Physics}
\dissertationAuthorName      {Divyajyoti}
\dissertationDateOfBirth     {9 March, 1997}
\dissertationDegree          {Doctor of Philosophy}
\dissertationTitle           {Synergies in analysing binary black hole mergers}
\dissertationSubtitleA{Effect of orbital eccentricity, spin-precession, and}
\dissertationSubtitleB{non-quadrupole modes}
\dissertationMonth           {July}
\dissertationYear            {2024}
\dissertationCertificateDate {4 July 2024}
\dissertationType            {Thesis}
\dissertationGuides          {Dr. Chandra Kant Mishra}
\dissertationCommitteeName   {Doctoral Committee}
\newglossaryentry
    {latex}
    {
        name        = {\LaTeX},
        description = {{\LaTeX} is a popular macro package for document preparation},
    }

\newglossaryentry
    {math}
    {
        name        = {Mathematics},
        description = {A very versatile language},
    }

\newglossaryentry
    {phy}
    {
        name        = {Physics},
        description = {Trying to understand the world by describing it using mathematics},
    }

\newglossaryentry
    {csprog}
    {
        name        = {Computer Science and Programming\ \ },
        description = {An art form that is commonly associated with bringing silicon to life},
    }

\newacronym{bbh} {BBH} {Binary Black Hole}
\newacronym{hm} {HM} {Higher Mode}
\newacronym{gwtc} {GWTC} {Gravitational Wave Transient Catalog}
\newacronym{lvk} {LVK} {LIGO-Virgo-KAGRA}
\newacronym{nr} {NR} {Numerical Relativity}
\newacronym{pe} {PE} {Parameter Estimation}
\newacronym{snr} {SNR} {Signal-to-Noise Ratio}
\newacronym{isco} {ISCO} {Innermost Stable Circular Orbit}
\newacronym{O1}{O1}{First observing run. Similarly for O2, O3, and O4}
\newacronym{nsbh}{NSBH}{Neutron-Star Black Hole}
\newacronym{bns}{BNS}{Binary Neutron Star}
\newacronym{siqm}{SIQM}{Spin-Induced Quadrupole Moment}
\newacronym{gw}{GW}{Gravitational Wave}
\newacronym{gr}{GR}{General theory of Relativity}
\newacronym{imr}{IMR}{Inspiral-Merger-Ringdown}
\newacronym{pn}{PN}{Post-Newtonian formalism}
\newacronym{ff}{FF}{Fitting Factor}
\newacronym{srf}{SRF}{Signal Recovery Fraction}
\newacronym{psd}{PSD}{Power Spectral Density}
\newacronym{ligo}{LIGO}{Laser Interferometer Gravitational-wave Observatory}
\newacronym{kagra}{KAGRA}{KAmioka GRAvitational-wave detector}
\newacronym{lisa}{LISA}{Laser Interferometer Space Antenna}
\newacronym{et}{ET}{Einstein Telescope}
\newacronym{ce}{CE}{Cosmic Explorer}
\newacronym{aLIGO}{aLIGO}{Advanced LIGO}
\newacronym{grb}{GRB}{Gamma-Ray Burst}
\newacronym{cmb}{CMB}{Cosmic Microwave Background}
\newacronym{pta}{PTA}{Pulsar Timing Array}
\newacronym{pycbc}{\texttt{PyCBC}}{Python-based toolkit for Compact Binary Coalescence signals}
\newacronym{bilby}{\texttt{Bilby}}{Bayesian inference library}
\newacronym{mcmc}{MCMC}{Markov Chain Monte Carlo}
\newacronym{far}{FAR}{False Alarm Rate}
\newacronym{em}{EM}{Electro-Magnetic}
\newacronym{tgr}{TGR}{Test of General Relativity}
\newacronym{eco}{ECO}{Exotic Compact Object}
\newacronym{bh}{BH}{Black Hole}
\newacronym{eob}{EOB}{Effective-One Body formalism}
\newacronym{lalsuite}{\texttt{LALSuite}\ \ \ }{LSC Algorithm Library Suite}
\newacronym{lsc}{LSC}{LIGO Scientific Collaboration}
\newacronym{sxs}{SXS}{Simulating eXtreme Spacetimes catalog}
\newacronym{rit}{RIT}{Rochester Institute of Technology catalog}
\newacronym{lso}{LSO}{Last Stable Orbit}
\newacronym{tb}{TB}{Template Bank}
\makeglossaries

\setglossarystyle{alttree}

\graphicspath{{./future-dets/}}

\usepackage{auxhook}
\begin{document}
\thispagestyle{empty} \pagenumbering{roman}
        \pagestyle{empty}
        \renewcommand*{\chapterpagestyle}{empty}
        \newgeometry{
    left    = 1.25in,
    right   = 0.5in,
    top     = 1in,
    bottom  = 1in,
    foot    = 0.67in,
}
%\doCoverTitleBorder
%\input zz-imp/pre-imp-cover
%\cleardoublepage

\ifdississynopsis\else
    \doCoverTitleBorder
    % \newgeometry{
%     left    = 1.25in,
%     right   = 0.5in,
%     top     = 1in,
%     bottom  = 1.34in,
%     foot    = 0.67in,
% }
%
% \AddToShipoutPictureBG{%
%     \color{\dissertationTitleTapeColor}
%     \AtPageLowerLeft{%
%         \rule{\dimexpr 1in + \oddsidemargin\relax}{\paperheight}%
%     }%
% }

\begin{minipage}{0.44\textwidth}
    \quad\includegraphics[width=0.5\textwidth]{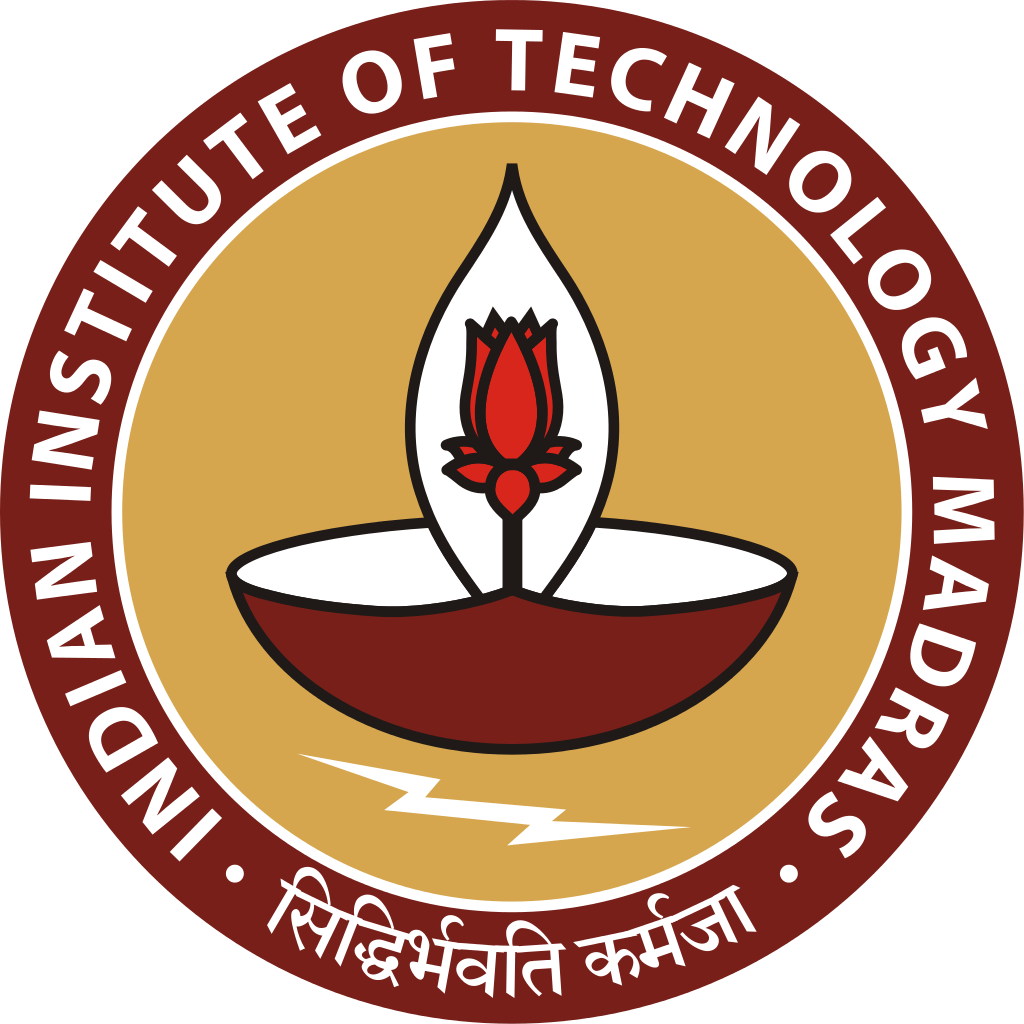}
\end{minipage}
\hfill
\begin{minipage}{0.55\textwidth}
    \setstretch{1}
    \textupper{\asserttoken{dissertationDepartment}{Department of {\TeX} Engineering}}
    \par
    \textupper{Indian Institute of Technology Madras}
    \par
    \textupper{Chennai -- 600036}
\end{minipage}
\par
\makebox[\textwidth][l]{\rule{\paperwidth}{1mm}}

\begin{center}
\vspace*{\parskip}
\setlength{\fboxrule}{2pt}
\setlength{\fboxsep}{10pt}
% \fbox{%
        \begin{minipage}{0.75\textwidth}
        \begin{center}
            {\bfseries\huge\asserttoken{dissertationTitle}{The Title of the Thesis Will Appear Here if you Set it}}\\
            \vspace{-0.25em}
            \textit{{\bfseries\large\asserttoken{dissertationSubtitleA}{The subtitle will appear here if you set it}}}\\
            \vspace{-0.75em}
            \textit{{\bfseries\large\asserttoken{dissertationSubtitleB}{The subtitle will appear here if you set it}}}
        \end{center}
        \end{minipage}
%      }
\end{center}

\if\relax\the\dissertationImage\relax%
    \def\diss@image{nodissimageprovided}
\else
    \def\diss@image{images/\the\dissertationImage}
\fi

\vspace{\parskip}

\IfFileExists
    {\diss@image}
    {\begin{center}
        \includegraphics[width=0.225\textheight]{\diss@image}
     \end{center}}
    {\vspace{0.1\textheight}}

\vspace{\dimexpr2\parskip}

\begin{center}
    \begin{minipage}[c]{0.9\textwidth}
        \begin{center}
            \begingroup
                \itshape
                A \asserttoken{dissertationType}{Whatever}
                \par
                \ifdississynopsis
                To be \else\fi
                Submitted by
            \endgroup
            \par
            \textupper{{\bfseries\asserttoken{dissertationAuthorName}{Firstname Lastname}}}
        \end{center}
    \end{minipage}
\end{center}

\vspace{\dimexpr2\parskip}

\begin{center}
    \begin{minipage}[c]{0.9\textwidth}
        \begin{center}
            \begingroup
                \itshape
                For the award of the degree
                \par
                Of
            \endgroup
            \par
            \textupper{{\bfseries\asserttoken{dissertationDegree}{Whatever Degree}}}
            \par
            \asserttoken{dissertationMonth}{January}\ \asserttoken{dissertationYear}{2022}
        \end{center}
    \end{minipage}
\end{center}

% \newpage
%
% \restoregeometry
% \ClearShipoutPictureBG

    \vspace*{\fill}
    \hskip 1em \textcopyright\ \asserttoken{dissertationYear}{2024}\  Indian Institute of Technology Madras
\fi

\restoregeometry
% \ClearShipoutPictureBG
\cleardoublepage

      \chapter*{}
\vspace*{\fill}

\begin{center}
\begin{minipage}{0.7\textwidth}
\itshape
    %Outside the ship, the thin membrane of space-time rippled with the gravitational waves, like a placid lake surface disturbed by a night breeze. The judgment of death for both worlds spread across the cosmos at the speed of light.

    ``As the smaller black hole plunged towards the larger one, it knew its death wouldn't be in vain. A billion years in the future, scientists on a tiny blue planet would listen to the song it sang before it went down for the final sleep...''
\end{minipage}
\end{center}
%\hspace*{0.75\textwidth} {\bfseries -- Liu Cixin}

\vspace*{\fill}
\newpage

      \chapter*{}
\vspace*{\fill}

\begin{center}
\begin{minipage}{0.75\textwidth}
\itshape
\centering
    To Mukesh, my partner in work and in life...
\end{minipage}
\end{center}

\vspace*{\fill}
\newpage

      %\printCertificate
      \chapter*{List of Publications}
\thispagestyle{empty}
\begin{enumerate}[{\bfseries I.}, leftmargin=1cm,itemindent=0em,]

    \itemhead{Publications in refereed journals appearing in thesis}

\begin{enumerate}

\item \textbf{Divyajyoti}, Sumit Kumar, Snehal Tibrewal, Isobel M. Romero-Shaw, Chandra Kant Mishra\\
\textit{Blind spots and biases: the dangers of ignoring eccentricity in gravitational-wave signals from binary black holes}\\
\textbf{Physical Review D 109, 043037 (2024)}. arXiv:2309.16638. \citep{Divyajyoti:2023rht}

\item \textbf{Divyajyoti}, N. V. Krishnendu, Muhammed Saleem, Marta Colleoni, Aditya Vijaykumar, K. G. Arun, Chandra Kant Mishra\\
\textit{Effect of double spin-precession and higher harmonics on spin-induced quadrupole
moment measurements}\\
\textbf{Physical Review D 109, 023016 (2024)}. arXiv:2311.05506. \citep{Divyajyoti:2023izl}

\item \textbf{Divyajyoti}, Preet Baxi, Chandra Kant Mishra, K. G. Arun\\
\textit{\small Detectability of gravitational higher modes in the third-generation era}\\
\textbf{Physical Review D 104, 084080 (2021)}. arXiv:2103.03241. \citep{Divyajyoti:2021uty}

\item Abhishek Chattaraj, Tamal RoyChowdhury, \textbf{Divyajyoti}, Chandra Kant Mishra, Anshu Gupta\\
\textit{\small High accuracy PN and NR comparisons involving higher modes for eccentric BBHs and a dominant mode eccentric IMR model}\\
\textbf{Physical Review D 106, 124008 (2022)}. arXiv:2204.02377. \citep{Chattaraj:2022tay}

\item The LIGO Scientific Collaboration (incl. \textbf{Divyajyoti}), the Virgo Collaboration, the KAGRA Collaboration\\
\textit{\small Tests of General Relativity with GWTC-3}\\
\textbf{Accepted for publication in Physical Review D (2022)}. arXiv:2112.06861. \citep{LIGOScientific:2021sio}
\end{enumerate} \par

%     \itemhead{Preprints appearing in thesis}

% \begin{enumerate}
%     \item \textbf{Divyajyoti}, Sumit Kumar, Snehal Tibrewal, Isobel M. Romero-Shaw, Chandra Kant Mishra:\\
% \textit{Blind spots and biases: the dangers of ignoring eccentricity in gravitational-wave signals from binary black holes}\\
% \textbf{Submitted to Physical Review D}. arXiv:2309.16638. \citep{Divyajyoti:2023rht}
% \end{enumerate} \par

    \itemhead{Publications in refereed journals (others)}
\begin{enumerate}

\item Mukesh Kumar Singh, \textbf{Divyajyoti}, Shasvath J Kapadia, Md Arif Shaikh, Parameswaran Ajith\\
\textit{Improved early-warning estimates of luminosity distance and orbital inclination of compact binary mergers using higher order modes of gravitational radiation}\\ 
\textbf{Monthly Notices of the Royal Astronomical Society 513 (3), 3798–3809 (2022)}. arXiv:2202.05802. \citep{Singh:2022tlh}

\item The LIGO Scientific Collaboration (incl. \textbf{Divyajyoti}), the Virgo Collaboration, the KAGRA Collaboration\\
\textit{GWTC-2.1: Deep Extended Catalog of Compact Binary Coalescences Observed by LIGO and Virgo During the First Half of the Third Observing Run}\\
\textbf{Physical Review D 109, 022001 (2024)}. arXiv:2108.01045. \citep{LIGOScientific:2021usb}
\end{enumerate} %\par

\end{enumerate}
\cleardoublepage

%%
%%
%%
%%%%%%%%%%%%%%%%%%%%%%%%%%%%%%%%%%%%%%%%%%%%%%%%%%%%%%%%%%%%%%%%%%%%%%%%%%%%%%%
% A place to keep maths symbol and unit definitions                           %
%                                                                             %
% If you want to add numbers, please do so in result_numbers.tex.             %
% If you want to add acronyms or abbreviations, use acronyms.tex.             %
%                                                                             %
% Symbols should follow https://dcc.ligo.org/LIGO-T2000185                    %    
%                                                                             %
%%%%%%%%%%%%%%%%%%%%%%%%%%%%%%%%%%%%%%%%%%%%%%%%%%%%%%%%%%%%%%%%%%%%%%%%%%%%%%%

%%%%%%%%%%%%%%%%%
%%%%% units %%%%%
%%%%%%%%%%%%%%%%%

\newcommand{\Msun}{\ensuremath{{M}_\odot}}
\newcommand\Mpcyr{\ensuremath{\mathrm{Mpc}^{3}\,\mathrm{yr}}}
\newcommand\Gpcyr{\ensuremath{\mathrm{Gpc}^{3}\,\mathrm{yr}}}
\newcommand\perMpcyr{\ensuremath{\mathrm{Mpc}^{-3}\,\mathrm{yr}^{-1}}}
\newcommand\perGpcyr{\ensuremath{\mathrm{Gpc}^{-3}\,\mathrm{yr}^{-1}}}

%%%%%%%%%%%%%%%%%%%
%%%%% symbols %%%%%
%%%%%%%%%%%%%%%%%%%

% mass parameters
\newcommand{\massone}{\ensuremath{m_1}}
\newcommand{\masstwo}{\ensuremath{m_2}}
\newcommand{\Mc}{\ensuremath{\mathcal{M}}}
\newcommand{\Mtot}{\ensuremath{M}}
\newcommand{\Mf}{\ensuremath{M_\mathrm{f}}}
\newcommand{\massratio}{\ensuremath{q}}
\newcommand{\Erad}{\ensuremath{E_\mathrm{rad}}}

% spin parameters
\newcommand{\chieff}{\ensuremath{\chi_\mathrm{eff}}}
\newcommand{\chip}{\ensuremath{\chi_\mathrm{p}}}
\newcommand{\chif}{\ensuremath{\chi_\mathrm{f}}}
\newcommand{\spintilt}[1]{\ensuremath{\theta_{{LS}_{#1}}}}
\newcommand{\spinone}{\ensuremath{\chi_1}}
\newcommand{\vecspinone}{\ensuremath{\vec\chi_1}}
\newcommand{\spintwo}{\ensuremath{\chi_2}}
\newcommand{\vecspintwo}{\ensuremath{\vec\chi_2}}
\newcommand{\LNewton}{\ensuremath{\hat{L}_\mathrm{N}}}
\newcommand{\thetaJN}{\ensuremath{\theta_{JN}}}

% distance
\newcommand{\DL}{\ensuremath{D_\mathrm{L}}}
\newcommand{\DC}{\ensuremath{D_\mathrm{c}}}
\newcommand{\redshift}{\ensuremath{z}}

% PE quantities
\newcommand{\PEprob}{\ensuremath{p}}%
\newcommand{\PEparameter}{\ensuremath{\vec{\theta}}}%
\newcommand{\PEdata}{\ensuremath{{d}}}%
\newcommand{\PEprior}{\ensuremath{\PEprob(\PEparameter)}}%
\newcommand{\PEposterior}{\ensuremath{\PEprob(\PEparameter | \PEdata)}}%
\newcommand{\PElikelihood}{\ensuremath{\PEprob(\PEdata | \PEparameter)}}%
\newcommand{\PEmodelh}{\ensuremath{h_{\mathrm{M}}}}% 

% PE setting
\newcommand{\flow}{\ensuremath{f_\mathrm{low}}}
\newcommand{\fhi}{\ensuremath{f_\mathrm{high}}}
\newcommand{\fsamp}{\ensuremath{f_\mathrm{s}}}
\newcommand{\fNyq}{\ensuremath{f_\mathrm{Nyquist}}}
\newcommand{\alphaRoll}{\ensuremath{\alpha^\mathrm{roll\text{-}off}}}

% Cosmology symbols
\newcommand{\HzeroSymbol}{\ensuremath{H_{0}}}
\newcommand{\WmSymbol}{\ensuremath{\Omega_{\mathrm{m}}}}

% KL divergences
\newcommand{\DKLchip}{\ensuremath{D_\mathrm{KL}^{\chi_\mathrm{p}}}}
\newcommand{\DKLchieff}{\ensuremath{D_\mathrm{KL}^{\chi_\mathrm{eff}}}}

% search pipeline parameters
\newcommand{\rankstat}{\ensuremath{x}}

% others 
\newcommand{\VT}{\ensuremath{\langle VT \rangle}}
\newcommand{\pastro}{\ensuremath{p_{\mathrm{astro}}}}
\newcommand{\pterr}{\ensuremath{p_{\mathrm{terr}}}}
\newcommand{\pbbh}{\ensuremath{p_{\mathrm{BBH}}}}
\newcommand{\pbns}{\ensuremath{p_{\mathrm{BNS}}}}
\newcommand{\pnsbh}{\ensuremath{p_{\mathrm{NSBH}}}}
\newcommand{\comovingv}{\ensuremath{V_\mathrm{c}}}
\newcommand{\comovingvt}{\ensuremath{\langle VT_\mathrm{c} \rangle}}
\newcommand{\injspinmax}{\ensuremath{\chi_\mathrm{max}}}

  \begin{FrontMatter}
      \printGlossaryAndAbbreviations
      \chapter{List of waveform models used}

\begin{table}[h]
\small
\def\arraystretch{1.1}
\begin{tabular}{|p{0.2\linewidth}|p{0.2\linewidth}|p{0.51\linewidth}|}
\hline
\multicolumn{1}{|c|}{\textbf{Waveform name}} & \multicolumn{1}{c|}{\textbf{Modes ($\bm{\ell, |m|}$)}} & \multicolumn{1}{c|}{\textbf{Features}} \\ \hline
\texttt{IMRPhenomD} & (2,2) & \citep{Khan:2015jqa} Aligned spin \\ \hline
\texttt{IMRPhenomPv2} & (2,2) & \citep{Hannam:2013oca} Single-spin precession, based on \texttt{IMRPhenomD} \\ \hline
\texttt{IMRPhenomHM} & (2,2), (2,1), (3,3), (3,2), (4,4), (4,3) & \citep{London:2017bcn} Higher modes, aligned spin, based on \texttt{IMRPhenomD} \\ \hline
\texttt{IMRPhenomXAS} & (2,2) & \citep{Pratten:2020fqn} Aligned spin, multi-banding \\ \hline
\texttt{IMRPhenomXHM} & (2,2), (2,1), (3,3), (3,2), (4,4) & \citep{Garcia-Quiros:2020qpx} Higher modes, aligned spin, multi-banding, based on \texttt{IMRPhenomXAS} \\ \hline
\texttt{IMRPhenomXPHM} & (2,2), (2,1), (3,3), (3,2), (4,4) & \citep{Pratten:2020ceb} Higher modes, double-spin precession, multi-banding, based on \texttt{IMRPhenomXHM} \\ \hline
\texttt{IMRPhenomXP} & (2,2) & \citep{Pratten:2020ceb} Double spin-precession, multi-banding, limiting case of \texttt{IMRPhenomXPHM} \\ \hline
\end{tabular}
\caption[List of quasi-circular, frequency-domain, phenomenological waveforms]{Quasi-circular, frequency-domain, phenomenological waveforms, modelling the entire inspiral-merger-ringdown regime. %Waveforms in \texttt{IMRPhenom} family are for compact binaries at comparable masses, whereas those in \texttt{IMRPhenomX} family for compact binaries at comparable and extreme mass ratios. 
The columns denote the name of the waveform approximant coded up in \texttt{LALSuite}, the spherical harmonic modes (in the co-precessing frame for waveforms with spin-precession), and main features of the waveform along with the most relevant publication.}
\end{table}
%\vspace{-1em}
\begin{table}[h]
\small
\def\arraystretch{1.1}
\begin{tabular}{|p{0.2\linewidth}|p{0.74\linewidth}|}
\hline
\multicolumn{1}{|c|}{\textbf{Waveform name}} & \multicolumn{1}{c|}{\textbf{Features}} \\ \hline
\texttt{TaylorF2Ecc} & \citep{Moore:2016qxz} Inspiral-only, frequency-domain, eccentricity corrections in phase, quasi-circular aligned spins \\ \hline
\texttt{EccentricFD} & \citep{Huerta:2014eca} Inspiral-only, frequency domain, eccentricity corrections in both amplitude and phase, non-spinning \\ \hline
\texttt{TEOBResumS}\tablefootnote{For \texttt{TEOBResumS}, the higher mode version \texttt{TEOBiResumS\_SM} \citep{Nagar:2020pcj} is also available, but we have used the dominant mode version in this thesis.} & \citep{Chiaramello:2020ehz} Inspiral-merger-ringdown, time-domain, eccentricity corrections in both amplitude and phase, aligned-spin \\\hline
\texttt{ENIGMA} & \citep{Huerta:2017kez} Inspiral-merger-ringdown, time-domain, eccentricity corrections in both amplitude and phase, non-spinning \\\hline
\end{tabular}
\caption[List of eccentric waveform models]{Eccentric, dominant mode, waveform models.}
\end{table}
      \printNotation
  \end{FrontMatter}
  \begin{MainMatter}
      \chapter{Introduction}

Gravity! A force as old as time itself! Since the beginning of civilisation, mankind has tried to make sense of nature's fundamental forces, to peek under the matrix and get answers to questions immemorial. Over time, as empires rose and fell, our understanding of the universe, fundamental forces and particles, physics and mathematics evolved, but one constant remained: \textit{curiosity}. It is this curiosity that drives scientists today to look beyond the realms of the known, to look for objects which are beyond our imagination, and to look back in time in an attempt to uncover the greatest mysteries of the universe: how it was formed, how it evolved, and what will be its end. Gravitational wave astronomy (GWA) is one such branch of physics that attempts to answer these questions by looking for signatures of perturbations in the very fabric of spacetime. Standing on the shoulders of giants, in this \textbf{thesis}, I attempt to answer some of the questions which plague the gravitational wave community today.

\section{History of gravitational waves}
\label{sec:intro-history}

Modern mathematical description of gravitational force goes back as early as the 16\textsuperscript{th} and 17\textsuperscript{th} century when a number of scientists, including Grimaldi and Riccioli \citep{Heilbron+1979}, Robert Hooke \citep{Hooke:1674abc}, Bullialdus and Borelli \citep{1645ibap.book.....B, 1666tmpe.book.....B}, and Isaac Newton, \citep{1687pnpm.book.....N} claim to have contributed towards the formulation of the inverse square law nature of the gravitational force. Only after Einstein published his work on the General Theory of Relativity \citep{Einstein:1916vd} did physicists perceive gravity as the curvature of spacetime instead of a force. The nature of this distortion of spacetime can be understood by looking at Einstein’s field equations, which are a set of 10 coupled, non-linear partial-differential equations written compactly as
\begin{equation}
    R_{\alpha\beta} - \frac{1}{2} R g_{\alpha\beta} = \frac{8\pi G}{c^4} T_{\alpha\beta}
    \label{eq:intro-einstein_coupled}
\end{equation}
The left-hand side of Eq.~\eqref{eq:intro-einstein_coupled} describes the geometry/curvature of the spacetime around the gravitating body, and the right-hand side denotes the energy and momentum of the source. In Eq.~\eqref{eq:intro-einstein_coupled}, $R_{\alpha\beta}$ is the Ricci tensor, which represents how a volume in curved space differs from a volume in Euclidean space; $R$ is the Ricci scalar which characterizes the curvature of spacetime; $g_{\alpha\beta}$ is the metric tensor; $G$ and $c$ are universal gravitational constant and speed of light in vacuum respectively; and $T_{\alpha\beta}$ is the stress-energy tensor that describes the density and flux of energy and momentum in spacetime, and is the source of the gravitational field. In 1905, Henri Poincaré suggested that accelerated masses in the relativistic field theory of gravity should produce gravitational waves, similar to an accelerating electric charge producing electromagnetic waves \citep{Poincare1906}. Einstein pursued the idea that wave-like solutions were possible for his equations.

Far from the source (in the weak field regime), the metric $g_{\alpha \beta}$ appearing in Eq.~\eqref{eq:intro-einstein_coupled} can be written as a linear perturbation to the Minkowski (flat) metric $\eta_{\alpha\beta}$ \citep{DensonHill:2016upp} as:
\begin{equation}
    g_{\alpha \beta} = \eta_{\alpha \beta} + h_{\alpha \beta}
\end{equation}
where $h_{\alpha \beta}$ is a small metric perturbation ($|h_{\alpha \beta}| \ll |\eta_{\alpha\beta}| $). 
%The linearized equations of motion can be written in a compact form by defining \citep{Maggiore:2007ulw}:
%
% \begin{equation}
%     \overline{h}_{\alpha \beta} = h_{\alpha \beta} - \frac{1}{2} \eta_{\alpha \beta} \eta^{\mu \nu} h_{\mu \nu}
% \end{equation}
%
%which can be inverted as:
%
% \begin{equation}
%     h_{\alpha \beta} = \overline{h}_{\alpha \beta} - \frac{1}{2} \eta_{\alpha \beta} \eta^{\mu \nu} \overline{h}_{\mu \nu}.
% \end{equation}
% %
% Thus, using Reimann tensor
% %
% \begin{equation}
%     R_{\alpha \beta \gamma \delta} = \frac{1}{2} (\partial_\beta \partial_\gamma h_{\alpha \delta} + \partial_\alpha \partial_\delta h_{\beta \gamma} - \partial_\alpha \partial_\gamma h_{\beta \delta} - \partial_\beta \partial_\delta h_{\alpha \gamma})
% \end{equation}
% %
% and Einstein tensor,
% %
% \begin{equation}
%     G_{\alpha \beta} = R_{\alpha \beta} - \frac{1}{2} g_{\alpha \beta} R,
% \end{equation}
%
This leads to the linearized form of Einstein equations [Eq.~\eqref{eq:intro-einstein_coupled}] which can be written as \citep[see][for complete derivation]{Maggiore:2007ulw}:
\begin{equation}
    \mathlarger{\Box} \overline{h}_{\alpha \beta} + \eta_{\alpha \beta}\partial^\mu \partial^\nu \overline{h}_{\mu\nu} - \partial^\mu \partial_\beta \overline{h}_{\alpha\mu} - \partial^\mu \partial_\alpha \overline{h}_{\beta\mu} = \frac{16\pi G}{c^4}T_{\alpha \beta},
\label{eq:intro-einstein-linearized}
\end{equation}
where
\begin{equation}
    \overline{h}_{\alpha \beta} = h_{\alpha \beta} - \frac{1}{2} \eta_{\alpha \beta} \eta^{\mu \nu} h_{\mu \nu},
\end{equation}
and $\mathlarger{\mathlarger{\Box}} \equiv \eta_{\alpha \beta} \partial^\alpha \partial^\beta = \partial_\alpha \partial^\alpha$ is the flat space d'Alembertian in the context of linearized theory. Choosing the harmonic gauge:
\begin{equation}
    \partial^\alpha \overline{h}_{\alpha \beta} = 0,
\end{equation}
Eq.~\eqref{eq:intro-einstein-linearized} reduces to:
\begin{equation}
    \mathlarger{\Box} \overline{h}_{\alpha\beta} = -\frac{16\pi G}{c^4}T_{\alpha \beta}
\end{equation}
and outside the source ($T_{\alpha \beta} = 0$) gives:
\begin{equation}
    \mathlarger{\Box} \overline{h}_{\alpha \beta} = 0.
\label{eq:intro-gws}
\end{equation}
which is a system of relativistic wave equations. Since $\mathlarger{\mathlarger{\Box}} = -(1/c^2)\partial^2_t + \vec{\nabla}^2$, Eq.~\eqref{eq:intro-gws} implies that gravitational waves travel at the speed of light. Further, choosing the \textit{transverse-traceless gauge}, or TT gauge \citep[see][for complete derivation]{Maggiore:2007ulw}:
\begin{equation}
    h^{0\beta} = 0,\ \ \ \ \ \ \ h^i_{~i} = 0,\ \ \ \ \ \ \ \partial^j h_{ij} = 0
\end{equation}
Eq.~\eqref{eq:intro-gws} has plane wave solutions:
\begin{equation}
    h^\text{TT}_{ab}(t,z) = \begin{pmatrix}
    h_+ & h_\times \\
    h_\times & -h_+ 
    \end{pmatrix}_{ab} \cos[\omega(t-z/c)]
    \label{eq:polarizations}
\end{equation}
where $z$ is the direction of propagation, $\omega$ is the frequency of the wave, and $a,b=1,2$ are indices in the transverse ($x,y$) plane. Thus, GWs have two polarizations with $h_+$ and $h_\times$ denoting the amplitudes of the "plus" and "cross" polarizations respectively.

In the years (decades) to come, there was much debate on the existence and propagation of gravitational waves \citep{Eddington:1922ds, 10.1063/1.2117822, 1937FrInJ.223...43E, Pirani:1956tn, Cervantes-Cota:2016zjc}. More importantly, ideas were discussed on whether these waves were tangible enough to be detected, and whether they even carried energy \citep{Pirani:1956tn, aip_1995, 2011rgp..rept.....R}. Feynman with his ``sticky bead argument'' had a significant role in convincing many of the scientific minds at the Chapel Hill Conference 1957 that gravitational waves carry energy \citep{2011rgp..rept.....R, Bondi:1957dt, Feynman_1961}. Joseph Weber, also present at the conference, was inspired by the discussions to build a detector to detect gravitational wave signals \citep{Weber:1960zz}.

\begin{figure}[t]
    \centering
    \includegraphics[width=0.7\linewidth]{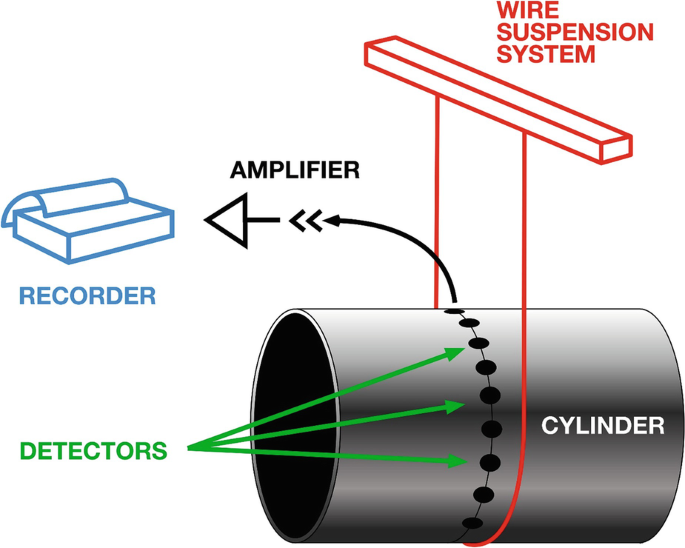}
    \caption[A sketch of Weber's cylinder detector]{\takenfrom{\cite{Beckman2021}} A sketch of Weber's cylinder detector.}
    \label{fig:intro-weber-det}
\end{figure}

He built a detector that measured the vibrations induced in a large metal cylinder due to the passing of gravitational waves \citep{Weber:1966zz}. A steel wire suspended the cylinder from a support built to isolate vibrations of its environment, and the whole setup was placed inside a vacuum chamber (Fig.~\ref{fig:intro-weber-det}). A belt of piezoelectric crystals placed around the cylinder measured the vibrations and converted them into electrical signals. Two such detectors were built, one at the University of Maryland and another at Argonne National Laboratory, 950 km apart, to eliminate spurious local signals \citep{Cervantes-Cota:2016zjc}. In 1969, Weber's team published the results claiming the detection of gravitational waves \citep{Weber:1969bz}. The following year, Weber claimed to detect multiple signals originating from the center of the Milky Way \citep{Weber:1970mx}. While Weber maintained his original position on the subject \citep{Weber_1971}, the other scientific groups showed that the signals were unlikely to be gravitational wave signals \citep{Sciama:1969zz, Kafka_1972}. Regardless of whether the scientific minds agreed on the credibility of these detections, several groups started a hunt for gravitational wave signals, building their own detectors and improving upon the design of the cylindrical "antennas" \citep{Th300}. In the years to come, with multiple detectors collecting data, there was a growing consensus among the community that no gravitational wave signals were being captured \citep{1978A&A....70...97K, 1975Natur.254..498H, 1975NCimL..12..111B, Drever:1973xf, Endal:1977ef, Douglass:1975qs, Billing:1976us, Giffard:1976nq, Hirakawa:1975qf, Levine:1974pa, Moss:1971ocz, 2005PhT....58l..62C}. Just when gravitational wave searches were starting to lose hope, the detection of a binary pulsar system by Hulse and Taylor in 1974 \citep{1979Natur.277..437T} redoubled the efforts towards building broadband detectors to detect gravitational wave signals and added fuel to the idea of the interferometric method of gravitational wave detection \citep{Gertsenshtein:1962kfm, Vladimir_1966, Weiss_1976, Th300, Allen:1975sn}.

Interferometric gravitational wave detectors (which eventually led to the direct detection of gravitational waves in 2015) exploit the fact that a gravitational wave signal passing through matter stretches it in one direction while compressing it in the perpendicular direction. We have touched upon how this is done in the section dedicated to current and future detectors (Sec.~\ref{sec:intro-dets}). Robert L. Forward, in 1971, published the design for the first interferometric gravitational wave detector \citep{Forward:1978zm}, acknowledging the contribution of Weber towards the idea \citep{Moss:1971ocz}. Parallely, in the years that followed, Rainer Weiss, along with David Shoemaker, made several efforts towards solidifying an interferometric gravitational wave detector and received funding from the National Science Foundation (NSF) \citep{Weiss_1976, Shoemaker:1987ft, Collins_1976, DeSabbata:1977wn}. Several other groups, either in collaboration or independently, worked on their own designs for interferometric detectors \citep{1985magr.meet.....S, 1985magr.meet.....W, Weiss_2000, 1986RALR....1.....H, Leuchs_1987, Hough:1989yv, 1985AnPh...10..219B, 2012JInst...7.3012A} but most were not materialised due to a lack of funding. 

In the summer of 1975, Weiss and Kip Thorne discussed various ideas for ground-based and space-based laser interferometers \citep{Weiss_1976, Levin_2017}, and settled on a ground-based laser interferometer. To that effect, in 1978, Thorne offered the job of building the detector to Ronald Drever at Caltech, and in late 1979, the NSF granted funds to Caltech and MIT to build the detectors (albeit on a much smaller scale compared to the current detectors). In the meantime, Weiss was corresponding with the director of the NSF Gravitational Physics division, Richard Isaacson, to build detectors which had arms at the scale of kilometers. Since funding both the MIT and Caltech projects was not deemed feasible by NSF, a single project named "Laser Interferometer Gravitational-Wave Observatory" (LIGO) was started jointly by Caltech-MIT and was approved to be funded by the NSF \citep{Levin_2017, Weiss_1983, Ruthen_1992} in 1988. 
Due to tensions between Weiss and Drever \citep{Cervantes-Cota:2016zjc, Collins+2004+546+557, Collins_2003, Niels_bohr_library_2020}, the project underwent a series of changes and delays, and finally, Barry C. Barish was appointed the director. Barish put an organisational structure to the project and a step-wise upgrade plan for the detectors \textit{viz.} initial-LIGO (iLIGO) and advanced-LIGO (aLIGO). Two LIGO detectors, one in Hanford and another in Livingston, were proposed to be built, and the construction ended in 1997. The LIGO laboratory and LIGO Scientific Collaboration were also founded to take care of the administration of laboratories, and scientific \& technological research, respectively.

In parallel, with the help of funding from German and British agencies, and donation of land by the University of Hanover and the State of Lower Saxony, construction of a 600 m detector (GEO~600) started in September 1995 \citep{max_planck_2005} in Germany. The detector started its operation in December 2002. While GEO~600 was being planned, the Virgo project, a collaboration between the French and Italian group of scientists, was approved by the French CNRS (Le Centre National de la Recherche Scientifique) in 1993 and Italian INFN (Istituto Nazionale di Fisica Nucleare) in 1994 \citep{1985AnPh...10R...1B}. The construction for Virgo started in 1996 but faced several hiccups between 1996-1999 \citep{1985magr.meet.....S}. To continue the project in a more organized manner, the European Gravitational Observatory (EGO) consortium was formed by CNRS and INFN. In June 2003, the construction of the initial Virgo detector was completed \citep{Levin_2017, ego_2003}.

Over the next decade or so, the alliance between LIGO, GEO~600, and Virgo detectors continued, improving the detector technologies. LIGO Scientific Collaboration (LSC) grew to become a worldwide collaboration with scientists from numerous countries working on various aspects of gravitational wave science and detection. The first observation run for initial LIGO was from 2002 to 2010, where it did not detect any GW signals. Between 2010 and 2015, various upgrades were made to the detectors, and aLIGO was formed. 

It was 13\textsuperscript{th} September 2015, and there were still four days until the start of the official observation run of Advanced LIGO, but nature was getting impatient. On the morning of 14\textsuperscript{th} September 2015, a gravitational wave signal, originating from the merger of two inspiralling black holes 1300 million light years from Earth, passed through the LIGO detectors and created history. The signal was detected in both detectors, and calculations confirmed the time delay between the two triggers to be consistent with the general theory of relativity. The event was named GW150914\footnote{GW followed by the detection date in YYMMDD format.} and, after rigorous checks and analyses, was published by the LIGO Scientific Collaboration and Virgo Collaboration in February 2016 \citep{Abbott:2016blz}. The following year, on 3\textsuperscript{rd} October 2017, Rainer Weiss, Barry C. Barish, and Kip S. Thorne were awarded the Nobel prize ``for decisive contributions to the LIGO detector and the observation of gravitational waves'' \citep{nobel_2017}.

\section{Sources}
\label{sec:intro-gw-types-sources}

\begin{figure}[p!]
    \centering
    \includegraphics[width=0.85\linewidth]{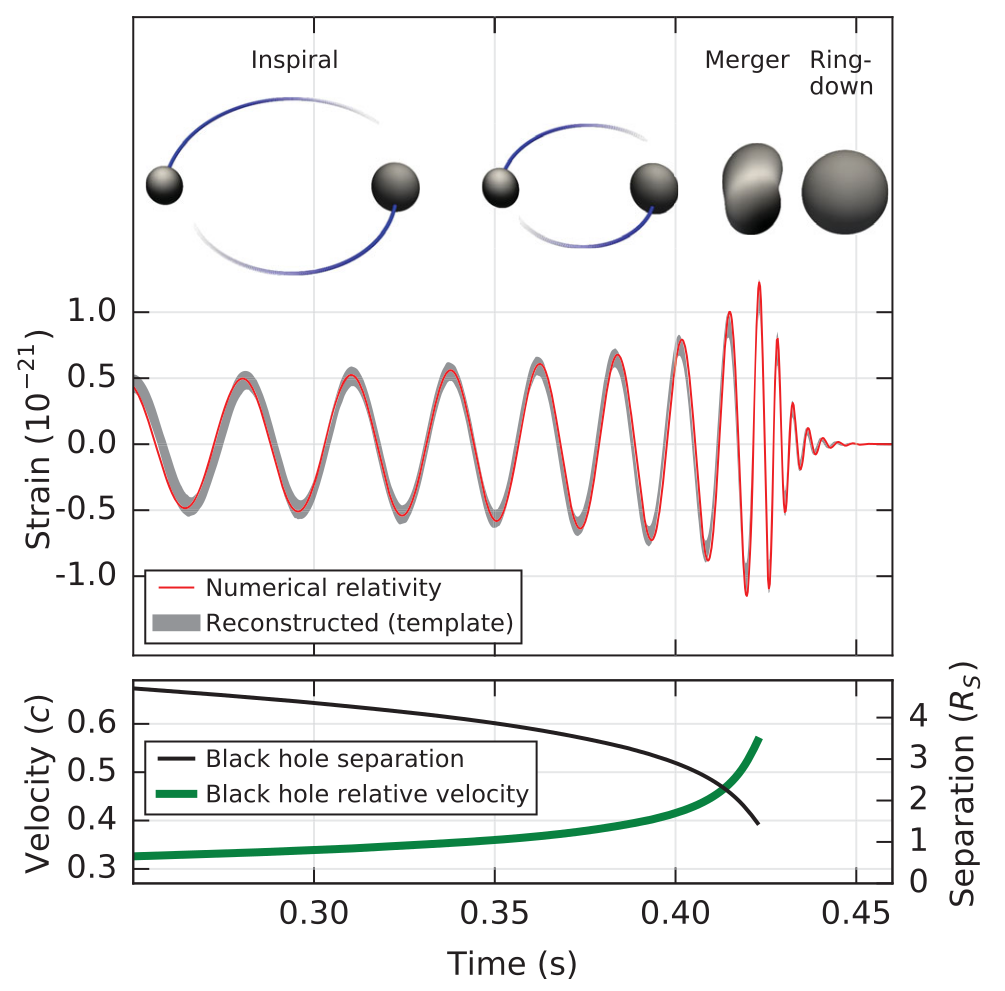}
    \includegraphics[width=\linewidth]{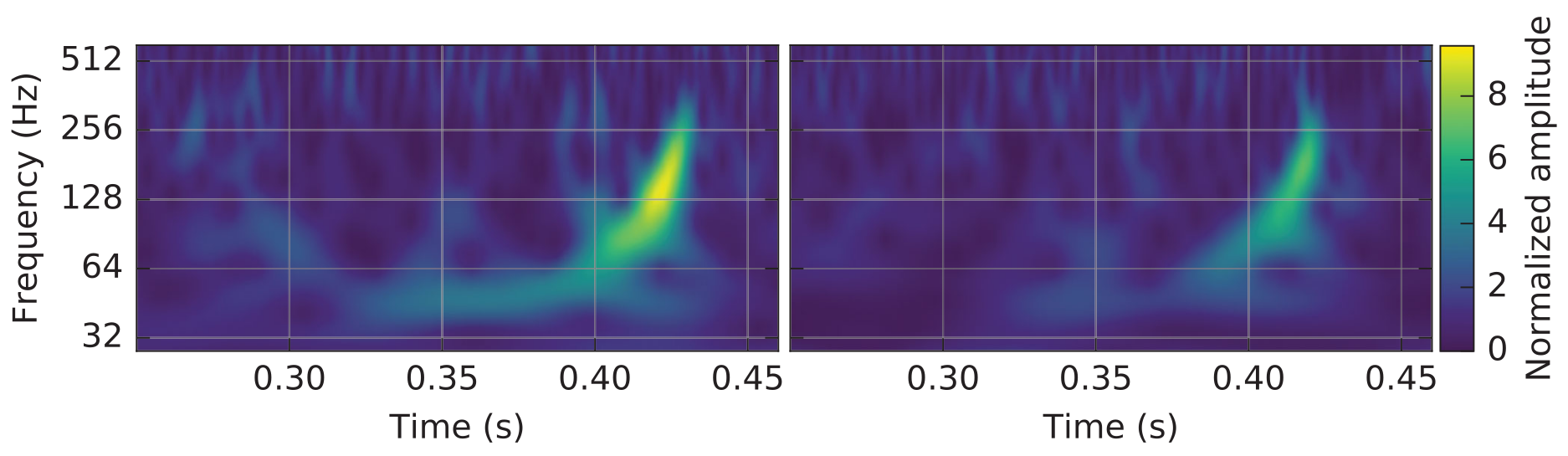}
    \caption[Gravitational wave chirp from GW150914]{\takenfrom{\cite{Abbott:2016blz}} Top: Estimated gravitational-wave strain amplitude from GW150914 projected onto H1. The inset images show numerical relativity models of the black hole horizons as the black holes coalesce. Middle row: The Keplerian effective black hole separation in units of Schwarzschild radii ($R_S = 2GM/c^2$) and the effective relative velocity given by the post-Newtonian parameter $v/c = (GM\pi f/c^3)^{1/3}$, where $f$ is the gravitational-wave frequency calculated with numerical relativity and M is the total mass ($65 \Msun$). Bottom row: A time-frequency representation \citep{Chatterji:2004qg} of the strain data in H1 (left) and L1 (right) detectors, showing the signal frequency increasing over time.}
    \label{fig:intro-chirp}
\end{figure}

Gravitational waves can be represented as a time-dependent perturbation to the metric tensor for flat spacetime, produced by a time-varying quadrupole moment at the leading order. In other words, all time-dependent non-spherical movements will produce gravitational waves. The sources of GWs are generally classified into four categories: chirps, continuous, burst, and stochastic \citep{ligo_sources, ligo_sources_caltech}.

\subsection{Chirps}
\label{subsec:intro-types-chirps}

These are signals that are generally produced by inspiralling compact binary systems that are about to merge and are composed of neutron star and/or black holes. The two compact objects in a binary revolve around a common center of mass and lose energy and (angular) momentum due to the emission of gravitational radiation. As a consequence, the orbit of the binary system shrinks, and binary's orbital period decreases. The two objects come closer and closer, finally colliding at relativistic speeds to form a remnant. Depending on the mass of the compact objects, the final few cycles of this process up to the merger can be captured in the current ground-based gravitational wave detectors (Fig.~\ref{fig:intro-chirp}). Since the signal's frequency and amplitude rapidly increase as it approaches the merger, it creates a chirping sound when converted to audio. Hence, these inspiralling gravitational waves are also known as chirps. Chirps are not only the strongest type of signals but are also the cleanest ones to model which (as we discuss below) is a prerequisite for their detection. Thus, it is no surprise that GW events detected by the LIGO-Virgo detectors till date are all chirp signals \citep{LIGOScientific:2018mvr, LIGOScientific:2020ibl, LIGOScientific:2021usb, LIGOScientific:2021djp}. One such signal is shown in Fig.~\ref{fig:intro-chirp} which was recorded by the LIGO Livingston and Hanford detectors for the event GW150914 (first direct detection of gravitational waves.)

\subsection{Continuous Gravitational Waves}
\label{subsec:intro-types-continuous}

\begin{figure}[t]
    \centering
    \includegraphics[width=0.6\linewidth]{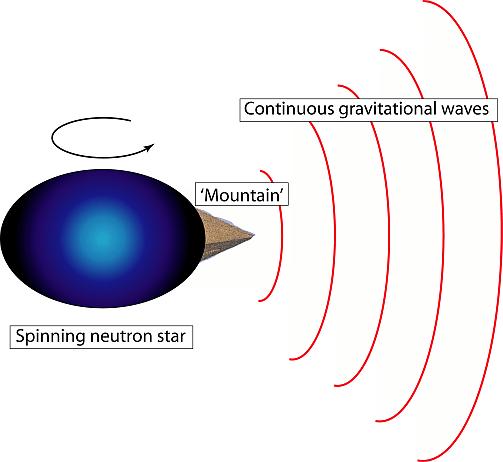}
    \caption[Sketch of a rotating neutron star emitting continuous GWs]{\takenfrom{\cite{Drummond_2018}} Sketch of a rotating neutron star with a mountain on the surface (not to scale) emitting continuous gravitational waves.}
    \label{fig:intro-rotating NS}
\end{figure}

\begin{figure}[p!]
    \centering
    \includegraphics[trim=130 120 100 120, clip, width=\linewidth]{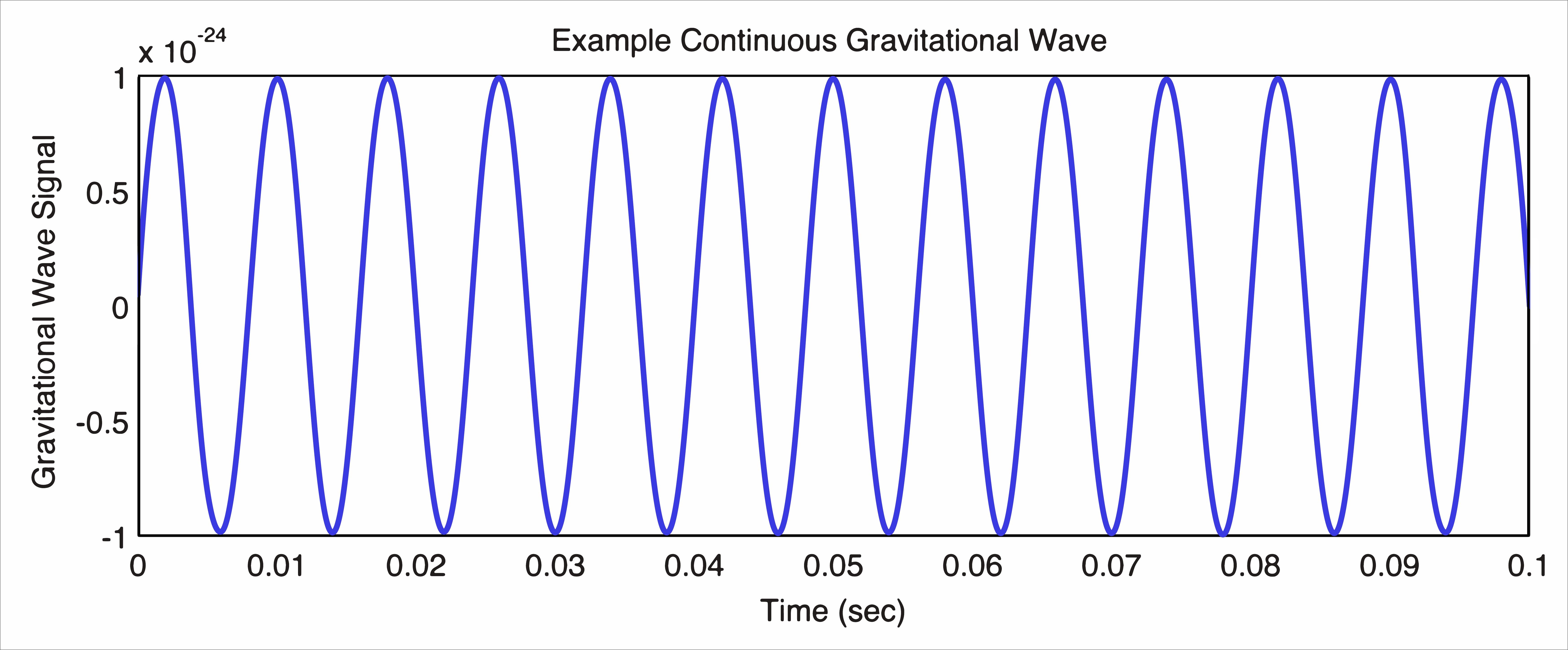}
    \includegraphics[width=0.85\linewidth]{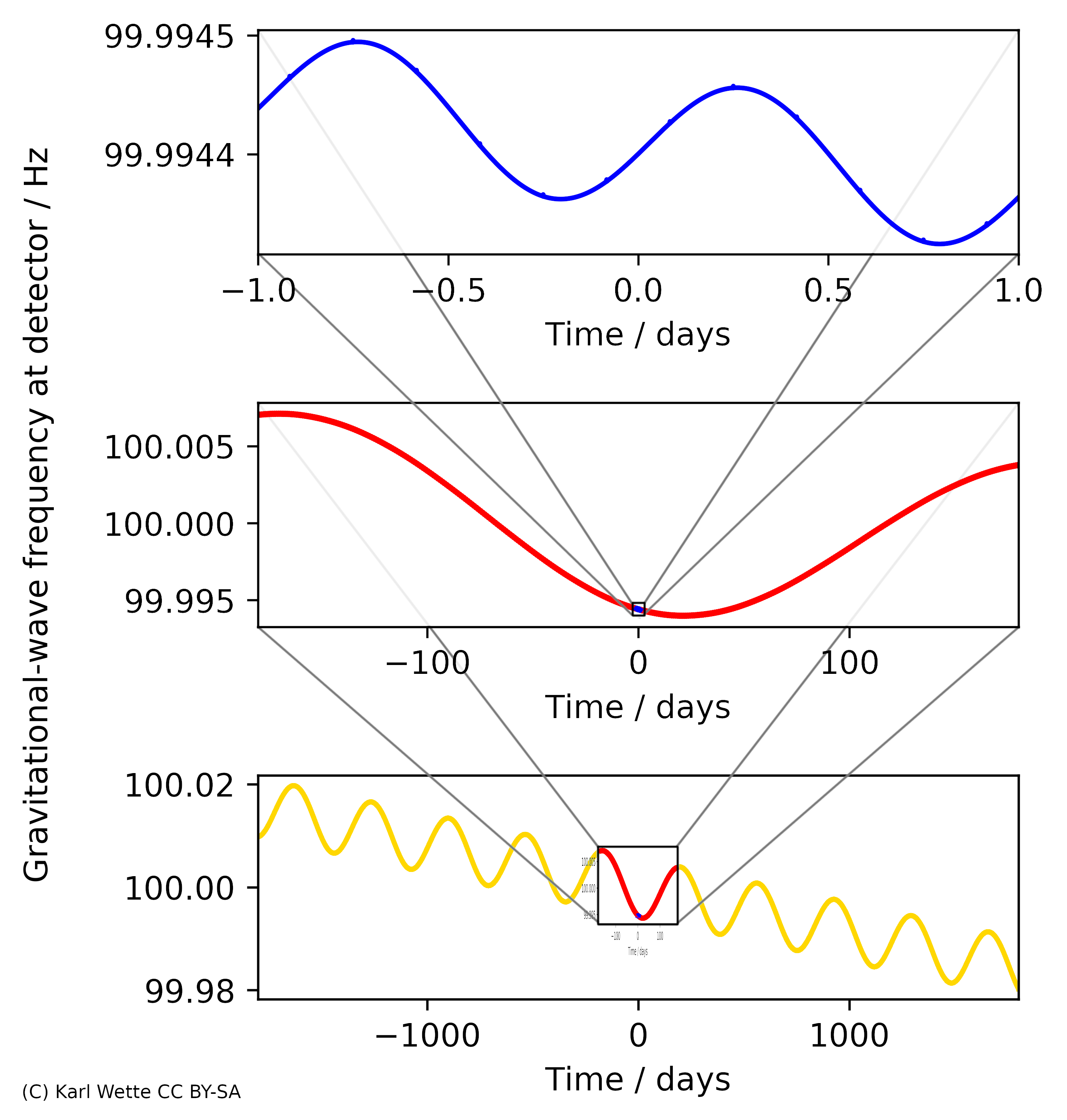}
    \caption[Frequency evolution of a continuous wave GW signal]{\takenfrom{\cite{ligo_continuous}} From top to bottom row: 1. A short snippet from an example continuous gravitational wave signal, 2. Changing frequency of the signal due to the daily rotation of Earth about its axis, 3. Change in the signal frequency due to Earth's motion around the Sun, 4. Decrease in the frequency due to spin-down of the neutron star emitting the signal. Image credits: A. Stuver, K. Wette.}
    \label{fig:intro-continuous-waves}
\end{figure}

When a system emits a train of gravitational waves at nearly fixed frequencies (monochromatic), they are known as continuous gravitational waves (shown in the top panel of Fig.~\ref{fig:intro-continuous-waves}). Unlike chirps, these are not transient in nature. Some of the sources for these kinds of GWs can be long inspirals of compact binary systems where the merger is so far out in the future that the frequency remains nearly constant (although the change in frequency can be detectable on longer time frames) in the detector band, or spinning compact stars with deformations such as a spinning neutron star with a ``mountain'' on the surface (see Fig.~\ref{fig:intro-rotating NS}). The amplitude of continuous gravitational waves can be an order of magnitude (or more) weaker than chirp signals. The current ground-based GW detectors are searching for continuous waves from spinning neutron stars in the Milky Way and hope to monitor these signals (once detected) over long periods of time. While the frequency of these signals remains constant for a short duration, it changes over longer periods due to the spin-down of the neutron star as well as the rotation and revolution of Earth (Fig.~\ref{fig:intro-continuous-waves}). Spin-down, or slowing of the rotation of the neutron star, is expected as it loses energy through gravitational and electromagnetic waves. See \cite{{Riles:2022wwz}} for a review on the search for continuous-wave gravitational radiation.

\subsection{Gravitational Wave Bursts}
\label{subsec:intro-types-burst}

Burst signals are short-lived gravitational wave signals which are expected to appear in the GW detectors similar to a `pop' or `crackle'. The potential sources for these can be asymmetric supernovae explosions, but little is known with certainty. Since these are unmodelled signals, the gravitational wave search algorithms used to look for these signals are often model-independent. See \cite{KAGRA:2021tnv} for an all-sky search for burst signals in the third gravitational wave observing run.

\subsection{Stochastic Gravitational Wave Background}
\label{subsec:intro-types-stochastic}

\begin{figure}[t]
    \centering
    \includegraphics[width=\linewidth]{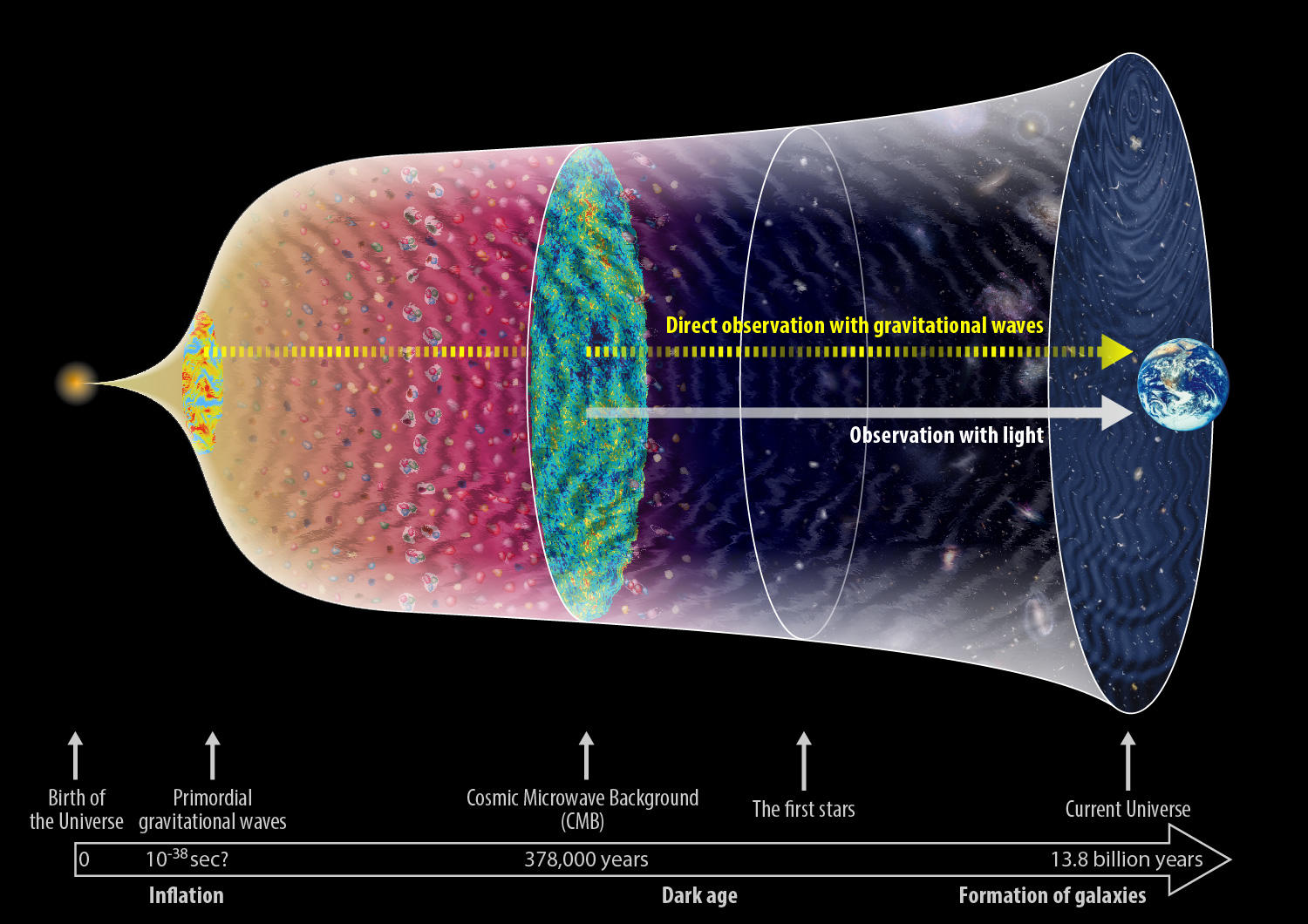}
    \caption[Diagram showing different epochs since the Big Bang and when gravitational waves and the CMB came into existence.]{\takenfrom{\cite{NAOJ_universe}} Diagram showing different epochs in the universe's evolution since the Big Bang and when gravitational waves and the CMB came into existence.}
    \label{fig:intro-bigbang}
\end{figure}

Stochastic gravitational wave signal is expected to be a mixture of random low amplitude gravitational wave signals originating from different kinds of sources. This kind of signal is also expected from the early evolution of the universe, much like the Cosmic Microwave Background (CMB) radiation. While the CMB comes from about 300,000 years after the Big Bang, stochastic gravitational waves can give us an idea about the universe as early as $10^{-36}$ to $10^{-32}$ seconds after the Big Bang (see Fig.~\ref{fig:intro-bigbang}). Another candidate which can produce such a gravitational wave background is the overlap of numerous signals from inspirals of supermassive binary black holes in the nanohertz regime [see, for instance, a recent publication \cite{NANOGrav:2023gor} by the International Pulsar Timing Array Collaboration \citep{ipta}]. The pulsar timing array constitutes a set of galactic pulsars that are monitored to look for correlations between the pulses arriving on Earth and to infer gravitational wave background \citep{Lynch:2015iua}.

%%%%%%%%%%%%%%%%%%%%%%%%%%%%%%%%%%%%%%%%%%

\section{Detectors}
\label{sec:intro-dets}

In Sec.~\ref{sec:intro-history}, we briefly described the resonant mass antennas which were the first detectors constructed to capture GW signals. These have been pursued in parallel with the interferometric GW detectors and include room-temperature bar antennas (such as Weber bar and GEOGRAV), cryogenically cooled bar antennas (such as AURIGA, NAUTILAS, EXPLORER, ALLEGRO, NIOBE, ALTAIR), and cryogenically cooled spherical antennas (GRAIL, TIGA, SFERA, and Graviton). But, since the past few decades, due to lack of interest and funding, the projects have not gained as much momentum as the interferometric GW detectors. See \cite{Aguiar:2010kn} for a review on resonant-mass GW detectors. This section gives an overview of interferometric GW detectors. We begin by outlining the basic working principle for these detectors and subsequently discuss the current and future detectors.

A gravitational wave passing through a ring of freely falling masses distorts the assembly over one time period as shown in Fig.~\ref{fig:intro-mass-rings}. The top row of figures denotes the `plus' (+) polarization whereas the bottom row shows the `cross' ($\times$) polarization which is $45^{\circ}$ apart \citep{Bergmann:1987tm}. Hence, the changes produced in the distance of a perpendicularly placed pair of particles should indicate the passing of a gravitational wave. Alternatively, one can measure the time taken for the light to travel between these particles.% since the speed of light is constant and unaltered by gravitational waves. 
Thus, an ``L'' shaped detector with arms of equal length and freely falling test masses (mirrors) at the corners and end points of the ``L'' is built. To quantify the strength of a gravitational wave, a dimensionless strain `$h$' can be defined, which denotes the maximum change in the detector's arms per unit length. If $l$ is the separation between the two mirrors along the arms of the detector and $\Delta l$ is the change in the distance between the two mirrors due to the passage of the gravitational wave, then $\Delta l = h\,l/2$ in each arm of the detector. Thus, the total difference in the length between the two arms per unit length is \citep{Kokkotas:2007zz}
\begin{equation}
    h = \frac{\Delta l}{l}.
    \label{eq:intro-strain-diff-length}
\end{equation}
As it can be seen, for a fixed strain $h$, increasing the length $l$ of the detector will lead to an increase in $\Delta l$ leading to a better detection. This motivates the design of the current laser interferometers used to detect gravitational waves. 

\begin{figure}[t]
    \centering
    \includegraphics[width=\linewidth]{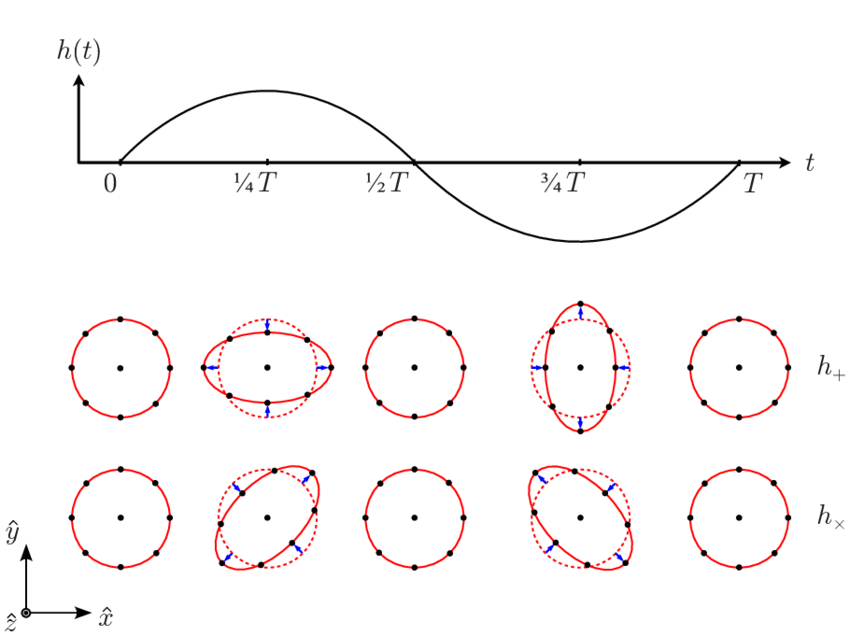}
    \caption[A monochromatic gravitational wave of pulsation $\omega = 2\pi/T$ propagates along the z- direction]{\takenfrom{\cite{LeTiec:2016sgy}} A monochromatic gravitational wave of pulsation $\omega = 2\pi/T$ propagates along the z- direction. The lower panel shows the effects of the plus and cross polarizations on a ring of freely falling particles, in a local inertial frame.}
    \label{fig:intro-mass-rings}
\end{figure}

The current interferometric gravitational wave detectors are ``L'' shaped detectors with mirrors at the ends of the arms that reflect light to create an interference pattern. A photo-detector senses the interference pattern, converting the light into an electrical signal, which can then be analyzed. A single laser beam is split at the intersection of the two arms. Half of the laser light is transmitted into one arm while the other half is reflected into the second arm. Mirrors are suspended as pendula at the end of each arm and near the beam splitter. Laser light in each arm bounces back and forth between these mirrors, and finally returns to the intersection, where it interferes with light from the other arm. If the lengths of both arms have remained unchanged, then the two combined light waves should completely subtract each other (destructive interference), and there will be no light observed at the output of the detector.  However, if a gravitational wave were to stretch one arm and compress the other slightly (about 1/1000 the diameter of a proton), the two light beams would no longer completely subtract each other, yielding light patterns at the detector output.

%-----------------------------------

\subsection{Response of a ground-based interferometric detector}
\label{subsec:intro-det-response}

\begin{figure}[t!]
    \centering
    \includegraphics[width=\linewidth]{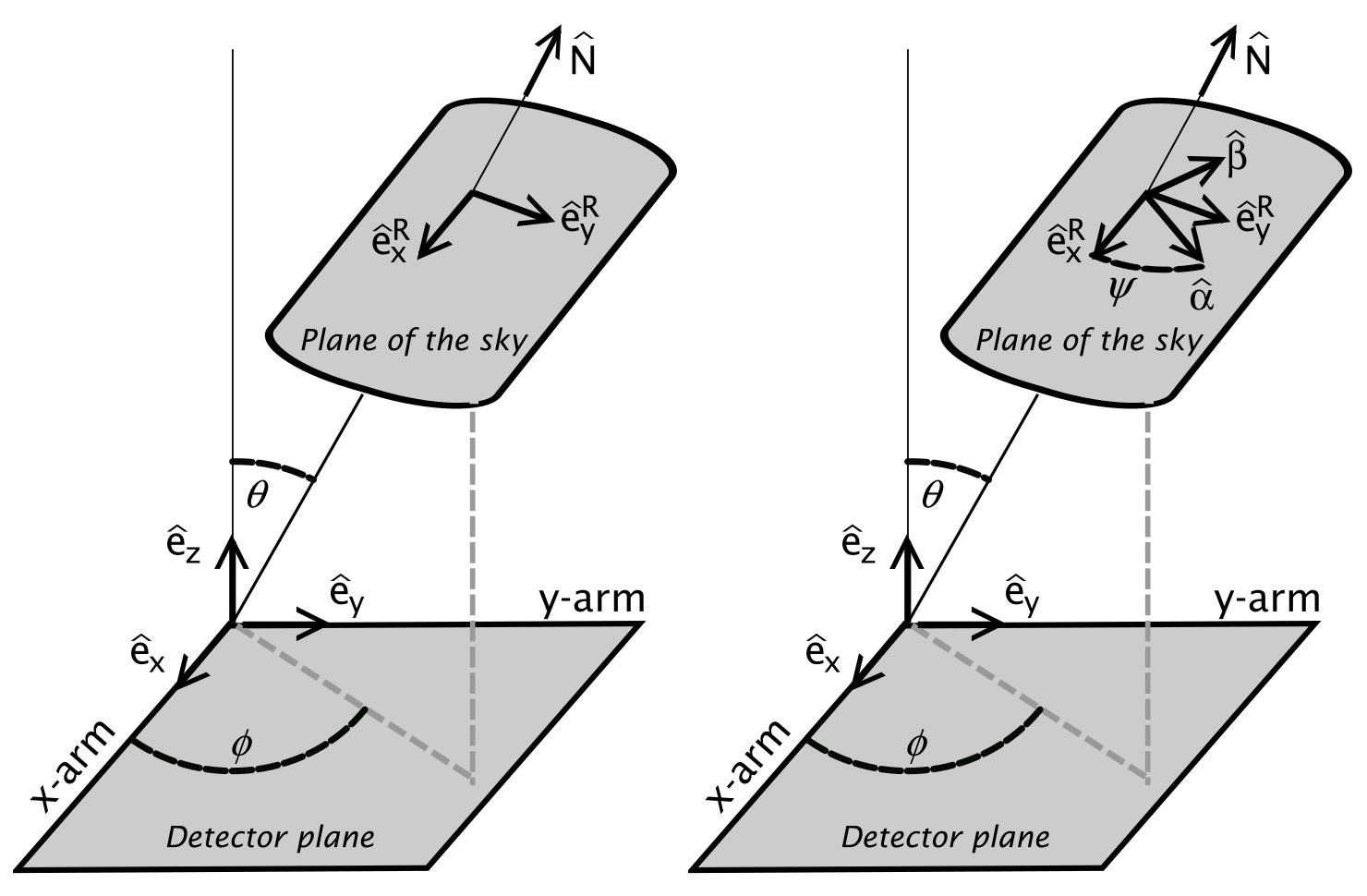}
    \caption[The relative orientation of the sky and detector frames]{\takenfrom{\cite{Sathyaprakash:2009xs}} The relative orientation of the sky and detector frames (left panel) and the effect of a rotation by the angle $\psi$ in the sky frame (right panel).}
    \label{fig:intro-det-tensor-diagram}
\end{figure}

The interferometer's response to gravitational waves, known as the \textit{antenna pattern response}, varies based on the detector's geometry and the gravitational wave's direction and polarization. Let $\boldsymbol{\hat{N}}$ be the direction of GW propagation, with radiation vectors $\boldsymbol{\hat{e}}_x^R$ and $\boldsymbol{\hat{e}}_y^R$, such that $\boldsymbol{\hat{e}}_x^R$ lies in the plane formed by the wave propagation direction and one arm of the GW detector. Here, we assume that the arm of the detector lies along $x$- axis of the detector plane, and its unit vector is $\boldsymbol{\hat{e}}_x$. Thus, the wave amplitude $h^+$ [see Eq.~\eqref{eq:polarizations}] has $\boldsymbol{\hat{e}}_x^R$ and $\boldsymbol{\hat{e}}_y^R$ as the axes of its ellipse. The full metric perturbation due to a GW coming from the direction $\boldsymbol{\hat{N}}$ can written as \citep{Yunes:2013dva, Nishizawa:2009bf}\footnote{We denote tensors with $\Tilde{\circ}$ on top.}:
\begin{equation}
    \Tilde{\boldsymbol{h}}(t) = h^+(t)\Tilde{\boldsymbol{e}}^{\,+} + h^\times(t) \Tilde{\boldsymbol{e}}^{\,\times},
\end{equation}
where $\Tilde{\boldsymbol{e}}^{\,+}$ and $\Tilde{\boldsymbol{e}}^{\,\times}$ are polarization tensors. If the direction of propagation for the gravitational wave is $z$- direction, these can be written as:
\begin{equation}
    \Tilde{\boldsymbol{e}}^{\,+} = 
    \begin{pmatrix}
        1 & 0 & 0 \\
        0 & -1 & 0 \\
        0 & 0 & 0
    \end{pmatrix},
    \Tilde{\boldsymbol{e}}^{\,\times} = 
    \begin{pmatrix}
        0 & 1 & 0 \\
        1 & 0 & 0 \\
        0 & 0 & 0
    \end{pmatrix}.
\label{eq:intro:z-direction-tensor}
\end{equation}
Now, with the help of Fig.~\ref{fig:intro-det-tensor-diagram}, we will explicitly write these above vectors and tensors in terms of the angles $(\theta, \phi, \psi)$ for a GW propagating in an arbitrary direction $\boldsymbol{\hat{N}}$. Let us take the detector plane unit vectors as:
\begin{equation}
\boldsymbol{\hat{e}}_x = 
    \begin{pmatrix}
        1 \\ 0 \\ 0
    \end{pmatrix},
\boldsymbol{\hat{e}}_y = 
    \begin{pmatrix}
        0 \\ 1 \\ 0
    \end{pmatrix},
\boldsymbol{\hat{e}}_z = 
    \begin{pmatrix}
        0 \\ 0 \\ 1
    \end{pmatrix}.
\end{equation}
Then, looking at left panel of Fig.~\ref{fig:intro-det-tensor-diagram}, the plane of the sky is rotated with respect to the detector plane by angles ($\theta, \phi$). Hence, the radiation basis vectors can be written as:
\begin{equation}
\boldsymbol{\hat{e}}_x^R = 
    \begin{pmatrix}
        \cos{\theta}\cos{\phi} \\
        \cos{\theta}\sin{\phi} \\ 
        -\sin{\theta}
    \end{pmatrix},
\boldsymbol{\hat{e}}_y^R = 
    \begin{pmatrix}
        -\sin{\phi} \\ 
        \cos{\phi} \\ 
        0
    \end{pmatrix},
\boldsymbol{\hat{e}}_z^R = 
    \begin{pmatrix}
        \sin{\theta}\cos{\phi} \\ 
        \sin{\theta}\sin{\phi} \\ 
        \cos{\theta}
    \end{pmatrix}.
\end{equation}
Now, we can rotate the sky plane with respect to the GW propagation direction $\boldsymbol{\hat{N}}$ (right panel of Fig.~\ref{fig:intro-det-tensor-diagram}) by a polarization angle $\psi$. Hence, in general, the new basis vectors $\boldsymbol{\hat{\alpha}}$ and $\boldsymbol{\hat{\beta}}$ can be written as:
\begin{subequations}
\begin{align}
    \boldsymbol{\hat{\alpha}} &= \boldsymbol{\hat{e}}_x^R \cos{\psi} + \boldsymbol{\hat{e}}_y^R \sin{\psi} \\
    \boldsymbol{\hat{\beta}} &= -\boldsymbol{\hat{e}}_x^R\sin{\psi} + \boldsymbol{\hat{e}}_y^R\cos{\psi}.
\end{align}
\end{subequations}
Now, we can re-write Eq.~\eqref{eq:intro:z-direction-tensor} using the generalized basis vectors as:
\begin{subequations}
    \begin{align}
        \Tilde{\boldsymbol{\epsilon}}^{\,+} &= \boldsymbol{\hat{\alpha}} \otimes \boldsymbol{\hat{\alpha}} - \boldsymbol{\hat{\beta}} \otimes \boldsymbol{\hat{\beta}} \\
        \Tilde{\boldsymbol{\epsilon}}^{\,\times} &= \boldsymbol{\hat{\alpha}} \otimes \boldsymbol{\hat{\beta}} + \boldsymbol{\hat{\beta}} \otimes \boldsymbol{\hat{\alpha}}.
    \end{align}
\end{subequations}
For detectors with perpendicular arms, antenna pattern response functions can be written as the difference of projection of the polarization tensor onto each of the interferometer arms:
\begin{subequations}
    \begin{align}
        F^+ &= \frac{1}{2}(\boldsymbol{\hat{e}}_x \otimes \boldsymbol{\hat{e}}_x - \boldsymbol{\hat{e}}_y \otimes \boldsymbol{\hat{e}}_y)^{ij}\Tilde{\boldsymbol{\epsilon}}^{\,+}_{ij} \\
        \nonumber &= \frac{1}{2}(\boldsymbol{\hat{e}}_x^{\,i} \boldsymbol{\hat{e}}_x^{\,j} - \boldsymbol{\hat{e}}_y^{\,i} \boldsymbol{\hat{e}}_y^{\,j})(\boldsymbol{\hat{\alpha}}_i \boldsymbol{\hat{\alpha}}_j - \boldsymbol{\hat{\beta}}_i \boldsymbol{\hat{\beta}}_j), \\
        \nonumber \,\, \\
        F^\times &= \frac{1}{2}(\boldsymbol{\hat{e}}_x \otimes \boldsymbol{\hat{e}}_x - \boldsymbol{\hat{e}}_y \otimes \boldsymbol{\hat{e}}_y)^{ij}\Tilde{\boldsymbol{\epsilon}}^{\,\times}_{ij} \\
        \nonumber &= \frac{1}{2}(\boldsymbol{\hat{e}}_x^{\,i} \boldsymbol{\hat{e}}_x^{\,j} - \boldsymbol{\hat{e}}_y^{\,i} \boldsymbol{\hat{e}}_y^{\,j})(\boldsymbol{\hat{\alpha}}_i \boldsymbol{\hat{\beta}}_j + \boldsymbol{\hat{\beta}}_i \boldsymbol{\hat{\alpha}}_j).
    \end{align}
\end{subequations}
Substituting the values for the matrices defined above, we obtain:
\begin{subequations}
    \begin{align}
        F^+ &= \frac{1}{2}(1+\cos^2\theta)\cos{2\phi}\cos{2\psi} - \cos{\theta}\sin{2\phi}\sin{2\psi} \\
        F^\times &= \frac{1}{2}(1+\cos^2\theta)\cos{2\phi}\sin{2\psi} + \cos{\theta}\sin{2\phi}\cos{2\psi},
    \end{align}
\end{subequations}
and Eq.~\eqref{eq:intro-strain-diff-length} can be written as:
\begin{equation}
    h(t) = F^+(\theta, \phi, \psi) h^+(t) + F^\times(\theta, \phi, \psi) h^\times(t).
\end{equation}
These antenna pattern functions dictate the amplitude of the GW received in the detector, depending on the orientation of the detector with respect to the source. The maximum value of either $F^+$ or $F^\times$ is 1. If the arms of the interferometer are not perpendicular to each other, the detector-plane coordinates $x$ and $y$ are defined so that the bisector of the angle between the arms aligns with the bisector of the angle between the coordinate axes. When averaged over all sky angles, these response functions take a value of 2/5 \citep[see also][for calculation of the rms values of antenna pattern functions]{Sathyaprakash:2009xs}.

%-----------------------------------

\subsection{Current interferometric detectors}
\label{subsec:intro-current-dets}

Currently, 5 ground-based gravitational wave detectors are operational. These include the two LIGO \citep{LIGOScientific:2014pky, 2020PhRvD.102f2003B} detectors, both of which are located in the USA, one in Livingston, and the other in Hanford, Virgo~\citep{Virgo:2014yos, PhysRevLett.123.231108} detector in Italy, GEO~600 \citep{Grote:2010zz} in Germany, and KAGRA (Kamioka Gravitational Wave Detector) in Japan \citep{KAGRA:2020tym, Aso:2013eba} which has recently joined the network. LIGO detectors have arms with 4 km length, but this is still not enough to detect the change in the length produced by a typical GW signal. Thus, in order to increase the length, Fabry Perot cavities are included in each of the arms, with an additional mirror in each arm near the beam splitter (Fig.~\ref{fig:intro-basic-interferometer}). After entering the instrument via the beam splitter, the laser in each arm bounces between these two mirrors about 300 times before being merged with the beam from the other arm. In addition to increasing the arm length from 4 km to $\sim 1200$ km, this also builds up the laser light within the interferometer, which increases sensitivity (since more photons keep track of the lengths of the arms) \citep{ligo_info}. To increase the sensitivity, in addition to the long length of the detectors' arms, the detector must operate at power ($\sim 750$ kW) much higher than what enters the detector ($\sim 40 $W). In order to achieve this, Power Recycling mirrors are introduced in the detectors (see Fig.~\ref{fig:intro-basic-interferometer}). Inside the interferometer, light from the laser passes through the transparent side of a power recycling mirror to the beam splitter and is directed down the arms of the interferometer. The instrument's alignment and mirror coatings ensure that nearly all of the laser light entering the arms follows a path back to the reflective side of the power recycling mirror before it exits to the photo-detector. As laser power constantly enters the interferometer from the laser itself, the power recycling mirror continually reflects the laser light that has already travelled through the instrument back into the interferometer. This process greatly boosts the power of the laser light inside the Fabry Perot cavities without generating such a powerful laser beam at the outset.
The boost in power generated by power recycling results in a sharpening of the interference fringes that appear when the two beams are superimposed.

\begin{figure}[t!]
    \centering
    \includegraphics[width=\linewidth]{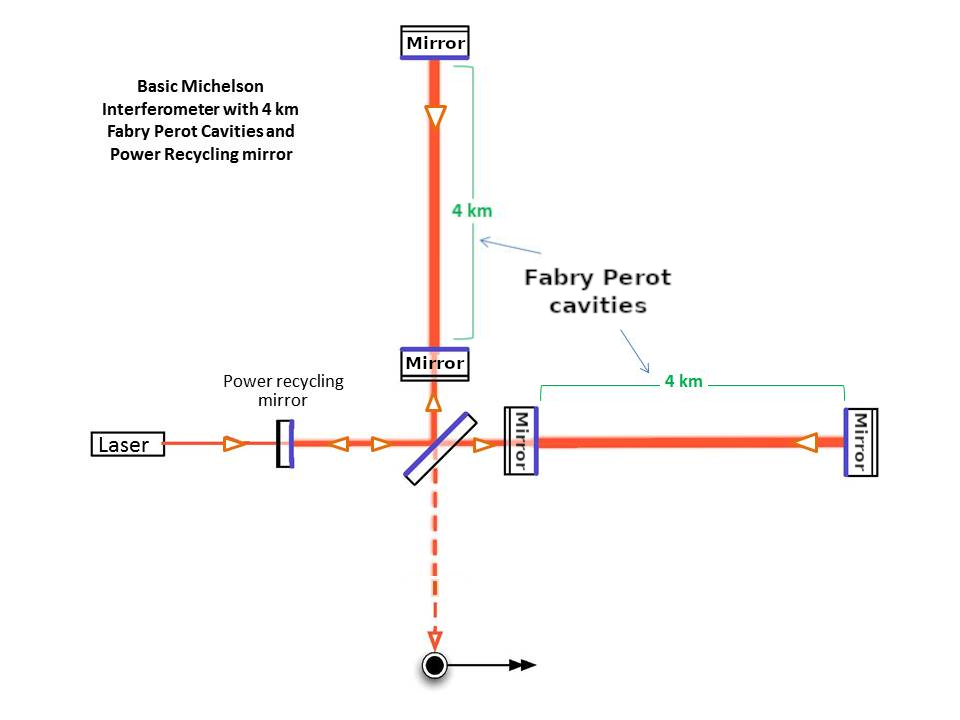}
    \caption[Basic Michelson interferometer with Fabry Perot cavities and Power Recycling mirror]{\takenfrom{\cite{ligo_ifo}} Basic Michelson with Fabry Perot cavities and Power Recycling mirror. LIGO's interferometers use multiple power recycling mirrors, but for simplicity, only one is shown. Image Credit: Caltech/MIT/LIGO Lab.}
    \label{fig:intro-basic-interferometer}
\end{figure}

These detectors also possess signal recycling mirrors, which further enhance the signal that is received by the photo-detector. Moreover, the interferometers have a seismic isolation system that dampens out unwanted vibrations (noise), making it easier for the instruments to sense the vibrations caused by gravitational waves. Virgo and KAGRA detectors are also similar in design but with arm lengths of 3 km. Further details about detector technologies can be seen at \cite{ligo_tech} for LIGO, \cite{virgo_det} for Virgo, and \cite{KAGRA:2020tym} for KAGRA.

In addition to the interferometric detectors, a set of galactic pulsars forming a Pulsar Timing Array (PTA) is monitored and analysed to search for correlated signatures in the pulse arrival times on Earth. This array of millisecond pulsars is used to detect and analyse long wavelength gravitational-wave background such as that originating from the inspirals of supermassive black hole binaries \citep{Lynch:2015iua}. There are several pulsar timing array projects ongoing around the world, and they have been collaborating under the title of International Pulsar Timing Array \citep{ipta}.

%-----------------------------------

\subsection{Future detectors}
\label{subsec:intro-future-dets}

\begin{figure}[t!]
\centering
\includegraphics[trim=10 10 10 10, clip, width=\linewidth]{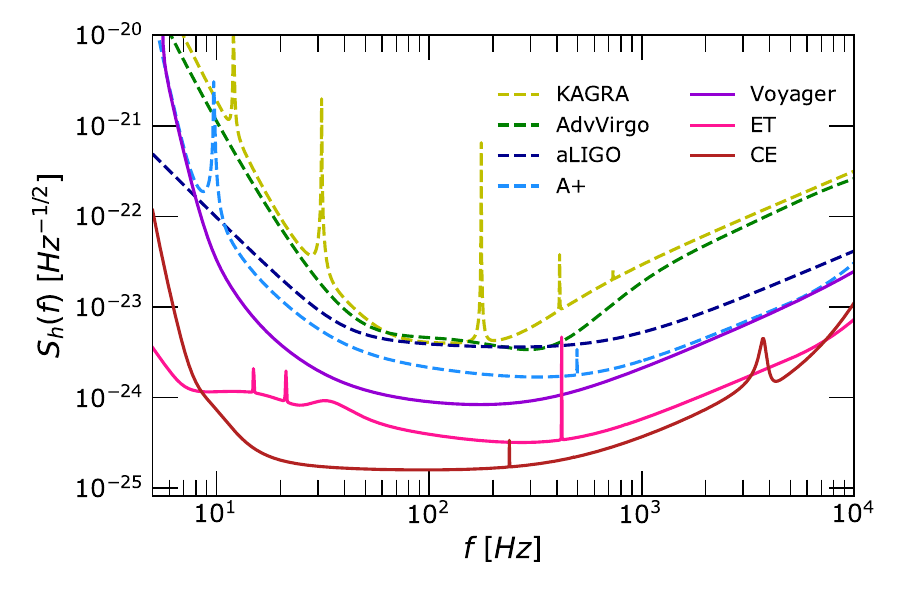}
\caption[Detector sensitivity curves for various ground-based detectors]{\takenfrom{\cite{Divyajyoti:2021uty}} Detector sensitivity curves for various ground-based detectors. The dashed lines denote current detectors (or their upgrades), and the solid lines denote future detectors.}
\label{fig:intro-PSDs}
\end{figure}

There are multiple proposals for future ground-based as well as space-based GW detectors. LIGO-India \citep{LI, Saleem:2021iwi, Unnikrishnan:2023uou} is one such detector which was approved recently, and the construction has begun. Further, there are planned upgrades for the current LIGO detectors to A+ \citep{KAGRA:2013rdx}, A\# \citep{A_sharp}, and Voyager \citep{LIGO:2020xsf, LIGO_Voyager} sensitivities. Going to the next generation, Cosmic Explorer (CE) \citep{Evans:2021gyd, LIGOScientific:2016wof, Reitze:2019iox} and Einstein Telescope (ET) \citep{Punturo_2010, Hild:2010id, Punturo:2010zz, ET-0106C-10} are among the leading proposals for the next generation of ground-based detectors. CE will be similar in layout to the current LIGO detectors, with two arms at a right angle to each other, forming an L shape. The length of these two arms is proposed to be 40\,km each, which is 10 times longer than the advanced LIGO detector. ET, on the other hand, will have a different layout. It will consist of three arms forming an equilateral triangle\footnote{The proposal for Einstein telescope is one of the most favoured proposals at the moment.}. Each arm will have a length of 10\,km, and the whole setup is underground. Both detectors are expected to achieve a sensitivity that is roughly an order of magnitude better than the current 2G detectors (aLIGO), on average, and a low-frequency sensitivity in the range 1-5\,Hz \citep{Chamberlain:2017fjl}. 

Going to the space-based detectors, the Laser Interferometer Space Antenna (LISA) \cite{Babak:2021mhe} is already under construction. The detector will be in the shape of an equilateral triangle with three spacecrafts on the three vertices of the triangle. Each spacecraft will contain two telescopes, two lasers, and two test masses, and the arrangement will be placed at the same distance from Sun as the Earth's orbit, but with an angle of $20^\circ$ lagging the Earth. The plane of the triangle formed by the spacecrafts will be tilted at $60^\circ$ compared to the Ecliptic. With arm lengths of $\sim 2.5$ million kms, it will open the gravitational wave window to a range of low frequencies from 0.1 mHz to 100 mHz.

Other detector proposals include a number of space based GW detectors such as 
\begin{itemize}
    \item DECi-hertz Interferometer Gravitational-wave Observatory (DECIGO) \citep{Kawamura:2020pcg}: Japanese space-based GW observatory expected to be most sensitive in the 0.1 to 10 Hz frequency band, filling in the gap between the sensitive bands of LIGO and LISA.
    \item Big Bang Observer (BBO) \citep{Harry:2006fi}: proposed successor to LISA by the European Space Agency with the primary scientific goal of observing the gravitational waves from the time shortly after the Big Bang.
    \item TianQin \citep{TianQin:2015yph}: proposed Chinese space-borne gravitational-wave observatory consisting of three spacecrafts in Earth orbit. The TianQin project is being led by Professor Luo Jun, President of Sun Yat-sen University.
    \item TianGO \citep{Kuns:2019upi}: proposed space-based, decihertz gravitational-wave detector.
    \item Lunar Gravitational-Wave Antenna (LGWA) \citep{LGWA:2020mma}: detector to monitor the vibrations of Moon with the potential to reveal gravitational waves in the mHz band.
\end{itemize}

\begin{sidewaysfigure}[h!]
    \centering
    \includegraphics[width=0.8\linewidth]{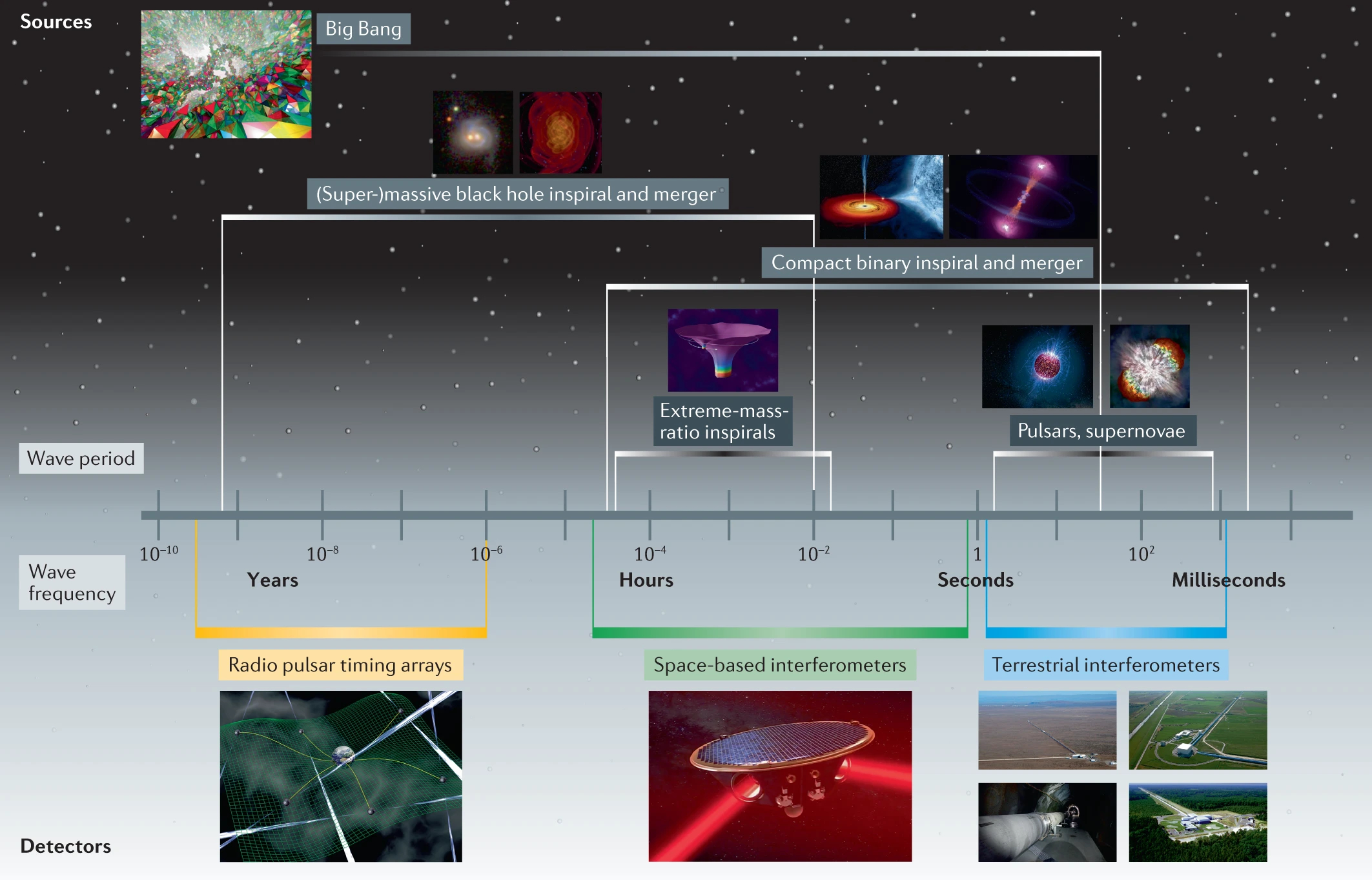}
    \caption[The gravitational-wave spectrum probed by strain-sensitive gravitational-wave detectors, ranging from $10^{-9}$ Hz to more than $1000$ Hz]{\takenfrom{\cite{Bailes:2021tot}} The gravitational-wave spectrum probed by strain-sensitive gravitational-wave detectors, ranging from $10^{-9}$ Hz to more than $1000$ Hz.}
\end{sidewaysfigure}
\clearpage

%%%%%%%%%%%%%%%%%%%%%%%%%%%%%%%%%%%%%%%%

\section{Observed events and their implications}
\label{sec:intro-detected-events}

\begin{sidewaysfigure}
    \centering
    \includegraphics[width=0.9\linewidth]{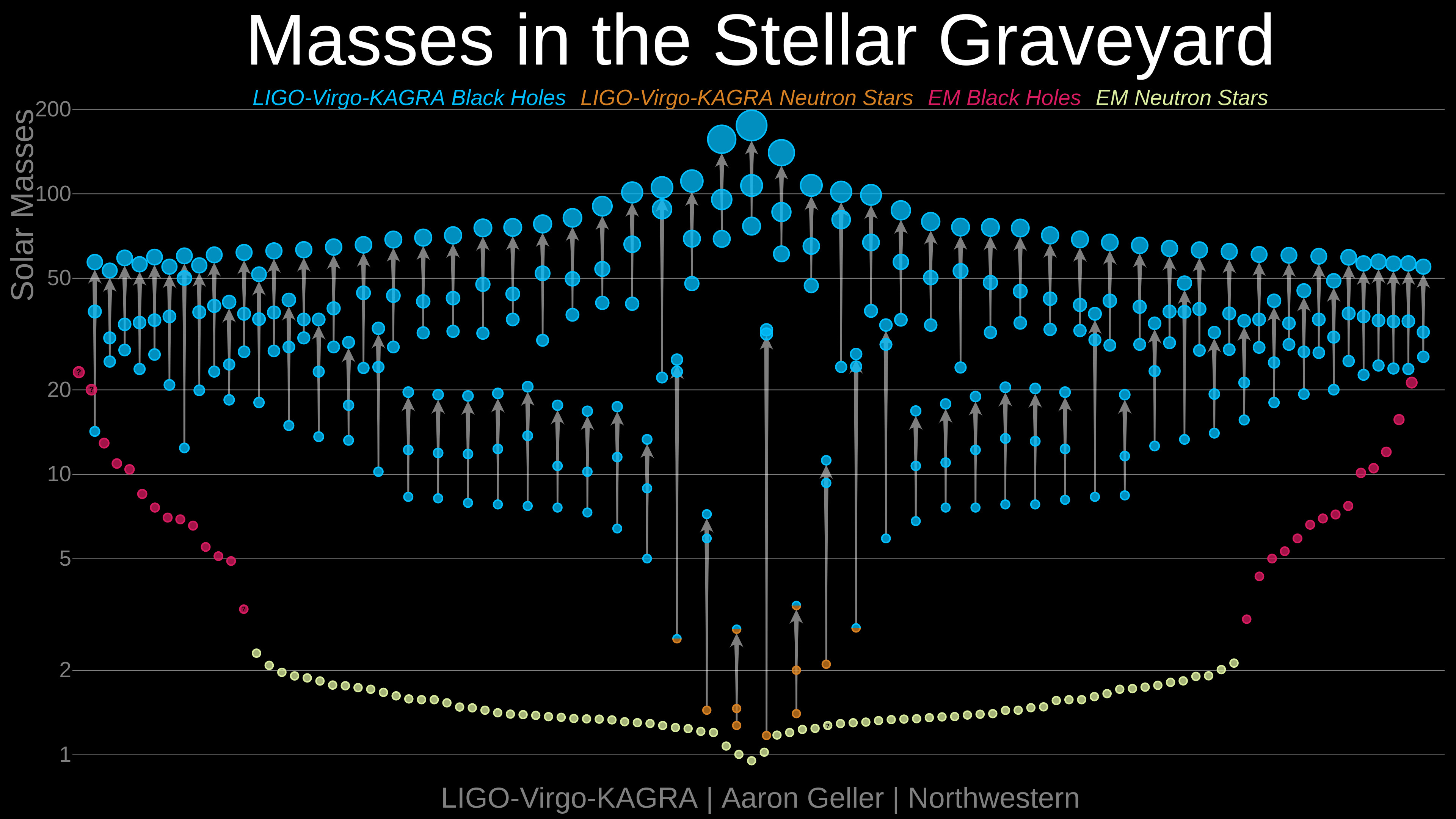}
    \caption[Masses of announced gravitational-wave detections, and black holes and neutron stars previously constrained through electromagnetic observations]{\takenfrom{\cite{ligo_gwtc3_events}} Graphic of masses of announced gravitational-wave detections, and black holes and neutron stars previously constrained through electromagnetic observations. This contains all GW events through the end of O3 with $p_\text{astro} > 0.5$. Image Credit: LIGO-Virgo / Aaron Geller / Northwestern University.}
    \label{fig:intro-events}
\end{sidewaysfigure}

Since the detection of the first gravitational wave event, GW150914, there have been three observation runs (O1-O3) of the current interferometric gravitational wave detectors, and the candidate events have been cataloged into four gravitational wave transient catalogs: GWTC-1 \citep{LIGOScientific:2018mvr}, GWTC-2 \citep{LIGOScientific:2020ibl}, GWTC-2.1 \citep{LIGOScientific:2021usb}, and GWTC-3 \citep{LIGOScientific:2021djp}. The first observing run was from 12\textsuperscript{th} September 2015 to 19\textsuperscript{th} January 2016, and gravitational waves from three binary black hole (BBH) mergers were reported. The second observing run, which took place between 30\textsuperscript{th} November 2016 and 25\textsuperscript{th} August 2017, reported the first binary neutron star (BNS) merger and seven BBHs. These events, which all have a false alarm rate (FAR) of $<1$ per year, are included in GWTC-1 along with 14 GW candidate events with $\text{FAR}<1$ per 30 days. The first half of the third observing run (O3a) was from 1\textsuperscript{st} April 2019 to 1\textsuperscript{st} October 2019, and 39 additional GW candidates with $\text{FAR} < 2$ per year were added to the list (GWTC-2). Most of these events were BBHs, while some signals were consistent with BNS or Neutron Star-Black Hole (NSBH) systems, but the classification could not be made unambiguously. GWTC-2 was accompanied by GWTC-2.1, which is a deep extended catalog of CBCs from O3a with 1201 candidates that pass $\text{FAR} < 2$ per day. Finally, GWTC-3 adds 35 CBC candidates with the probability of astrophysical origin $p_\text{astro}>0.5$ \citep[see Sec.~IV C and appendix D 7 of][for details on $p_\text{astro}$ calculation]{LIGOScientific:2021djp} identified during the second half of the third observing run (O3b) which was from 1\textsuperscript{st} November 2019 to 27\textsuperscript{th} March 2020. These include signals from BBH and NSBH systems, along with 1048 subthreshold candidates which meet the threshold of $\text{FAR}<2$ per day but do not meet the $p_\text{astro}$ criteria. Overall, 90 GW candidates meet the $p_\text{astro}$ criteria from the three observing runs to date. These are shown in Fig.~\ref{fig:intro-events}. We now discuss a few events which have a special place in our hearts. The event nomenclature is GWYYMMDD (for events from O1-O2) or GWYYMMDD\_hhmmss (for events from O3)\footnote{The last six digits indicating the time in addition to the date were added because the detector sensitivities increased in O3 such that multiple triggers were recorded per day.}, which is GW followed by the time stamp (UTC) of the signal. For example, GW200220\_124850 occurred on 2020-02-20 at 12:48:50.

\subsection{Exceptional events}
\label{subsec:intro-exceptional-events}

\subsubsection{GW150914}

This was the first direct detection of gravitational waves, which was observed with a false alarm rate of 1 per 203000 years or 5.1$\sigma$ significance and a signal-to-noise ratio (SNR) of 24 \citep{LIGOScientific:2016vbw, LIGOScientific:2016vlm}. The signal was strong enough to be detected in the filtered strain data without using any waveform models. It was consistent with the merger of two black holes with masses $36^{+5}_{-4} \Msun$ and $29^{+4}_{-4} \Msun$ at a redshift of $z=0.09^{+0.03}_{-0.04}$. The final black hole had a mass of $62^{+4}_{-4} \Msun$ with approximately $3^{+0.5}_{-0.5} \Msun$ of mass emitted in the form of gravitational wave energy. We list here some of the astrophysical implications of this event \citep{TheLIGOScientific:2016htt}:
\begin{itemize}
    \item Confirmed the prediction of Einstein's general theory of relativity in very strong-field regime.
    \item Provided the \textbf{first direct evidence for the formation and merger of binary black holes} within the age of the universe. 
    \item Both black holes were more massive than the previously inferred black holes from X-ray binary observations.
    \item Increased the lower limits on the merger rate of such events, ruling out certain theoretical models which predicted lower rates. 
    \item Confirmed that the speed of gravitational waves was consistent with the speed of light with better constraints than earlier and improved the constraints on the graviton mass.
\end{itemize}

\subsubsection{GW151226}

This was a binary black hole merger of $14.2^{+8.3}_{-3.7} \Msun$ and $7.5^{+2.3}_{-2.3} \Msun$ component masses \citep{LIGOScientific:2016sjg} consistent with inferred masses of black holes found in x-ray binaries. The dimensionless spin magnitude of at least one companion black hole was greater than 0.2. This was the \textbf{first GW event with spinning black hole(s)}. Because of the signal’s smaller
strain amplitude and time-frequency morphology, in other words, the signal was buried deep inside the noise, the generic transient searches that initially identified GW150914 did not detect GW151226. Hence, the initial identification of this signal was confirmed by performing two independent offline matched-filter searches \citep{Usman:2015kfa, Messick:2016aqy, LIGOScientific:2016vbw} that used the waveform models in \cite{Taracchini:2013rva, Purrer:2015tud}.

\subsubsection{GW170817}

This was the first detection of a binary neutron star merger \citep{TheLIGOScientific:2017qsa, LIGOScientific:2018hze} and the only gravitational wave event to date with a confirmed electromagnetic counterpart \citep{GBM:2017lvd}. It was nearly a 100-second signal observed in the GW detectors of inspiralling pair of neutron stars, which ended in the merger \citep{LIGOScientific:2017zic}. Nearly 1.7 s after the GW signal ended, there was a gamma-ray burst (GRB) detected by FERMI and INTEGRAL spacecraft. It was confirmed that this burst came from the same source and \textbf{confirmed the hypo\textbf{thesis} that binary neutron star mergers can produce short GRBs}. The aftermath of the signal was followed up by 70 observations by various telescopes on the ground and in space, and the emission was reported for days (months) after the GW event, in the whole range of electromagnetic spectrum from Gamma rays to radio waves. We list here some of the major astrophysical implications of this event:

\begin{itemize}
    \item Event with the best sky localization measurement of just 16 sq. deg.
    \item A BNS event like this is expected to result in kilonova, followed by radioactive decay of heavy r-process nuclei, responsible for synthesizing heavy elements like gold and platinum found in our universe.
    \item The detection of electromagnetic counterpart, in addition to gravitational waves, started a new era in multi-messenger astronomy; nearly 100 preprints were submitted within a day of the announcement of this event.
    \item The difference between the speed of light and the speed of gravitational waves was improved by 14 orders of magnitude from its previous bound.
    \item An event like this can be used as a standard siren to make independent measurements of the Hubble constant. With more events like GW170817 in the future, the bounds on the Hubble constant can be tight enough to resolve the Hubble tension.
    \item This led to several new tests of GR in the strong field regime and ruled out several alternate theories of gravity.
    \item It provided novel opportunity to probe the properties of matter at extreme conditions such as those found in the interior of neutron stars, and constraints were placed on the neutron star radii and equations of state \citep{LIGOScientific:2018cki}.
\end{itemize}

\subsubsection{GW190412}

This was the first event with considerable asymmetry in the component masses with the primary ($\sim 30 \Msun$) being nearly 3.75 times more massive than the secondary ($\sim 8 \Msun$) \citep{LIGOScientific:2020stg}. Because of asymmetry, in addition to the dominant mode of gravitational radiation, a \textbf{subdominant mode was also detected for the first time}. This provided great advantage in the analysis with waveform models that include subdominant modes and resulted in the breaking of degeneracy between various parameters such as distance and inclination angle (orientation with respect to the line-of-sight) of the binary system.

\subsubsection{GW190814}

This is another event \citep{LIGOScientific:2020zkf}, like GW190412, where a subdominant mode was detected. In fact, the power in the subdominant mode was even higher than it was in GW190412. This was also the event where the spin magnitude of the primary is restricted to 0 with extremely high precision (upper bound of 0.07). The mass of the secondary lies entirely inside (including the 90\% error bars) the lower mass gap between the mass of neutron stars and black holes, making it either the \textbf{heaviest neutron star or the lightest black hole} detected. There is still some debate about the nature of the secondary object.

\subsubsection{GW190521}

This is a merger of two black holes with $85^{+21}_{-14} \Msun$ and $66^{+17}_{-18} \Msun$ masses \citep{Abbott:2020tfl, Abbott:2020mjq} where the primary black hole mass lies in the \textbf{mass gap produced by pulsational pair-instability supernova} processes, indicating alternative formation channels for this kind of binary black hole system. The mass of the remnant was also high enough ($142_{-16}^{+28}$) to be considered in the intermediate-mass black hole range. This event has been shown to be consistent with a merger of a system with spin-induced orbital precession, and high orbital eccentricity, among other scenarios.

\subsubsection{Honorable mentions}
\begin{itemize}
    \item GW170104: BBH with the effective spin parameter showing significance in the negative region indicating that black hole spins show a preference for system aligned away from the orbital angular momentum \citep{LIGOScientific:2017bnn}.
    
    \item GW170814: First 3-detector observation of GW signal, thus constraining the sky localization to just $60$ sq. deg. This enabled the test of GW polarizations and, subsequently, a new class of phenomenological tests of GR \citep{LIGOScientific:2017ycc}.
    
    \item GW191219\_163120: This event shows support for the highest mass ratio ($\sim 26.5$) till date, also making it a potential NSBH candidate, but the uncertainty in $p_\text{astro}$ which is dependent on the pipelines, and the mass ratio inference itself which is outside the calibration limit of the waveform models makes it hard to infer its properties with confidence \citep{LIGOScientific:2021djp}.
    
    \item GW200105\_162426 and GW200115\_042309: These were the first two neutron star - black hole merger events reported, and neither had a confirmed electromagnetic counterpart. Including these events in the catalog improved the estimates for the merger rate of NSBH events significantly \citep{LIGOScientific:2021qlt}. 
\end{itemize}

As we have discussed in the sections above, gravitational waves have opened a new window to the universe, allowing probes into new sectors of astrophysics and fundamental physics. Now, we focus on two important implications of gravitational wave astrophysics considering only the current detected events by the LIGO-Virgo detectors: population inference of compact binary systems and tests of general relativity possible with the GW detections.

\subsection{Population of merging compact binaries}

With a total of about 90 compact binary coalescences detected in the first three observing runs, it is possible to put constraints on the populations and merger rates of BBH, NSBH, and BNS systems. Although the detected events included in the catalogs meet some $p_\text{astro}$ threshold, a more stringent constraint\footnote{A criteria of FAR $<1$ per year is imposed on BBH events whereas an even stricter constraint of FAR $<0.25$ per year is imposed for all analyses considering NS systems, due to the lower number of observations.} \citep{KAGRA:2021duu} is put on the events when inferring the merger rates and population distributions in order to avoid contamination from signals of non-astrophysical origin. This reduces the total number of events to 76. Here, we list some of the key features of population distributions inferred from the detected GW events \citep{KAGRA:2021duu}.

\begin{itemize}
    \item NSBH binaries form a distinct population separate from BNS and BBH populations, and the merger rates of NSBH binaries are substantially larger than the BBH merger rates.

    \item There seems to be a relative dearth of observations of binaries which have component masses in the range 3-5 $M_\odot$.

    \item While the Galactic BNS favour a peak at 1.35 $M_\odot$ \citep{Shao:2021dbg, Mapelli:2020xeq, Olejak:2022zee}, the inferred mass distribution for BNS mergers from GWTC-3 does not exhibit a peak at this value.

    \item The observed BBH mass distribution seems to be clumped with multiple major and minor peaks showing in the chirp mass distribution. The primary mass distribution peaks near $\sim 10 M_\odot$ which suggests that globular clusters contribute subdominantly to the detected population \citep{Fishbach:2020ryj, Mapelli:2018wys, Regimbau:2011rp}. Dynamical formations in young clusters are also disfavoured \citep{Romano:2016dpx, Christensen:2018iqi, Callister:2020arv}.

    \item The spin distribution of BBH population exhibits low overall spins, and the effective inspiral spin parameter ($\chi_\text{eff}$) distribution is concentrated near 0.

    \item Considering the other peaks in the mass distribution and combining them with the spin distribution leads to the conclusion that BBH binaries originate from the initial BH mass function or are produced by different populations formed by separate physical processes or formation channels.

    \item The analysis of BBH rate evolution with redshift strongly disfavours the possibility that the merger rate does not evolve with redshift.    
\end{itemize}

In this \textbf{thesis}, we utilize the population distributions inferred by the gravitational wave events to construct populations of black hole binary systems for our injection studies. For instance, we have used two different mass distributions taken from \cite{LIGOScientific:2020kqk} to construct the population of sources studied in Chapter \ref{chap-hm}. Moreover, in Chapter \ref{chap:ecc}, we use astrophysically motivated population models for redshift \citep{Ng:2020qpk} and eccentricity \citep{Zevin:2021rtf, Kremer:2019iul} to create the population of eccentric binary black hole mergers.

%-----------------------------------------

\subsection{Tests of general relativity}
\label{subsec:intro-tgr}

Gravitational waves in modified theories of gravity may differ from general relativity in generation, propagation, and polarization. Thus, the tests of general relativity using GWs can be broadly classified into two categories: consistency tests and parametrized tests. Consistency tests check the consistency of the observed GW signal with the predicted GR waveform. These may be consistency tests of the signal morphology or the overall consistency of the GR signal with the data. Parametrized tests, on the other hand, introduce specific deviation parameters in the models to check for particular effects that may arise due to the violation of GR. We briefly describe the tests which were performed under these two categories for the events in GWTC-3 catalog \citep{LIGOScientific:2021sio}.

\subsubsection{Consistency tests}

\begin{itemize}
    \item \textit{Residuals test}: The random noise in different detectors can be taken to be incoherent. Detecting consistent noise in the network, even after eliminating the gravitational wave signal from the data, suggests a discrepancy between the signal present in the data and the GR template employed. The residual analysis is designed to detect such discrepancies in the data with GR \citep{LIGOScientific:2016lio, LIGOScientific:2019fpa, Abbott:2020mjq}.

    \item \textit{Inspiral–merger–ringdown consistency test}: checks the consistency of the mass and spin of the remnant BH inferred from the low- and high-frequency parts of the signal \citep{Ghosh:2016qgn, Ghosh:2017gfp}.
\end{itemize}

\subsubsection{Parametrized tests}

\begin{itemize}
    \item \textit{Parametrized tests of GW generation}: This test introduces deviations at different post-Newtonian orders since post-Newtonian coefficients are sensitive to different physical effects. Measuring a non-zero value in these deviation parameters can point to limitations of the waveform models, which are based on GR \citep{Blanchet:1994ez, Blanchet:1994ex, Arun:2006hn, Arun:2006yw, Yunes:2009ke, Mishra:2010tp, Li:2011cg, Li:2011vx}.
    
    \item \textit{Modified GW dispersion relation}: In general relativity, it is predicted that gravitational waves propagate non-dispersively. This characteristic is described by the dispersion relation $E^2 = p^2 c^2$, where $E$ and $p$ represent the energy and momentum of the wave, respectively. Observing the dispersion of gravitational waves can be regarded as an indicative signature of alterations to GR. In this test, a parameterized model \citep{Mirshekari:2011yq, Will:1997bb} for dispersion of GWs is used that aids in investigating the potential existence of dispersion in the data without explicitly relying on the specifics of the modified theory.
    
    \item \textit{Polarization test}: The assessment of the polarization content of gravitational waves serves as a means to limit potential deviations from GR. This limitation arises because the theory permits only two out of the six polarization states predicted by generic metric theories of gravity \citep{Eardley:1973br, Eardley:1973zuo}.
    
    \item \textit{Echoes searches}: The observed mergers of massive compact objects may involve not only black holes as predicted by classical GR but also different types of compact objects governed by exotic physics, collectively known as exotic compact objects (ECOs). This category encompasses entities such as firewalls \citep{Almheiri:2012rt}, fuzzballs \citep{Lunin:2001jy}, gravastars \citep{Mazur:2004fk}, boson stars \citep{Liebling:2012fv}, AdS black bubbles \citep{Danielsson:2021ykm}, and dark matter stars \citep{Giudice:2016zpa}. A shared characteristic among these objects is the absence of a horizon, leading to the reflection of ingoing gravitational waves (such as those generated during a merger) multiple times off effective radial potential barriers. This process results in wave packets leaking out to infinity, potentially exhibiting regular intervals, and these phenomena are termed "echoes" \citep{Cardoso:2016oxy, Cardoso:2016rao, Cardoso:2017cqb}. Searches for echoes serve as constraints on the existence of recurring ringdown signals \citep{Cardoso:2016rao, Cardoso:2017cqb, Abedi:2016hgu, Ashton:2016xff, Westerweck:2017hus, Uchikata:2019frs}, which are anticipated in specific categories of Exotic compact objects (ECOs).

    \item \textit{Ringdown test}: The gravitational radiation emitted from the highly distorted black hole remnant resulting from a merger is commonly known as ringdown. The waveform after the onset of the ringdown phase is a combination of quasi-normal modes (QNM) characterized by a complex frequency \citep{Vishveshwara:1970cc, Cunningham:1978zfa}. According to GR, the frequency and damping times for astrophysical black holes (BHs) are entirely dictated by the mass and spin of the resulting BH remnant \citep{Hansen:1974zz, Carter:1971zc, Gurlebeck:2015xpa, Penrose:1969pc}. The connection between frequency and remnant parameters establishes that the detection of multi-mode ringdown signals provides a distinctive assessment of the BH nature of the merger remnant \citep{Dreyer:2003bv, Berti:2006wq}. This approach holds the potential to differentiate among various classes of ECOs \citep{Berti:2015itd}.

    \item \textit{Spin-induced multipole moment test}: Spinning objects exhibit contributions to the multipole decomposition of their gravitational field, including quadrupole and higher-order terms arising from their rotational deformations. In accordance with the no-hair conjecture, the spin-induced multipole moments assume unique values for black holes based on their mass and spin characteristics \citep{Hansen:1974zz, Carter:1971zc, Gurlebeck:2015xpa}. Gravitational waveforms that describe spinning compact binary systems encapsulate information regarding the effects of spin-induced multipole moments. Measuring the deviation from this spin-induced multipole parameter can point to the non-black hole nature of the components in a binary system. We discuss further details of this test in Sec.~\ref{subsec:intro-siqm} and subsequent extensions using double spin-precessing and higher mode waveform models {in Chapter \ref{chap:siqm}}.
\end{itemize}

%%%%%%%%%%%%%%%%%%%%%%%%%%%%%%%%%%%%%%%%%%%%%%%%%%

\section{Searches and parameter estimation}
\label{sec:intro-searches-matched-filtering}

The data collected at a gravitational wave detector has (if at all) signals which are buried deep within the noise of the detector. In order to search for such signals, most searches employed today are based on a technique known as \textbf{Matched Filtering}, which critically depends on our prior understanding of signal morphology. Matched filtering (also known as Weiner filtering) \citep{Th300, Helstrom68, schutz_1991} is a technique where data is cross-correlated against a linear filter to find the signal buried within the noise. If we can construct a "template" which describes the form of the signal buried in the data, we can use that as a "filter" to extract the signal. The template is essentially a model based on theory (say general relativity), which is expected to accurately describe the signal in the data, or in other words, is our best guess of the signal waveform from theory. A bank of templates (template bank) is used with varying parameters, such as different values of masses and spins, in order to search for the filter which best describes the signal (since the form of the signal depends on these parameters and we don't know the binary parameter values \textit{a priori}) \citep{Allen:2005fk}.

Let us consider the GW data stream represented by a time series $s(t)$. We assume that the data contains stationary Gaussian noise $n(t)$ with zero expectation value, and a signal $g(t)$. Then, the data can be written in the form
\begin{subequations}
\begin{align}
    s(t) &= n(t) + g(t) \\
    \langle {n(t)} \rangle &= 0.
\end{align}
\end{subequations}
Now, we hope to use a template $h(t,\theta_i)$ that can extract the signal from the data. The template is dependent on parameters $\theta_i$, that characterize the source. Modelled searches perform matched filtering between signal templates $\Tilde{h}(f)$ and the detector data $\Tilde{s}(f)$ in the Fourier domain to apply the matched filters for different times of arrival efficiently. 

For Gaussian and stationary noise, the one-sided power spectral density (PSD), $S_n(f)$, of noise $n(t)$ can be defined as 
\begin{equation}
    \langle \Tilde{n}(f) \Tilde{n}^*(f') \rangle \equiv \frac{1}{2} \delta(f-f') S_n(f),
\end{equation}
where $\Tilde{n}(f)$ is the Fourier transform of $n(t)$ given by
\begin{equation}
    \Tilde{n}(f) = \int_{-\infty}^{\infty} n(t) {\rm{e}}^{-2\pi i f t} \text{d}t.
\end{equation}
Using the definition for PSD, a noise weighted inner product between two functions $a$ and $b$ can be defined as \citep[see][for detailed derivation]{Maggiore:2007ulw},
\begin{equation}
    \langle a|b \rangle = 4\int_0^\infty \frac{\Tilde{a}^*(f)\Tilde{b}(f)}{S_n(f)} \text{d}f,
    \label{eq:inner-product}
\end{equation}
which is nothing but the noise-weighted cross-correlation between $a$ and $b$. Hence the filtered signal-to-noise ratio (SNR) (or simply matched filter SNR) can be defined as \citep{Owen:1998dk}
\begin{equation}
    \frac{S}{N} = \frac{\langle g|h \rangle}{\text{rms}\langle n|h \rangle} = \frac{\langle g|h \rangle}{\sqrt{\langle h|h \rangle}}.
\end{equation}
Now, an "optimal filter" which best describes the signal will be the signal itself, and so, if one assumes that the template has exactly the same form as the signal in the detector data, one can define the "optimal" SNR ($\rho$) ~\citep{Cutler:1994ys, Cutler:1992tc, Th300} as
\begin{equation}
\rho^2 = \langle h|h \rangle.
\label{eq:opt-snr}
\end{equation}
The SNRs from individual detectors can be combined into the SNR for the network \citep{LIGOScientific:2021djp, Pai:2000zt} as
\begin{equation}
    \rho_\text{network} = \sqrt{\rho_1^2 + \rho_2^2 + \rho_3^2 + ...}
\end{equation}
where $\rho_i$ denotes the SNR in the $i$\textsuperscript{th} detector. Hence for a network consisting of $k$ detectors with similar sensitivities, we will have $\rho_\text{network} \propto \sqrt{k}$.

\subsection{Search pipelines}

Currently, gravitational wave searches are performed in two stages: online searches and offline searches. Online searches are performed in (near) real-time as the data is being collected and allow for the rapid release of public alerts associated with candidates to enable the search for multi-messenger counterparts. On the other hand, offline searches are done using the cleaned dataset, and get the advantage of improved background statistics, extensive data calibration, vetting and conditioning  \citep{LIGOScientific:2021djp}. In the latest (third) gravitational wave transient catalog (GWTC-3) \citep{LIGOScientific:2021djp}, five search pipelines were used for online searches: Gstreamer LIGO Algorithm Library (\texttt{GstLAL}) \citep{Messick:2016aqy, Sachdev:2019vvd, Cannon:2020qnf, Hanna:2019ezx}, Multi-Band Template Analysis (\texttt{MBTA}) \citep{Adams:2015ulm, Aubin:2020goo}, Python-based toolkit for Compact Binary Coalescence signals (\texttt{PyCBC}) \citep{Allen:2005fk, Allen:2004gu, DalCanton:2014hxh, Usman:2015kfa, Nitz:2017svb, Davies:2020tsx}, Summed Parallel Infinite Impulse Response (\texttt{SPIIR}) \citep{Chu:2020pjv}, and coherent WaveBurst (\texttt{cWB}) \citep{Klimenko:2015ypf, Klimenko:2004qh, Klimenko:2011hz}. While the first four pipelines search based on CBC templates, \texttt{cWB} searches for burst signals with minimal assumptions about the sources. Except \texttt{SPIIR}, the other four pipelines were also used to conduct offline searches. The template banks for modelled searches were constructed in the parameter space which included component masses ($m_1, m_2$) of the binary system and corresponding dimensionless spins ($\chi_1, \chi_2$) aligned with the orbital angular momentum. We discuss the binary parameters in detail in Sec.~\ref{subsec:intro-bbh-params}.

%-------------------------------------------

\subsection{Binary Black Hole parameter space}
\label{subsec:intro-bbh-params}

A typical gravitational wave signal from a binary black hole merger observed in ground-based detectors such as LIGO and Virgo can be characterized by a set of at least 15 parameters. These include the component masses ($m_1$ and $m_2$), spin vector components ($S_\text{1x}$, $S_\text{1y}$, $S_\text{1z}$, $S_\text{2x}$, $S_\text{2y}$, $S_\text{2z}$) corresponding to the two spin vectors ($\Vec{S_1}$ and $\Vec{S_2}$), source's luminosity distance ($d_L$), inclination angle ($\iota$) of the binary, time of coalescence ($t_c$), phase of coalescence ($\phi_c$), right ascension ($\alpha$), declination ($\delta$), and polarization angle ($\psi$). In addition to these, if the binary is on an elliptical orbit, the parameter space is extended by including additional orbital parameters such as orbital eccentricity ($e$) at a reference epoch (or equivalently at a reference frequency), and a parameter describing the orientation of the ellipse at the same epoch, e.g., the mean anomaly ($l$) \citep{Ramos-Buades:2023yhy}. Moreover, if the source of the GW signal is not a binary black hole system, other parameters may be required to infer the source properties. For instance, for a binary neutron star system, tidal deformability parameters may be used to describe the deformation of the components and infer information about the equation of state of the neutron stars \citep[see, for instance,][where bounds on $\Tilde{\Lambda}$ were obtained for the GW event GW170817]{LIGOScientific:2018hze}.

While performing parameter estimation of a GW signal, the parameters listed above can suitably be replaced by an equivalent set obtained by combining two or more parameters from the above set. This is usually done for several reasons: the combination of parameters may be better measured compared to the individual parameters, there exist degeneracies between the parameters such that all of them may not be accurately measured simultaneously, there is a reduction in computation time for a certain choice of parameters depending on the sampling algorithms, and so on. For instance, chirp mass \citep{Poisson:1995ef, Finn:1992xs, Blanchet:1995ez} which is defined as
\begin{equation}
    \mathcal{M} = \frac{(m_1 m_2)^{3/5}}{(m_1+m_2)^{1/5}},
\end{equation}
is one of the best-measured parameters for a binary coalescence, and often used in combination with mass ratio, $q=m_1/m_2$, or symmetric mass ratio $\eta = m_1 m_2/(m_1+m_2)^2$ in the parameter space instead of sampling directly in $m_1$ and $m_2$. Similarly, the individual spins of the binary components may not be measured with high precision due to various degeneracies between parameters. Hence, while the spin vector components (in cartesian or spherical polar coordinate system) may be used to sample the parameter space, often combinations of spins and mass ratio are used to infer the in-plane and out-of-plane contributions of the spin vectors. We describe these combinations in detail in Sec.~\ref{sec:intro-spins} where we discuss the spin-induced orbital precession of a binary. Moreover, one or more deviation parameters are often introduced in the waveform models which, depending on the test of GR, help to constrain the deviation from a GR signal. One such deviation parameter ($\delta\kappa$) has been explored in detail in Chapter \ref{chap:siqm}. We provide an exhaustive list of all the parameters used in this \textbf{thesis} at the beginning of this document (Notation).

%-------------------------------------------

\subsection{Bayesian inference}
\label{subsec:intro-bayesian}

\begin{figure}[t!]
    \centering
    \includegraphics[width=\linewidth]{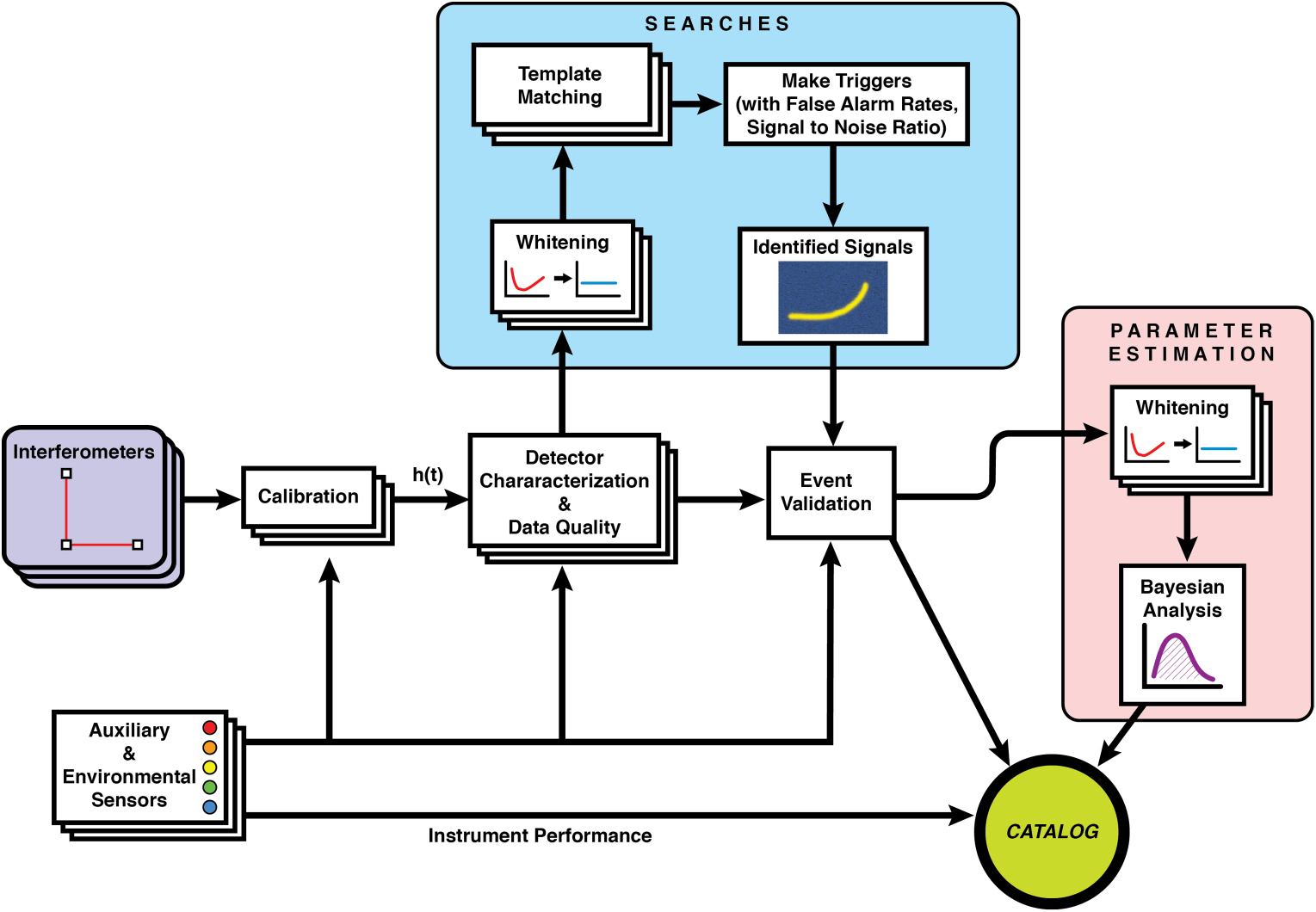}
    \caption[A simplified schematic summarizing the main steps in LIGO–Virgo data processing]{\takenfrom{\cite{LIGOScientific:2019hgc}} A simplified schematic summarizing the main steps in LIGO–Virgo data processing, from the output of the data to the results reported in a catalog of transient events.}
    \label{fig:intro-noise-removal}
\end{figure}

After detecting a gravitational wave signal, various techniques (such as whitening\footnote{The goal of a whitening procedure is to make the sequence of data delta-correlated (characterized by a normal or Gaussian distribution and a correlation between adjacent samples that approaches a delta function), removing all the correlation of the noise \citep{Cuoco:2000gv}. In other words, it normalizes the power at all frequencies so that excess power at any frequency is visible.}) are used to remove noise from the data (see Fig.~\ref{fig:intro-noise-removal} showing a simplified schematic of data processing). While this process is far from perfect, it usually yields a signal which can then be analysed to infer the parameters which describe the properties of the source. In this \textbf{Thesis}, we use Bayesian methods to infer the properties of both simulated and real gravitational wave signals.

The posterior probability for a parameter $\Vec{\theta}$, given the data $d$ and model $\mathcal{H}$, is given by the Bayes theorem as \citep{1763RSPT...53..370B}
\begin{equation}
  p(\Vec{\theta}|d,\mathcal{H}) = \frac{\mathcal{L}(d|\Vec{\theta},\mathcal{H}) \pi(\Vec{\theta}|\mathcal{H})}{\mathcal{Z}(d|\mathcal{H})}\,,
\end{equation}
where $\mathcal{L}(d|\Vec{\theta},\mathcal{H})$ denotes the likelihood, $\pi(\Vec{\theta}|\mathcal{H})$ is the prior, and $\mathcal{Z}(d|\mathcal{H})$ represents the evidence. Prior is chosen such that it includes any \textit{a priori} information we have about the parameters. An example of this is taking an appropriate range of values of the masses of the binaries that we expect to be detected by the current detectors. Likelihood function ($\mathcal{L}(d|\Vec{\theta, \mathcal{H}})$) shows the probability of detectors measuring the data $d$, for a signal described by the model hypothesis $\mathcal{H}$ (say a signal model based on GR) and source properties $\Vec{\theta}$. Now, since the total probability of posterior when integrated over the entire range of the parameters $\Vec{\theta}$ gives one, this leads us to the definition of evidence as:
\begin{subequations}
    \begin{align}
        \int p(\Vec{\theta}|d, \mathcal{H}) \text{d}\Vec{\theta} &= \bigintssss \frac{\mathcal{L}(d|\Vec{\theta},\mathcal{H}) \pi(\Vec{\theta}|\mathcal{H})}{\mathcal{Z}(d|\mathcal{H})} \text{d}\Vec{\theta} = 1 \\
        \Rightarrow \mathcal{Z}(d|\mathcal{H}) &= \int \mathcal{L}(d|\Vec{\theta},\mathcal{H}) \pi(\Vec{\theta}|\mathcal{H}) \text{d}\Vec{\theta}
    \end{align}
\end{subequations}
which is the marginalized likelihood that serves as a measure of how well the data is modelled by the hypothesis $\mathcal{H}$. Although, it is just a normalization constant in parameter estimation, it serves a crucial role in Model selection. Consider two hypotheses $\mathcal{H}_A$ and $\mathcal{H}_B$ for some data $d$. The ratio between the evidence for both of these hypotheses yields the Bayes factor given by \citep{Isi:2022cii}
\begin{equation}
\mathcal{B}_{A/B} = \frac{\mathcal{Z}(d|\mathcal{H}_A)}{\mathcal{Z}(d|\mathcal{H}_B)} = \frac{\int \mathcal{L}(d|\Vec{\theta_A},\mathcal{H}_A) \pi(\Vec{\theta_A}|\mathcal{H}_A) \text{d}\Vec{\theta_A}}{\int \mathcal{L}(d|\Vec{\theta_B},\mathcal{H}_B) \pi(\Vec{\theta_B}|\mathcal{H}_B) \text{d}\Vec{\theta_B}},
\label{eq:bayes_factor}
\end{equation}
where the marginalization is done for different sets of parameters $\Vec{\theta_A}$ and $\Vec{\theta_B}$ for $\mathcal{H}_A$ and $\mathcal{H}_B$ respectively. The definitions of hypotheses include the choice of priors and any other assumptions that go into the likelihoods. In this \textbf{thesis}, we use Bayes factor in multiple studies to compare the models used for analyses. For instance, in Chapter \ref{chap:ecc}, we use it to investigate whether an eccentric waveform model is preferred over a quasi-circular model for the analysis of the injected GW signals. In Chapter \ref{chap:siqm}, Bayes factor is used to quantify the probability that the GW events considered there are from binary black hole mergers and not from other exotic objects. If the Bayes Factor $\mathcal{B}_{A/B}$ is considerably larger than 1 [or equivalently $\log(\mathcal{B}_{A/B})>0$], then it indicates that data prefers hypothesis $\mathcal{H}_A$ over $\mathcal{H}_B$. This can also be written in terms of the odds ratio, which is 
\begin{equation}
    \mathcal{O}_{A/B} = \frac{\pi(\mathcal{H}_A)}{\pi(\mathcal{H}_B)} \mathcal{B}_{A/B},
\end{equation}
which indicates the betting odds in favour of one model over another. It is a common practice though to choose the priors for both hypotheses as the same i.e. $\pi(\mathcal{H}_A) = \pi(\mathcal{H}_B)$ so that there is no \textit{a priori} preference for either hypothesis, in which case $\mathcal{O}_{A/B} = \mathcal{B}_{A/B}$.

In the standard likelihood function, which is commonly used for the data analysis of GW transients \citep{Finn:1992wt, Romano:2016dpx}, both model and data are expressed in the frequency domain. It has stationary Gaussian noise, which is a decent approximation in most cases \citep{Berry:2014jja, Abbott:2016wiq, LIGOScientific:2019hgc} except for when the instruments are affected by a glitch \citep{Pankow:2018qpo, Powell:2018csz} which, if they are, the Detector Characterization and data quality team informs the analysts that the signal may have data quality issues. Upon identifying a significant trigger through search pipelines, analysts typically examine multi-resolution time-frequency scalograms of the data surrounding the trigger, commonly referred to as Q-scans. Q-scans serve as qualitative checks that necessitate visual inspection. These scans are instrumental in revealing the presence of any pronounced glitches in the data, as exemplified by the case of the binary neutron star event GW170817 \citep{TheLIGOScientific:2017qsa}. Once the parameter estimation analyses have been run, a close examination of Q-scans of the residuals is conducted to identify any potential impact of unmodelled noise features on the analyses. For further details on noise mitigation techniques, kindly refer to \cite{LIGOScientific:2019hgc}. Fig.~\ref{fig:intro-gw170817-glitch} shows the glitch that was observed in LIGO Livingston during the detection of GW170817, and the steps that were taken to remove it.

\begin{figure}[p!]
    \centering
    \includegraphics[width=0.95\linewidth]{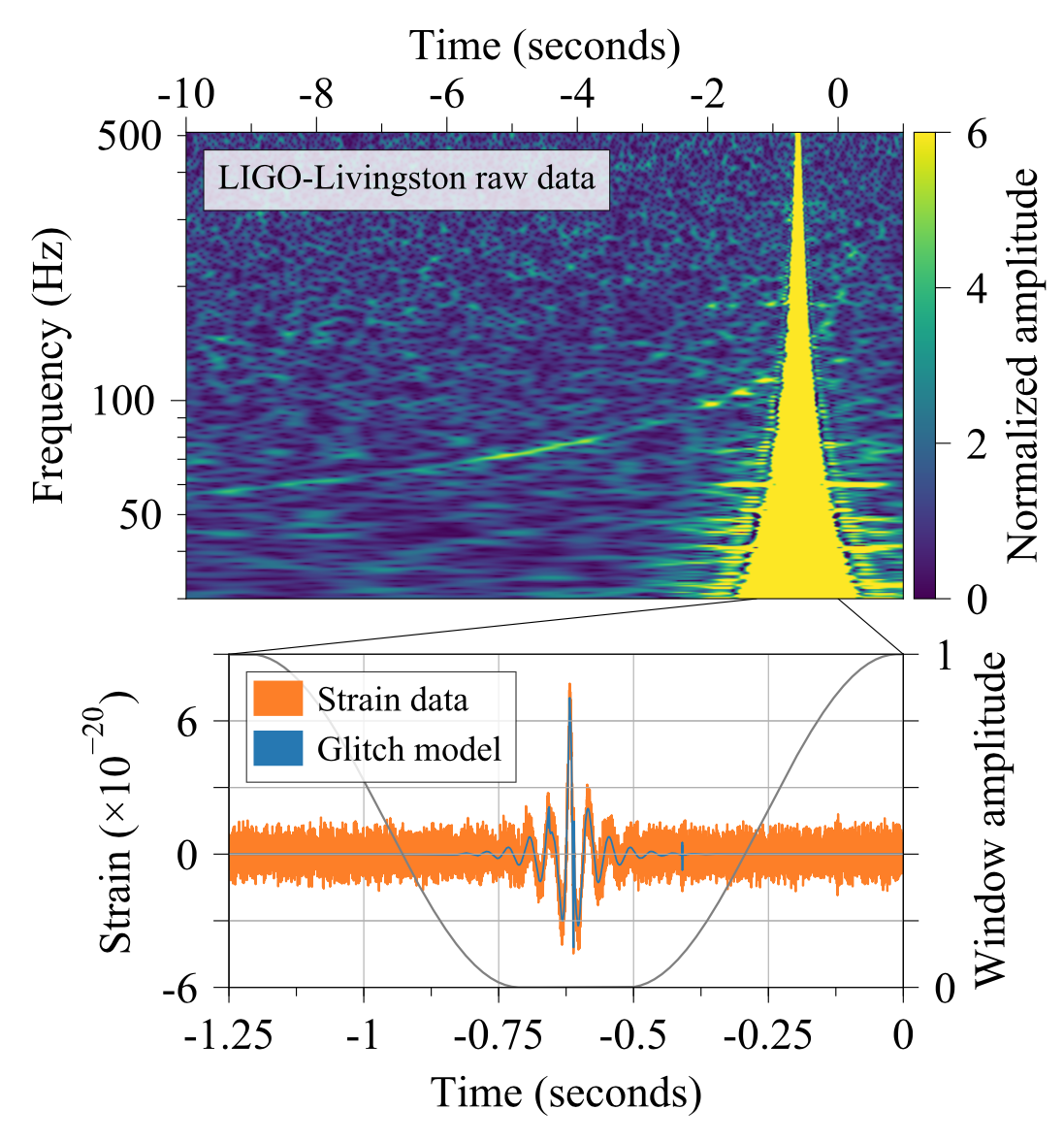}
    \caption[Mitigation of the glitch in LIGO-Livingston data for GW170817]{\takenfrom{\cite{TheLIGOScientific:2017qsa}} Mitigation of the glitch in LIGO-Livingston data for GW170817. Top panel: A time-frequency representation \citep{Chatterji:2004qg} of the raw LIGO-Livingston data used in the initial identification of GW170817. The coalescence time reported by the search is at time 0.4s in this figure, and the glitch occurs 1.1s before this time. The time-frequency track of GW170817 is clearly visible despite the presence of the glitch. Bottom panel: The raw LIGO-Livingston strain data (orange curve) showing the glitch in the time domain. To mitigate the glitch in the rapid reanalysis that produced the sky map, the raw detector data were multiplied by an inverse Tukey window (grey curve, right axis) that zeroed out the data around the glitch \citep{Usman:2015kfa}. To mitigate the glitch in the measurement of the source's properties, a model of the glitch based on a wavelet reconstruction \citep{Cornish:2014kda} (blue curve) was subtracted from the data. The gravitational-wave strain amplitude of GW170817 is of the order of $10^{-22}$ and so is not visible in the bottom panel.}
    \label{fig:intro-gw170817-glitch}
\end{figure}

%-------------------------------------------

\subsection{Samplers}
\label{subsec:intro-samplers}

There are several approaches to parameter inference; for instance, \texttt{BAYESTAR} \citep{Singer:2015ema, Singer:2016eax} performs rapid localization of GW sources by calculating probabilities on a multi-resolution grid of the sky, whereas \texttt{RapidPE} \citep{Pankow:2015cra} and \texttt{RIFT} \citep{Lange:2018pyp} use highly-parallelized grid-based methods. There also exist various schemes which use machine learning algorithms for calculating posterior probabilities \citep{George:2017pmj, Gabbard:2019rde}. However, in this \textbf{thesis}, we use packages like \texttt{LALInference} \citep{Veitch:2014wba}, \texttt{PyCBCInference} \citep{Biwer:2018osg}, and \texttt{bilby\_pipe} \citep{Romero-Shaw:2020owr} which employ \textit{stochastic sampling} methods. These methods such as Markov-Chain Monte Carlo (MCMC) \citep{Christensen:1998gf, Christensen:2001cr, Rover:2006bb, Rover:2006ni, vanderSluys:2007st, vanderSluys:2008qx} and \textit{Nested} sampling \citep{Veitch:2009hd, Veitch:2008ur} algorithms are implemented specifically for gravitational wave data analysis of ground-based detectors. Alternatively, there are algorithms which have been developed for other gravitational wave detectors, such as pulsar timing arrays \citep{Lentati:2013rla, Vigeland:2013zwa} and LISA \citep{MockLISADataChallengeTaskForce:2009wir, Babak:2008aa}. Additionally, methods like those shown in \cite{Zackay:2018qdy} use relative-binning \citep{Cornish:2010kf, Cornish:2020vtw} to reduce the cost of computing the likelihood, whereas \texttt{BayesWave} \citep{Cornish:2014kda, Littenberg:2014oda} uses trans-dimensional MCMC to fit an \textit{a priori} unknown number of sine-Gaussian wavelets to the data.

Numerous Monte Carlo sampling schemes, in general, calculate the posterior probability for every point in the parameter space. While this works for a low-dimensional parameter space [for instance, if binary components are assumed to be non-spinning \citep{Ajith:2009fz}], it becomes highly inefficient and computationally expensive for parameter space in higher dimensions. Since gravitational wave inference deals with a parameter space of 15 dimensions or more, stochastic samplers offer a reasonable solution to this problem. The stochastic samplers can be divided broadly into two (not mutually exclusive) categories: MCMC \citep{Hogg:2017akh, Metropolis:1953am, Hastings:1970aa} and nested sampling \citep{2004AIPC..735..395S, 2020MNRAS.493.3132S}. In both of these, independent samples are drawn stochastically (every point in the signal has a non-zero probability of being sampled, and the samples are non-uniformly spaced) from the posterior such that the number of samples in the range $(\Vec{\theta}, \Vec{\theta}+\Delta\Vec{\theta}) \propto p(\Vec{\theta}|d,\mathcal{H})\Delta\Vec{\theta}$.

In MCMC, posterior samples are generated by noting the positions of particles (or \textit{walkers}) undergoing a biased random walk\footnote{A random walk where the evolving variable jumps from its current state to one of the various potential new states, but, unlike a pure random walk, the probabilities of the potential new states are not equal.} through the parameter space where the transition probability of the Markov chain dictates the probability of moving to a new point in the space. Initial samples that are collected during the \textit{burn-in} period\footnote{The preliminary steps, during which the chain moves from its unrepresentative initial value to the modal region of the posterior. The goal of burn-in is to give the Markov chain time to reach its equilibrium distribution.} are discarded, and the sampling process stops once a user-defined condition is met, such as a threshold number of posterior samples have been collected which can provide an accurate representation of the posterior. Thus, in short the MCMC sampler has a goal of collecting a specified number of samples, and the posterior is constructed in the \textit{post-processing} stage. The goal for nested samplers, on the other hand, is slightly different. 

In nested sampling methods, posterior samples are generated as a by-product of calculating the evidence. The sampler starts with a fixed number of \textit{live points} (a set of samples that are used to explore the posterior distribution), which are drawn from the prior distribution. At each iteration, the live point with the lowest likelihood gets discarded, and in turn, another point is chosen from a higher likelihood region of the parameter space. The product between the likelihood of the discarded point and the difference in the prior volume between successive iterations, when summed over all iterations, yields the evidence approximated with the trapezoidal rule [see Eq.~(29) of \cite{Veitch:2014wba}]. The nested samples, when weighted by the posterior probability at that point in the parameter space, get converted to the posterior samples. The sampler stops the process once a pre-defined condition has been met, such as when the fraction of evidence that remains in the prior volume is smaller than 0.1 (say). In other words, the condition dictates that the evidence estimate would not change by more than a factor of 0.1 if all the remaining prior support were at the maximum likelihood. 

In this \textbf{thesis}, we have used nested samplers for the analyses. Specifically, we use \texttt{LALInferenceNest} \citep{Veitch:2014wba} for analysis done with \texttt{LALInference}, and \texttt{dynesty} \citep{2020MNRAS.493.3132S} for parameter estimation with \texttt{PyCBCInference} and \texttt{bilby\_pipe}.

%%%%%%%%%%%%%%%%%%%%%%%%%%%%%%%%%%%%%%%%%%%%

\section{Physical effects in GW signals from binary black hole mergers}

The current template based gravitational wave search pipelines rely on quasi-circular waveform models suitable\footnote{An eccentric compact binary system is expected to circularize by the time it reaches the late stages of inspiral due to emission of GWs \citep{PhysRev.131.435}.} for detecting GWs from compact binary mergers expected to be observed in ground-based GW detectors. While these models describe the complete evolution of binary through the inspiral, merger, and ringdown stages, they assume components with spins aligned with binary's orbital angular momentum (i.e.~non-precessing). Moreover, certain selection biases (we discuss some of these in the subsequent text) allow one to only model for the dominant quadrupole mode (i.e.~these models neglect the presence of other modes in observed data). Further, while the effect of non-quadrupole modes and spin-precession are included when performing parameter estimation studies, the binary is again assumed to be on quasi-circular orbit. As we shall see later in this \textbf{thesis}, this latter assumption leads to biases in parameter estimation studies (Chapter \ref{chap:ecc}). Here, we focus on all the three physical effects (orbital eccentricity, spin-precession, and
non-quadrupole/higher order modes), and describe the importance of these effects in searches and parameter estimation of binary black hole systems.

\subsection{Orbital eccentricity}
\label{sec:intro-ecc}

%Orbital eccentricity is the measure of the ellipticity of the orbit of a binary system. While modelling a gravitational wave signal for compact objects in quasi-elliptical orbits, parameters such as orbital eccentricity ($e$) and mean anomaly ($l$) are often used. \ckm{If you want to be specific, you must elaborate on parametrizations. Read Kaushik's paper} A diagrammatic view of various elements of an eccentric binary system is shown in Fig.~\ref{fig:ecc_diagram}.

While most of the detected signals are consistent with GW emission from inspiralling BBHs on quasi-circular orbits, a few events have been argued to be more consistent with coming from binaries with non-negligible orbital eccentricity at detection \citep[e.g.,][]{Romero-Shaw:2021ual, Romero-Shaw:2022xko, Iglesias:2022xfc}. In particular, GW190521~\citep{Abbott:2020mjq, Abbott:2020tfl} has been interpreted as coming from a moderate- to highly-eccentric BBH \citep{Romero-Shaw:2020thy, Gamba:2021gap, Gayathri:2020coq}. Non-negligible orbital eccentricity measured at detection in the LIGO-Virgo-KAGRA (LVK) sensitive frequency range (above $10$~Hz) implies that the radiating BBH was driven to merge by external influences: for example, as part of a field triple \citep[e.g.,][]{Antonini:2017ash}, in a densely-populated star cluster \citep[e.g.,][]{Rodriguez:2017pec}, or in the accretion disk of a supermassive black hole \citep[e.g.,][]{Tagawa:2020jnc}. 

\begin{figure}[t!]
    \centering
    \includegraphics[width=0.6\linewidth]{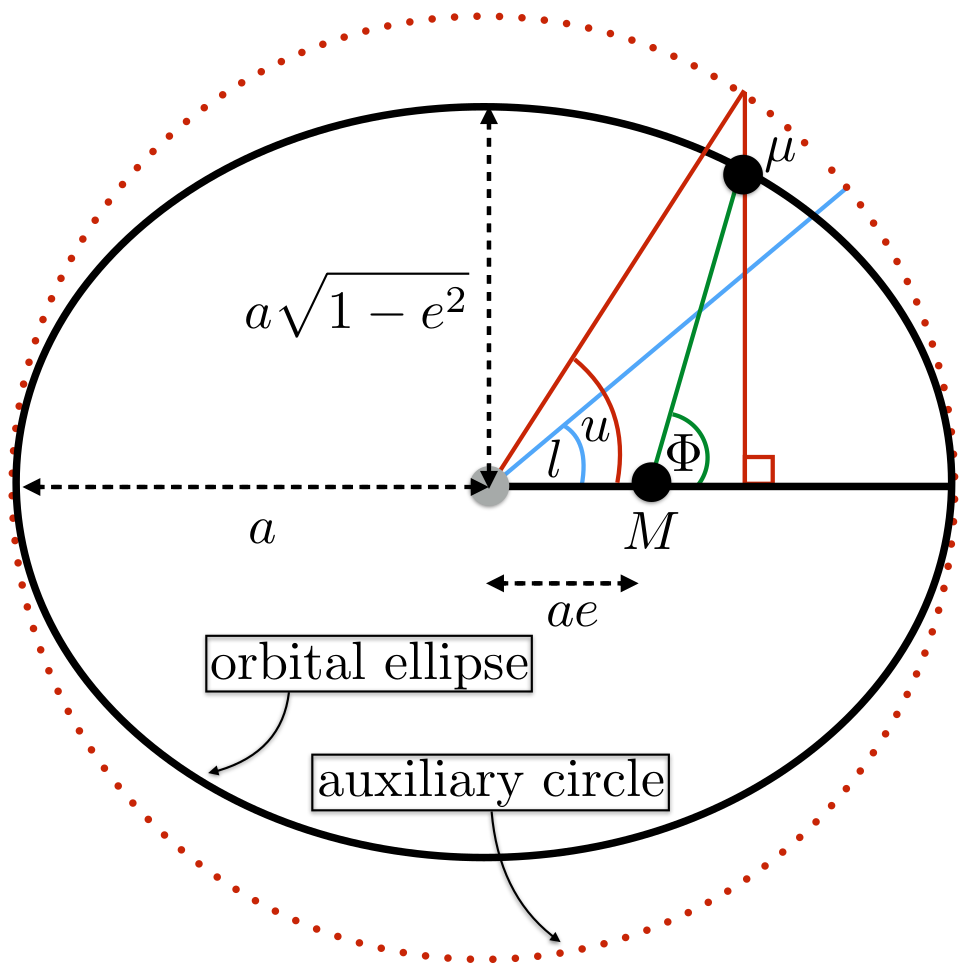}
    \caption[A diagram illustrating the relationship between the various angular elements in a binary system with orbital eccentricity $e$, reduced mass $\mu$, and total mass $M$.]{\takenfrom{\cite{Taylor:2015kpa}} A diagram illustrating the relationship between the various angular elements in a binary system with orbital eccentricity $e$, reduced mass $\mu$, and total mass $M$. The semi-major and semi-minor axes are $a$ and $a \sqrt{1-e^2}$, respectively. If we measure the angles from the moment of periapsis, then $\Phi$ is the true anomaly\footnotemark, $l$ is the mean anomaly, and $u$ is the eccentric anomaly. The auxiliary circle has a radius equal to the orbital semi-major axis.}
    \label{fig:ecc_diagram}
\end{figure}

Search pipelines based on matched-filtering methods use quasi-circular waveform templates motivated by the expected efficient circularisation via GW emission of compact binary orbits during the late stages of their evolution~\citep{PhysRev.136.B1224}.
However, binaries formed through dynamical processes in dense stellar environments \citep{PortegiesZwart:1999nm, 2010MNRAS.407.1946D, Rodriguez:2016kxx, Banerjee:2017mgr, DiCarlo:2019pmf, Mapelli2020, Mandel:2018hfr} or through Kozai-Lidov processes \citep{Kozai:1962zz, LIDOV1962719} in field triples \citep{Martinez:2020lzt, 2016ARA&A..54..441N}, may be observed in ground-based detectors such as advanced LIGO and Virgo with residual eccentricities $\gtrsim0.1$ \citep[e.g.,][]{2011A&A...527A..70K, Lower:2018seu, Antonini:2017ash, Samsing:2017rat, Rodriguez:2017pec, Zevin:2021rtf, Tagawa:2020jnc}. While pipelines employing quasi-circular templates should be able to detect the majority of systems with eccentricities $e_{10}\lesssim 0.1$ at a GW frequency of $10$ Hz \citep{Favata:2021vhw} if observed with current LIGO-Virgo detectors, binaries with larger eccentricities would require constructing template banks for matched-filter searches including the effect of eccentricity \citep[e.g.,][]{Brown:2009ng, Zevin:2021rtf}. \footnotetext{The symbol $\Phi$ has been used for other angles throughout the \textbf{thesis}, and this is the \emph{only} instance where it is used to indicate the true anomaly.}
Moreover, the presence of even small eccentricities ($e_{10}\sim0.01-0.05$) can induce bias in extracting source properties~\citep{Favata:2013rwa, Abbott:2016wiq, Favata:2021vhw, Ramos-Buades:2019uvh, Narayan:2023vhm, Guo:2022ehk, Saini:2022igm, Bhat:2022amc}. As the detectors upgrade to the LIGO A+/Voyager configurations with improved low-frequency sensitivity, neglecting eccentricity in detection and parameter estimation pipelines may lead to failure in identifying the presence of eccentric signals in data, and/or incorrect inference of source properties. This problem is likely to be amplified in detections made with next-generation ground-based instruments such as Cosmic Explorer and Einstein Telescope since their sensitivity to frequencies $\sim 1-5$~Hz and above should enable them to frequently observe systems with detectable eccentricities~\citep{Lower:2018seu, Chen:2020lzc}.

\begin{figure}[t!]
    \centering
    \includegraphics[width=\linewidth]{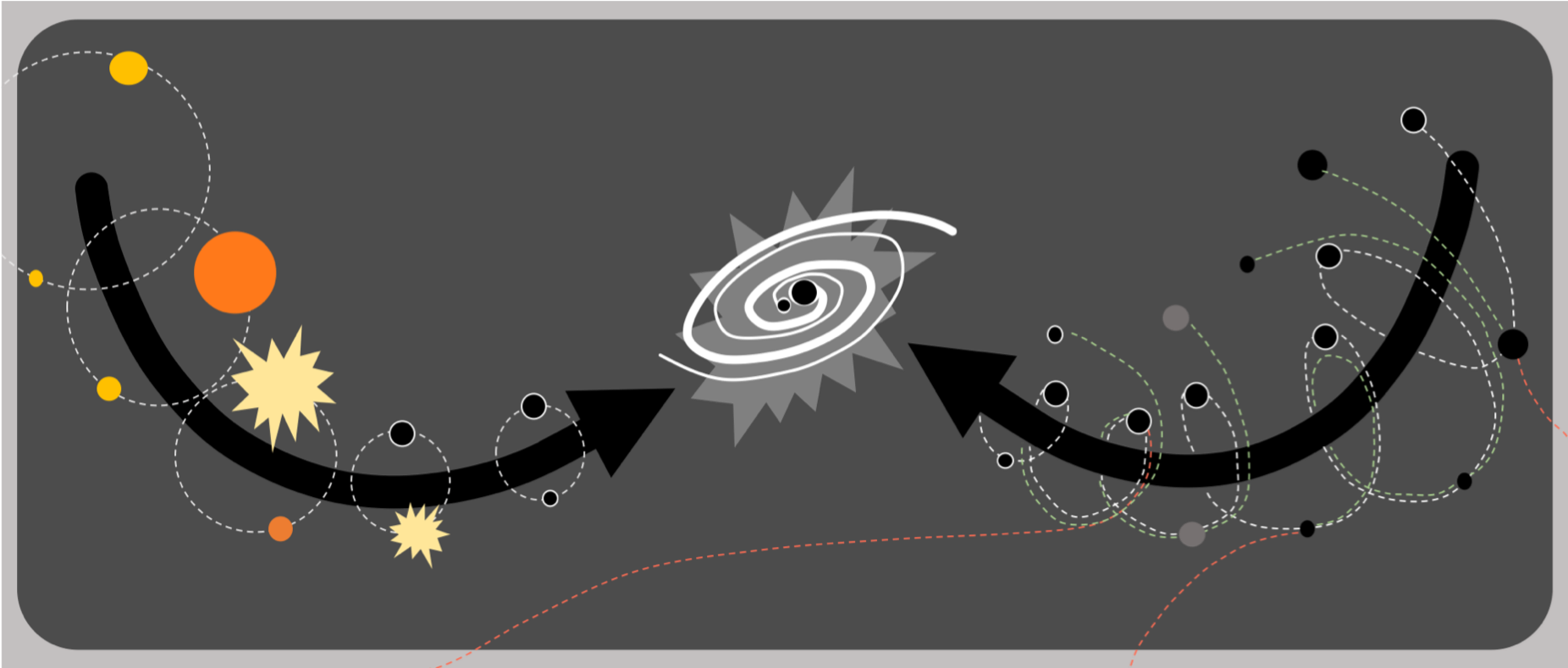}
    \caption[Artist’s depiction of two black holes falling into a dance as they spiral towards each other and eventually collide]{\takenfrom{\cite{ecc_news_article}} Artist’s depiction of two black holes falling into a dance as they spiral towards each other and eventually collide. Image Credit: Isobel Romero-Shaw.}
    \label{fig:ecc-bh-formation}
\end{figure}

In searches for compact binary coalescence signals, computation time and availability of accurate waveform models play crucial roles. In recent years, using the standard search pipelines such as \texttt{cWB} and \texttt{PyCBC}, there have been some targeted searches for eccentric systems \citep{Cheeseboro:2021rey, Pal:2023dyg, Lenon:2021zac, Cokelaer:2009hj, Wang:2021qsu, Ravichandran:2023qma, Ramos-Buades:2020eju, LIGOScientific:2019dag, LIGOScientific:2023lpe, Taylor:2015kpa, Martel:1999tm, Tai:2014bfa}, and upper limits have been set in the absence of detection. For instance, with data from the first two observing runs of LIGO and Virgo detector network, \cite{Nitz:2019spj} provided $90\%$ credible upper limits for binary neutron stars as $\sim 1700$ mergers Gpc$^{-3}$ Yr$^{-1}$ for eccentricities $\leq 0.43$, for dominant mode frequency at $10$~Hz. Using data from the third Gravitational wave transient catalog (GWTC-3), the LVK collaboration placed an upper limit for the merger rate density of high-mass binaries with eccentricities $0 < e \leq 0.3$ at $0.33$ Gpc$^{-3}$ yr$^{-1}$ at 90\% confidence level \citep{LIGOScientific:2023lpe}. For sub-solar mass binaries, \cite{Nitz:2021vqh} provided $90\%$ credible upper limits for $0.5-0.5$ ($1.0-1.0$)~M$_\odot$ binaries to be $7100$ ($1200$) Gpc$^{-3}$ Yr$^{-1}$.

As shown in \cite{PhysRev.136.B1224}, since most of the binary systems detected in the frequency band of current ground-based detectors are expected to be circularized, modelling eccentric waveforms has not always been required for accurate detection and parameter estimation. Further, including additional parameters has often led to difficulties in developing optimal template placement strategies to search for eccentric binaries. Moreover, modelling of eccentric binaries can often be challenging due to additional time scales involved along with frequency-dependent modulations of amplitude and phase due to eccentric nature of the binary orbits. Methods currently in use for eccentric searches \citep{Klimenko:2005xv,Tiwari:2015gal, Coughlin:2015jka, Lower:2018seu, Salemi:2019uea} with little or no dependence on signal model are sensitive only to high masses \citep[ideally $\gtrsim 70M_{\odot}$][]{LIGOScientific:2019dag,LIGOScientific:2023lpe}. Nevertheless, one can infer the presence of orbital eccentricity in signals detected by standard searches tuned to quasi-circular BBH by employing available eccentric waveform models in parameter estimation studies \citep[e.g.,][]{Lower:2018seu, Romero-Shaw:2019itr, Romero-Shaw:2020thy, Romero-Shaw:2020aaj, Lenon:2020oza, Romero-Shaw:2021ual, Romero-Shaw:2022xko, Romero-Shaw:2022fbf, Iglesias:2022xfc, Gamba:2021gap, Bonino:2022hkj}. Alternatively, one can also compare the GW strain data directly to numerical relativity waveform simulations of GW from eccentric BBH through marginal likelihood computation \citep{Gayathri:2020coq} via direct Monte Carlo integration \citep{Lange:2018pyp} over a set of parameters. 

While inspiral-only models for GW signals from eccentric compact binary systems are sufficiently accurate and are rapid enough to generate for use in direct parameter estimation via Bayesian inference \citep{Konigsdorffer:2006zt, Yunes:2009yz, Klein:2010ti, Mishra:2015bqa, Moore:2016qxz, Tanay:2016zog, Klein:2018ybm, Boetzel:2019nfw, Ebersold:2019kdc, Moore:2019xkm, Klein:2021jtd, Khalil:2021txt, Paul:2022xfy, Henry:2023tka}, their use may be limited to low mass (typically $\lesssim25M_{\odot}$)~\citep{LIGOScientific:2011jth} eccentric binaries due to the absence of merger and ringdown in the signal model. 
Waveform models describing BBH evolution through inspiral, merger, and ringdown (IMR) stages are under development and/or available for use \citep[e.g.,][]{Hinder:2017sxy, Huerta:2017kez, Cao:2017ndf, Chiaramello:2020ehz, Liu:2023dgl} but these are generally slower to generate than their quasi-circular counterparts, and Bayesian inference using these models has usually required reduction of accuracy conditions \citep[e.g.,][]{OShea:2021faf}, using likelihood reweighting techniques \citep[e.g.,][]{Romero-Shaw:2019itr}, or utilising highly computationally expensive parallel inference on supercomputer clusters \citep[e.g.,][]{pBilby, Romero-Shaw:2022xko}. Further, eccentric versions of effective-one-body (EOB) waveforms \citep{Liu:2019jpg} including higher modes \citep{Ramos-Buades:2021adz, Nagar:2021gss, Chattaraj:2022tay, Iglesias:2022xfc} and an eccentric numerical relativity (NR) surrogate model \citep{Islam:2021mha} are also available, and are expected to be even slower to generate as additional modes are to be computed together with the dominant mode. A caveat to all of the Bayesian inference studies mentioned above is the absence of spin-induced precession \citep{PhysRevD.49.6274} in the waveform model employed. Since both eccentricity and misaligned spins introduce modulations to the gravitational waveform \citep{OShea:2021faf, Romero-Shaw:2020thy}, it may be critical to account for both spin precession and eccentricity while aiming to measure either or both of the two effects, particularly for GWs from high-mass BBHs \citep{Romero-Shaw:2022fbf, Ramos-Buades:2019uvh}. Even though the currently available eccentric waveform models may not include the effect of spin-induced precession\footnote{While the model in \cite{Klein:2021jtd} does include the effect of spin-precession and eccentricity, it is an inspiral-only prescription and thus might only be suitable for long inspiral signals such as those observed by LISA \citep{Babak:2021mhe}.}, these are still useful for studying systematic errors incurred due to the neglect of orbital eccentricity in waveform models used in detecting and analysing events included in LVK catalogues \citep[e.g.,][]{LIGOScientific:2021djp}.

Chapter \ref{chap:ecc} of this \textbf{thesis} focuses on the effect of orbital eccentricity on the detection and parameter estimation of binary black hole systems.

%----------------------------------------

\subsection{Spin-precession}
\label{sec:intro-spins}

In a compact binary system, in general, one or both the objects may be spinning such that their spin vectors are not aligned with the orbital angular momentum vector ($\Vec{L}$). In such a case where the total spin vector, $\Vec{S} = \Vec{S_1} + \Vec{S_2}$, is misaligned with $\Vec{L}$ (see Fig.\ref{fig:intro-spin-prec}), the orbital plane of the binary precesses about the total angular momentum $\Vec{J}$ \citep{PhysRevD.49.6274}. This spin-induced precession gets imprinted on the gravitational wave signal in the form of modulations in amplitude and phase \citep{Cutler:1992tc, PhysRevD.49.6274}. Ideally, in order to get complete information on the spins of the binary system, one would like to get good constraints on all the six component spins (three for each spin vector), but due to degeneracies, these may not always be well constrained. Thus, when performing parameter estimation, we often use a combination of spin magnitudes and angles to get information about the spins of a binary.

\begin{figure}[t]
    \centering
    \includegraphics[trim=0 30 30 0, clip, width=0.6\linewidth]{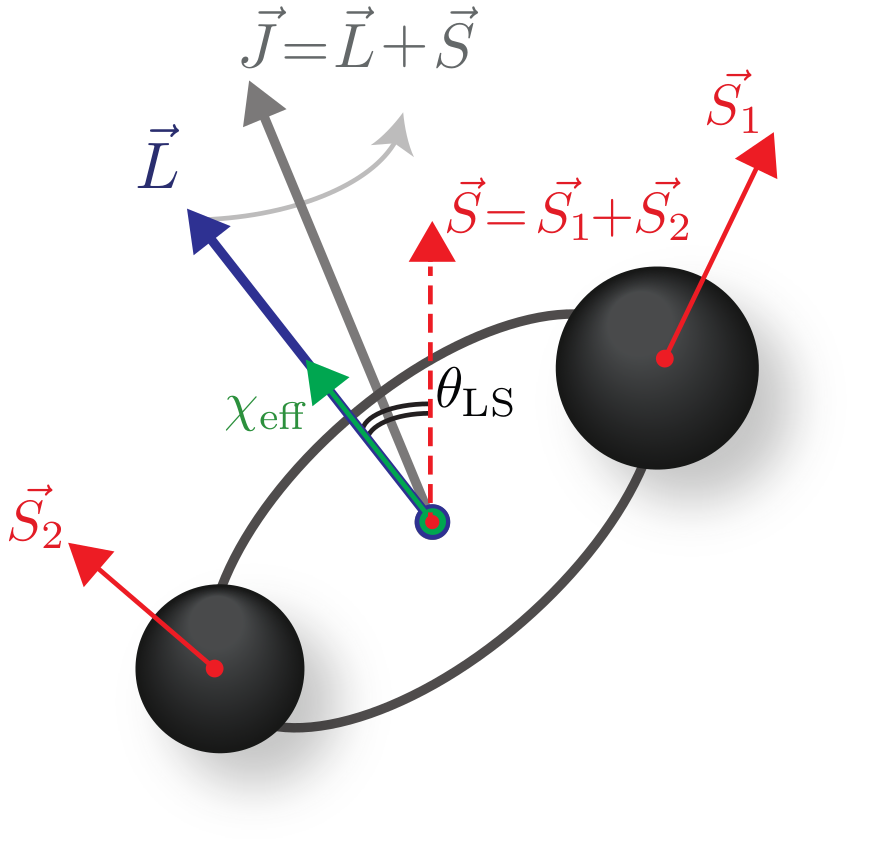}
    \caption[Diagram of the vectors and angles that define the spinning BBH problem]{\takenfrom{\cite{Rodriguez:2016vmx}} Diagram of the vectors and angles that define the spinning BBH problem. For any system where $\Vec{S}$ and $\Vec{L}$ are misaligned, orbital plane will precess about the total angular momentum, $\Vec{J}$.}
    \label{fig:intro-spin-prec}
\end{figure}

The two spin parameters, one for the out-of-plane spin effects - the effective spin parameter ($\chi_\mathrm{eff}$) \citep{Ajith:2009bn, Santamaria:2010yb}, and another for the in-plane spin effects - the spin-precession parameter ($\chi_\mathrm{p}$) \citep{Schmidt:2012rh, Hannam:2013oca, Schmidt:2014iyl} are among the best-measured spin parameters in the parameter estimation of a GW signal. It has been demonstrated that $\chi_\mathrm{eff}$ captures the spin effects along the direction of the angular momentum axis~\citep{Ajith:2009bn} and $\chi_\mathrm{p}$ measures the spin effects in the orbital plane of the binary~\citep{Schmidt:2014iyl}. The effective spin parameter for a binary with dimensionless spin components, $\chi_i = (\vec{S_{i}} \dotproduct \hat{L})/ m_i^2$, can be defined as
\begin{equation}
    \chi_{\rm{eff}}=\frac{\chi_1 m_1 + \chi_2 m_2}{m_1+m_2}.
    \label{eq:chieff}
\end{equation}
Here $\vec{S_i}$ is the individual spin angular momentum vector of the compact object in the binary with mass $m_{i}$, and $\hat{L}$ represents the unit vector along the angular momentum axis of the binary. The value of $\chi_\text{eff}$ varies from -1 (maximally spinning system where spin vectors are anti-aligned with the orbital angular momentum), 0 (non-spinning system, or a combination of aligned and anti-aligned spin vectors), to 1 (maximally spinning system where both spin vectors are aligned with the orbital angular momentum). 

On the other hand, in terms of the perpendicular spin vectors, $S_{i\perp}=|\hat{L}\times (\vec{S_i}\times \hat{L})|$, the effective spin-precession parameter can be written as,
\begin{equation}
    {\chi_\mathrm{p}}=\frac{1}{A_{1}m_{1}^2} \max(A_{1}S_{1\perp}, A_{2} S_{2\perp}),
    \label{eq:chip}
\end{equation}
where, $A_{1}=2+(3/2q)$ and $A_{2}=2+(3q/2)$ are mass parameters defined in terms of the mass ratio $q=m_1/m_2>1$. $\rm{\chi_p}$ varies from 0 to 1, indicating the degree of misalignment with $\Vec{L}$.

Measuring the spins, in particular the orientation of the spin vector, of a binary black hole system can provide vital information about the formation channels \citep{LIGOScientific:2018jsj, Rodriguez:2016vmx, Safarzadeh:2020jsc, Farr:2017gtv, Rodriguez:2019huv, Mapelli2020, Mandel:2018hfr, Belczynski:2001uc}. Binaries formed in isolation are expected to predominantly have aligned spins \citep{1992A&AS...96..653D, 1993MNRAS.260..675T, 2003MNRAS.342.1169V, Kalogera:2006uj, Dominik:2012kk, Dominik:2013tma, Belczynski:2016obo, Gerosa:2018wbw, Bavera:2020inc, Belczynski:2017gds, Olejak:2020oel}, whereas those formed through dynamical evolution \citep{PortegiesZwart:1999nm, Antonini:2017ash, Rodriguez:2016kxx, PortegiesZwart:2004ggg, Samsing:2013kua, Antonini:2012ad, Ziosi:2014sra, Antonini:2017tgo, Fragione:2018vty, Fragione:2019poq, Safarzadeh:2020mlb, Antonini:2016gqe, Park:2017zgj} or hierarchical mergers \citep{Safarzadeh:2020qrc, Hamers:2020huo, Fragione:2020aki, Hamers:2020fiv, Kimpson:2016dgk, Lim:2020cvm} are more likely to show signs of spin-precession. Additionally, it has been shown through numerical simulations of BBHs that the remnant's kick velocity may depend significantly on BH spin orientations \citep{LISA:2017pwj, Campanelli:2007cga, Gonzalez:2007hi, Lousto:2011kp, Merritt:2004xa, Komossa:2008as, sesana:2007zk, Gonzalez:2006md}. In this \textbf{thesis}, we are interested in investigating the effect of spin-precession on tests of binary black hole mimickers such as spin-induced quadrupole moment test (Chapter \ref{chap:siqm}) as well as the interplay of spin-precession and orbital eccentricity (Chapter \ref{chap:ecc}).

\subsection{Non-quadrupole modes}
\label{sec:intro-hm}

Another effect which plays an important role in the modelling of gravitational waveforms is the presence of non-quadrupole modes (also referred to as subdominant modes or higher modes) in signals from compact binary systems which are asymmetric (unequal mass components) and/or whose orbital planes are not optimally inclined towards Earth (face-off binaries). Further, each mode can also have eccentricity-induced subdominant contaminant modes. The multipolar structure of the radiation field guarantees relatively weaker strengths of these modes compared to the dominant mode. In other words, we are more likely to detect the dominant mode from a near-equal mass/face-on binary than non-quadrupole modes from an unequal mass/face-off system. This detection bias makes it difficult to detect higher-order modes (HMs) in observed sources. Including higher modes in the waveforms proves to be very useful when extracting source properties and finds numerous implications in astrophysics ~\citep{LIGOScientific:2020stg, LIGOScientific:2020zkf, Arun:2007hu, Arun:2014ysa, Babak:2008bu, Bustillo:2015qty, Bustillo:2016gid, Chatziioannou:2019dsz, Graff:2015bba, Kalaghatgi:2019log, Kumar:2018hml, OShaughnessy:2014shr, Purrer:2019jcp, Shaik:2019dym, Trias:2007fp, VanDenBroeck:2006ar, Varma:2016dnf, Varma:2014jxa}, cosmology~\citep{Arun:2007hu, Babak:2008bu, Borhanian:2020vyr} and fundamental physics \citep{Purrer:2019jcp, Shaik:2019dym}.

%Multipolar decomposition of the gravitational waveform is a convenient tool for representing gravitational radiation from systems like compact binary mergers~\citep{Thorne:1980ru}. It helps immensely in handling the nonlinearities of GR in the perturbative approaches to GR such as PN theory \citep[see][for a detailed review]{Blanchet:2013haa}. Symmetric trace-free (STF) tensors and spin-weighted spherical harmonics provide two equivalent bases for such a decomposition \citep[see for instance][]{Blanchet:2008je, Kidder:2007rt}. The latter has been more popular recently due to its extensive use by the numerical relativity community, as it provides a natural basis for extracting the waveform from numerical simulations \citep[see for instance][]{Mroue:2013xna}.

The GW strain can be expressed as a linear combination of different modes defined using a basis of spin-weighted spherical harmonics of weight $-2$ as follows \citep{Goldberg:1966uu}
\begin{equation}
    h_+ - i h_\times = h(t,\overrightarrow\lambda,\Theta,\Phi) = \sum_{\ell \geq 2}\sum_{-\ell \leq m \leq \ell} h^{\ell m}(t,\overrightarrow\lambda) Y^{\ell m}_{-2}(\Theta, \Phi).
\label{eq:hlm-sph-time-domain}
\end{equation}

Here, $t$ denotes the time coordinate, the intrinsic parameters like masses and spins are denoted by $\overrightarrow\lambda$, and ($\Theta$, $\Phi$) are the spherical angles in a source-centered coordinate system with total angular momentum along the z-axis. To relate the spin $[-2]$ weighted spherical harmonics ($Y^{\ell m}_{-2}$) with the usual spherical harmonic basis ($Y^{\ell m}$), we reproduce Eqs.~(2.3)-(2.5) of \cite{Blanchet:2008je} here: 
\begin{subequations}
\begin{align}
Y^{\ell m}_{-2} &= \sqrt{\frac{2\ell+1}{4\pi}}\,d^{\,\ell
m}_{\,2}(\Theta)\,e^{i \,m \,\Phi}\,,\\d^{\,\ell m}_{\,2} &=
\sum_{k=k_1}^{k_2}\frac{(-)^k}{k!}
\frac{\sqrt{(\ell+m)!(\ell-m)!(\ell+2)!(\ell-2)!}}
{(k-m+2)!(\ell+m-k)!(\ell-k-2)!}\left(\cos\frac{\Theta}{2}\right)^{2\ell+m-2k-2}
\!\!\!\left(\sin\frac{\Theta}{2}\right)^{2k-m+2}\,
\end{align}
\end{subequations}
where $k_1=\mathrm{max}(0,m-2)$ and $k_2=\mathrm{min}(\ell+m,\ell-2)$. The separate modes $h^{\ell m}$ can be obtained from the surface integral using the orthonormality properties of these harmonics
\begin{equation}
h^{\ell m} = \int d\Omega \,\Bigl[h_+ - i h_\times\Bigr] \,\overline{Y}^{\,\ell m}_{-2} (\Theta,\Phi)\,,
\end{equation}
where the bar or overline denotes the complex conjugate. Several waveforms, both numerical and phenomenological, have been developed which include higher modes \citep{Blackman:2017pcm, Cotesta:2018fcv, Cotesta:2020qhw, Foucart:2020xkt, Garcia-Quiros:2020qpx, Khan:2018fmp, Khan:2019kot, Liu:2021pkr, London:2017bcn, Mehta:2017jpq, Mehta:2019wxm, Nagar:2021gss, Nagar:2020xsk, Nagar:2020pcj, Ossokine:2020kjp, Pratten:2020ceb, Rifat:2019ltp, Varma:2019csw, Varma:2018mmi}. Many of these waveforms are incorporated in the LSC Algorithm Library Suite (\texttt{LALSuite}) \citep{lalsuite}.
These waveforms make use of the analytical and semi-analytical treatment of the compact binary dynamics within the PN~\citep{Kidder:2007rt, Arun:2008kb, Arun:2004ff, Arun:2009pq, Blanchet:1996wx, Blanchet:2006gy, Blanchet:2004ek, Blanchet:1995fg, Blanchet:1995ez, Blanchet:2001ax, Blanchet:2008je, Blanchet:2001aw, Blanchet:1996pi, Bohe:2012mr, Buonanno:2012rv, Kidder:1995zr, Marsat:2012fn, Marsat:2013caa, Mishra:2016whh, Porto:2010zg, Porto:2008jj, Porto:2008tb} and effective-one-body~\citep{Cotesta:2018fcv, Cotesta:2020qhw, Buonanno:1998gg, Buonanno:2000ef, Damour:2000we, Damour:2001tu, Damour:2015isa, Goldberger:2004jt, Sennett:2019bpc} frameworks as well as of numerical relativity (NR) simulations [see \cite{Boyle:2019kee} for a recent update on NR waveform catalog by the SXS collaboration, \cite{Jani:2016wkt} for Georgia Tech catalog, \cite{Ferguson:2023vta} for second MAYA catalog following the Georgia Tech catalog, and \cite{Healy:2017psd} for the RIT catalog]. A comparison between different numerical relativity schemes leading to simulations of BBH spacetimes can be found in \cite{Ajith:2012az, Hinder:2013oqa, Brugmann:2008zz}.

Methods used for detecting the presence of these higher-order modes have been discussed in \cite{Mills:2020thr, Ghonge:2020suv, OBrien:2019hcj, Roy:2019phx}. The effect of HMs on the detection and parameter estimation of binary black hole mergers has been studied extensively \citep[see for instance,][]{Kalaghatgi:2019log, Lange:2018pyp, Payne:2019wmy, Tagoshi:2014xsa, Varma:2016dnf, Varma:2014jxa, Krishnendu:2021fga, Mills:2020thr, Bustillo:2015qty}.
%On the other hand, $h^{\ell m}$ directly be related to the multipole moments $U_L$ and $V_L$ as
%
%\begin{equation}
%h^{\ell m} = -\frac{G}{\sqrt{2}\,R\,c^{\ell+2}}\left[U^{\ell m}-\frac{i}{c}V^{\ell m}\right]\,,
%\end{equation}
%
%where $U^{\ell m}$ and $V^{\ell m}$ are the radiative mass and current moments in standard (non-STF) guise \cite{Kidder:2007gz}. These are related to the STF moments by
%
%\begin{subequations}
%\begin{align}
%U^{\ell m} &= \frac{4}{\ell!}\,\sqrt{\frac{(\ell+1)(\ell+2)}{2\ell(\ell-1)}}\,\alpha_L^{\ell m}\,U_L\,,\\ V^{\ell m} &= -\frac{8}{\ell!}\,\sqrt{\frac{\ell(\ell+2)}{2(\ell+1)(\ell-1)}}\,\alpha_L^{\ell m}\,V_L\,.
%\end{align}
%\end{subequations}
%
%Here $\alpha_L^{\ell m}$ is the STF tensor which connects the usual basis of spherical harmonics $Y^{\ell m}$ to the set of STF tensors $N_{\langle L\rangle}=N_{\langle i_1}\cdots N_{i_\ell\rangle}$ (where the brackets indicate the STF projection). These are related by
%
%\begin{subequations}
%\begin{align}
%N_{\langle L\rangle}(\Theta,\Phi) &= \sum_{m=-\ell}^{\ell} \alpha_L^{\ell m}\,Y^{\ell m}(\Theta,\Phi)\,,\\Y^{\ell m}(\Theta,\Phi) &= \frac{(2\ell+1)!!}{4\pi \ell!}\,\overline{\alpha}_L^{\ell m}\,N_{\langle L\rangle}(\Theta,\Phi)\,,
%\end{align}
%\end{subequations}
%
%with the STF tensorial coefficient being
%
%\begin{equation}
%\alpha_L^{\ell m} = \int d\Omega\,N_{\langle L\rangle}\,\overline{Y}^{\,\ell m}\,.
%\end{equation}
%

\subsubsection{Implications of higher-order modes}
\label{subsec:hm-implications}

One of the most important consequences of including higher order modes into the gravitational waveforms can be linked to their sensitivity to frequencies that are inaccessible through the dominant (quadrupole) mode. Typically, including higher order modes into the waveforms will extend the GW spectrum to higher frequencies. For instance, inspiral for the dominant (quadrupole) mode ($\ell$=2, $|m|$=2 or simply the 22 mode) can be assumed to terminate at twice the orbital frequency at the last stable orbit ($f_{\rm LSO}$), while the same for a higher mode waveform including the $k$th harmonic will be visible until the GW frequency becomes $k\,f_{\rm LSO}$. A direct consequence of this is the increase in the mass reach of broadband detectors \citep{Arun:2007qv, VanDenBroeck:2006qu}. 

The higher order modes, through amplitude corrections to gravitational waveforms, also bring in new dependencies in terms of mass ratio, component spins, and inclination angle into the gravitational waveforms; see for instance \cite{VanDenBroeck:2006ar} (nonspinning case), and \cite{Arun:2008kb} (for spinning case). By including them into the waveforms, one can break the degeneracies present in the waveform, such as those between inclination angle and luminosity distance~\citep{Usman:2018imj}, and that between mass ratio and spins \citep{Hannam:2013uu, Ohme:2013nsa}. While better measurements of the luminosity distance allow putting tighter bounds on cosmological parameters such as the Hubble constant \citep{Borhanian:2020vyr, Sathyaprakash:2009xt}, improved inclination angle estimates can lead to better modeling of off-axis gamma ray bursts \citep{Arun:2014ysa}. 

Further, the use of higher modes has been shown to improve the efficiency of parametrized tests of GR~\citep{Mishra:2010tp, Yunes:2009ke} and massive graviton tests~\citep{Arun:2009pq}. A new test of GR based on the consistency of different modes of the gravitational waveform was proposed~\citep{Dhanpal:2018ufk,Islam:2019dmk, Shaik:2019dym} and performed on a few selected events from GWTC-2 \citep{Capano:2020dix}. A multipolar null test of GR was also proposed in \cite{Kastha:2018bcr,Kastha:2019brk} which would measure the contribution to the gravitational waveforms from various multipoles and test their consistency with the predictions of GR.
 Recently, it was shown that the detection of higher modes can improve the early warning time and localization of compact binary mergers, especially NS-BH systems\citep{Kapadia:2020kss, Singh:2022tlh, Singh:2020lwx}. 

In this \textbf{thesis}, we show results for higher mode eccentric injections in Chapter \ref{chap:ecc} and study the effect of higher-order modes on the measurement of spin-induced quadrupole moments, along with spin-precession in Chapter \ref{chap:siqm}. In Chapter \ref{chap-hm}, we focus on the detectability of higher modes in the third generation detectors, commenting on the parameter space in which they are most relevant and showing that several of GWTC-2 events would've shown detection of higher modes if 3G detectors were operational at the time.
 
%----------------------------------------

\subsection{Impact of spin-precession and higher modes on SIQM test}
\label{subsec:intro-siqm}

Various methods exist for investigating a true nature of the compact object in a binary system \citep{Laarakkers:1997hb, PhysRevD.55.6081, Poisson:1997ha, Pacilio:2020jza, Gurlebeck:2015xpa, LeTiec:2020spy, Mendes:2016vdr, Uchikata:2016qku, Johnson-Mcdaniel:2018cdu, JimenezForteza:2018rwr, Abdelsalhin:2018reg, Datta:2019euh, Hartle:1973zz, Chatziioannou:2016kem, Datta:2020gem}. An analysis based on the spin-induced multipole moments is one among them. Spin-induced multipole moments arise due to the spins of individual compact objects in the binary \citep{Laarakkers:1997hb}. Typically, the evolution of an inspiralling compact binary can roughly be divided into three stages: an early-inspiral, late-inspiral \& merger, and the final ringdown. During the early inspiral stage, the separation between the compact objects in the binary is large, and hence, their evolution can be modelled as a perturbation series in the velocity parameter. The post-Newtonian (PN) theory provides an analytic expression for the inspiral phase incorporating various physical effects such as the spin-orbit effects, self-spin effects, cubic and higher order spin-effects, spin-precession effects, orbital eccentricity effects, etc \citep[see for instance,][]{Blanchet:2002av, Mishra:2016whh, Henry:2022dzx}. On the other hand, one needs to invoke numerical relativity techniques to model the highly non-linear relativistic merger phase \citep[see][for a review on numerical relativity modelling techniques]{Lehner:2001wq}. Further, the ringdown part can again be modelled perturbatively using the BH perturbation theory techniques~\citep{Sasaki:2003xr, Pretorius:2007nq}, although all numerical relativity simulations evolve the BBH spacetimes through this final ringdown stage.

In the inspiral phase, the effect of spin-induced quadrupole moment (SIQM) starts to appear at 2PN, together with the other spin-spin terms. More precisely, the leading order PN coefficient is of the schematic form $Q=-\kappa \chi^2 m^3$ \citep{Poisson:1995ef}, where the negative sign indicates the \emph{oblate} deformation due to the spinning motion. The proportionality constant, $\kappa$, can take different values for different compact objects. For isolated black holes, $\kappa_\text{BH}$ is 1.\footnote{For BBHs, $\kappa_\text{BH}$ can deviate from 1, but the effect is negligible compared to the measurements being made here \citep{Buonanno:2012rv}.} Slowly spinning neutron stars can have $\kappa$ values in the range $\rm{\kappa_{NS}}\sim 2-14$~\citep{Laarakkers:1997hb, Pappas:2012ns, Pappas:2012qg}. In contrast, for exotic stars like boson stars, this range can be $\rm{\kappa_{BS}\sim}10-100$~\citep{PhysRevD.55.6081} depending on internal composition. There also exist gravastar proposals where the value $\rm{\kappa_{GS}}$ can match the BH value but also allows for negative values and prolate deformations~\citep{Uchikata:2015yma}. Measuring the SIQM parameter ($\kappa$) from GW observations can thus provide unique information about the nature of the compact object. 

In principle, the BH nature of the binary components can be probed by measuring their individual SIQM coefficients $\kappa_1$ and $\kappa_2$, parametrized as deviations away from unity $\delta\kappa_1$ and $\delta\kappa_2$. However, for the stellar-mass compact binaries accessible to LIGO and Virgo, it is often difficult to simultaneously constrain $\delta\kappa_1$ and $\delta\kappa_2$ due to the strong degeneracies between the two parameters, and with other binary parameters like the spins and masses \citep{Krishnendu:2017shb, Krishnendu:2018nqa}. Hence, a symmetric combination of these deviation parameters may be used for probing the binary black hole nature of GW source, choosing the anti-symmetric combination to be zero.

The method for measurement of deviation on $\kappa$ has been explored in detail~\citep{Krishnendu:2017shb, Krishnendu:2018nqa, Krishnendu:2019tjp, Krishnendu:2019ebd, Saleem:2021vph} and applied to observed GW events from the first three observing runs of advanced LIGO-Virgo detectors \citep{LIGOScientific:2020tif, LIGOScientific:2021sio}. Further, it has been demonstrated in the past that it
is possible to measure spin-induced multipole moments for intermediate mass-ratio \citep{Brown:2006pj, Rodriguez:2011aa} and extreme mass-ratio inspirals \citep{Barack:2006pq, Babak:2017tow}. This parameter can also be constrained through electromagnetic observations of active galactic nuclei \citep[see for instance][]{Laine:2020dnr} and supermassive BHs \citep{EventHorizonTelescope:2019dse}. Moreover, the possibility of measuring spin-induced quadrupole using future detectors, and simultaneous measurement of spin-induced quadrupole and octupole moment parameters have also been studied \citep{Krishnendu:2018nqa, Saini:2023gaw}. A template bank for binaries of exotic compact object searches has recently been developed, accounting for the spin-induced quadrupole moment and tidal effects~\citep{Chia:2022rwc}.  Moreover, \cite{LaHaye:2022yxa} came up with a fully precessing waveform implementation of SIQM test for low-mass binaries, focusing on binaries in which at least one object is in the lower mass-gap ($< 3 M_\odot$) \citep{Lyu:2023zxv}. While ~\cite{Krishnendu:2017shb, LIGOScientific:2020tif, LIGOScientific:2021sio} report the SIQM measurements using a phenomenological waveform model, \texttt{IMRPhenomPv2} \citep{Husa:2015iqa, Khan:2015jqa, Hannam:2013oca}, containing only the dominant modes ($\ell=2, |m|=2$) in the co-precessing frame and an effective spin parameter, in this \textbf{thesis} we extend the test for waveforms in the \texttt{IMRPhenomX} family, thus accounting for double spin-precession and higher modes. Specifically, we use three waveforms from this family: \texttt{IMRPhenomXHM} which is an aligned spin model with higher modes, \texttt{IMRPhenomXP} which contains only the dominant modes along with double spin-precession, and \texttt{IMRPhenomXPHM} which contains the higher modes as well as double spin-precession. We apply the extended test on spin-precessing binary black hole signals and present the results in Chapter \ref{chap:siqm}. We find that waveform with double spin-precession gives better constraints for $\delta\kappa$, compared to waveform with single spin-precession. We also revisit earlier constraints on the SIQM-deviation parameter for selected GW events observed through the first three observing runs (O1-O3) of LIGO-Virgo detectors. The effects of higher-order modes on the test are also explored for various mass-ratio and spin combinations by injecting simulated signals in zero-noise. Our analyses indicate that binaries with mass ratios greater than three and significant spin-precession may require waveforms that account for spin-precession and higher modes to perform parameter estimation reliably.

%---------------------------------------

\begin{figure}[p!]
    \centering
    \includegraphics[width=\linewidth]{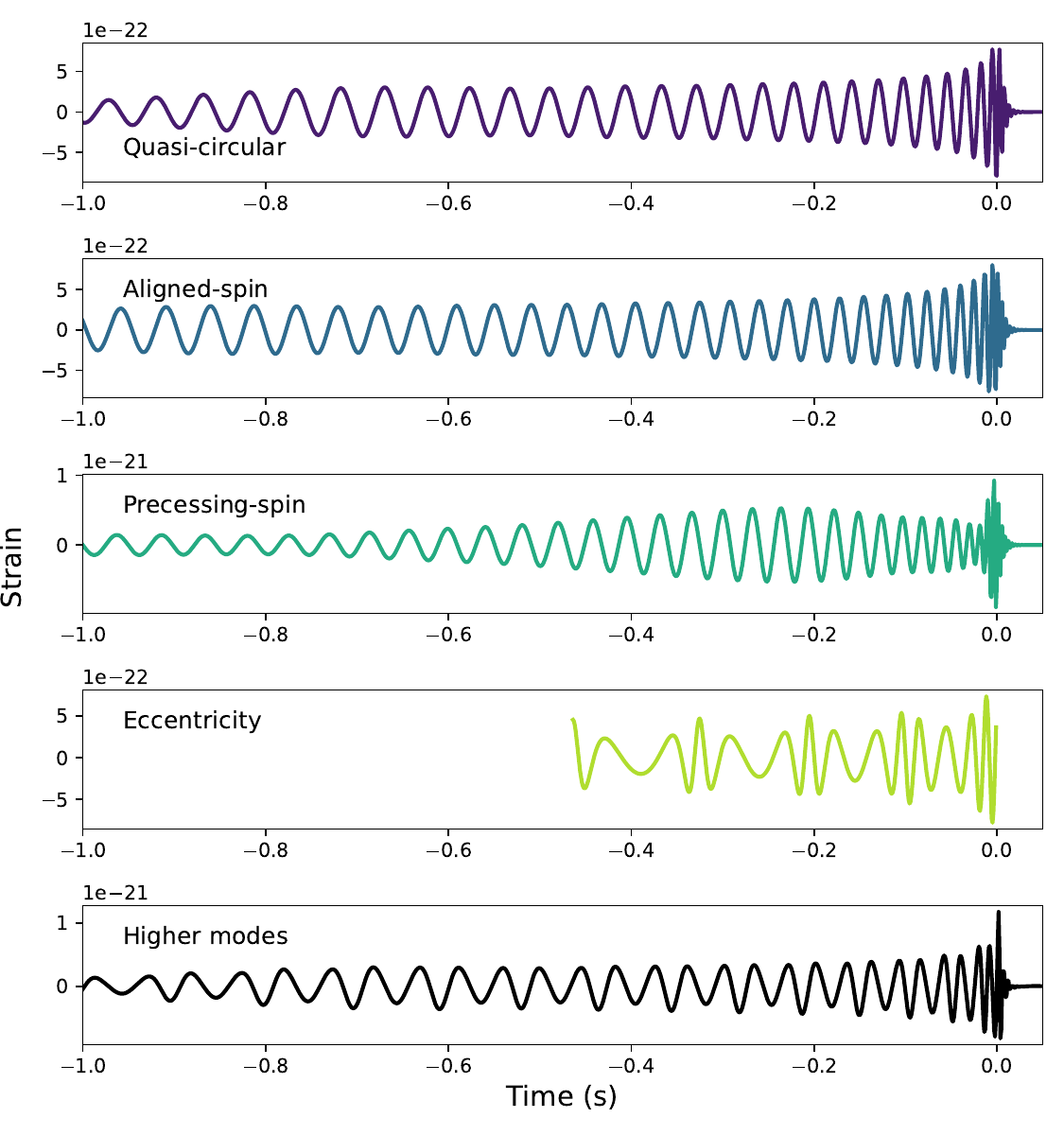}
    \caption[Diagrams showing the effect of including higher modes, spins, and orbital eccentricity in a binary black hole gravitational waveform]{Diagrams showing the effect of including (one at a time) higher modes, spins, and orbital eccentricity in a binary black hole gravitational waveform. The waveform models used for generating the data are \texttt{IMRPhenomXPHM} \citep{Garcia-Quiros:2020qpx, Pratten:2020ceb} for quasi-circular (all except 4\textsuperscript{th} row), and \texttt{InspiralENIGMA} \citep{Huerta:2017kez, Paul:2024abc} for eccentric waveforms.}
    \label{fig:intro-wf_effects}
\end{figure}

\subsection{Interplay of orbital eccentricity, spin-precession, and higher modes}
\label{subsec:intro-interplay-of-effects}

As discussed in the sections above, orbital eccentricity, spin-precession, and higher-order modes play a crucial role in understanding some of the fundamental questions related to the inference of gravitational wave data and formation channels of gravitational wave sources. Each effect presents as oscillations in the gravitational wave signal, as shown in Fig.~\ref{fig:intro-wf_effects} where the \textit{plus} polarization strain in the time domain has been plotted. We have used the model \texttt{IMRPhenomXPHM} \citep{Garcia-Quiros:2020qpx, Pratten:2020ceb} to plot the quasi-circular waveforms (all but the 4\textsuperscript{th} row), and \texttt{InspiralENIGMA} \citep{Huerta:2017kez, Paul:2024abc} for the eccentric waveform (4\textsuperscript{th} row). The plots are generated, with a starting frequency of 20 Hz, for a black hole binary system with $(m_1, m_2) = (60, 15) M_\odot$, inclined at an angle of $\iota=60^\circ$, placed at a distance of $d_L = 400$ Mpc. In the top row, no additional effect has been added and it portrays a non-spinning, quasi-circular BBH system with $(\ell, |m|) = (2,2)$ modes only. The next row labelled "Aligned-spin" introduces non-zero values for spin vector components aligned with the orbital angular momentum as ($S_\text{1z}=0.45$, $S_\text{2z}=0.2$) so that the resultant $\chi_\text{eff}=0.4$. Compared to the non-spinning case, this signal is longer in length. For the "precessing-spin" case shown in the 3\textsuperscript{rd} row, all the six spin components have non-zero values ($S_\text{1x}=0.65$, $S_\text{1y}=0.26$, $S_\text{1z}=0.45$, $S_\text{2x}=0.6$, $S_\text{2y}=0.3$, $S_\text{2z}=0.2$). This results in the spin-precession parameter to be $\rm{\chi_p}=0.7$, while keeping $\chi_\text{eff}=0.4$. Here, we see modulations in the waveform envelope due to the spin-induced precession of the binary's orbit. The next row shows an eccentric waveform signal with an eccentricity of 0.3 at 20 Hz. And the last row corresponds to a non-spinning, quasi-circular waveform which contains the $(\ell, |m|) = (2,2), (3,3), (4,4), (2,1), (3,2)$ modes. It can be seen from the last three panels that higher-order modes, orbital eccentricity, and spin-precession all induce oscillations in the GW signal and it is important to understand the interplay of these effects in order to characterize the sources correctly. The interplay of these effects has been studied in various combinations \citep{Puecher:2022sfm, Krishnendu:2021cyi, Fairhurst:2019srr, Fairhurst:2019vut, Romero-Shaw:2022fbf, Xu:2022zza, Hegde:2023hfv, Iglesias:2022xfc, Moore:2019vjj}. In this \textbf{thesis}, we focus extensively on how the detection and parameter estimation of binary black hole systems is affected by one or more of these effects. In Chapter \ref{chap:ecc}, we focus on orbital eccentricity, specifically on the detectability and parameter estimation of eccentric systems and how the analysis of eccentric signals with precessing spin waveforms affects the inference of source parameters \citep{Divyajyoti:2023rht}. We also analyse eccentric binary black hole signals containing higher modes \citep{Chattaraj:2022tay}. Next, in Chapter \ref{chap:siqm}, we observe the effect of spin-precession and higher modes on the spin-induced quadrupole moment test (see Sec.~\ref{subsec:intro-siqm}) and comment on the regions of parameter space where these effects are dominant \citep{Divyajyoti:2023izl}. Finally, in Chapter ~\ref{chap-hm}, we explore the detectability of higher modes in the future ground-based GW detectors \citep{Divyajyoti:2021uty}.
      \chapter{Detection and parameter estimation of eccentric binary black hole mergers}
\label{chap:ecc}

\section{Introduction}
\label{sec:intro}

As discussed earlier, orbital eccentricity is one of the most important effects in understanding the formation channels of a compact binary system. Since there hasn't been a confident detection of a gravitational wave signal from an eccentric binary system, it is natural to ask whether our search pipelines are competent to search for these signals. Moreover, we need to have a comprehensive understanding of the systematics arising due to waveform models when analysing eccentric signals so as to avoid misinterpretation of the results. This forms the theme of this chapter.

We assess the impact of employing a circular template bank for GW searches when the source population may exhibit eccentricity. To achieve this, we simulate diverse source populations, by varying binary black hole masses and eccentricities. In Sec.~\ref{sec:gw_searches}, we determine the detection efficiency of the population by comparing inherent signal strength to the values obtained from the circular template bank. We also investigate the regions in the parameter space where the largest loss in signal strength occurs due to the difference between injection and recovery waveform model or due to non-inclusion of eccentricity.
Next, in order to explore the effect of eccentricity on various inferred parameters of a BBH system, in Sec.~\ref{sec:pe}, we perform parameter estimation (PE) of injected non-spinning and aligned spin eccentric signals generated using numerical relativity (NR) codes, and recover them using multiple waveform models with different combinations of spins and eccentricity. Our aim is to observe and quantify the biases incurred due to absence of eccentricity in the recovery waveform when analyzing eccentric signals and to verify that those biases can be corrected to a certain extent by using the currently available eccentric waveforms. 
To this effect, we perform multiple sets of injections which include synthetic GW signals consistent with non-spinning (Secs.~\ref{subsec:non-spin-inj} and \ref{sec:noisy_injs}) and aligned-spin (Sec.~\ref{subsec:align-spin-inj}) BBHs in eccentric orbits, and recover these injections with a variety of either quasi-circular or eccentric waveform models, including no spin, aligned spins, or misaligned (precessing) spins. We also perform non-spinning higher mode eccentric injections and recover them with quasi-circular, higher mode, IMR waveforms (Sec.~\ref{sec:hm_injs}). Due to a lack of available waveform models, we are unable to perform injections using models with both spin-precession and eccentricity. We observe that the state-of-the-art inspiral-merger-ringdown (IMR) quasi-circular waveforms of the \texttt{IMRPhenomX} \citep{Pratten:2020fqn, Pratten:2020ceb} family do not recover the true chirp mass of an eccentric signal. 
%Regardless of the spin configuration (non-spinning, spin-aligned, or spin-precessing) taken for these quasi-circular recovery waveforms, the bias of the chirp mass and spin posteriors when the injected signal is non-spinning and eccentric is similar relative to the non-biased recovery of the quasi-circular injection. 
%Therefore, we conclude that for low-mass (total mass of $35$~M$_\odot$), long-inspiral, non-spinning, eccentric binary systems, parameter estimation with a precessing-spin waveform model will not lead to a false positive value of precession. This supports the conclusions drawn in \citet{Romero-Shaw:2022fbf}, where the authors demonstrate that eccentricity and spin-precession are distinguishable for signals with more than a few cycles in-band. 
Further, analyzing these same signals using a computationally efficient inspiral-only eccentric waveform results in significant reduction of biases in the posteriors on intrinsic parameters of the binary, leading to the recovery of true values within the $90\%$ credible bounds. 
%This implies that for GWs from low-mass and aligned- or low-spin BBHs, inspiral-only waveforms that are readily available within the \texttt{lalsuite} framework are adequate for accurate recovery of the source parameters. We also see clear correlation between chirp mass and eccentricity posteriors for both the non-spinning and aligned spin eccentric injections, in agreement with the previous studies \citep{Favata:2021vhw, Ramos-Buades:2019uvh, Ramos-Buades:2023yhy}, in addition to a mild correlation of eccentricity with the effective inspiral spin parameter $\chi_\text{eff}$ when the injection is aligned-spin and eccentric, consistent with the correlations seen in \citet{OShea:2021ugg}.

While an inspiral-only eccentric waveform model should be better for analyzing eccentric signals than a full IMR quasi-circular model, it is likely that biases can be further reduced and posteriors be better constrained if an IMR eccentric waveform model is used for recovery. However, our results demonstrate that even eccentric waveform models with limited physics (no merger or ringdown, no spin-precession, no higher modes) can reduce errors in the inference of BBH parameters, and may, therefore be a useful stepping stone towards analysis with full IMR models including eccentricity, spin-precession, and higher modes. We comment on the implications of these biases and conclude in Sec.~\ref{sec:ecc-concl}.

\section{Reduced GW search sensitivity to eccentric binaries}
\label{sec:gw_searches}

This section explores the fraction of events that might be missed due to the neglect of eccentricity in template banks used for matched-filter based searches such as \texttt{PyCBC} \citep{Usman:2018imj} or \texttt{GSTLAL} \citep{Messick:2016aqy}. GW searches may miss a fraction of signals because of the following reasons:

\begin{enumerate}
    \item The templates may not be accurate representations of the real signal, especially if these signals include additional physics (e.g., eccentricity, misaligned spins, higher modes) not included in the template bank.
    \item Template banks are discrete in nature.
\end{enumerate}

To quantify the loss of sensitivity of the search, one can define the \textit{match} between the template $h(\theta_i, \Phi)$ and a signal $g$ in terms of the overlap between them as:
\begin{equation}
    m(g,h(\theta_i)) = \text{max}(\langle g|h(\theta_i, \Phi)\rangle), 
    \label{eqn:match}
\end{equation}
where $h(\theta_i, \Phi)$ is a template with intrinsic parameters $\theta_i$ and extrinsic parameters $\Phi$, and the RHS in Eq. \eqref{eqn:match} is maximised over all the extrinsic parameters. The fitting factor $FF(g)$, for a signal $g$ is defined as \citep{Dhurkunde:2022aek}:
\begin{equation}
    FF(g) = \text{max}(m(g, h(\theta_i))), \label{eqn:fitting_factor}
\end{equation}
where RHS is maximized over all the templates $h(\theta_i)$. In order to get fraction of recovered signals relative to an optimal search ($FF=1$), we use the metric described in \cite{Buonanno:2002fy}, which takes into account the intrinsic SNR of the signal to calculate the signal recovery fraction ($SRF$), defined as \citep{Dhurkunde:2022aek}:
\begin{equation}
    SRF \equiv \frac{\sum_{i=0}^{n_s - 1} FF^3_\text{TB}(s_i) \sigma^3(s_i)}{\sum_{i=0}^{n_s-1}\sigma^3(s_i)} ,\label{eqn:signal_fraction_recovery}
\end{equation}
\noindent
where $FF_\text{TB}$ is the fitting factor for a volumetric distribution of $n_s$ sources using a template bank, and $\sigma(s_i)$ is the intrinsic loudness of each signal $s_i$. We calculate the $SRF$ for a given distribution of sources and a template bank.

\subsection{Reduced detectability of eccentric systems}
\label{subsec:detection}

We use a reference population distributed uniformly in source frame masses $m^{\rm source}_{1,2} \in [5, 50]$ and redshift up to 3. For eccentricity parameter ($e$), log-uniform distribution is used while the sources are distributed uniformly in sky. In this section, we consider non-spinning population to quantify the effects of eccentricity.
The population is generated with three different waveform models: 

\begin{enumerate}
    \item \textbf{\texttt{IMRPhenomD}} \citep{Husa:2015iqa, Khan:2015jqa}: This is used to create a population of non-spinning, quasi-circular signals.\footnote{While \texttt{IMRPhenomD} is an aligned-spin quasi-circular waveform model, in our study we have restricted to a non-spinning population of binary black holes.} Since we use the same waveform for recovery, this serves as the optimal search and we expect the maximal recovery of the injected signals.

    \item \textbf{\texttt{TaylorF2Ecc}} \citep{Moore:2016qxz, Kim:2019abc}: This is used to create a population of non-spinning, eccentric signals where the eccentricity distribution is taken as ${e_{10}} \in [10^{-7},0.3]$ for eccentricity defined at $10$ Hz ($e_{10}$). Note: \texttt{TaylorF2Ecc} does not include eccentricity corrections in amplitude of the signal.

    \item \textbf{\texttt{EccentricFD}} \citep{Huerta:2014eca}: We use the same eccentricity distribution as \texttt{TaylorF2Ecc} but the signals are generated with \texttt{EccentricFD} which includes eccentricity corrections in both the phase and amplitude of the GW signal.
\end{enumerate}

% \begin{itemize}
%     \item \textbf{Zero eccentricity distribution:} We use the \texttt{IMRPhenomD} \citep{Husa:2015iqa, Khan:2015jqa} waveform model to inject non-spinning, quasi-circular signals and use a template bank with the same waveform for recovery. This serves as the optimal search and we expect the maximal recovery of the injected signals.
%     \item \textbf{Non-zero eccentricity distribution:} For these sets of injections, we use two waveform models: we generate one set with \texttt{TaylorF2Ecc} \citep{Moore:2016qxz, Kim:2019abc}, and the other with \texttt{EccentricFD} \citep{Huerta:2014eca}. We use the same component masses as above, and a log-uniform eccentricity distribution in the range $\log_{10}{e_{10}} \in [-7,-0.3]$ defined at $10$ Hz. We choose this upper limit so that we stay within the regions of validity of the waveform models.
% \end{itemize}

For recovery, we use the `quasi-circular' template bank (non-spinning) constructed with \texttt{IMRPhenomD}. We use stochastic placing algorithms \citep{Harry:2009ea, Babak:2008rb} implemented in \texttt{PyCBC} to generate template bank for component masses in the range $m_{1,2} \in [3, 200]~M_{\odot}$, using the minimal match criteria of 0.98. We use this template bank to quantify the fraction of lost signals if the intrinsic population has some eccentricity distribution. We calculate the optimal SNR for each injection using the template bank and then estimate the $FF$ for the set of injections. We use the low-frequency cutoff of $10$~Hz and detector sensitivities for i) advanced LIGO \citep{PyCBC-PSD:aLIGO}, and ii) $A+$ \citep{PyCBC-PSD:Aplus}. 

\begin{figure}[p!]
    \centering
    \includegraphics[trim=0 0 40 20, clip, width=0.56\linewidth]{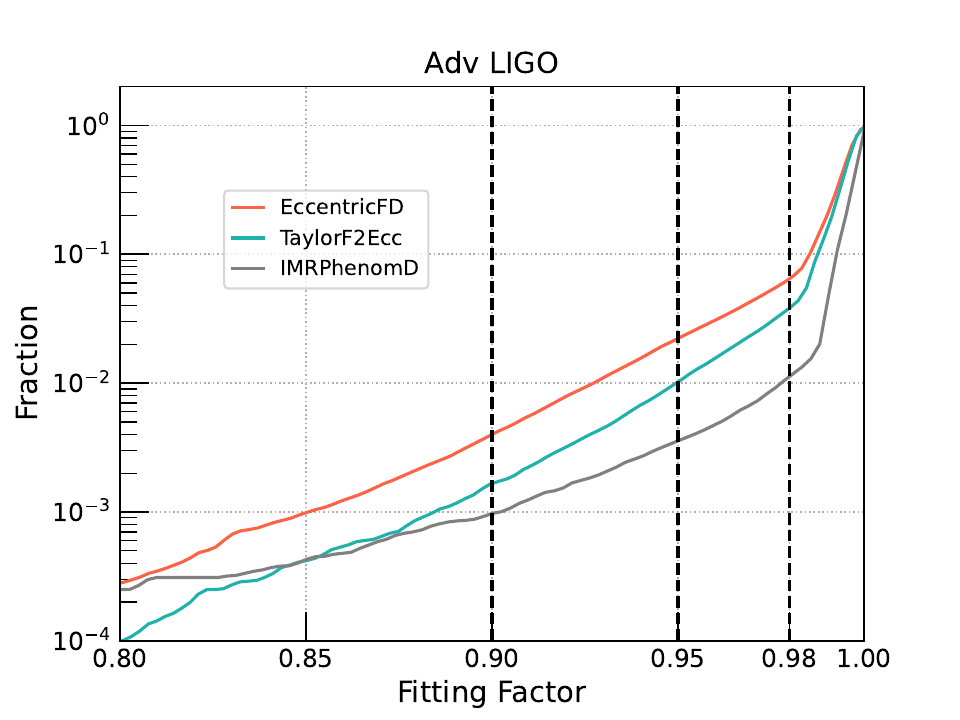}
    \includegraphics[trim=0 0 40 20, clip, width=0.56\linewidth]{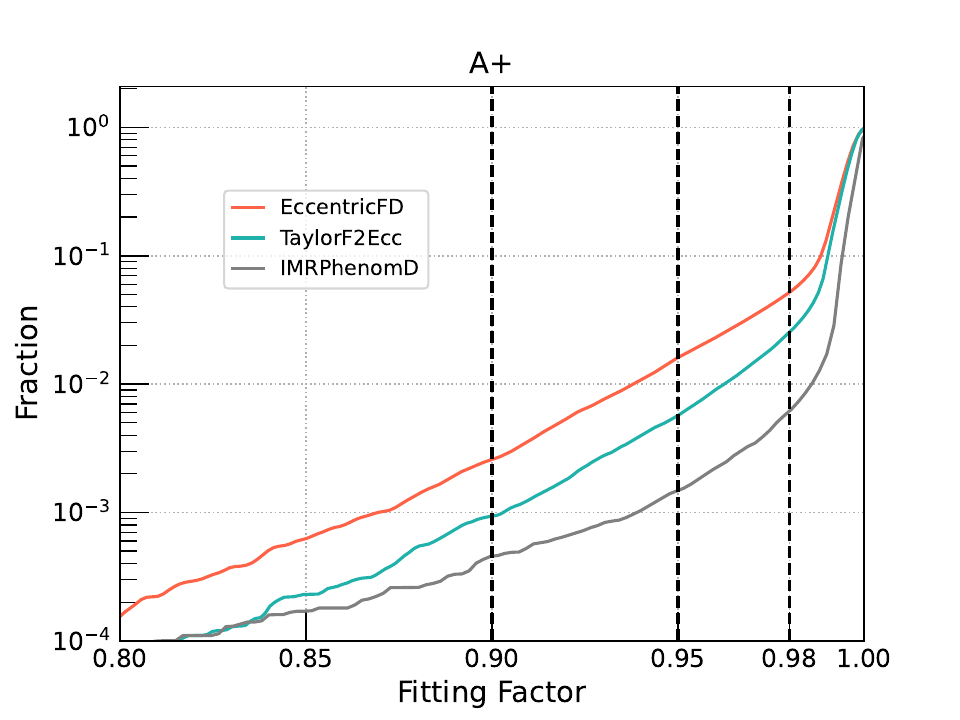}

    \caption[Cumulative fraction of events above a given Fitting Factor ($FF$) for various populations]{Cumulative fraction of events above a given Fitting Factor ($FF$) for various populations, distributed uniformly in masses and log-uniformly in eccentricity (measured at $10$~Hz) with the match calculated against the standard template bank, shown here for detector sensitivities: Adv-LIGO and $A+$. The grey curves show the fraction recovered for the reference population with no eccentricity, while the green and red curves show the fraction recovered for the eccentric population represented by \texttt{TaylorF2Ecc} and \texttt{EccentricFD} models respectively. Three vertical dashed black lines show the fitting factor values of $0.9$, $0.95$, and $0.98$ increasing in value from the left. These plots show that if we use the quasi-circular template bank to search for a population which contains a log-uniform distribution of eccentricities, we fail to detect a higher fraction of signals in searches. E.g. in the top panel (Adv LIGO sensitivity): eccentric population constructed with \texttt{TaylorF2Ecc} (\texttt{EccentricFD}) waveform has $\approx$ 1 percent ($\approx$ 2.2 percent) events with fitting factor less than 0.95. $FF$ for baseline model \texttt{IMRPhenomD} is $\approx$ 0.4 percent. In the bottom panel ($A+$): eccentric population constructed with \texttt{TaylorF2Ecc} (\texttt{EccentricFD}) waveform has $\approx$ 0.6 percent ($\approx$ 1.6 percent) events with $FF$ less than 0.95. $FF$ for baseline model \texttt{IMRPhenomD} is $\approx$ 0.1 percent.}
    \label{fig:cumulative_histogram_ff}
\end{figure}

\begin{figure}[t!]
    \centering
    \includegraphics[trim=20 0 40 20, clip, width=0.49\linewidth]{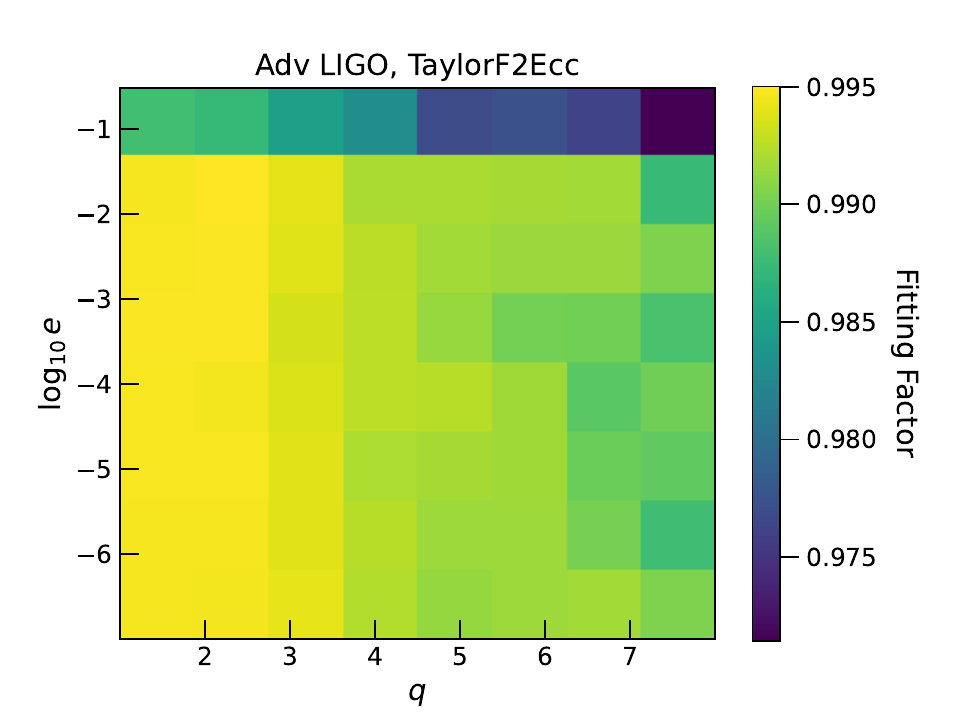} \hspace{3pt}
    \includegraphics[trim=40 0 20 20, clip, width=0.49\linewidth]{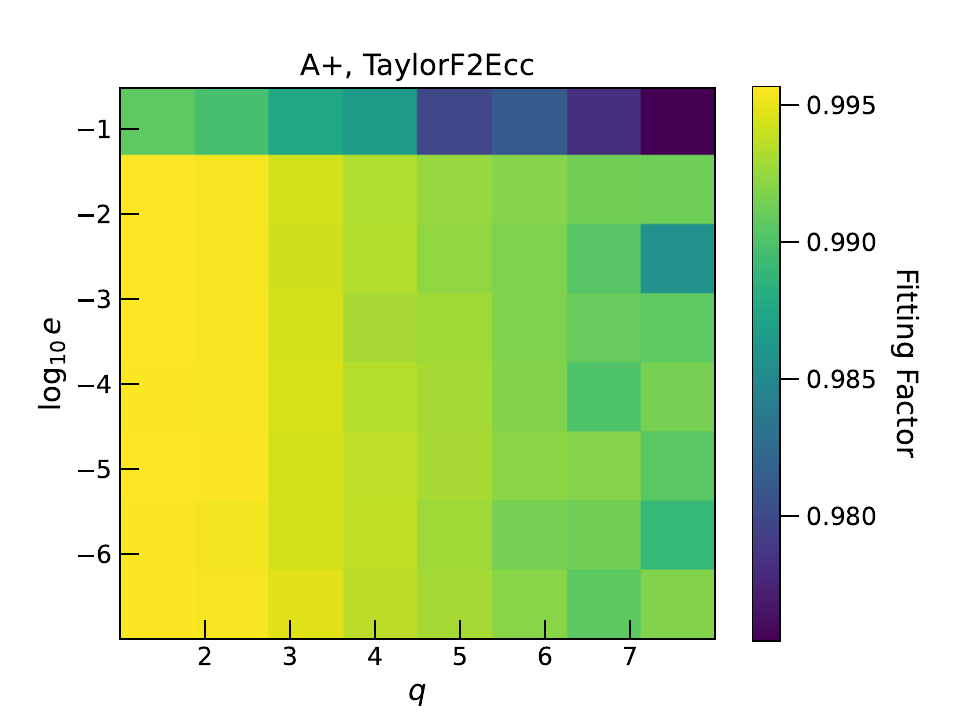}
    \includegraphics[trim=20 0 40 20, clip, width=0.49\linewidth]{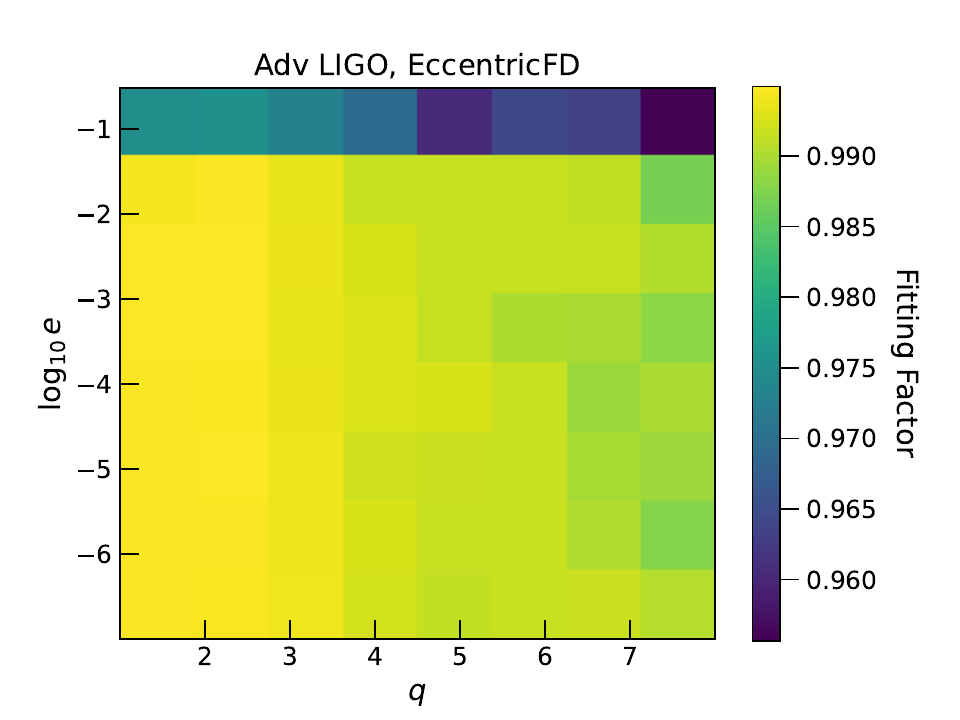} \hspace{3pt}
    \includegraphics[trim=40 0 20 20, clip, width=0.49\linewidth]{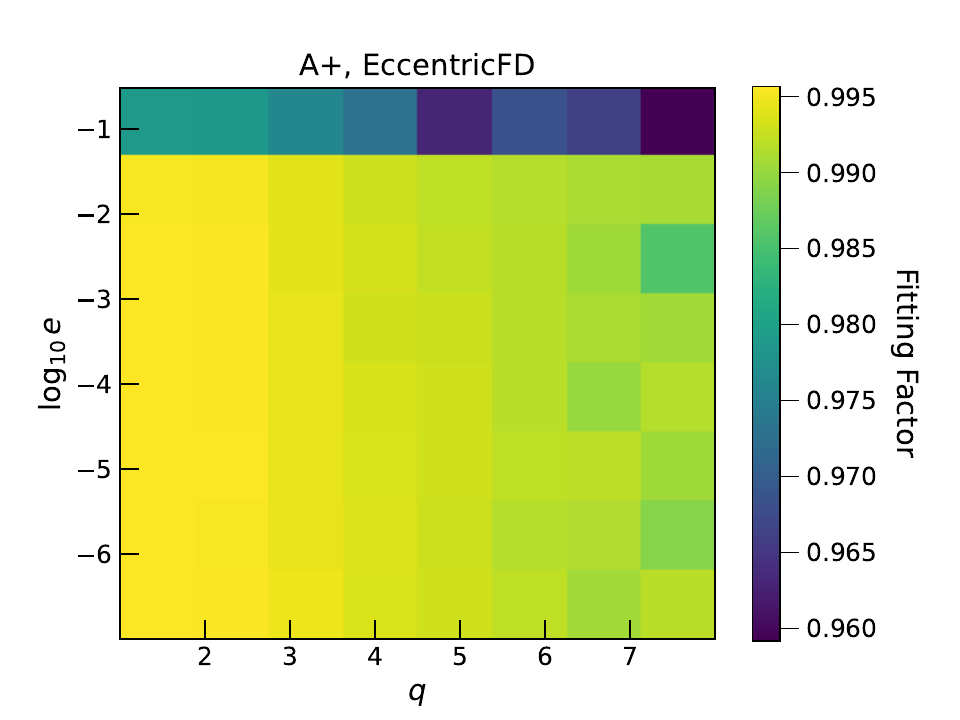}
    \caption[The Fitting Factor ($FF$) varying with mass ratio ($q$) and eccentricity measured at 10 Hz ($\log_{10}{(e)}$)]{The Fitting Factor ($FF$) varying with mass ratio $q$ and $\log_{10}{(e)}$ for a population uniform in component masses and log-uniform in eccentricity $e_{10}$ measured at 10 Hz. Top row represents the population generated with \texttt{TaylorF2Ecc} and bottom row represents the population generated with \texttt{EccentricFD}. For all plots, we use the recovery template bank generated with \texttt{IMRPhenomD}. The left column shows results assuming the detector sensitivity of advLIGO and the right column shows results assuming the detector sensitivity of $A+$. The maximal loss in fitting factor occurs for high mass ratios ($q=m_1/m_2$) and high eccentricity regimes, with high eccentricity values playing dominating role. We use only the inspiral part of the waveform, up to frequency corresponding to innermost stable circular orbit (ISCO), to calculate the $FF$.}
    \label{fig:ff_ecc_q}
\end{figure}
%\clearpage

In Fig.~\ref{fig:cumulative_histogram_ff}, we show the fitting factors for all three injection sets considered above. As expected, the quasi-circular injection set generated with \texttt{IMRPhenomD} and recovered with the quasi-circular template bank gives us the maximum $FF$. The upper cut-off frequency, for each system, is chosen to be the frequency corresponding to the innermost stable circular orbit ($f_\text{ISCO}$) for a test particle orbiting a Schwarzschild black hole. We use the lower frequency cut-off of $10$~Hz to calculate the match via Eq. \eqref{eqn:match}. The loss of $FF$ is visible with both eccentric injection sets. In Fig.~\ref{fig:ff_ecc_q}, we explore the parameter regime where the reduction in $FF$ is maximum. As expected, larger eccentricity values ($e_{10}>0.01$) give us lower $FF$. We also notice that the combination of high mass ratio and high eccentricity gives us maximum loss in the $FF$. We propose that more extreme mass ratios lead to a larger reduction in $FF$ for the same value of $e_{10}$ because binaries with more extreme $q$ have longer GW signals in-band, and therefore have more inspiral cycles over which the mismatch due to eccentricity accumulates. 

Figures \ref{fig:cumulative_histogram_ff} and \ref{fig:ff_ecc_q} show that for a given population of eccentric signals, there will be loss of $FF$ for signals with eccentricity $e_{10} > 0.01$. This trend becomes more prominent for more extreme values of mass ratio $q$. The extent of the overall search volume loss depends on the proportion of high-eccentricity signals in the population. In order to include eccentricity in GW searches, we require i) efficient eccentric waveform models and ii) a low computational cost in comparison to the gain in the search volume. 

\begin{table}[t!]
\small
\begin{tabular}{|c|c|c|c|c|}
\hline
  \textbf{\begin{tabular}[c]{@{}c@{}}Injection Waveform\end{tabular}} &
  \hspace{5pt}$\textbf{\begin{tabular}[c]{@{}c@{}}$\bm{SRF}$ \\ Full Range \end{tabular}}$\hspace{5pt} &
  \hspace{5pt}$\textbf{\begin{tabular}[c]{@{}c@{}}$\bm{SRF}$ \\ ($\bm{e_{10}>0.01}$) \end{tabular}}$\hspace{5pt} &
  \hspace{5pt}$\textbf{\begin{tabular}[c]{@{}c@{}}$\bm{SRF}$ \\ ($\bm{q>3}$) \end{tabular}}$\hspace{5pt} & \hspace{5pt}$\textbf{\begin{tabular}[c]{@{}c@{}}$\bm{SRF}$ \\ ($\bm{e_{10}>0.01}$) \& \\ ($\bm{q>3}$) \end{tabular}}$\hspace{5pt} \\ \hline
  \texttt{IMRPhenomD}     & 0.992 (0.992) & -     & 0.986 (0.997) & -  \\ \hline
  \texttt{TaylorF2Ecc} & 0.989 (0.987) & 0.973 (0.97) & 0.923 (0.978) & 0.923 (0.948)   \\ \hline
  \texttt{EccentricFD} & 0.989 (0.987) & 0.969 (0.963)   &  0.923 (0.979)  & 0.918 (0.944) \\ \hline
\end{tabular}
\caption[$SRF$ for various injection sets quantifying the reduced detectability of eccentric signals when circular template bank is used]{The $SRF$, as described in Eq.~\eqref{eqn:signal_fraction_recovery}, is calculated for injection sets described in the text to quantify the reduced detectability of eccentric signals when circular template bank is used. For recovery, we use a template bank designed for non-eccentric searches using \texttt{IMRPhenomD} waveform model. For each column, two numbers are shown: one for Advanced LIGO search sensitivity and the numbers in the bracket are quoted for A+ search sensitivity. The $SRF$ for the optimal search (injection with \texttt{IMRPhenomD}) indicates the maximum. For eccentric injections, the loss in the $SRF$ is maximum in the parameter space ($e_{10}>0.01, q>3$) which is affected most due to loss in $FF$.}
\label{table:SRF}
\end{table}

The $SRF$ depends on the intrinsic source population under consideration. If the fraction of signals with high eccentricity ($>0.01$) is large, we expect to fail to recover a higher fraction of them. In order to estimate SRF, we choose a network of three detectors: HLV, with two LIGO detectors H and L at Hanford and Livingston respectively, and the Virgo (V) detector in Italy. For the full population described above, we estimate $SRF$ to be 0.992 for optimal search (injection and recovery done with \texttt{IMRPhenomD}) for both the Advanced LIGO and A+ detector sensitivities. We kept the same sensitivity for Virgo \citep{PyCBC-PSD:AdvVirgo} in both the networks. With eccentric injections using \texttt{EccentricFD}, the $SRF$ for the full population is estimated to be 0.989 (0.987) for Advanced LIGO (A+) detector sensitivity. For the other set of eccentric injections generated using \texttt{TaylorF2Ecc}, the estimated $SRF$ is 0.989 (0.987) for Advanced LIGO (A+) detector sensitivity. This indicates that the presence of eccentricity in GW signal reduces the overall $SRF$ if quasi-circular recovery models are used. Moreover, if the population has a significant number of events from the parameter space which is responsible for most loss in $FF$, the $SRF$ is further reduced, indicating failure of recovering a comparatively large fraction of events in that parameter region. For a targeted region in parameter space of non-negligible eccentricity ($e_{10}>0.01$) and high mass ratio ($q>3$), we summarize the results in Table \ref{table:SRF}. In this targeted region, we can get the value of $SRF$ as low as $\sim$0.918 compared to the $SRF$ of $\sim$0.99 for the optimal pipeline.

To gain insights into a realistic population, we create another injection set. This set incorporates a power-law distribution of source frame masses, consistent with GWTC-3 population analysis \citep{KAGRA:2021duu}, and an eccentricity distribution drawn from simulations outlined in \cite{Kremer:2019iul, Zevin:2021rtf}. Figure 1 of \cite{Zevin:2021rtf} describes the eccentricity distribution at 10 Hz for detectable BBH mergers.  We limit the source frame mass distribution to the range $[5, 50] M_{\odot}$, aligning with the template banks we generated. We use the same HLV detector network with two sensitivities for LIGO detectors. For this injection set, the $SRF$ for the baseline model was calculated at $\sim 0.986$ for both Advanced LIGO and A+ sensitivities. Focusing on targeted regions ($e_{10}>0.01, q>3$), the $SRF$ is found to be $\sim 0.944\ (0.946)$ for \texttt{EccentricFD} (\texttt{TaylorF2Ecc}) injections with Advanced LIGO design sensitivity, and $\sim 0.927\ (0.928)$ for \texttt{EccentricFD} (\texttt{TaylorF2Ecc}) injections with A+ design sensitivity.

While we might detect eccentric signals via either quasi-circular template-based searches or unmodelled searches increasing the true fraction of the underlying eccentric population that will enter into our catalogues, bias could be introduced in the subsequently inferred parameters. For this reason, quantifying the bias introduced by analysing eccentric signals with quasi-circular waveform models is necessary. We turn our attention to this in the following section.

%%%%%%%%%%%%%%%%%%%%%%%%%%%%%%%%%%%%

\section{Mischaracterizing eccentric binaries with parameter estimation}
\label{sec:pe}

In this section, we present results from injection analyses. We assess waveform systematics due to the neglect of eccentricity in PE studies employing quasi-circular waveform models to recover injections into detectors with zero noise (i.e., the detector response to the signal is accounted for, but no additional Gaussian noise is added to the power spectral density representing the detector's sensitivity). We also perform injections into simulated Gaussian noise and find them to be consistent with zero noise injection analyses. We perform two sets of injections: one with non-spinning simulations based on NR \citep{Chattaraj:2022tay, Hinder:2017sxy}, and one using an EOB-based IMR signal model (\texttt{TEOBResumS}) for aligned-spin injections \citep{Nagar:2018zoe, Chiaramello:2020ehz, Mora:2002gf, Nagar:2021xnh, Placidi:2021rkh, Albanesi:2022ywx, Albanesi:2022xge, Placidi:2023ofj}.

Complete inspiral-merger-ringdown waveforms are constructed by matching PN and NR waveforms for individual modes in a region where the PN prescription closely mimics the NR data following the method of \cite{Varma:2016dnf}. These are traditionally referred to as "hybrids" and are used as targets for modelling and data analysis purposes. One such hybrid is shown in Fig.~\ref{fig:ecc-hyb-22}. The blue dotted line marks the beginning of NR waveform, and the shaded grey region $t \in (1000M, 2000M)$ shows the matching window where hybridization was performed. Overlapping hybrid and NR waveforms on the left of the matching window hint at the quality of hybridization performed here. For non-spinning injections, we use hybrids developed in \cite{Chattaraj:2022tay}. We analyse non-spinning quasi-circular and eccentric signals with mass ratios $q=(1, 2, 3)$ and a fixed total mass of $M=35 M_\odot$ (for dominant mode, zero-noise analyses) and $M=40 M_\odot$ (for Gaussian noise injections and HM analyses). For $q=1$ injection, since the mass ratio prior is restricted to $q \geq 1$, almost the entire posterior lies above the injected value, skewing the posteriors for the other (correlated) parameters; this has been discussed in detail in Sec.~\ref{appendix:q_1}. Hence, for dominant mode, aligned-spin eccentric injections, we choose $q=(1.25, 2, 3)$ with the total mass $M=35M_\odot$ (same as dominant mode, non-spinning injections), and drop the $q=1$ case. We employ state-of-the-art quasi-circular waveform models (with and without spins) to recover the injections via Bayesian parameter estimation (PE). We also perform PE with an approximate inspiral eccentric waveform \texttt{TaylorF2Ecc}~\citep{Moore:2016qxz, Kim:2019abc}. The approximate eccentric model used here for PE does not include contributions from spin corrections associated with eccentricity and is based on a \texttt{Taylor} approximant~\citep{Damour:2002kr, Buonanno:2009zt} different from the one used in \texttt{TEOBResumS}. We assume that our sources are at a luminosity distance of $410$~Mpc and inclined at an arbitrary angle of $30^\circ$ to the line of sight. The sky location angles have been set arbitrarily as ($\alpha \sim 164^\circ,\ \delta=60^\circ,\ \psi=60^\circ$), and the geocent time ($t_\text{gps}$) is taken to be 1137283217~s. Since the SNR of a GW signal depends on extrinsic parameters in addition to intrinsic parameters like mass and eccentricity, different extrinsic parameters may lead to different SNRs, changing the widths of the posteriors presented here. 

\begin{figure}[t!]
    \centering
    \includegraphics[width=\linewidth]{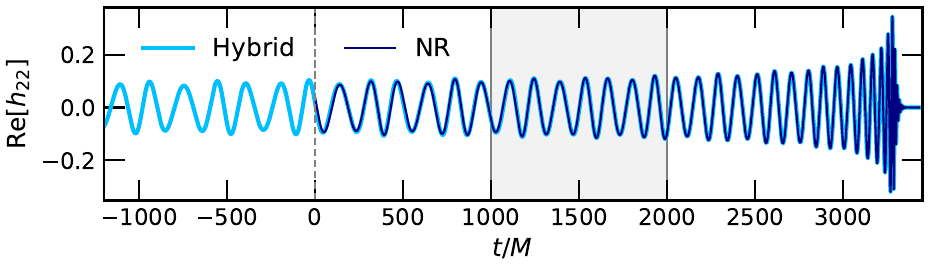}
    \caption[PN-NR hybrid waveform corresponding to NR simulation SXS:BBH:1364]{\takenfrom{\cite{Chattaraj:2022tay}} PN-NR hybrid waveform corresponding to NR simulation SXS:BBH:1364, an asymmetric mass binary with mass ratio $q=2$. The initial eccentricity of the constructed hybrid is $e_0 = 0.108$ at $x_\text{low} = 0.045$. The eccentric inspiral waveform used for constructing the target hybrids is presented in \cite{Tanay:2016zog}.}
    \label{fig:ecc-hyb-22}
\end{figure}

\begin{table}[t!]
\centering
\def\arraystretch{1.3}
\begin{tabular}{|c|c|c|}
\hline
\textbf{Parameter} & \textbf{Prior} & \textbf{Range} \\ \hline
$\mathcal{M}$ & \begin{tabular}[c]{@{}c@{}}Uniform in \\ component masses \end{tabular} & $5 \text{ - } 50~M_\odot$ \\ \hline
$q$ & \begin{tabular}[c]{@{}c@{}}Uniform in \\ component masses \end{tabular} & $1 \text{ - } 5$ \\ \hline
$d_L$ & Uniform radius & $100 \text{ - } 3000$ Mpc \\ \hline
$\iota$ & Uniform sine & $0 \text{ - } \pi$ \\ \hline
$t_c$ & Uniform & $t_\text{gps}+(-0.1 \text{ - } 0.1)$~s \\ \hline
$\phi_c$ & Uniform & $0 \text{ - } 2\pi$ \\ \hline
$\chi_{i\text{z}}$$^*$ & Uniform & $0 \text{ - } 0.9$ \\ \hline
$a_1$, $a_2$ & Uniform & $0 \text{ - } 0.9$ \\ \hline
$(S_i^\Theta + S_i^\Phi)^*$ & Uniform solid angle & \begin{tabular}[c]{@{}c@{}} $\Theta \in (0,\pi)$, \\ $\Phi \in (0,2\pi)$ \end{tabular} \\ \hline
$(\alpha + \delta)$ & Uniform sky & \begin{tabular}[c]{@{}c@{}} $\delta \in (\pi/2,-\pi/2)$, \\ $\alpha \in (0,2\pi)$ \end{tabular} \\ \hline
$e$ & Uniform & $0 \text{ - } 0.4$ \\ \hline
\end{tabular}
\caption[Priors for parameters used in various quasi-circular and eccentric recoveries]{Priors for parameters used in various quasi-circular and eccentric recoveries. Here, $\chi_{i\text{z}}$ is only used for aligned spin recoveries, whereas for precessing spin-recoveries, we have used $a_1, a_2, S_i^\Theta, S_i^\Phi$. The parameter $e$ is only included in the parameter space for eccentric recoveries. $^*$ - where $i=[1,2]$ refers to the binary components}
\label{table:ecc-priors}
\end{table}

\begin{table}[t!]
\centering
\def\arraystretch{1.3}
\begin{tabular}{|c|c|c|c|c|}
\hline
\textbf{S. No.} &
  \textbf{\begin{tabular}[c]{@{}c@{}}Injection Simulation ID / \\ Waveform\end{tabular}} &
  \hspace{5pt}$\mathbf{q}$\hspace{5pt} &
  \textbf{$\mathbf{e_{20}}$} &
  \textbf{$\mathbf{\chi_\text{eff}}$} \\ \hline
1  & SXS:BBH:1132     & 1 & 0     & -   \\ \hline
2  & HYB:SXS:BBH:1355 & 1 & 0.104 & -   \\ \hline
3  & HYB:SXS:BBH:1167 & 2 & 0.0   & -   \\ \hline
4  & HYB:SXS:BBH:1364 & 2 & 0.104 & -   \\ \hline
5  & HYB:SXS:BBH:1221 & 3 & 0.0   & -   \\ \hline
6  & HYB:SXS:BBH:1371 & 3 & 0.123 & -   \\ \hline
7  & \texttt{TEOBResumS}       & 1.25 & 0.0   & 0.3 \\ \hline
8  & \texttt{TEOBResumS}       & 1.25 & 0.1   & 0.3 \\ \hline
9  & \texttt{TEOBResumS}       & 2 & 0.0   & 0.3 \\ \hline
10  & \texttt{TEOBResumS}       & 2 & 0.1   & 0.3 \\ \hline
11  & \texttt{TEOBResumS}       & 3 & 0.0   & 0.3 \\ \hline
12 & \texttt{TEOBResumS}       & 3 & 0.1   & 0.3 \\ \hline
\end{tabular}
\caption[List of non-spinning, eccentric NR hybrid simulations \citep{Chattaraj:2022tay} and injections based on aligned-spin eccentric EOB model \texttt{TEOBResumS} \citep{Nagar:2018zoe}]{List of non-spinning, eccentric NR hybrid simulations (constructed in \cite{Chattaraj:2022tay}) and injections based on aligned-spin eccentric EOB model \texttt{TEOBResumS}. Columns include a unique hybrid ID for each simulation [SXS IDs are retained for identification with SXS simulations \citep{Boyle:2019kee, Buchman:2012dw, Chu:2009md, Hemberger:2013hsa, Scheel:2014ina, Blackman:2015pia, SXS:catalog, Lovelace:2011nu, Lovelace:2010ne, Mroue:2013xna, Mroue:2012kv, Lovelace:2014twa, Kumar:2015tha, Lovelace:2016uwp, LIGOScientific:2016sjg, Hinder:2013oqa, Abbott:2016apu, Chu:2015kft, Abbott:2016wiq, Varma:2018mmi, Varma:2018aht, Varma:2019csw, Varma:2020bon, Islam:2021mha, Ma:2021znq} used in constructing the hybrids] and the name of the waveform model used for generating injections, information concerning the mass ratio ($q=m_1/m_2$), eccentricity ($e_{20}$) at the reference frequency of $20$~Hz for a total mass of $M=35$ M$_\odot$, and effective spin $\chi_\text{eff}$ defined in Eq.~\eqref{eq:chieff} (only shown for spinning injections).} 
\label{table:hybrids}
\end{table}

To estimate parameters, we perform Bayesian inference (discussed in Sec.~\ref{subsec:intro-bayesian}) using \texttt{PyCBC Inference Toolkit}~\citep{Biwer:2018osg} and explore the parameter space that includes ($\mathcal{M}, q, t_c, d_L, \phi_c, \iota, \alpha, \delta$). For aligned spin recoveries, we use two additional parameters corresponding to the $\rm{z}$-components of the spin vectors \textit{viz.} $\chi_\text{1z}$ and $\chi_\text{2z}$. For recoveries with spin-precession, we use isotropic spin distribution sampling the six spin components in spherical polar coordinates: ($a_1, a_2, S_1^\Theta, S_2^\Theta, S_1^\Phi, S_2^\Phi$). For eccentric recoveries, we include an additional eccentricity ($e$) parameter in the parameter space. Table \ref{table:ecc-priors} contains the prior ranges on all the parameters. In our analysis, we marginalise over the polarization angle. We also calculate Bayes factors between recoveries with eccentric ($\mathcal{H}_E$) and quasi-circular ($\mathcal{H}_C$) models, given data $d$, defined as:
\begin{equation}
    \mathcal{B}_{E/C} = \frac{p(d|\mathcal{H}_E)}{p(d|\mathcal{H}_C)}.
\end{equation}
where $E$ and $C$ correspond to eccentric and quasi-circular recoveries respectively. The following subsections provide details of the specific injections, as well as various variations of recovery-waveform spin settings with which these injections have been recovered. While discussing the results in the following subsections, we make use of the term "recovery" to indicate a result in which the $90\%$ credible interval of the posterior includes the injected value, and the systematic bias (difference between the median value and injected value) in the posterior is less than the width of the posterior (at $90\%$ confidence). We judge that the result shows a significant bias if the injected value lies completely outside the $90\%$ credible interval of the posterior. As indicated earlier, these biases are dependent on SNRs which, in this study, fall in the range of typical SNRs observed in the GW event catalogs. We use the HLV network with design sensitivities of Advanced LIGO \citep{PyCBC-PSD:aLIGO} and Virgo \citep{PyCBC-PSD:AdvVirgo} detectors to perform all the parameter estimation analyses shown here.

\subsection{Dominant mode, non-spinning, eccentric injections}
\label{subsec:non-spin-inj}

We perform zero-noise injections using non-spinning, quasi-circular as well as quasi-elliptical GW waveforms for BBH mergers of total mass of $M=35$~M$_\odot$ with mass ratios $q=(1, 2, 3$). These injections include the dominant modes ($\ell=2, |m|=2$) of the eccentric and quasi-circular IMR hybrids constructed in \cite{Chattaraj:2022tay}, in addition to a quasi-circular SXS simulation (\textsc{SXS:BBH:1132}). Details of the simulations used in this study, including their eccentricity at a reference frequency of $20$~Hz, are shown in Table \ref{table:hybrids}. To calculate the eccentricity at $20$~Hz GW frequency, we have used the reference value ($e_0$) from Table I of \cite{Chattaraj:2022tay} which they have quoted for a dimensionless frequency of $x_0=0.045$. Using the following relation we compute the gravitational wave frequency corresponding to $x_0=0.045$ and total mass $35$ $M_\odot$:
\begin{equation}
    f_\text{GW} = \frac{x_0^{3/2}}{\pi M} = 17.62~\text{Hz}
\end{equation} 
where $M$ is the total mass taken in natural units (seconds). Now that we have $e_0$ at $f_\text{GW}$, we evolve it using Eq. (4.17a) of \cite{Moore:2016qxz} (eccentricity evolution for orbit averaged frequency) to get eccentricity value at $20$~Hz ($e_{20}$ shown in Table \ref{table:hybrids}). Note that the starting frequency for likelihood calculation is also 20 Hz.

\subsubsection{Quasi-circular, IMR recovery}

\begin{figure}[t!]
    \centering

    \includegraphics[trim=100 10 100 30, clip, width=\linewidth]{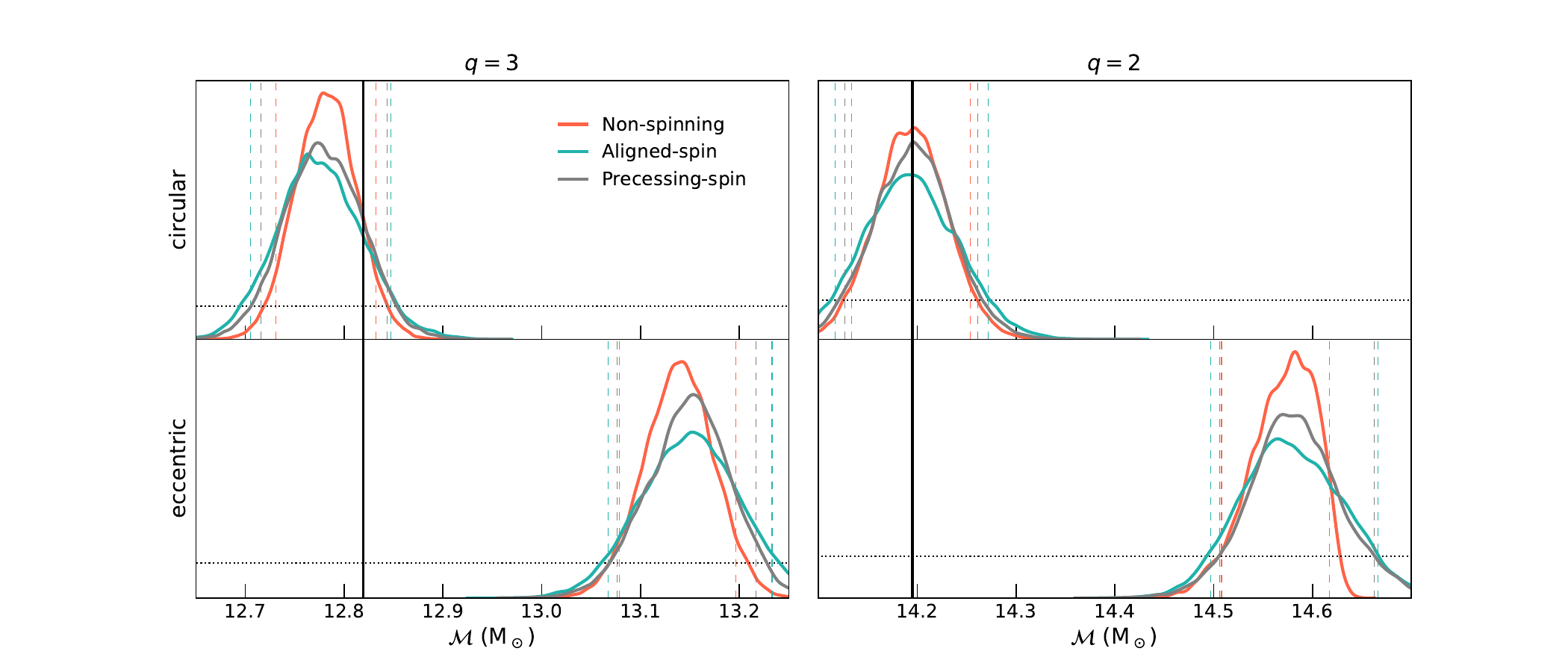}
    \caption[Chirp mass posteriors for non-spinning injections with mass ratios ($q=2,3$)]{Chirp mass posteriors for injections with mass ratios ($q=2,3$). The rows indicate the nature of the injections, with the top panel showing results for the quasi-circular injection, and the bottom panel showing results for the eccentric injection ($e_{20} \sim 0.1$). The colours correspond to different spin settings used during recovery. Recovery is performed using quasi-circular waveforms in all cases: \texttt{IMRPhenomXAS} is used for the non-spinning (red) and aligned spin (green) recoveries, and \texttt{IMRPhenomXP} is used for recovery allowing precessing spins (grey). The dashed vertical coloured lines of the same colours denote the $90\%$ credible interval of the corresponding recoveries, the solid black line shows the injected value of $\mathcal{M}$, and the dotted black curve indicates the prior which is same for all recoveries. The injected value is recovered within the $90\%$ credible interval for the quasi-circular injections, while it is not recovered for the eccentric injections. The slight shift of posteriors for the quasi-circular injection in the $q=3$ case may be attributed to systematic differences between the waveform models used for injection and recovery. The matched filter SNRs for $q=2$ and $q=3$ are $38$ and $33$ respectively.}
    \label{fig:hist-ns-as-ps}
\end{figure}

\begin{figure}[t!]
\centering
    \includegraphics[trim=20 10 30 30, clip, width=0.6\linewidth]{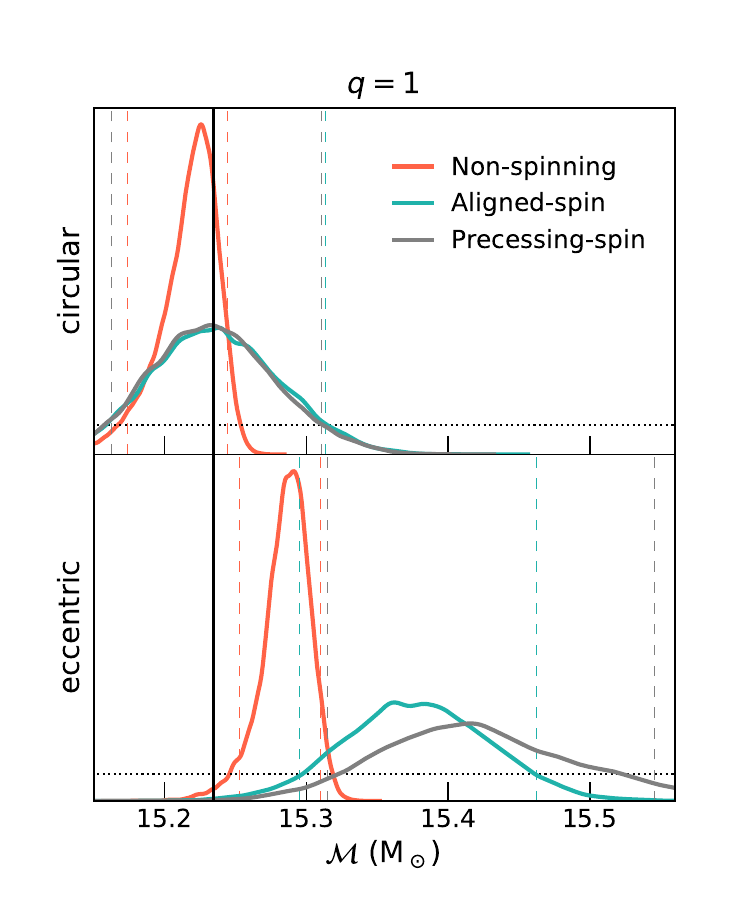}
    \caption[Posteriors for $q=1$ for non-spinning quasi-circular and eccentric injections]{Posteriors for $q=1$ for non-spinning quasi-circular (top) and eccentric (bottom) injections. Recovery with different waveform models is indicated with different colours, corresponding to quasi-circular waveform \texttt{IMRPhenomXAS} for non-spinning (red) and aligned spin (green), and \texttt{IMRPhenomXP} for precessing spins (grey). The injection value is shown with black line. The dotted line shows the prior used, which is same for the recovery of both quasi-circular and eccentric injections. The matched filter SNR for these injections is $41$.}
    \label{fig:hist_q_1}
\end{figure}

Here, we explore the bias introduced in the source parameters recovered via parameter estimation of GW events when eccentricity is ignored, i.e.~we inject an eccentric signal but do not use eccentric waveforms for recovery. For this exercise, we use the quasi-circular IMR phenomenological waveform models \texttt{IMRPhenomXAS} \citep{Pratten:2020fqn} and \texttt{IMRPhenomXP} \citep{Pratten:2020ceb}. 
In figures \ref{fig:hist-ns-as-ps} and \ref{fig:hist_q_1}, we plot the posterior probability distributions on chirp mass for non-spinning, quasi-circular and eccentric injections recovered using IMR quasi-circular waveform models. The injection is recovered using three spin setting configurations: non-spinning (NS), in which we restrict all spins to $0$; aligned-spin (AS), in which we restrict spin-tilt angles to $0$ and allow spin magnitudes to range between $0$ and $0.99$; and precessing-spin (PS), in which we allow all spin parameters to vary. By using different spin configurations, we investigate whether spurious measurements of spins occur for non-spinning eccentric injections. In addition to studying biases in spin parameters due to the presence of eccentricity, we also compare the effect of spin settings on the recovery of chirp mass in the presence of eccentricity, since eccentricity and chirp mass are known to be correlated parameters \citep[see for instance][]{Favata:2021vhw}. We use the waveform \texttt{IMRPhenomXAS} for non-spinning and aligned-spin recoveries, and \texttt{IMRPhenomXP} for the precessing-spin recovery. We display the corresponding corner plots in Figs.~\ref{fig:corner-ps-q-2}, \ref{fig:corner-ps-q-3}, and \ref{fig:corner-ps-q-1} for $\mathcal{M}$, $\chi_\text{eff}$, and $\chi_p$. Looking at Figs.~\ref{fig:hist-ns-as-ps} - \ref{fig:corner-ps-q-1} we make the following observations:

\begin{itemize}
    \item For all the values of mass ratio that we consider, the recovery of eccentric injections with quasi-circular waveform models results in a significant bias of the chirp mass posterior, such that the injected value falls outside the $90\%$ credible interval.
    
    \item The spin settings (non-spinning, aligned-spin, or precessing-spin) chosen for recovery do not affect the magnitude of shift in the chirp mass posterior for mass ratios 2 and 3, or in other words; the bias in the recovered chirp mass is the same regardless of assumptions about the spin magnitudes and tilt angles. 
    
    \item For $q=1$, the shift in the chirp mass posteriors for different spin configurations, seen in Fig.~\ref{fig:hist_q_1}, can partly be explained due to the prior railing of mass ratio leading to almost the entire posterior volume lying outside the injected value. This can lead to the prior railing in component masses and other correlated parameters. This is discussed in detail in the next sub-section.
        
    \item The $\chi_\text{eff}$ posterior is largely consistent with zero for both quasi-circular and eccentric injections. The same trend is seen for both the $q=2$ and $q=3$ cases. The slight deviation of $\chi_\text{eff}$ from $0$ for $q=1$ case can be explained by looking at the correlation between chirp mass and $\chi_\text{eff}$ (see Fig.~\ref{fig:corner-ps-q-1}). 
    
    \item The posterior for $\chi_\text{p}$ peaks towards 0 for the $q=2$ and $q=3$ cases, showing little to no evidence of spin-precession in the signal. For $q=1$, the $\chi_p$ posterior is uninformative. Both the uninformative posterior and posteriors that peak towards $0$ support the conclusion that eccentricity is not confused for spin-induced precession in long-duration signals from low-mass BBHs.
\end{itemize}

For the $q=3$ case, the slight deviation of posteriors from the injected value for the quasi-circular injection (top left panel of Fig.~\ref{fig:hist-ns-as-ps}) is most likely due to systematic differences between the injection and the recovery waveform. Even at a reasonably modest eccentricity of $e_{20}\sim0.1$, the chirp mass posterior is shifted enough that the injected value is not recovered within $90\%$ confidence. The shift in the chirp mass posteriors for non-spinning injections is consistent with the `\textit{effective chirp mass}' for eccentric binaries, defined in \cite{Bose:2021pcw} \citep[See also][for a similar definition of \textit{`effective chirp mass'} parameter]{Favata:2021vhw}. Further, for a non-spinning eccentric system with moderate total mass ($M=35 M_\odot$), the presence of eccentricity in the signal is not mimicked by a spin-precessing quasi-circular waveform. This is consistent with the findings of \citet{Romero-Shaw:2022fbf}, who find that eccentricity and spin-precession may be distinguished in signals with long inspirals coming from low-mass BBH due to the signal duration exceeding the timescale upon which modulations induced by eccentricity differ significantly from those induced by spin-precession. The fact that the spin posteriors are similar for both eccentric and quasi-circular injections also implies that a lack of spin can be confidently identified in low-mass systems regardless of their eccentricity.

\begin{figure}[t!]
    \centering
    \includegraphics[trim=10 0 0 0, clip, width=\linewidth]{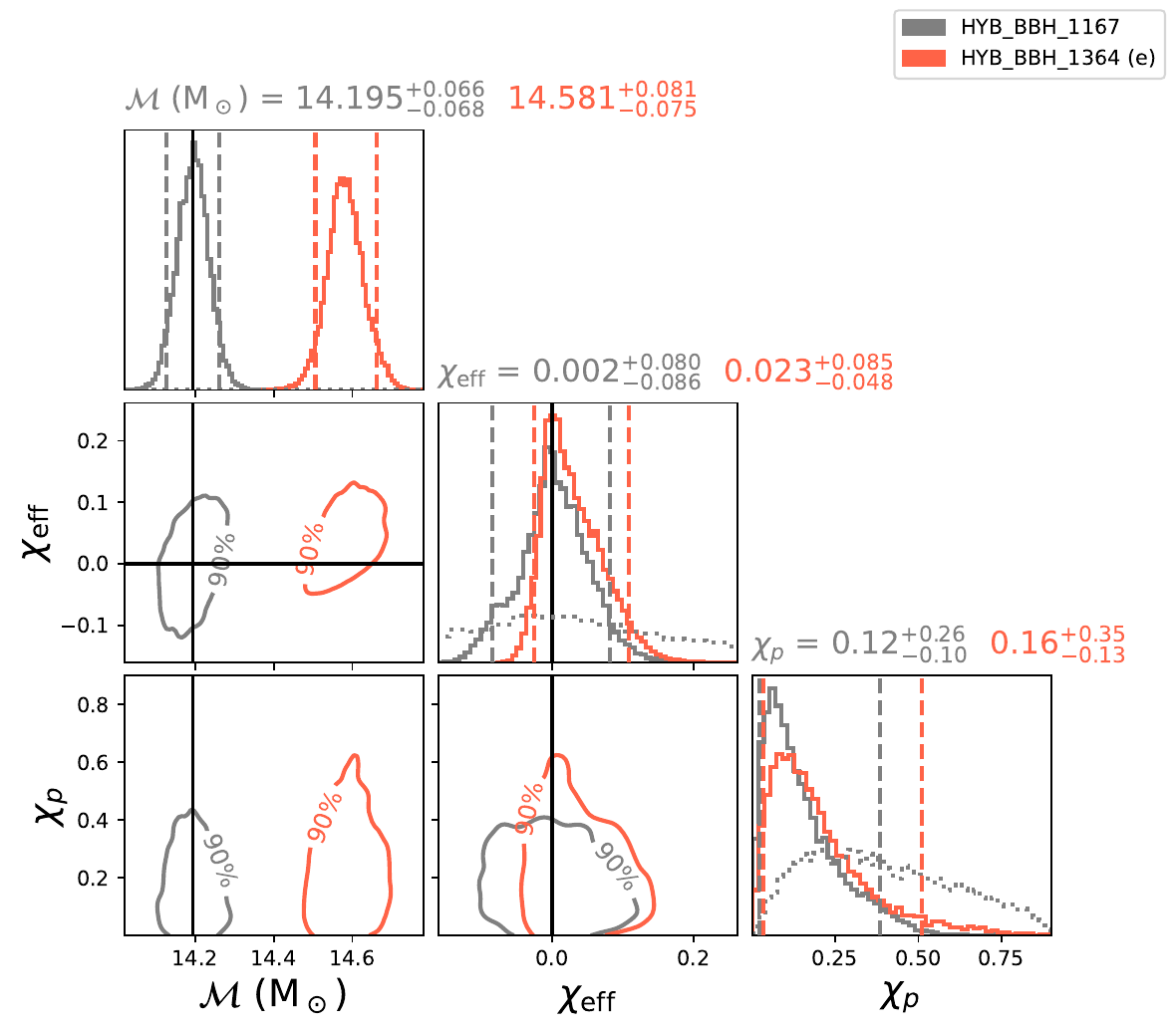}
    \caption[Corner plot for $q=2$ showing posteriors on $\mathcal{M}$, $\chi_\text{eff}$, and $\chi_\text{p}$, for precessing-spin recovery of both quasi-circular and eccentric, non-spinning injections]{The corner plot for $q=2$ injection showing posteriors on chirp mass ($\mathcal{M}$), effective spin ($\chi_\text{eff}$), and spin precession parameter ($\chi_\text{p}$), for precessing-spin recovery (performed using \texttt{IMRPhenomXP}) of both quasi-circular (grey) and eccentric (red), non-spinning injections. The injection values are shown with black lines. The histograms shown on the diagonal of the plot are 1D marginalized posteriors for the respective parameters with vertical dashed lines denoting $90\%$ credible intervals. The dotted curves in the 1D plots show the priors used, which are same for the recovery of both quasi-circular and eccentric injections. The prior height for $\mathcal{M}$ is too little compared to the posterior and is not visible in this plot. Therefore, we show the $\mathcal{M}$ prior in Fig.~\ref{fig:hist-ns-as-ps} instead.}
    \label{fig:corner-ps-q-2}
\end{figure}

\begin{figure}[t!]
    \centering
    \includegraphics[trim=10 0 0 0, clip, width=\linewidth]{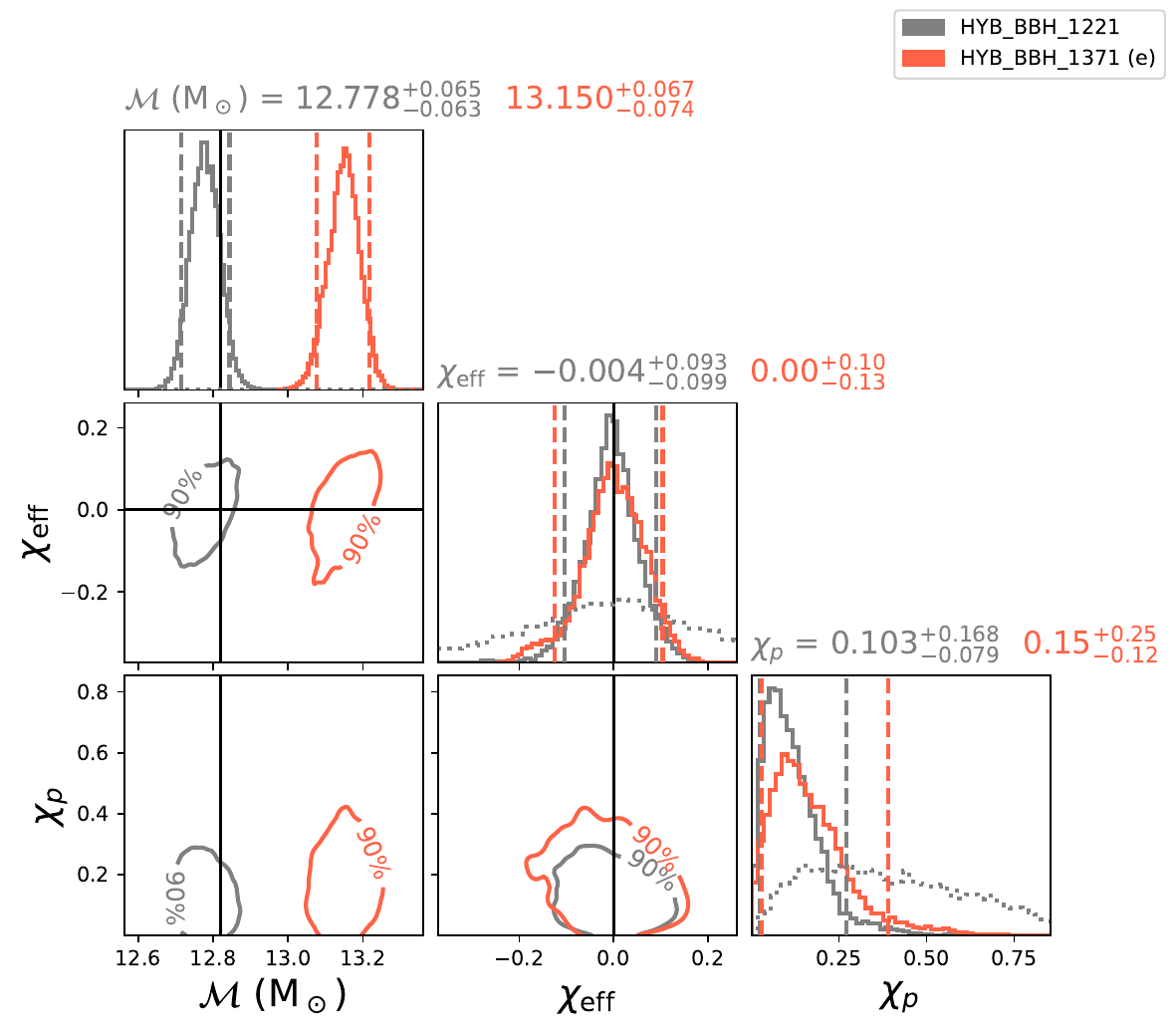}
    \caption[Corner plot for $q=3$ showing posteriors on $\mathcal{M}$, $\chi_\text{eff}$, and $\chi_\text{p}$, for the precessing-spin recovery of both quasi-circular and eccentric, non-spinning injections]{Same as Fig.~\ref{fig:corner-ps-q-2} but for $q=3$.}
    \label{fig:corner-ps-q-3}
\end{figure}

\begin{figure}
\centering
    \includegraphics[trim=10 0 0 0, clip, width=\linewidth]{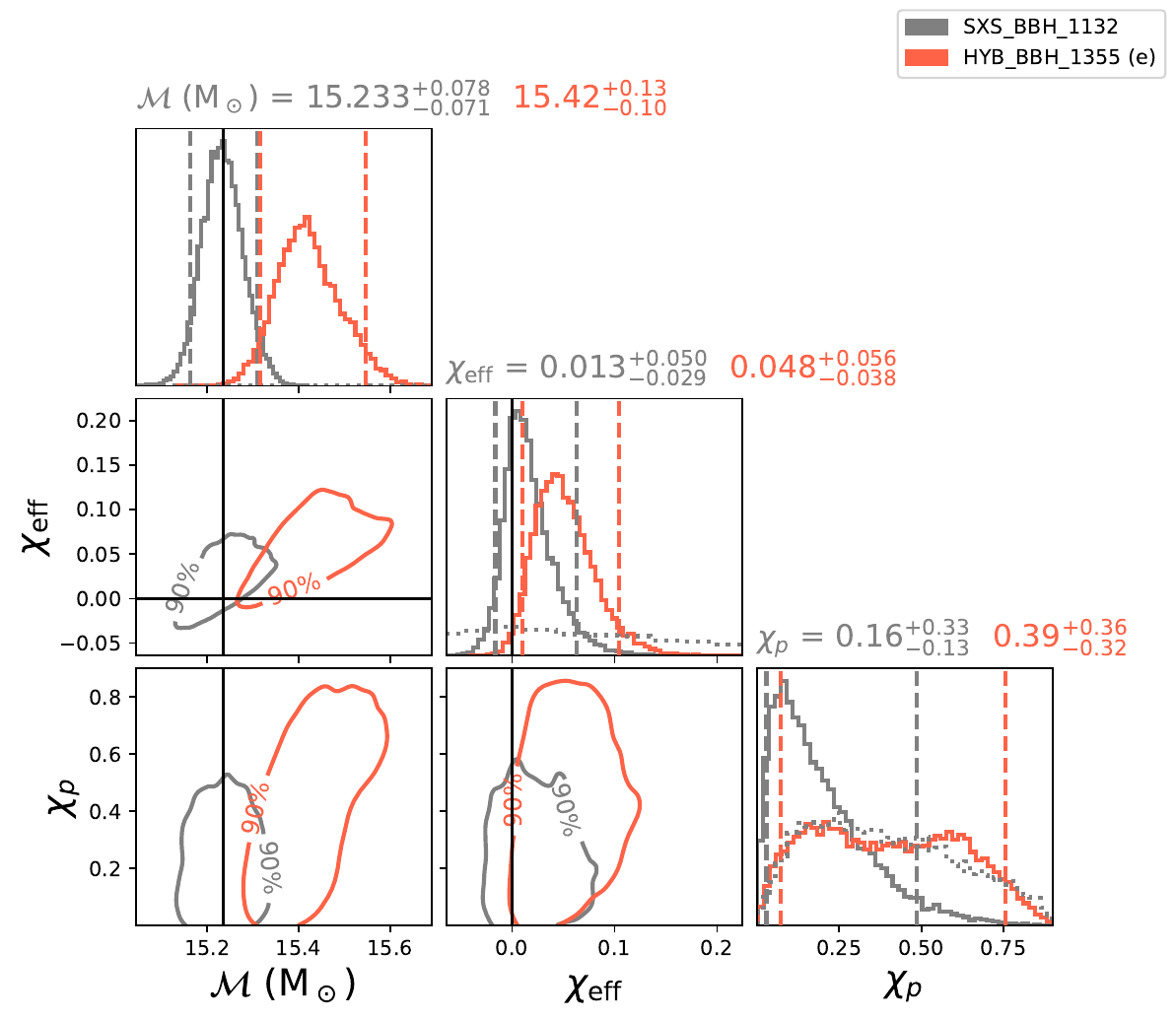}
    \caption[Corner plot for $q=1$ showing posteriors on $\mathcal{M}$, $\chi_\text{eff}$, and $\chi_\text{p}$, for the precessing-spin recovery of both quasi-circular and eccentric, non-spinning injections]{%The corner plot for $q=1$ showing posteriors on the chirp mass ($\mathcal{M}$), effective spin ($\chi_\text{eff}$), and spin precession parameter ($\chi_\text{p}$), for the precessing-spin recovery of both quasi-circular (grey) and eccentric (red), non-spinning injections. The injection values are shown with black lines. The histograms shown on the diagonal of the plot are 1D marginalized posteriors for the respective parameters with vertical dashed lines denoting $90\%$ credible intervals. The dotted curves in the 1D plots show the priors used, which are same for the recovery of both quasi-circular and eccentric injections.
    Same as Fig.~\ref{fig:corner-ps-q-2} but for $q=1$.}
    \label{fig:corner-ps-q-1}
\end{figure}
\clearpage

\subsubsection{Note on equal-mass case}
\label{appendix:q_1}

The $q=1$ case is different from the other mass ratio cases due to physical limits on the mass ratio prior ($q \geq 1$). The shift in the chirp mass posteriors for different spin configurations, seen in Fig.~\ref{fig:hist_q_1}, can partly be explained by the prior railing of mass ratio leading to a prior railing in component masses. Since the true value of injection is exactly $q=1$, parameters correlated with $q$ can become biased due to the entire posterior volume existing above the $q=1$ boundary. In order to confirm this, we carry out an identical baseline injection run where we inject ($\ell$=$2$, $|m|$=$2$) mode, non-spinning, quasi-circular signal into zero noise using \texttt{IMRPhenomXAS}, and recover it in the same three spin configurations (non-spinning, aligned spin, precessing spin) as used for the hybrids. We observe that the trends are identical to the ones observed using the hybrids. Hence we conclude that for $q=1$ case, the slight deviation from the usual trend is because of the mass ratio prior skewing the chirp mass posteriors.

\subsubsection{Eccentric, inspiral-only recovery}

\begin{figure}[p!]
    \centering
    \includegraphics[trim=10 0 10 10, clip, width=0.74\linewidth]{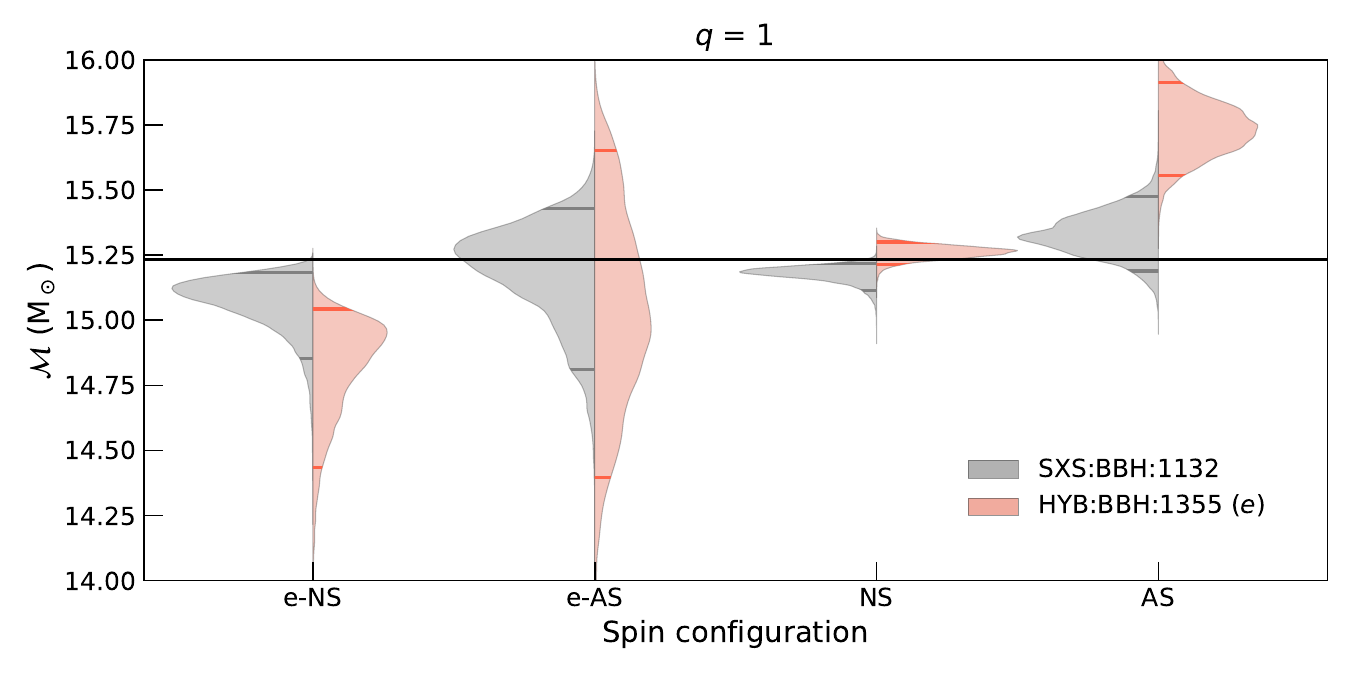}
    \includegraphics[trim=10 0 10 10, clip, width=0.74\linewidth]{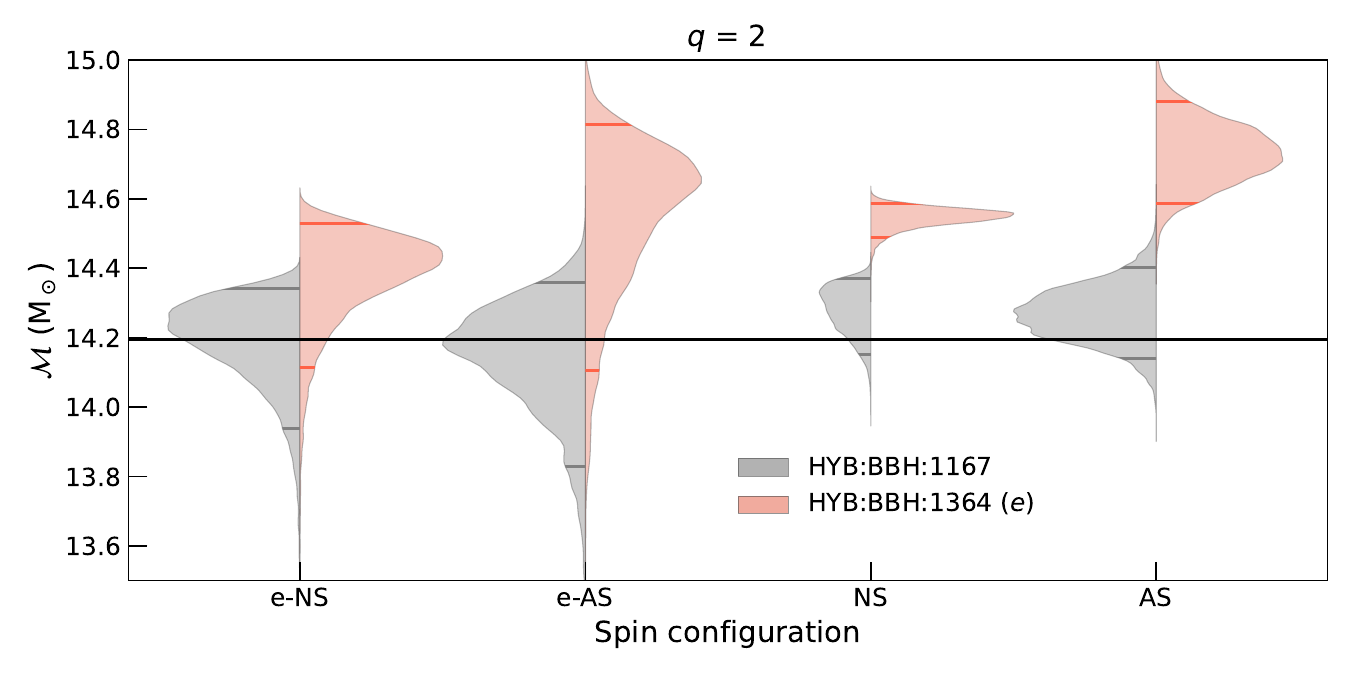}
    \includegraphics[trim=10 0 10 10, clip, width=0.74\linewidth]{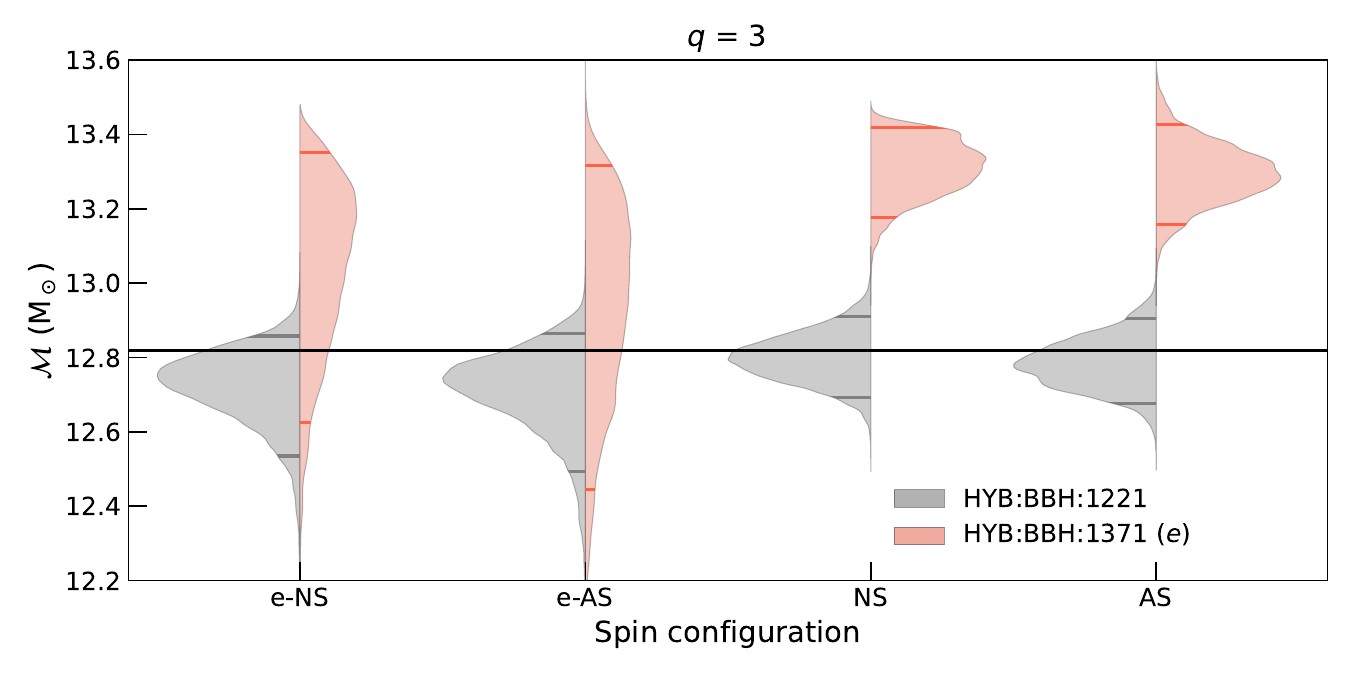}
    \caption[Recovery with \texttt{TaylorF2Ecc} for eccentric and quasi-circular injections in the form of violin plots]{We show the recovery with \texttt{TaylorF2Ecc} for eccentric and quasi-circular injections in the form of violin plots. The colours are used to distinguish the injections: red shows eccentric ($e_{20} \sim 0.1$) injections, while grey shows quasi-circular injections. The horizontal axis denotes the spin configuration in the recovery of posteriors. e-NS, e-AS, NS, and AS correspond to eccentric non-spinning, eccentric aligned spin, quasi-circular non-spinning, and quasi-circular aligned spin recoveries, respectively. The vertical axis corresponds to chirp mass $\mathcal{M}$ values. The black horizontal line indicates the injection value and coloured lines inside the shaded posteriors indicate the $90\%$ credible interval. The matched filter SNRs for $q=2$ and $q=3$ are $31$ and $28$ respectively.}
    \label{fig:violin-ns-inj}
\end{figure}

\begin{figure}[t!]
    \centering
    \includegraphics[width=0.85\linewidth]{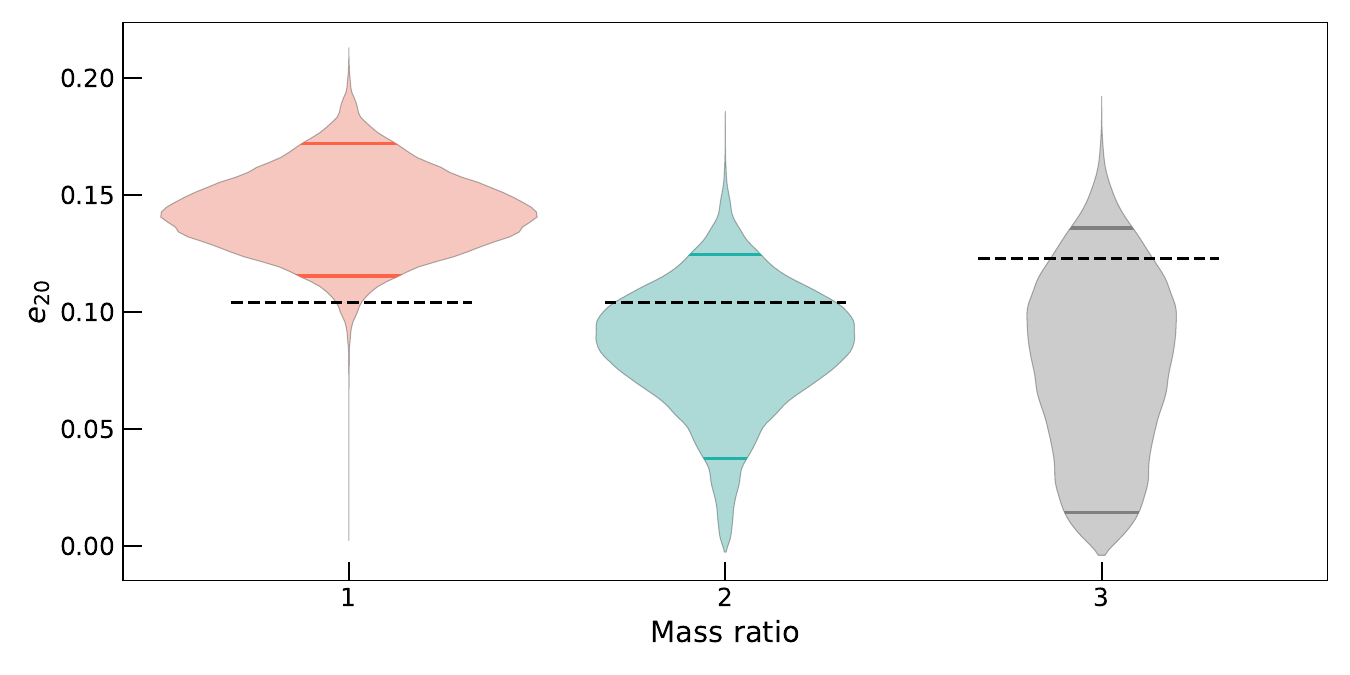}
    \caption[Eccentricity posteriors for $q$=(1, 2, 3) when the injections are non-spinning and eccentric, and the recovery is with \texttt{TaylorF2Ecc} in eccentric, non-spinning configuration]{Eccentricity posteriors for $q=(1, 2, 3)$ when the injections are non-spinning and eccentric, and the recovery is with \texttt{TaylorF2Ecc} in eccentric, non-spinning configuration. The coloured lines inside the shaded posteriors indicate 90\% credible interval whereas the black dashed lines denote the injected eccentricity values.}
    \label{fig:ecc_ns_inj_q_1_2_3}
\end{figure}

\begin{figure}[t!]
    \centering
    \includegraphics[trim=10 0 0 0, clip, width=\linewidth]{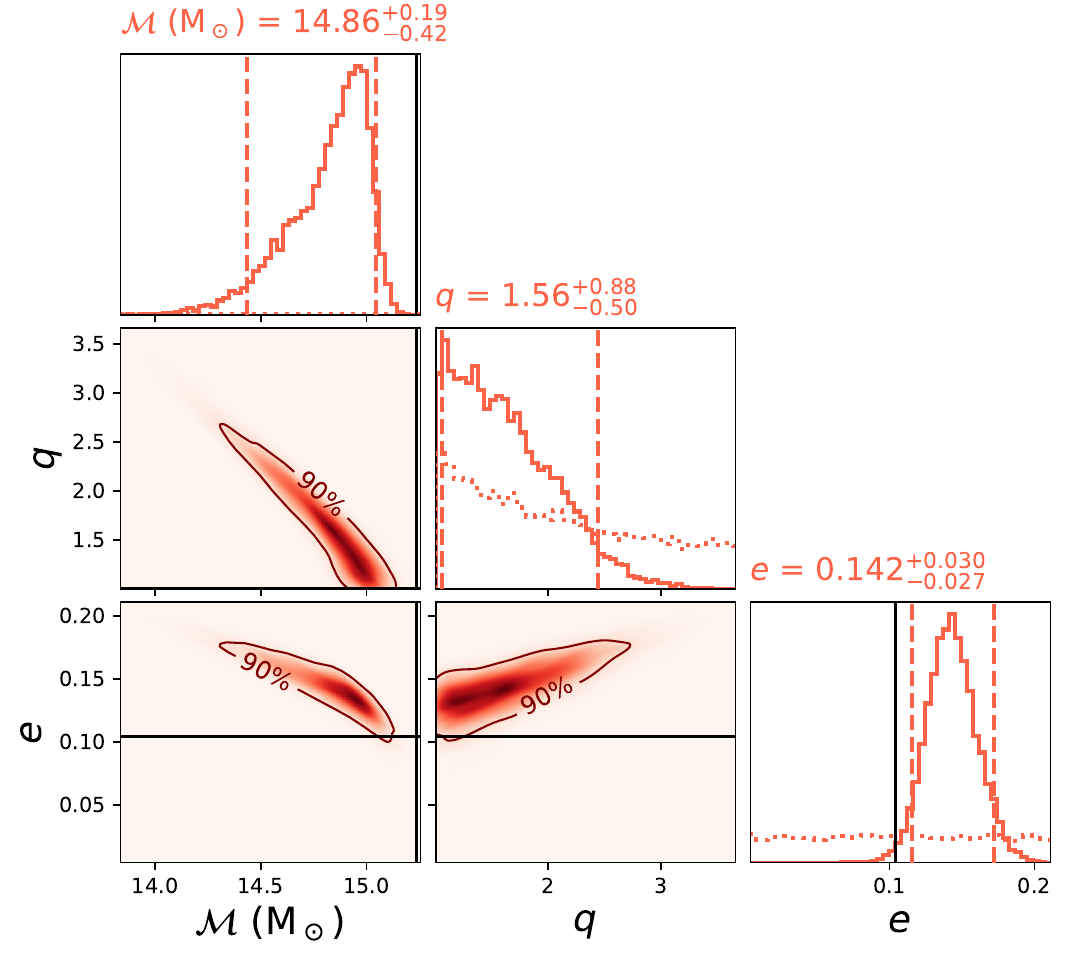}
    \caption[Corner plot showing $\mathcal{M}$, $q$, and $e_{20}$ posteriors, for $q=1$, non-spinning recovery from Fig.~\ref{fig:violin-ns-inj} performed using \texttt{TaylorF2Ecc}]{The corner plot for $q=1$ showing chirp mass ($\mathcal{M}$), mass ratio ($q$), and eccentricity ($e_{20}$) posteriors for the non-spinning recovery from Fig.~\ref{fig:violin-ns-inj} performed using \texttt{TaylorF2Ecc}. The histograms on the diagonal of the plot are 1D marginalized posteriors for the respective parameters with vertical dashed lines denoting $90\%$ credible intervals. The dotted curves show the priors used.}
    \label{fig:corner-ns-ecc-q-1}
\end{figure}

\begin{figure}[t!]
    \centering
    \includegraphics[trim=10 0 0 0, clip, width=0.8\linewidth]{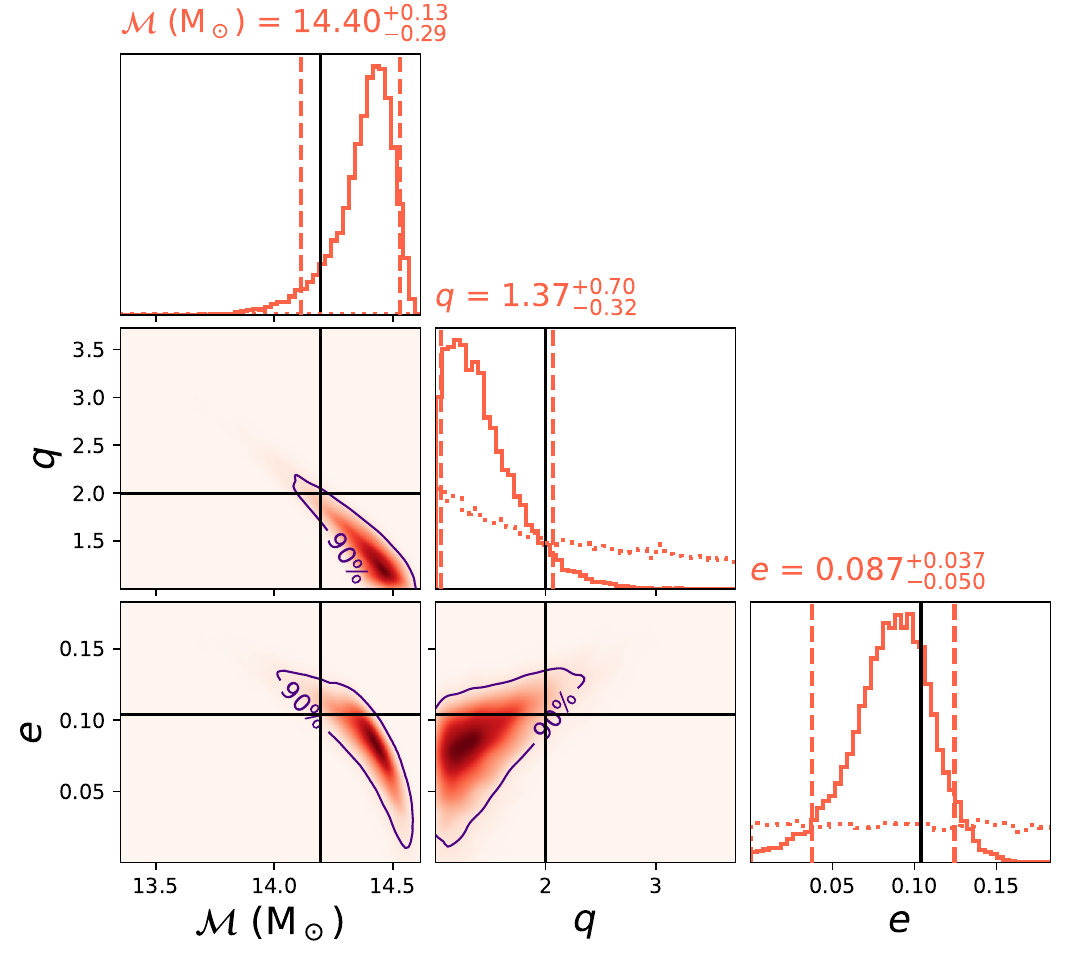}
    \includegraphics[trim=10 0 0 0, clip, width=0.8\linewidth]{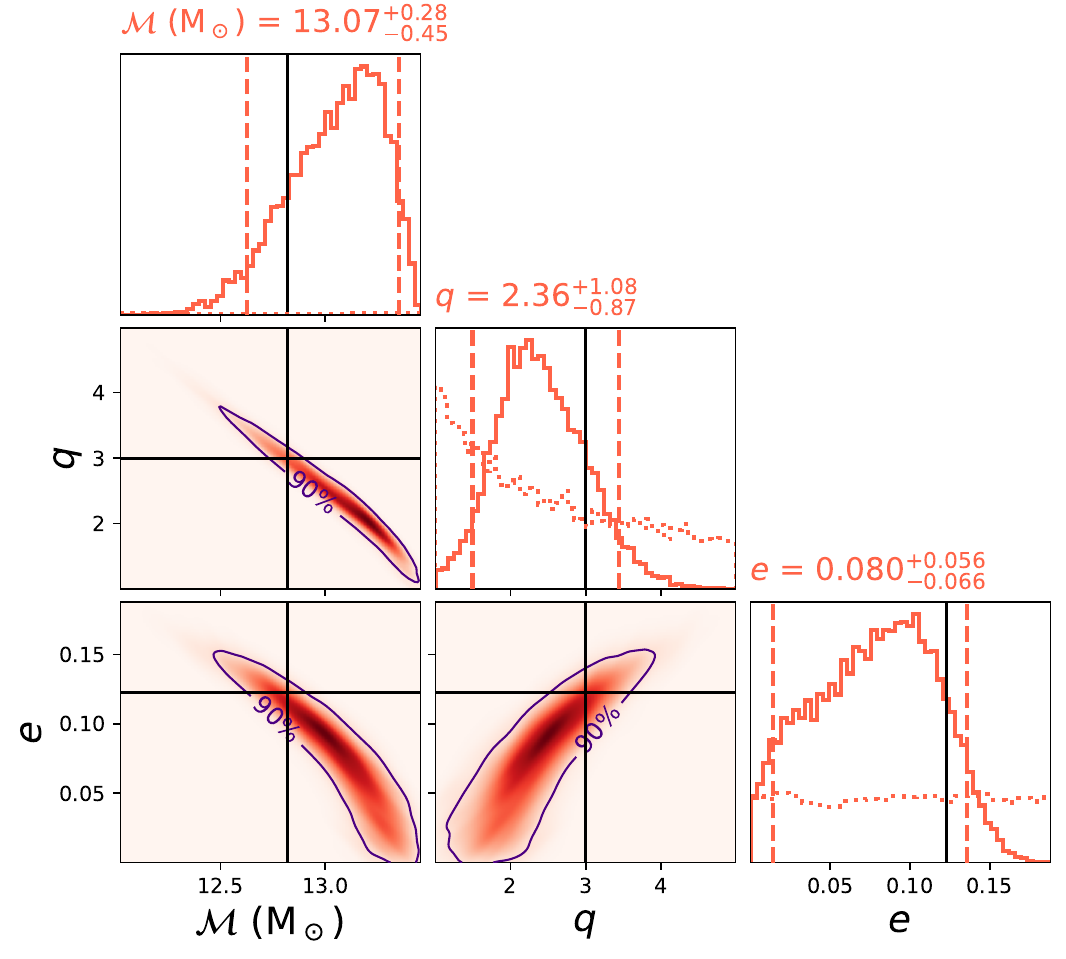}
    \caption{Same as Fig.~\ref{fig:corner-ns-ecc-q-1} but for $q=2$ (top) and $q=3$ (bottom).}
    \label{fig:corner-ns-ecc-q-2-3}
\end{figure}

We analyze the same injections as in the previous section with an eccentric inspiral-only waveform model \texttt{TaylorF2Ecc}. We present the results in figures \ref{fig:violin-ns-inj} and \ref{fig:ecc_ns_inj_q_1_2_3}. In Fig.~\ref{fig:violin-ns-inj}, we also show results obtained when the recovery is performed under the constraint $e_{20}=0$ with the same waveform, in order to account for any biases arising due to systematic differences between waveform model families and/or the lack of merger and ringdown in \texttt{TaylorF2Ecc}. Since \texttt{TaylorF2Ecc} is an inspiral-only model, we have truncated the likelihood integration using the quasi-circular waveform model to the same frequency as the eccentric recovery for a fair comparison and to get comparable SNRs. This frequency has been chosen to be $110$~Hz, close to the ISCO frequency for a $35$~M$_\odot$ system.

In the case of $q=(2,3)$ for eccentric injections (plotted in red in the Fig.~\ref{fig:violin-ns-inj}), a quasi-circular recovery excludes the injection value of chirp mass from the $90\%$ credible interval of the posterior when the recovery waveform has arbitrary spin constraints, whereas when eccentricity is included in the recovery waveform model and is sampled over, the injected value is recovered within the $90\%$ credible interval. We show the 1D posteriors for eccentricity in Fig.~\ref{fig:ecc_ns_inj_q_1_2_3}, and the 2D contours of $e_{20}$ with $\mathcal{M}$ and $q$ in Figs.~\ref{fig:corner-ns-ecc-q-1} and \ref{fig:corner-ns-ecc-q-2-3} which highlight the correlations between these parameters. We see that eccentricity shows negative and positive correlations with
$\mathcal{M}$ and $q$, respectively. The $\log$-Bayes factors between eccentric and quasi-circular recoveries performed with \texttt{TaylorF2Ecc} are close to $0$ for $q=2$ and $q=3$, but for $q=1$ it is $\sim11$, and thus favours recovery with an eccentric template when the injected waveform is eccentric. However, for $q=1$ even the quasi-circular recovery (NS) for quasi-circular injection (grey) is not recovered within $90\%$ confidence. Waveform systematics between the injected and recovered waveforms may partially cause this. Additionally, as shown in Fig.~\ref{fig:ecc_ns_inj_q_1_2_3}, the injected $e_{20}$ is not recovered within $90\%$ confidence for the eccentric injection. In addition to the waveform systematics, this is likely because the injected $q$ is at the lower boundary of the prior, so the entire $q$ posterior spans higher $q$ than injected, and (as shown in Fig.~\ref{fig:corner-ns-ecc-q-1}) higher $q$ correlates with higher $e_{20}$. To eliminate possible biases due to this prior effect, we use $q=1.25$ instead of $q=1$ in the following section.
\clearpage

\subsection{Dominant mode, aligned-spin, eccentric injections}
\label{subsec:align-spin-inj}

We inject an aligned-spin eccentric signal using the waveform model \texttt{TEOBResumS} \citep{Nagar:2018zoe} and recover in four configurations: quasi-circular aligned spin (AS), eccentric aligned spin (e-AS), eccentric non-spinning (e-NS), and quasi-circular precessing spin (PS). For the first three cases (AS, e-AS, and e-NS) we have used the waveform model \texttt{TaylorF2Ecc} for recovery, whereas for the last case (PS), we have used \texttt{IMRPhenomXP}. Since \texttt{TaylorF2Ecc} is an inspiral-only waveform, we truncate the likelihood calculation for recoveries with both waveforms at $110$~Hz in accordance with the choice of total mass as described above. As above, the total mass is $35 M_\odot$, and here we choose to inject signals with mass ratios $q=(1.25, 2, 3)$. The injected spin magnitudes are $\chi_\text{1z} = \chi_\text{2z} = 0.3$, hence $\chi_\text{eff} = 0.3$. The eccentric injections have $e_{20}=0.1$, consistent with injections in earlier sections, with the eccentricity defined at the orbit-averaged frequency of $20$ Hz \citep{TEOBResumS:bitbucket}. The chirp mass posteriors are plotted in the form of violin plots in Fig.~\ref{fig:violin-as-inj}. For arbitrary spin settings, the quasi-circular waveform models are not able to correctly recover the chirp mass of the eccentric injection, whereas the eccentric waveform model with aligned spins correctly recovers the injected value. 
This once again indicates that a quasi-circular precessing waveform model can cause biases in the recovered value of chirp mass if the signal is truly eccentric. Further, looking at figures \ref{fig:corner-ps-q-1.25-2-as-inj} and \ref{fig:corner-ps-q-3-as-inj}, we note that the $\chi_p$ posteriors peak at the same value for both eccentric and quasi-circular injections. This again suggests that an aligned spin eccentric signal when recovered with quasi-circular precessing model does not mimic a spin-precessing signal any more than a quasi-circular aligned-spin injection.

\begin{figure}[p!]
    \centering
    \includegraphics[trim=10 0 10 10, clip, width=0.81\linewidth]{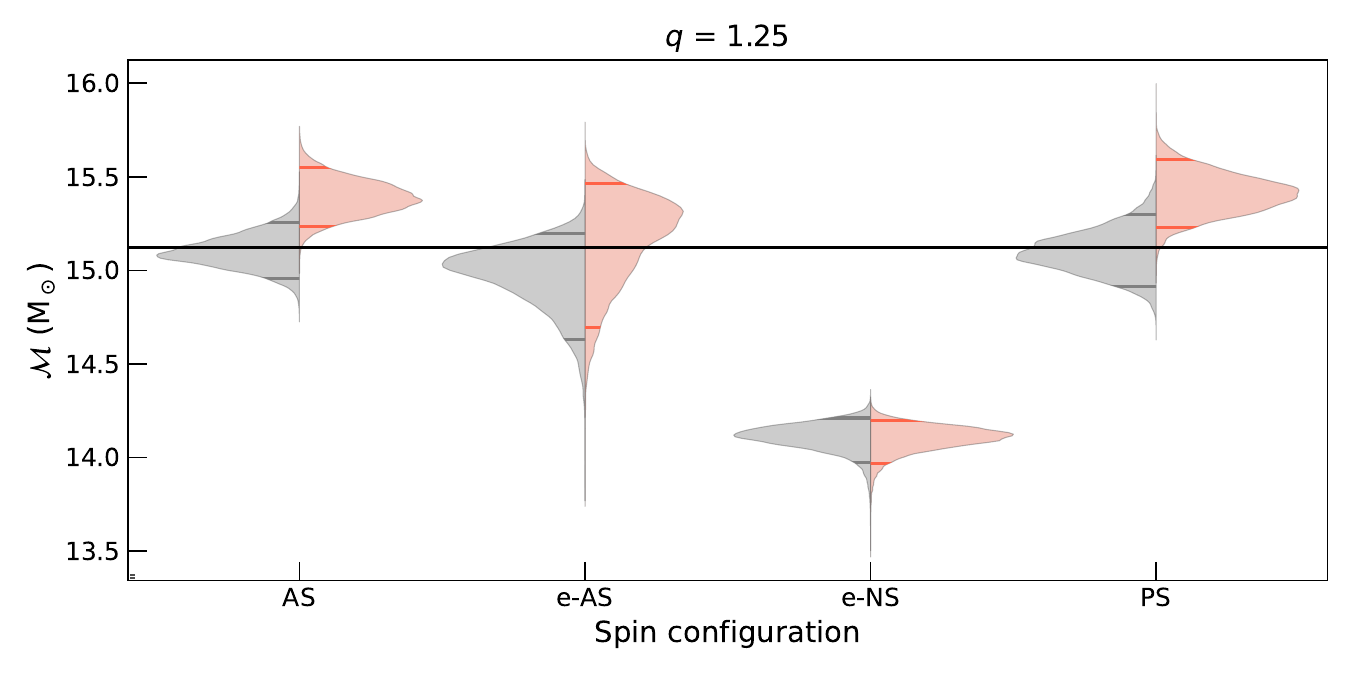}
    \includegraphics[trim=10 0 10 10, clip, width=0.81\linewidth]{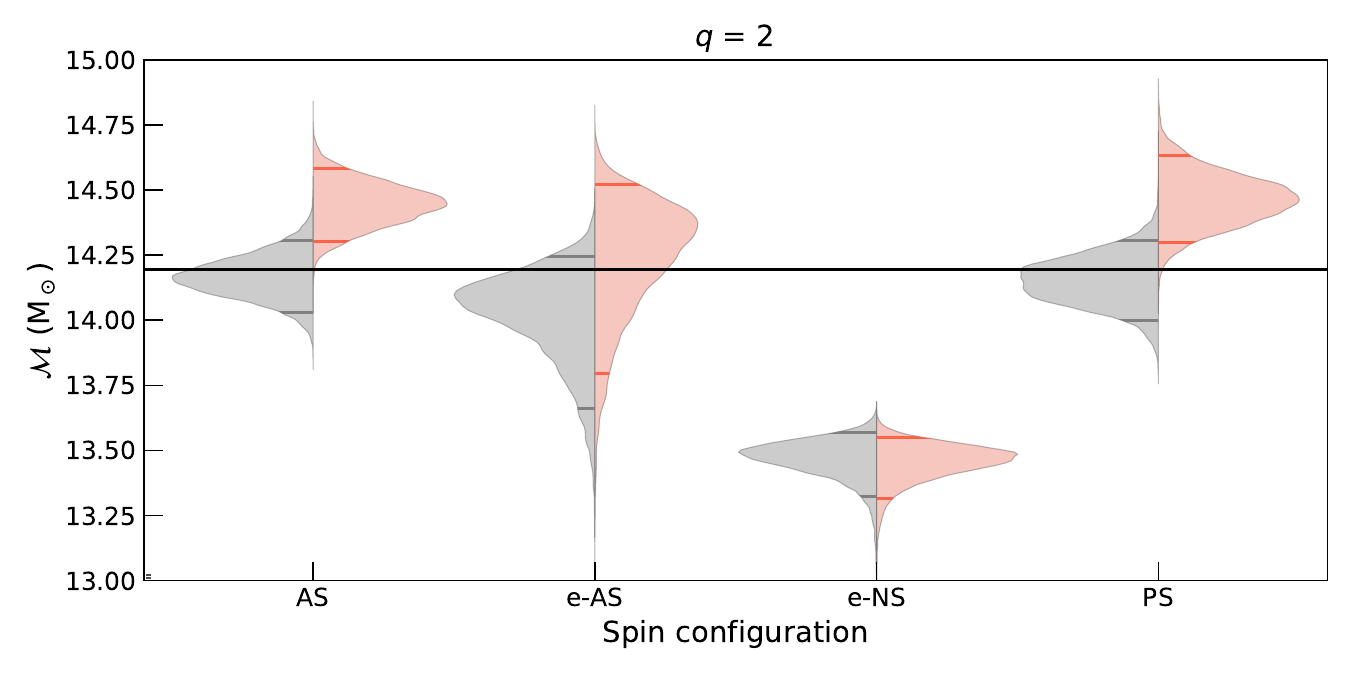}
    \includegraphics[trim=10 0 10 10, clip, width=0.81\linewidth]{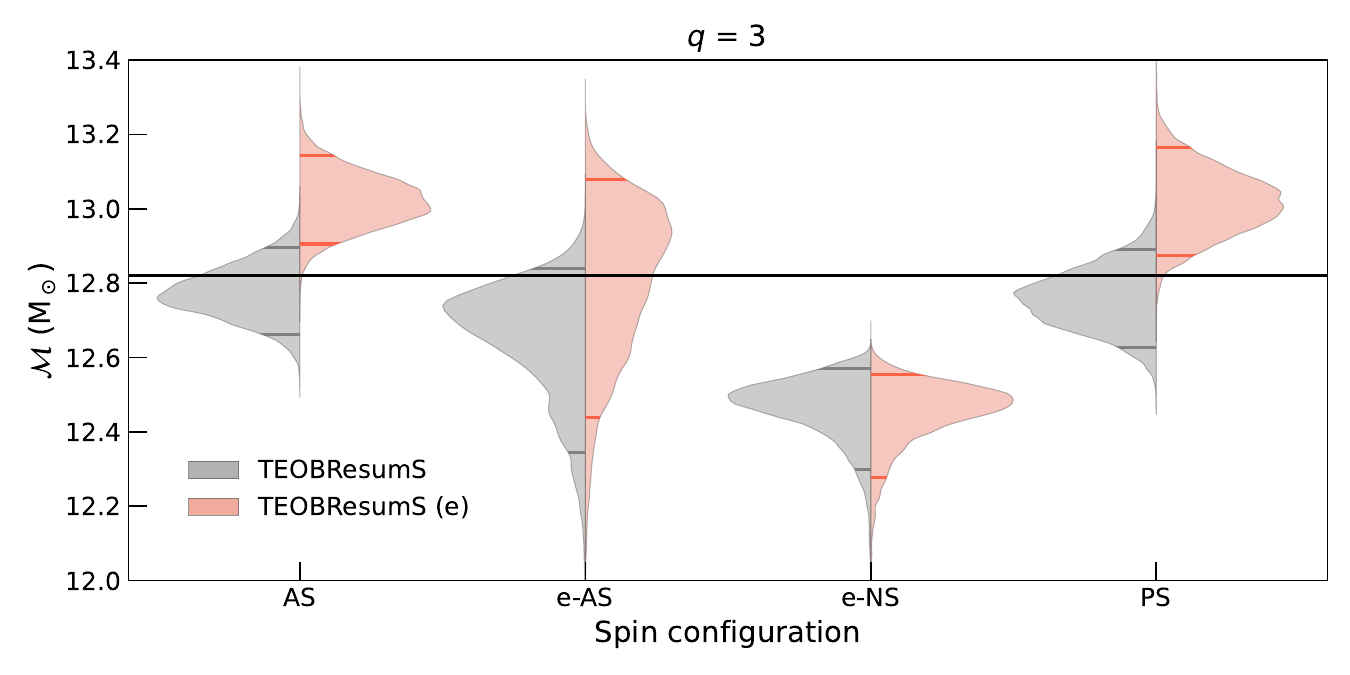}
    \caption[Eccentric and quasi-circular aligned-spin injections, recovered in quasi-circular aligned-spin (AS), eccentric aligned-spin (e-AS), and eccentric non-spinning (e-NS), and quasi-circular precessing-spin (PS) configurations]{Eccentric ($e_{20}=0.1$) (red) and quasi-circular (grey) aligned-spin injections generated with eccentric waveform model \texttt{TEOBResumS}, recovered with \texttt{TaylorF2Ecc} in quasi-circular aligned-spin (AS), eccentric aligned-spin (e-AS), and eccentric non-spinning (e-NS) configurations, and with quasi-circular waveform model \texttt{IMRPhenomXP} including precessing spin (PS). The aligned-spin eccentric injection is only correctly recovered using the e-AS configuration in the recovery waveform. The matched filter SNRs for $q=1.25$, $q=2$ and $q=3$ are $35$, $33$ and $30$ respectively.}
    \label{fig:violin-as-inj}
\end{figure}

\begin{figure}[t!]
    \centering
    \includegraphics[trim=10 0 0 0, clip, width=0.79\linewidth]{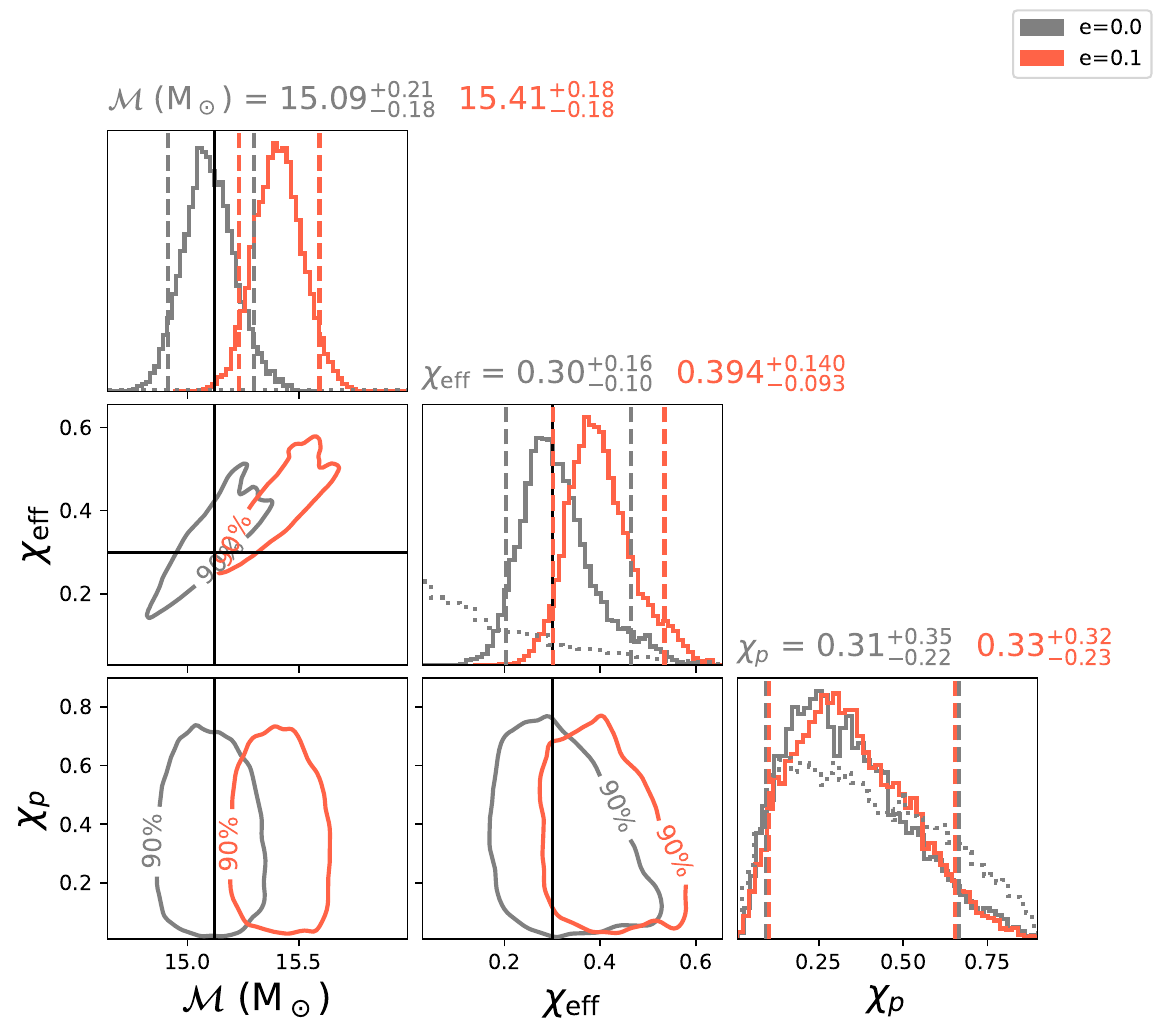}
    \includegraphics[trim=10 0 0 0, clip, width=0.79\linewidth]{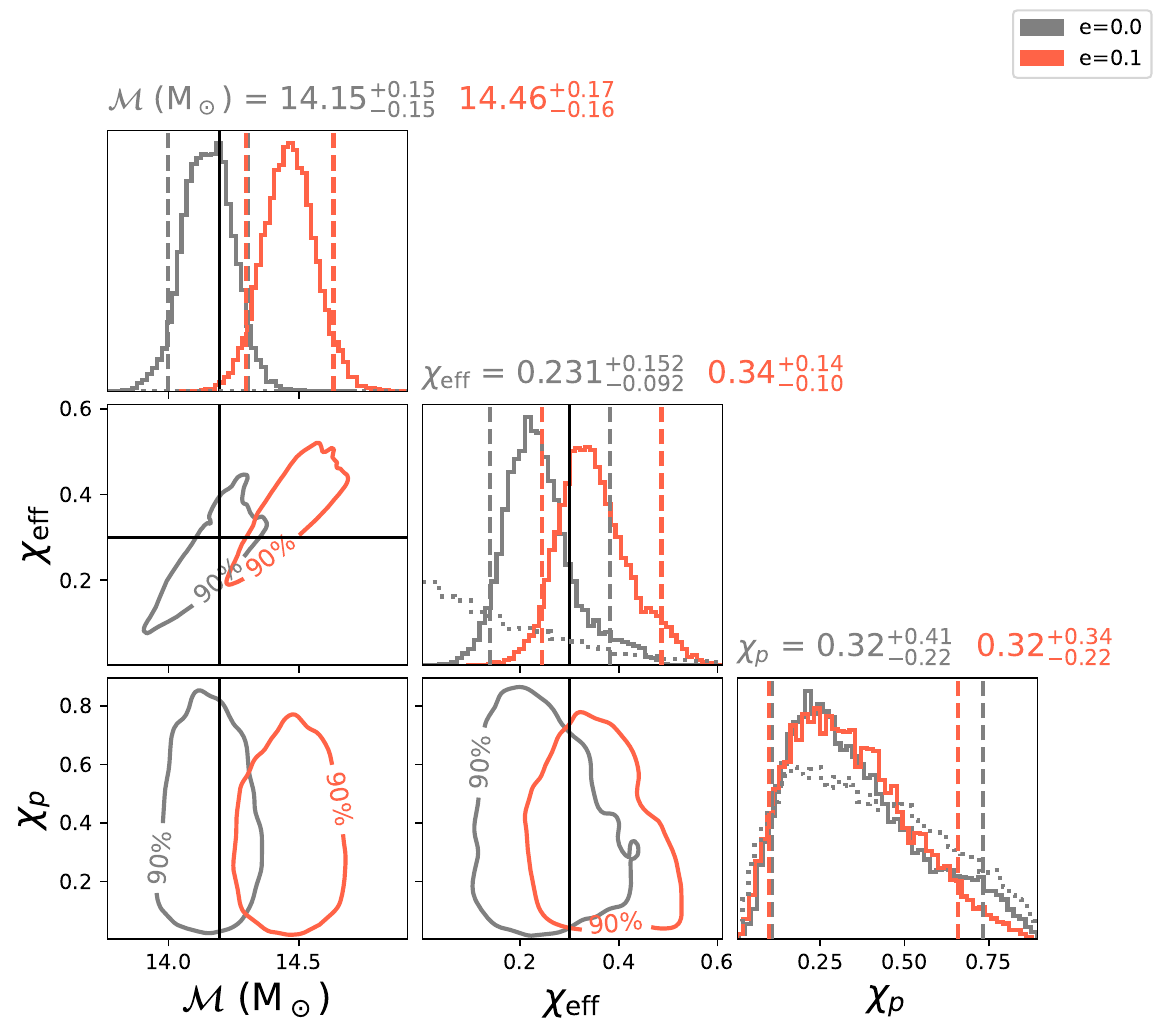}
    \caption{Same as Fig.~\ref{fig:corner-ps-q-2}, but for aligned-spin injections with mass ratios $q=1.25$ (top) and $q=2$ (bottom).}
    \label{fig:corner-ps-q-1.25-2-as-inj}
\end{figure}

\begin{figure}[t!]
    \centering
    \includegraphics[trim=10 0 0 0, clip, width=0.85\linewidth]{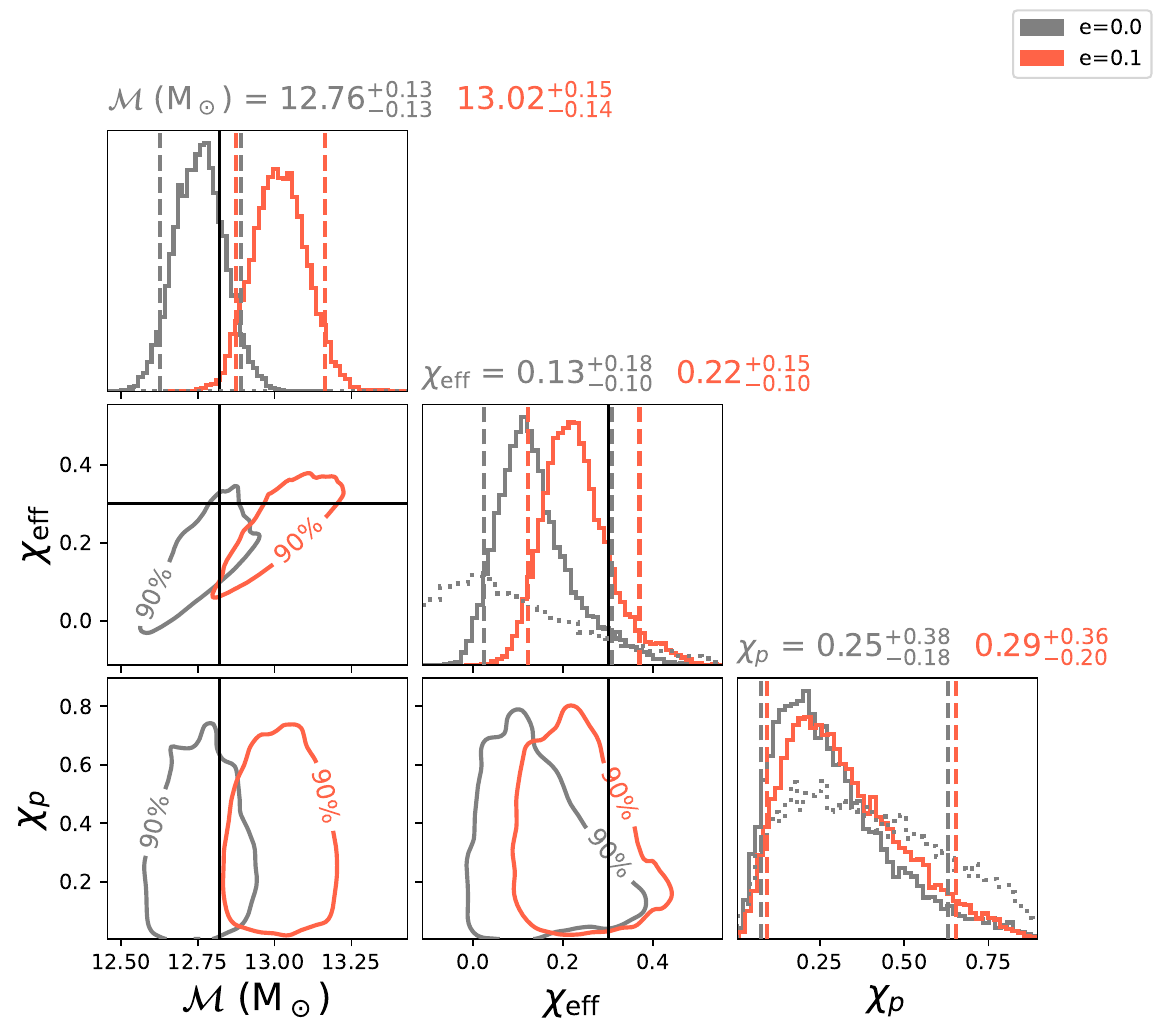}
    \caption{Same as Fig.~\ref{fig:corner-ps-q-1.25-2-as-inj} but for $q=3$.}
    \label{fig:corner-ps-q-3-as-inj}
\end{figure}

\begin{figure}[t!]
    \centering
    \includegraphics[width=\linewidth]{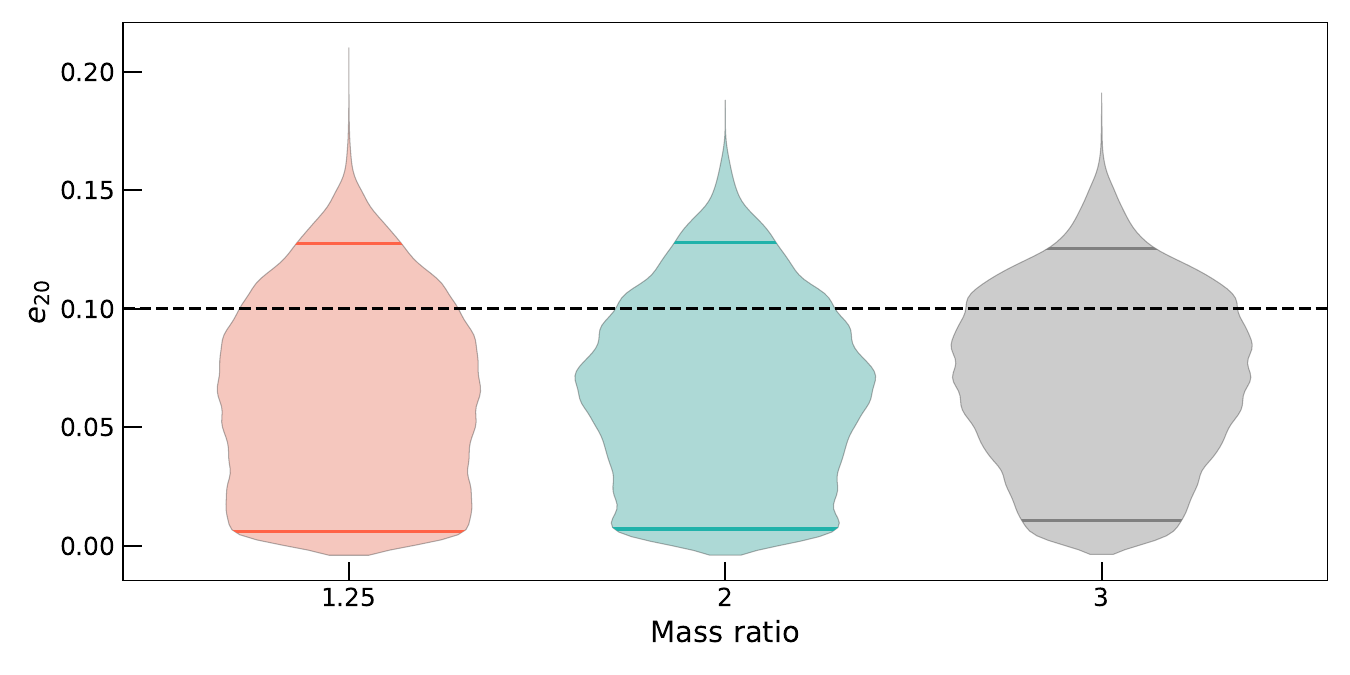}
    \caption[Eccentricity posteriors for $q=(1.25, 2, 3)$ when the injections are aligned-spin and eccentric, and the recovery is with \texttt{TaylorF2Ecc} in eccentric, aligned-spin configuration]{Eccentricity posteriors for $q=(1.25, 2, 3)$ when the injections are aligned-spin and eccentric, and the recovery is with \texttt{TaylorF2Ecc} in eccentric, aligned-spin configuration. The coloured lines inside the shaded posteriors indicate 90\% credible interval whereas the black dashed lines denote the injected eccentricity value.}
    \label{fig:ecc_as_inj_q_1.25_2_3}
\end{figure}

\begin{figure}[t!]
    \centering
    \includegraphics[trim=10 0 0 0, clip, width=\linewidth]{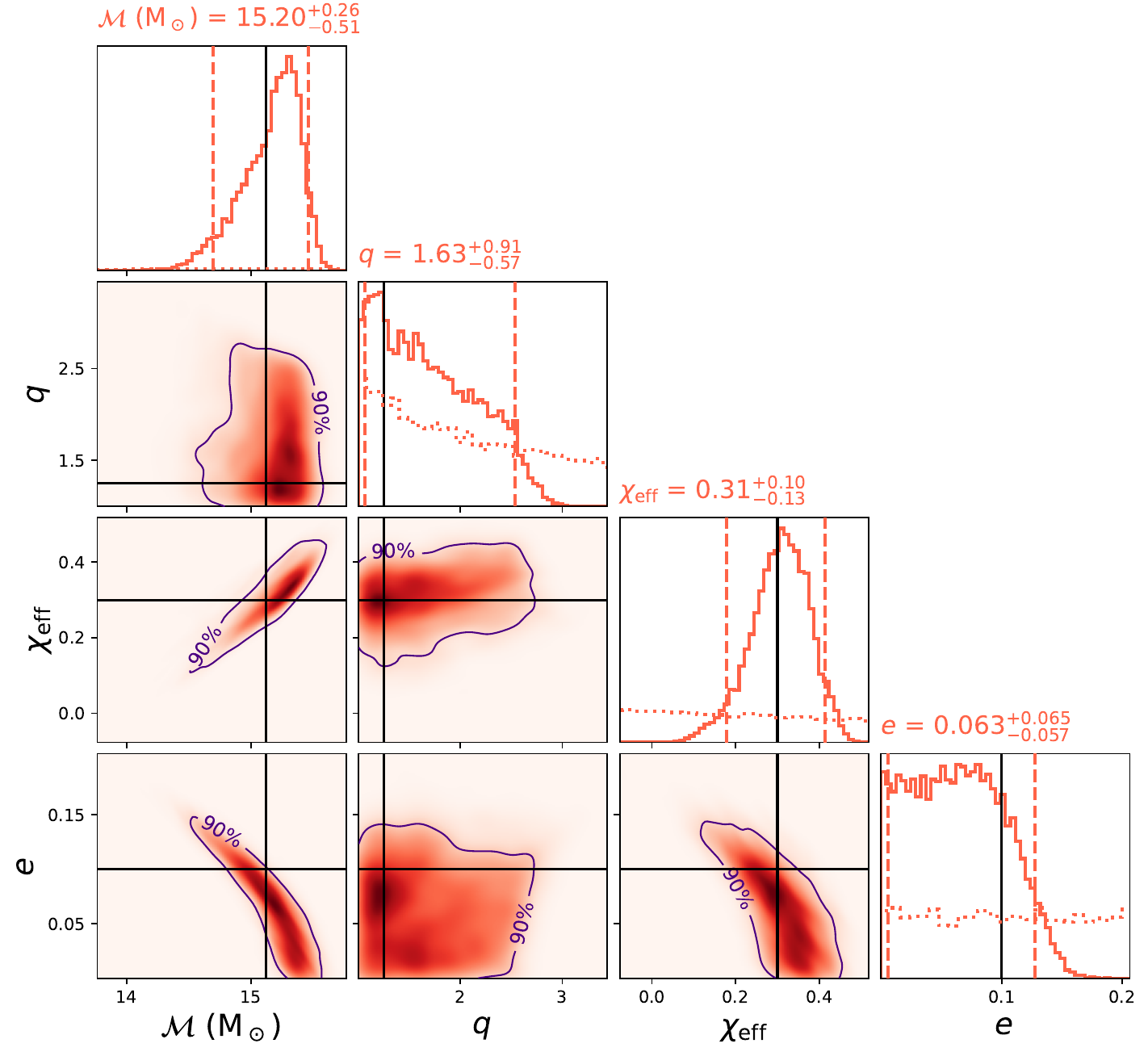}
    \caption[Corner plot of $\mathcal{M}$, $q$, $\chi_\text{eff}$, and $e$, for $q=1.25$ aligned spin recovery from Fig.~\ref{fig:violin-as-inj} performed using \texttt{TaylorF2Ecc}]{The corner plot of chirp mass ($\mathcal{M}$), mass ratio ($q$), effective spin parameter ($\chi_\text{eff}$), and eccentricity ($e$), for $q=1.25$ aligned spin recovery from Fig.~\ref{fig:violin-as-inj} performed using \texttt{TaylorF2Ecc}. The histograms on the diagonal of the plot are 1D marginalized posteriors for the respective parameters with vertical dashed lines denoting $90\%$ credible intervals. The dotted curves show the priors used.}
    \label{fig:corner-as-ecc-q-1.25}
\end{figure}

\begin{figure}[t!]
    \centering
    \includegraphics[trim=10 0 0 0, clip, width=0.8\linewidth]{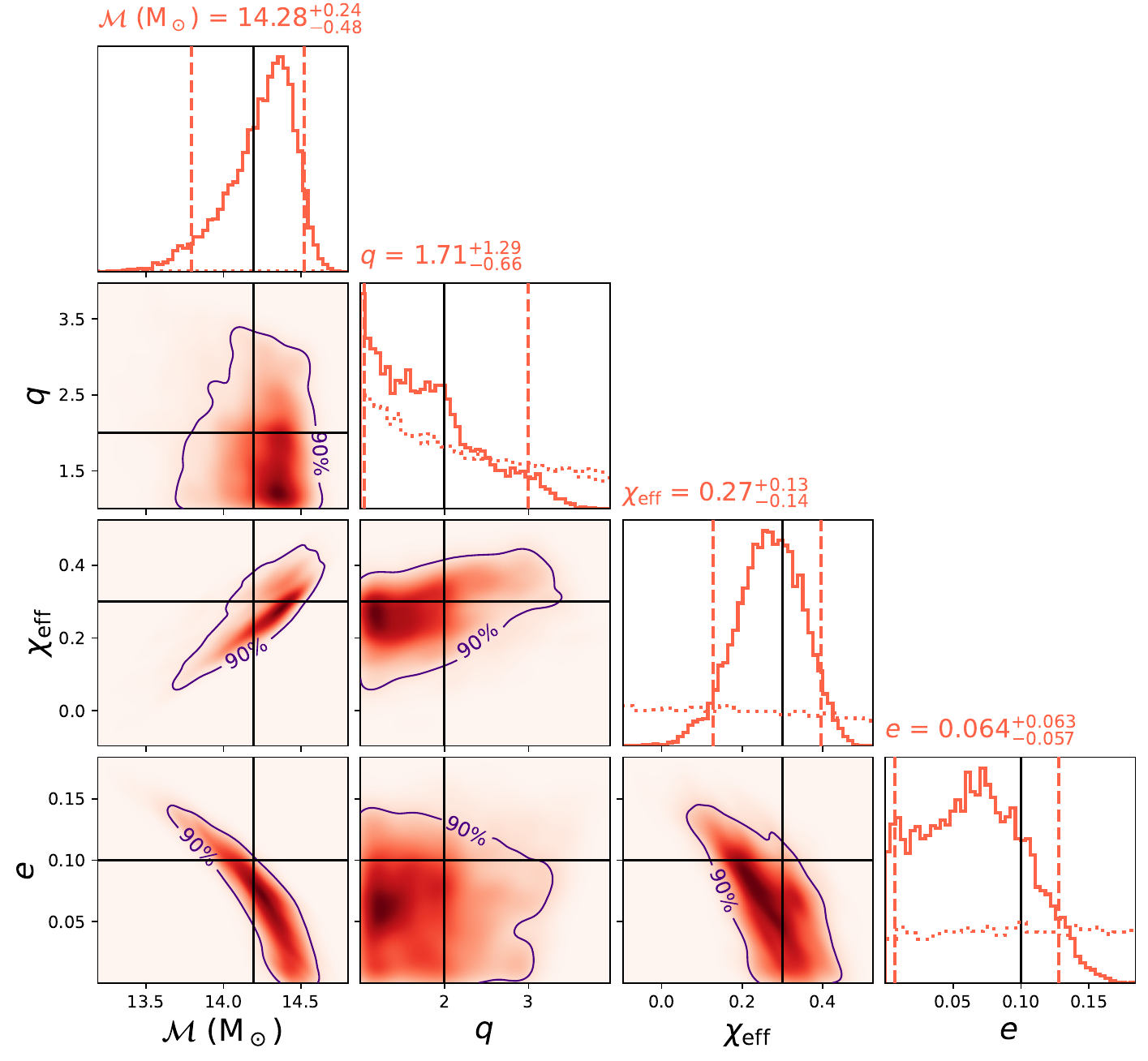}
    \includegraphics[trim=10 0 0 0, clip, width=0.8\linewidth]{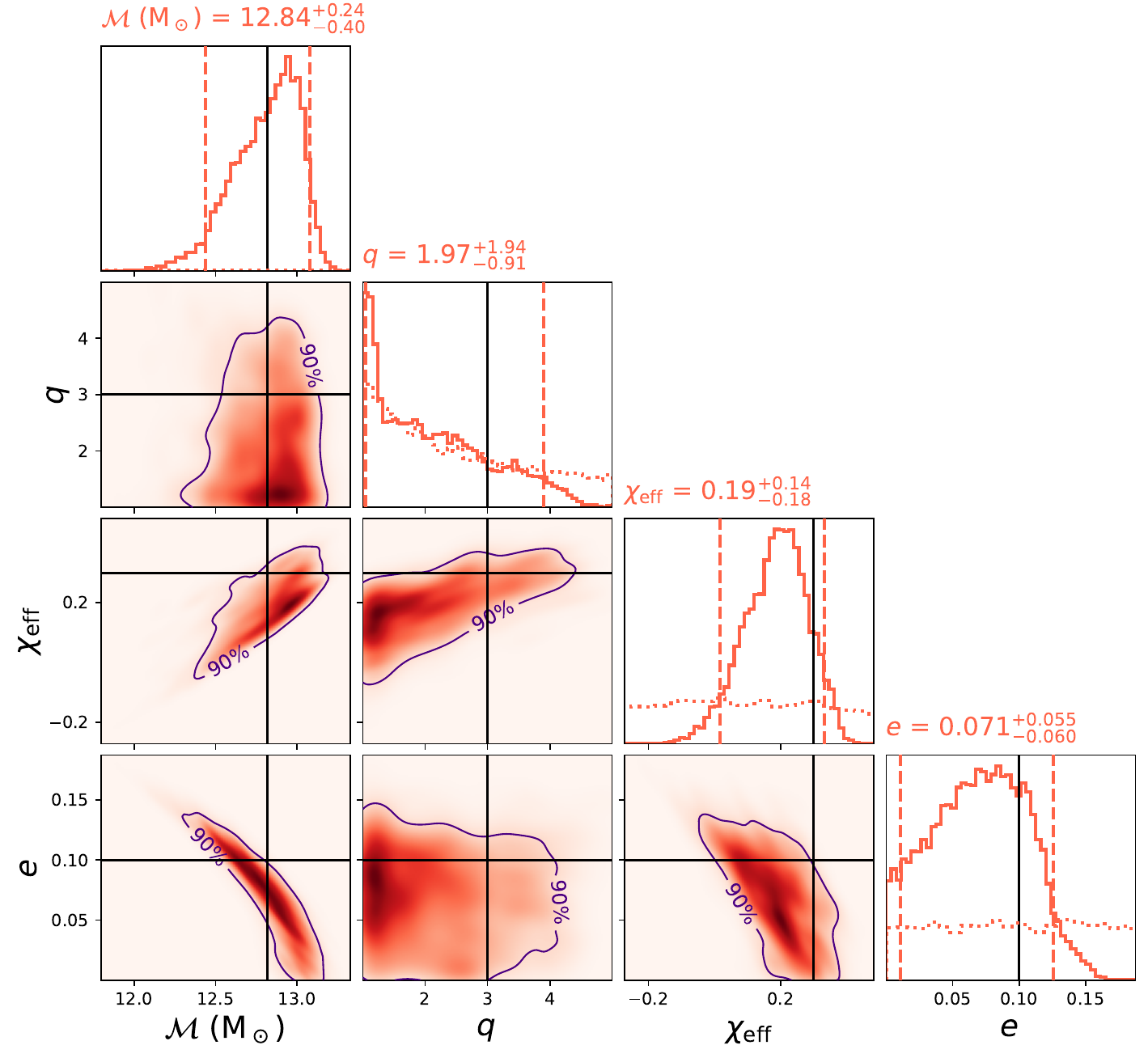}
    \caption{Same as Fig.~\ref{fig:corner-as-ecc-q-1.25} but for $q=2$ (top) and $q=3$ (bottom).}
    \label{fig:corner-as-ecc-q-2-3}
\end{figure}

We also analyse the same injection with the eccentric waveform model with zero spins (shown as e-NS in the Fig.~\ref{fig:violin-as-inj}). In this case, the posteriors for both the quasi-circular and eccentric injections are biased towards lower masses than injected. Noting that a positively aligned spin system has more cycles compared to its non-spinning counterpart (which is also the case for lower-mass systems), when an aligned-spin signal is recovered using a non-spinning waveform model, it is naturally biased towards lower masses. This bias in the chirp mass also causes a bias in the eccentricity posterior, resulting in a higher value of eccentricity due to a negative correlation with chirp mass. This can result in posteriors peaking at non-zero values of eccentricities when an aligned spin quasi-circular signal is analysed with a non-spinning eccentric model. Hence, an eccentric and spinning (aligned) system may be recovered with a positive bias in chirp mass when eccentricity is ignored, and a negative bias when the spins are ignored. Again, we provide the eccentricity posteriors in Fig \ref{fig:ecc_as_inj_q_1.25_2_3} as well as in the form of corner plots along with $\mathcal{M}$, $q$, and $\chi_\text{eff}$ parameters in figures \ref{fig:corner-as-ecc-q-1.25} and \ref{fig:corner-as-ecc-q-2-3}. We see a clear correlation between $e_{20}$ and $\mathcal{M}$, and a mild correlation between $\chi_\text{eff}$ and $e_{20}$, consistent with the findings of \citet{OShea:2021faf}. Also, as in Sec.~\ref{subsec:non-spin-inj}, we find that the Bayes factors ($\mathcal{B}_{E/C}$) for $q=1.25$, $2$, and $3$ are not high enough to indicate a clear preference for either the quasi-circular or eccentric waveform model for any injection.
\clearpage

\subsection{Simulated noise injections: dominant mode, non-spinning and eccentric}
\label{sec:noisy_injs}

\begin{figure}[t!]
    \centering
    \includegraphics[trim=40 0 30 20, clip, width=0.49\linewidth]{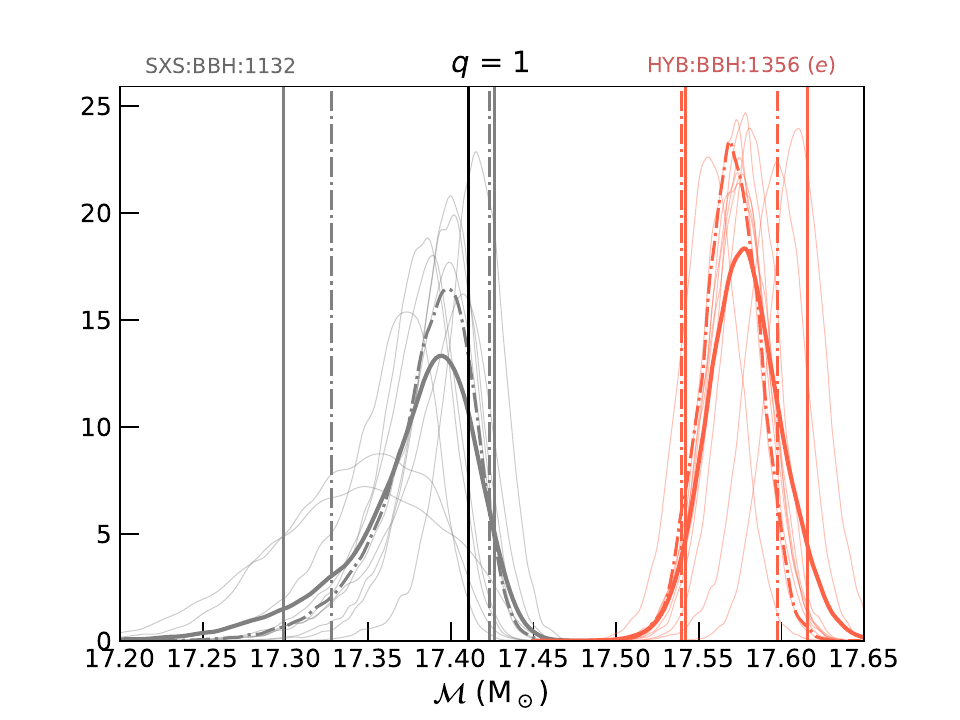}
    \includegraphics[trim=40 0 30 20, clip, width=0.49\linewidth]{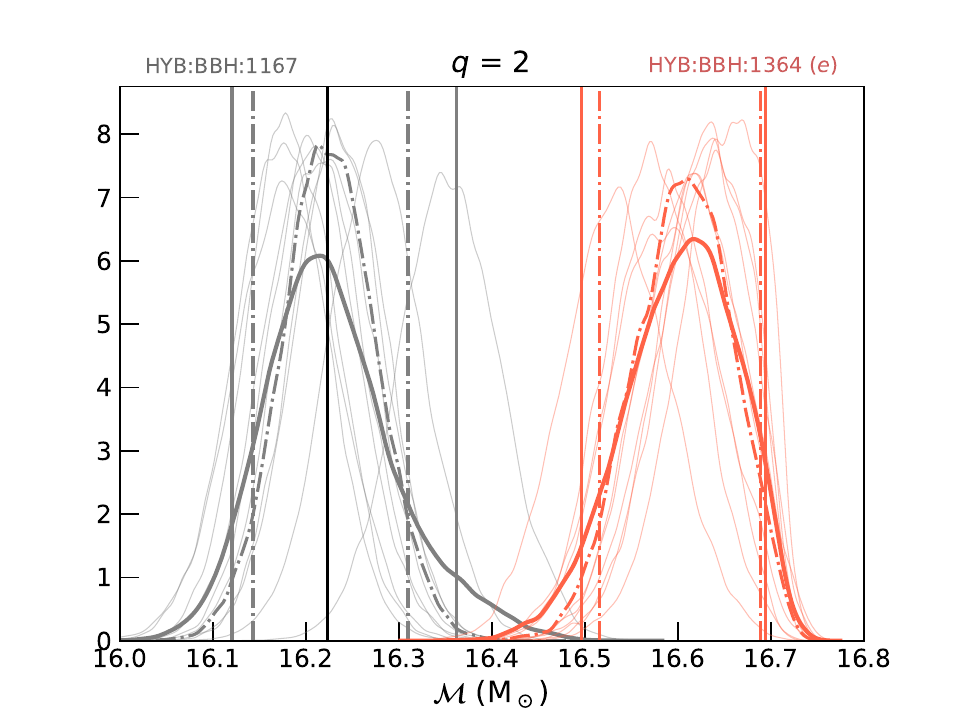}
    \includegraphics[trim=40 0 30 20, clip, width=0.49\linewidth]{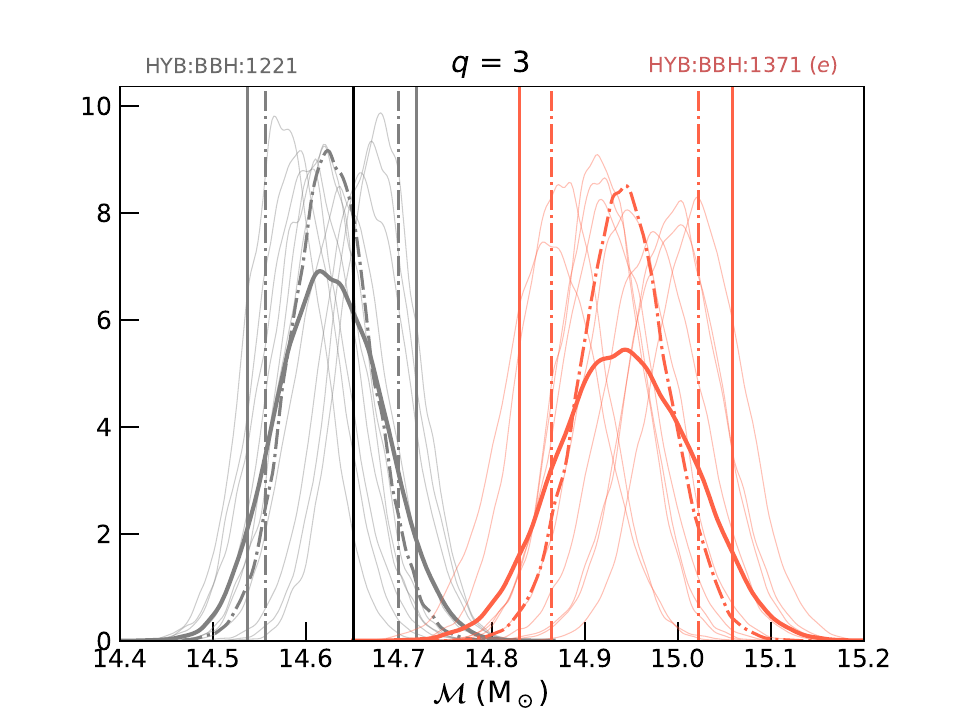}
    \caption[The chirp mass posteriors, for injections with mass ratios $q=1,2,3$, compared between eccentric and quasi-circular non-spinning injections, all recovered using \texttt{IMRPhenomXAS} with zero spins]{The chirp mass posteriors, for injections with mass ratios $q=1,2,3$, compared between eccentric (red) and quasi-circular (grey) non-spinning injections, all recovered using \texttt{IMRPhenomXAS} with zero spins. The faint lines depict injections in different noise realisations, and the dark solid curve is the combined posterior. We have also plotted the zero-noise injection as dashed curve. The coloured vertical lines represent the $90\%$ credible interval of the combined posterior of the same colour as the line, and the black line denotes the injected value.}
    \label{fig:noise-inj}
\end{figure}

We perform a set of injection recoveries with Gaussian noise simulated using the power spectral densities (PSDs) of the detectors. The total mass of the injected BBH systems is $40 M_\odot$ and the mass ratios are $q=1, 2, 3$, with the slightly heavier mass chosen to increase the computational efficiency of the analysis. These are recovered using quasi-circular waveform \texttt{IMRPhenomXAS} with zero spins. The results can be seen in Fig.~\ref{fig:noise-inj}. For each case, we have taken $10$ noise realizations, which each correspond to the posteriors shown by thin curves in the plot. An equal number of samples were taken from each of these runs and combined to form the average posterior shown by the thick coloured curve. We also perform a zero-noise injection for each mass ratio case, which is shown by the dot-dashed curve in the plot. The vertical coloured lines denote $90\%$ credible interval and the black line shows the injected value. As can be seen, the average posterior of all the noisy injections agrees well with the zero-noise curve for each case.

\subsection{Higher mode, non-spinning, eccentric injections}
\label{sec:hm_injs}

\begin{table}[t!]
\centering
\def\arraystretch{1.3}
\begin{tabular}{|c|c|c|c|}
\hline
\textbf{S. No.} &
  \textbf{\begin{tabular}[c]{@{}c@{}}Injection Simulation ID / \\ Waveform\end{tabular}} &
  \hspace{5pt}$\mathbf{q}$\hspace{5pt} &
  \textbf{$\mathbf{e_{20}}$} \\ \hline
1  & SXS:BBH:1132     & 1 & 0 \\ \hline
2  & HYB:SXS:BBH:1356 & 1 & 0.121 \\ \hline
3  & HYB:SXS:BBH:1167 & 2 & 0.0 \\ \hline
4  & HYB:SXS:BBH:1364 & 2 & 0.089 \\ \hline
5  & HYB:SXS:BBH:1221 & 3 & 0.0 \\ \hline
6  & HYB:SXS:BBH:1371 & 3 & 0.105 \\ \hline
\end{tabular}
\caption[List of non-spinning, eccentric NR hybrid simulations \citep{Chattaraj:2022tay} used for HM injections]{List of non-spinning, eccentric NR hybrid simulations (constructed in \cite{Chattaraj:2022tay}) used for HM injections. Columns include a unique hybrid ID for each simulation (SXS IDs are retained for identification with SXS simulations used in constructing the hybrids), information concerning the mass ratio ($q=m_1/m_2$), eccentricity ($e_{20}$) at the reference frequency of $20$~Hz for a total mass of $M=40$ M$_\odot$.} 
\label{table:hybrids-hm}
\end{table}

\begin{figure}[p!]
    \centering
    \includegraphics[width=0.81\linewidth]{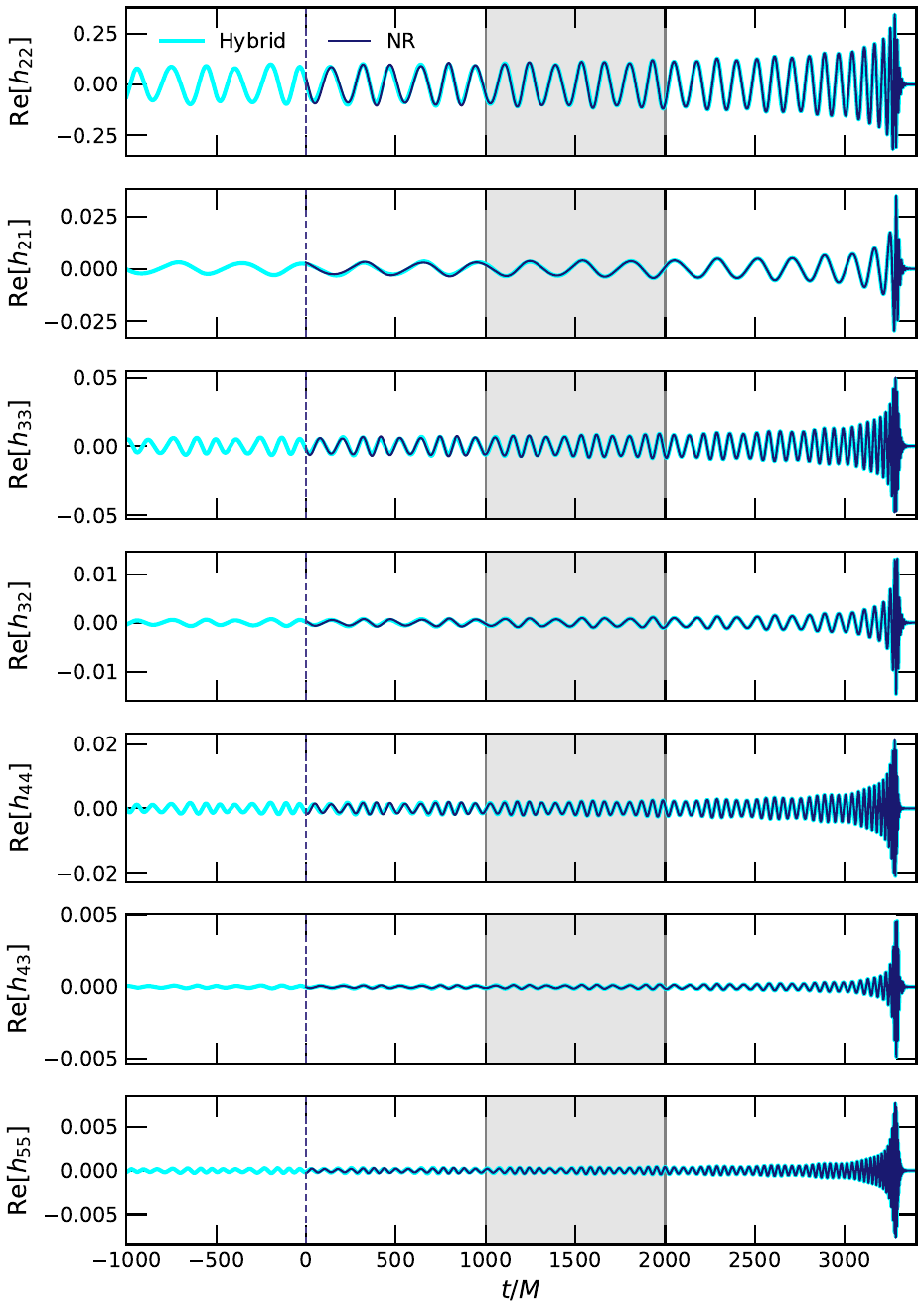}
    \caption[PN-NR hybrid waveform corresponding to NR simulation SXS:BBH:1364, an asymmetric mass binary with mass ratio q=2]{\takenfrom{\cite{Chattaraj:2022tay}} PN-NR hybrid waveform corresponding to NR simulation SXS:BBH:1364, an asymmetric mass binary with mass ratio $q=2$. The initial eccentricity of the constructed hybrid is $e_0 = 0.108$ at the dimensionless orbital averaged frequency of $x_0 = 0.045$. This frequency parameter, $x$, is related to the total mass ($M$) and gravitational wave frequency ($f_\text{GW}$) as $x = (\pi M f_\text{GW})^{2/3}$. The blue dotted line marks the beginning of the NR waveform and the shaded grey region $t \in (1000M, 2000M)$ shows the matching window where hybridization was performed. Overlapping hybrid and NR waveforms on the left of the matching window hint at the quality of hybridization performed here.}
    \label{fig:NR-hyb-EccTD}
\end{figure}

\begin{figure}[t!]
      \centering
      \includegraphics[trim=30 0 30 10, clip, width=0.49\linewidth]{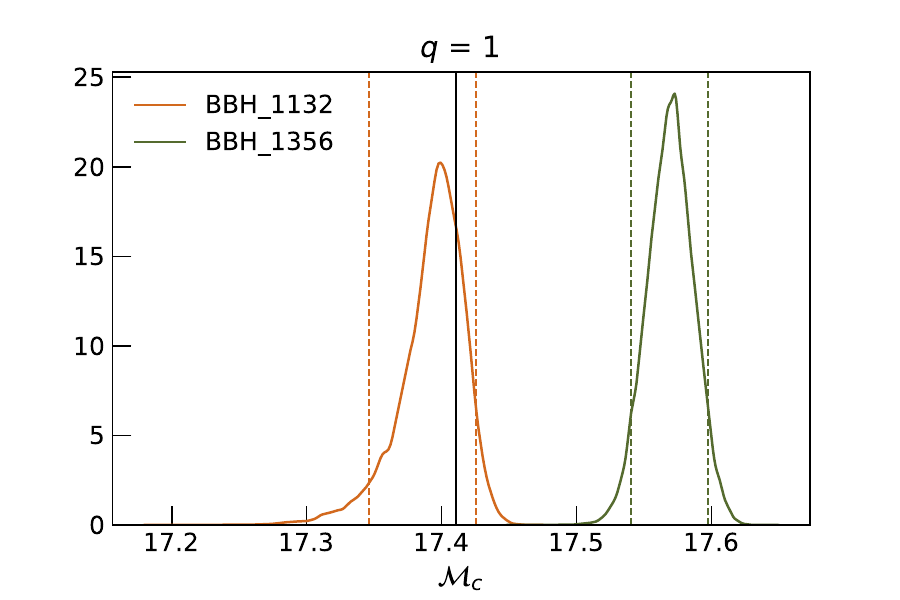}
      \includegraphics[trim=30 0 30 10, clip, width=0.49\linewidth]{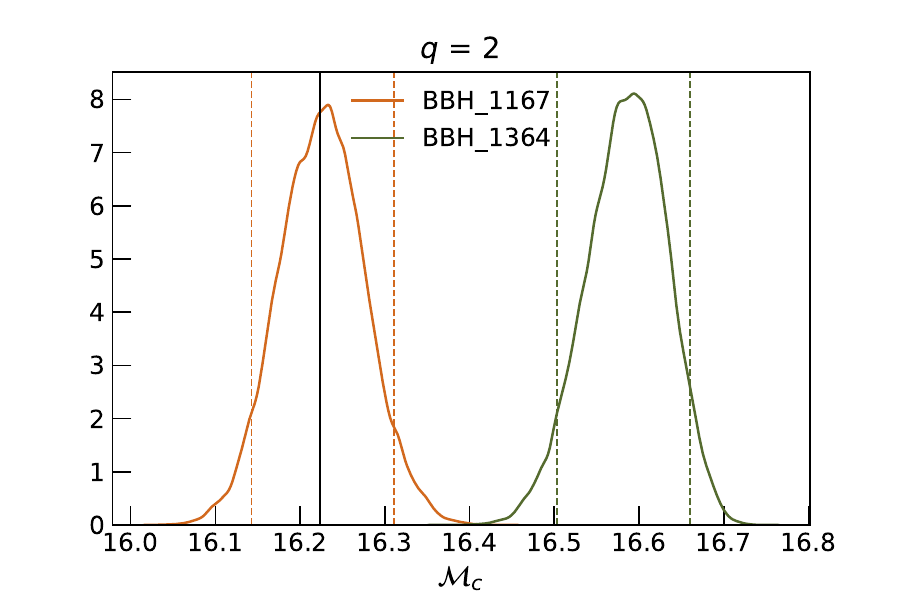}
      \includegraphics[trim=30 0 30 10, clip, width=0.49\linewidth]{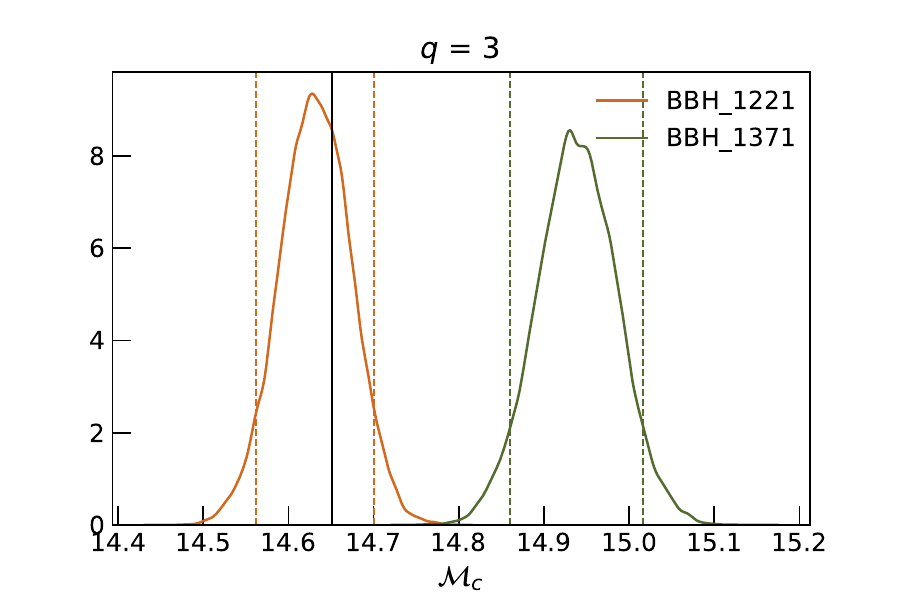}
      \caption[Chirp-mass recovery of circular and eccentric injections using quasi-circular higher mode waveforms]{Chirp-mass recovery of circular and eccentric injections using quasi-circular, higher mode waveforms. Thick black lines denote the injected value of the chirp-mass parameter and the dashed lines denote 90\% credible intervals. The orange (olive) posteriors denote measurement of the chirp mass for the injected circular (eccentric) signal. Both circular and eccentric injections correspond to a BBH of total mass $40 M_{\odot}$, and the eccentric simulations have an orbital eccentricity of $e_{20}\sim0.1$. Non-recovery of the injected value of the chirp mass for the eccentric case can be interpreted as the bias induced due to the neglect of eccentricity and associated higher modes.
      }
    \label{fig:hist-hm}
  \end{figure}

In this section, we perform injection studies using the higher modes of the eccentric, non-spinning hybrids (one such hybrid is shown in Fig.~\ref{fig:NR-hyb-EccTD}) to assess the biases that are introduced when higher mode, quasi-circular waveforms are used to recover these. 
Figure~\ref{fig:hist-hm} shows 90\% error bounds in the measurement of the binary's chirp mass for mass ratios $q=(1, 2, 3)$. The total mass of injected signals is assumed to be fixed at 40$M_{\odot}$. The value of eccentricity at 20 Hz (starting frequency of the analysis) is $e_{20} \sim 0.1$ (see Table \ref{table:hybrids-hm} for exact values). The injections here include ($\ell$, $|m|$)=(2, 2), (3, 3), (4, 4), (2, 1), and (3, 2) modes. These are then recovered using quasi-circular higher mode waveform \texttt{IMRPhenomXHM} \citep{Garcia-Quiros:2020qpx}, which contains all the above-mentioned modes. Additional $(4, 3)$ and $(5, 5)$ modes present in the hybrids have been dropped to have the same set of modes in injection and the recovery waveforms to avoid misinterpretation of the results. The aim here is to observe if the bias in the parameter estimates is purely due to the combined impact of eccentricity and associated higher-order modes.
Vertical black lines in the figure indicate the injected parameter values, while the recovery is shown by posteriors with 90\% error bounds (vertical dashed lines). The posteriors in orange denote injections with circular simulations (SXS:BBH:1132, HYB:BBH:1167, HYB:BBH:1221) while those in olive green denote injections with eccentric simulations (HYB:BBH:1356, HYB:BBH:1364, HYB:BBH:1371).

It can be seen in Fig.~\ref{fig:hist-hm} that eccentric injections are not recovered with the quasi-circular waveform. This indicates that the presence of residual eccentricity of the order $e_{20}\sim0.1$ and eccentricity-induced corrections to the higher modes in systems entering the ground-based detectors will lead to significant biases in recovering source parameters. This observation should motivate including the effect of orbital eccentricity in dominant and other higher modes in waveforms from compact binary mergers. Recovery of other relevant parameters by the means of corner plots for all three mass ratio cases is displayed in Figs.~\ref{fig:corner_hm_q_1}-\ref{fig:corner_hm_q_3}. This helps us understand correlations between different parameters.

We see correlations among various mass parameters ($\mathcal{M}_c$, $M$, and $\eta$) and strong correlations between luminosity distance $d_L$ and inclination angle ($\iota$).
Again, $\eta$ has an upper limit of $\eta\leq0.25$ (by definition), thus for $q$=$1$, since $\eta=0.25$, the posterior hits the prior boundary on the right. Because of the correlations between $\mathcal{M}_c$ and $\eta$, this translates into slight shifting of the posterior which results in the injection value not coinciding with the median of the posterior. This is observed both in Figs.~\ref{fig:hist-hm} and \ref{fig:corner_hm_q_1}. This feature also appears when we inject a signal using \texttt{IMRPhenomXHM} and recover using the same waveform, thus indicating that this is solely because of the correlations between $\eta$ and $\mathcal{M}_c$ and not because of differences in the injected and recovery waveform. The slight shift in $q=3$ posterior of circular injection (HYB:BBH:1221) though can be attributed to the slight differences in the injected and recovered signal. 

Another interesting feature of these plots is that, while the mass parameters show a shift in the posterior with eccentric injections, the distance and inclination posteriors are still able to recover the injected values, even for eccentric injections. This hints at the fact that, while the eccentricity parameter is strongly correlated with the mass parameters of the binary, it is not so with the case of extrinsic parameters like luminosity distance and inclination angle \citep[see also][]{OShea:2021faf}.

\begin{figure}[t!]
      \centering
      \includegraphics[trim=10 10 110 120, clip, width=\linewidth]{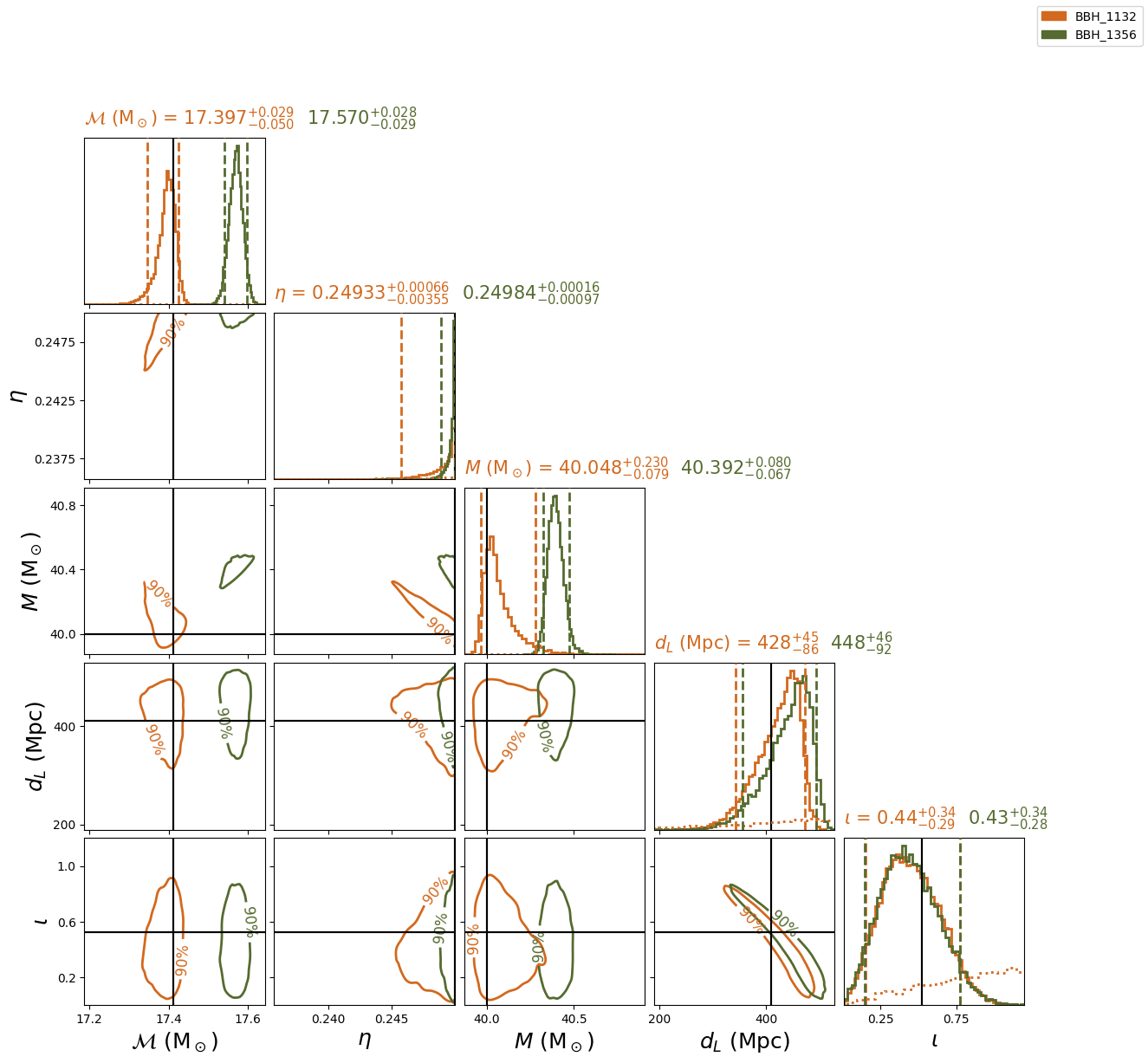}
      \caption[Corner plot for higher mode injections of $q=1$]{Corner plot for $q=1$, where circular (SXS:BBH:1132) simulation is shown in orange, and eccentric (HYB:BBH:1356) simulation is shown in dark green. Histograms on the diagonal show marginalized 1D posteriors, whereas the contours denote the joint 2D posteriors for various parameters. The vertical dashed lines in 1D histograms mark 90\% credible intervals, and dotted lines in orange show the prior. The black lines mark the injected values for various system parameters.}
    \label{fig:corner_hm_q_1}
  \end{figure}
  
\begin{figure}[t!]
      \centering
      \includegraphics[trim=10 10 110 120, clip, width=\linewidth]{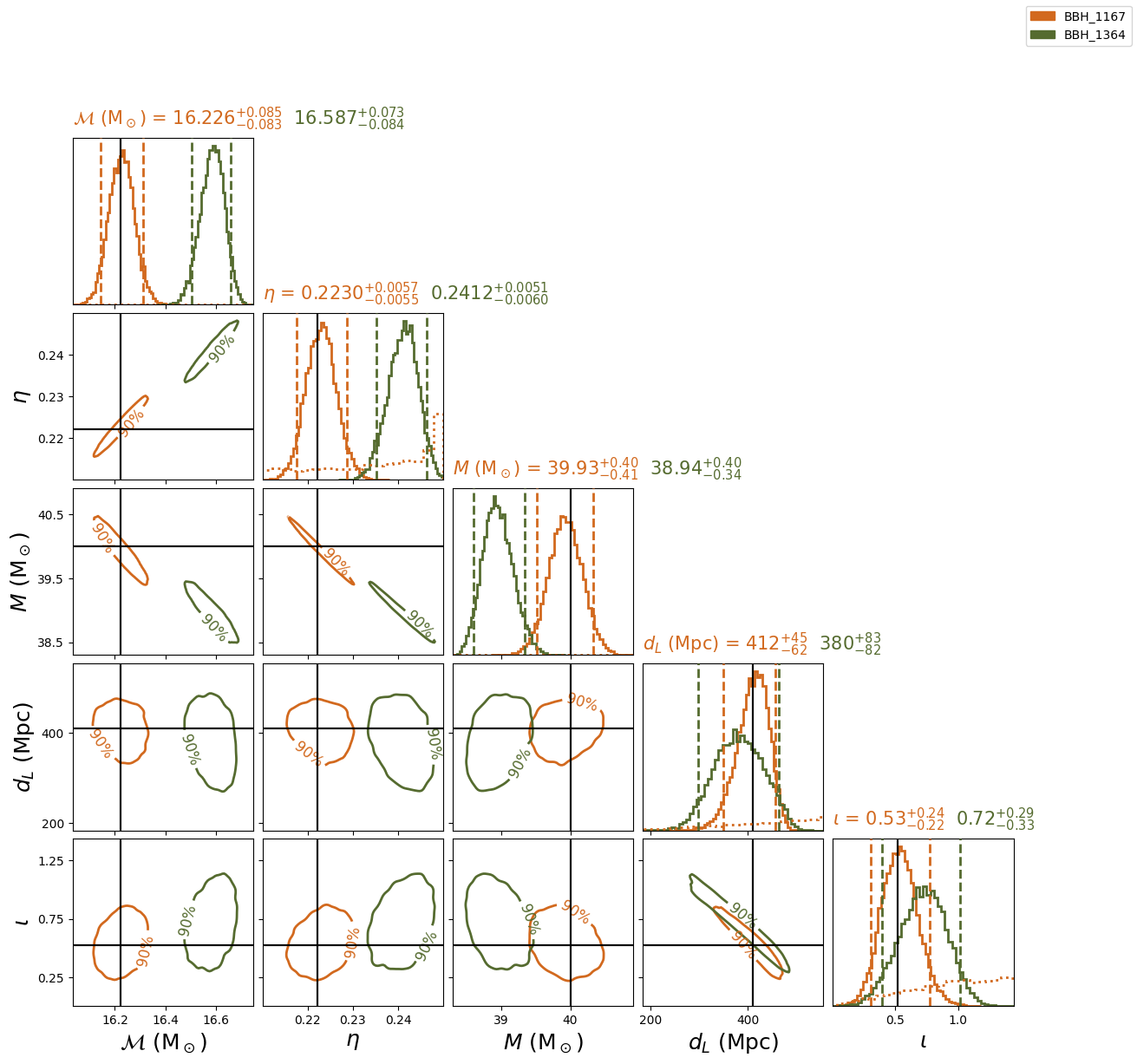}
      \caption[Same as Fig \ref{fig:corner_hm_q_1} but for $q=2$]{Same as Fig \ref{fig:corner_hm_q_1} but for $q=2$. The hybrids (HYB:BBH:1167) and (HYB:BBH:1364) are used for quasi-circular and eccentric injections respectively.}
    \label{fig:corner_hm_q_2}
  \end{figure}
  
\begin{figure}[t!]
      \centering
      \includegraphics[trim=10 10 110 120, clip, width=\linewidth]{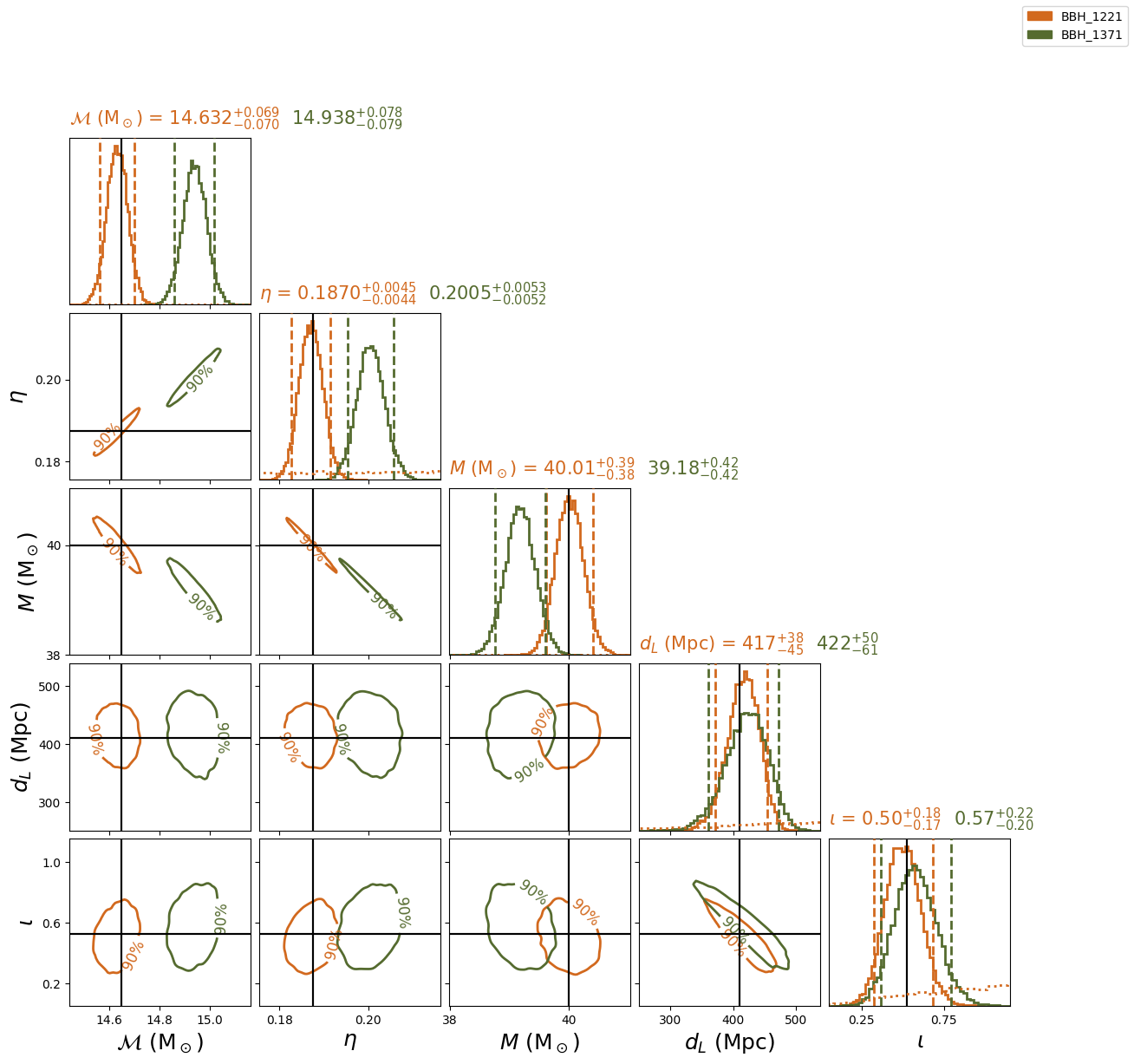}
      \caption[Same as Fig \ref{fig:corner_hm_q_1} but for $q=3$]{Same as Fig \ref{fig:corner_hm_q_1} but for $q=3$. The hybrids (HYB:BBH:1221) and (HYB:BBH:1371) are used for quasi-circular and eccentric injections respectively.}
    \label{fig:corner_hm_q_3}
  \end{figure}
\clearpage

%----------------------------------------------------------------------------------------------------------

\section{Summary and Conclusions} 
\label{sec:ecc-concl}
Measurable orbital eccentricity is a key indicator of BBH formation channels. However, catalogues of BBH detections, e.g. \citet{LIGOScientific:2021djp}, typically neglect this parameter and study all GW candidates using only quasi-circular signal models. Additionally, matched-filter searches for GW signals typically rely on quasi-circular waveform templates. In this work, we explore both the detectability of the eccentric signals when eccentricity is neglected from matched-filtering searches, and the biases that result from performing parameter estimation on eccentric GW signals using quasi-circular waveform models under a variety of spin assumptions. 

We find that there is a loss in the fitting factor ($< 0.95$) for eccentricities higher than $0.01$ at $10$~Hz in conjunction with high values of mass ratio ($q>3$). Further, we find that there's a loss in $SRF$ up to 6\% for the region in parameter space with $e_{10} > 0.01$ and mass ratio $q > 3$. While we restrict this calculation to the inspiral part of the signal, we argue that the loss in $FF$ would be similar for full IMR signals, since eccentricity is efficiently radiated away from an inspiralling system and so the binary should be close to circular before the merger and ringdown. 
The overall loss in the fraction of recovered signals depends on the fraction of events in the population that have a high-eccentricity and high-mass ratio. 
These population characteristics, in turn, depend on the formation channels contributing to the population.
For example, we would detect a higher fraction of binaries that formed in globular clusters (GCs) than those that formed in active galactic nuclei (AGN), since the eccentricity and mass ratio distributions expected from binaries formed in GCs are less extreme than those expected from AGN \citep[e.g.,][]{Zevin:2018kzq, Tagawa:2020qll, Tagawa:2020jnc, Yang:2019cbr}.
Therefore, with severity depending on the balance between the formation channels contributing to the observed population, missing eccentric binaries in searches can lead to errors in the inferred merger rate and underlying population characteristics.

Even if an eccentric signal is detected via a matched-template search based on quasi-circular waveform templates, the recovery of source parameters can be biased when the signal is analysed using a quasi-circular waveform model. We perform parameter estimation on non-spinning and aligned-spin eccentric injections, and recover them using various spin assumptions with quasi-circular and eccentric waveform models. We find that for $e_{20}\sim 0.1$, analyses with the quasi-circular waveform models are unable to recover the injected values of chirp mass within the $90\%$ credible interval. This holds true even for the injections which include both eccentricity and higher modes, which, when analysed with quasi-circular higher mode waveforms, still show the bias in the recovered chirp mass parameter. Further, we note that for the relatively low-mass BBH systems considered in this study, no spurious spin detections are made for non-spinning eccentric injections, and no spurious inferences of precession are made for any eccentric injections. This leads us to conclude that for relatively low mass systems, spin-precession does not mimic eccentricity. The spin parameter posteriors are similar for both quasi-circular and eccentric injections. 

Since both eccentricity and spin-precession can indicate that a binary formed in a dynamical environment, our results suggest that a non-spinning low-mass eccentric system if analyzed using quasi-circular waveform models only, may be mistaken for a binary that formed in isolation since the quasi-circular waveform models do not enable measurements of eccentricity and the spin posteriors show no additional signatures of dynamical formation. This may lead to miscategorization of such systems as uninteresting or "vanilla" binaries. Moreover, if we are routinely biased to higher masses even for a small subset of signals that include the influence of binary eccentricity, the population distribution of mass will gradually be shifted higher. Eventually, this could lead to incorrect inferences about, for example, the location of the pair-instability mass gap and the fraction of the population comprised of hierarchical mergers \citep[using hierarchical inference methods such as][]{Mould:2022ccw}. While the shifts in the inferred chirp mass for the low-mass and moderate-eccentricity injections studied here are relatively minor, for higher eccentricities and higher masses the bias would likely be worse [see, for instance, Eq. (1.1) in \cite{Favata:2021vhw}].
We also observe that for the eccentricity values chosen here ($e_{20}\sim 0.1$), even an inspiral-only eccentric waveform is able to recover the injected chirp mass within the $90\%$ confidence interval. Therefore, we conclude that for GW signals from relatively low-mass BBH, inspiral-only eccentric waveform models are adequate for identifying and quantifying orbital eccentricity.
      \DeclareRobustCommand{\SNAME}[1]{\IfEqCase{#1}{{GW200322A}{S200322ab}{GW200316A}{S200316bj}{GW200311B}{S200311bg}{GW200308A}{S200308bl}{GW200306A}{S200306ak}{GW200302A}{S200302c}{GW200225A}{S200225q}{GW200224A}{S200224ca}{GW200220A}{S200220ad}{GW200219A}{S200219ac}{GW200218A}{S200218al}{GW200216A}{S200216br}{GW200210A}{S200210ba}{GW200209A}{S200209ab}{GW200208B}{S200208am}{GW200208A}{S200208q}{GW200202A}{S200202ac}{GW200129A}{S200129m}{GW200128A}{S200128d}{GW200115A}{S200115j}{GW200112A}{S200112r}{GW200105A}{S200105ae}{GW191230A}{S191230an}{GW191222A}{S191222n}{GW191219A}{S191219ax}{GW191216A}{S191216ap}{GW191215A}{S191215w}{GW191213A}{S191213bb}{GW191204B}{S191204r}{GW191204A}{S191204h}{GW191129A}{S191129u}{GW191127A}{S191127p}{GW191126A}{S191126l}{GW191113A}{S191113q}{GW191109A}{S191109d}{GW191105A}{S191105e}{GW191103A}{S191103a}{200311A}{S200311ba}{200219B}{S200219bj}{200201A}{S200201bh}{200121A}{S200121aa}{191118A}{S191118ae}}}
\DeclareRobustCommand{\FULLNAME}[1]{\IfEqCase{#1}{{GW200322A}{GW200322\_091133}{GW200316A}{GW200316\_215756}{GW200311B}{GW200311\_115853}{GW200308A}{GW200308\_173609}{GW200306A}{GW200306\_093714}{GW200302A}{GW200302\_015811}{GW200225A}{GW200225\_060421}{GW200224A}{GW200224\_222234}{GW200220A}{GW200220\_061928}{GW200219A}{GW200219\_094415}{GW200218A}{GW200218\_100521}{GW200216A}{GW200216\_220804}{GW200210A}{GW200210\_092254}{GW200209A}{GW200209\_085452}{GW200208B}{GW200208\_222618}{GW200208A}{GW200208\_130117}{GW200202A}{GW200202\_154313}{GW200129A}{GW200129\_065458}{GW200128A}{GW200128\_022011}{GW200115A}{GW200115\_042309}{GW200112A}{GW200112\_155838}{GW200105A}{GW200105\_162426}{GW191230A}{GW191230\_180458}{GW191222A}{GW191222\_033537}{GW191219A}{GW191219\_163120}{GW191216A}{GW191216\_213338}{GW191215A}{GW191215\_223052}{GW191213A}{GW191213\_194013}{GW191204B}{GW191204\_171526}{GW191204A}{GW191204\_110529}{GW191129A}{GW191129\_134029}{GW191127A}{GW191127\_050227}{GW191126A}{GW191126\_115259}{GW191113A}{GW191113\_071753}{GW191109A}{GW191109\_010717}{GW191105A}{GW191105\_143521}{GW191103A}{GW191103\_012549}{200311A}{200311\_103121}{200219B}{200219\_201407}{200201A}{200201\_203549}{200121A}{200121\_031748}{191118A}{191118\_212859}{GW191129G}{GW191129\_134029}{GW191204G}{GW191204\_171526}{GW191216G}{GW191216\_213338}{GW200129D}{GW200129\_065458}{GW200225B}{GW200225\_060421}{GW200316I}{GW200316\_215756}{GW190412A}{GW190412\_053044}}}
\DeclareRobustCommand{\NNAME}[1]{\IfEqCase{#1}{{GW200322A}{GW200322\_091133}{GW200316A}{GW200316\_215756}{GW200311B}{GW200311\_115853}{GW200308A}{GW200308\_173609}{GW200306A}{GW200306\_093714}{GW200302A}{GW200302\_015811}{GW200225A}{GW200225\_060421}{GW200224A}{GW200224\_222234}{GW200220A}{GW200220\_061928}{GW200219A}{GW200219\_094415}{GW200218A}{GW200218\_100521}{GW200216A}{GW200216\_220804}{GW200210A}{GW200210\_092254}{GW200209A}{GW200209\_085452}{GW200208B}{GW200208\_222618}{GW200208A}{GW200208\_130117}{GW200202A}{GW200202\_154313}{GW200129A}{GW200129\_065458}{GW200128A}{GW200128\_022011}{GW200115A}{GW200115}{GW200112A}{GW200112\_155838}{GW200105A}{GW200105}{GW191230A}{GW191230\_180458}{GW191222A}{GW191222\_033537}{GW191219A}{GW191219\_163120}{GW191216A}{GW191216\_213338}{GW191215A}{GW191215\_223052}{GW191213A}{GW191213\_194013}{GW191204B}{GW191204\_171526}{GW191204A}{GW191204\_110529}{GW191129A}{GW191129\_134029}{GW191127A}{GW191127\_050227}{GW191126A}{GW191126\_115259}{GW191113A}{GW191113\_071753}{GW191109A}{GW191109\_010717}{GW191105A}{GW191105\_143521}{GW191103A}{GW191103\_012549}{200311A}{200311\_103121}{200219B}{200219\_201407}{200201A}{200201\_203549}{200121A}{200121\_031748}{191118A}{191118\_212859}}}
\DeclareRobustCommand{\MINIMALNAME}[1]{\IfEqCase{#1}{{GW200322A}{GW200322}{GW200316A}{GW200316}{GW200311B}{GW200311\_11}{GW200308A}{GW200308}{GW200306A}{GW200306}{GW200302A}{GW200302}{GW200225A}{GW200225}{GW200224A}{GW200224}{GW200220A}{GW200220}{GW200219A}{GW200219\_09}{GW200218A}{GW200218}{GW200216A}{GW200216}{GW200210A}{GW200210}{GW200209A}{GW200209}{GW200208B}{GW200208\_22}{GW200208A}{GW200208\_13}{GW200202A}{GW200202}{GW200129A}{GW200129}{GW200128A}{GW200128}{GW200115A}{GW200115}{GW200112A}{GW200112}{GW200105A}{GW200105}{GW191230A}{GW191230}{GW191222A}{GW191222}{GW191219A}{GW191219}{GW191216A}{GW191216}{GW191215A}{GW191215}{GW191213A}{GW191213}{GW191204B}{GW191204\_17}{GW191204A}{GW191204\_11}{GW191129A}{GW191129}{GW191127A}{GW191127}{GW191126A}{GW191126}{GW191113A}{GW191113}{GW191109A}{GW191109}{GW191105A}{GW191105}{GW191103A}{GW191103}{200311A}{200311\_10}{200219B}{200219\_20}{200201A}{200201}{200121A}{200121}{191118A}{191118}{GW191129G}{GW191129}{GW191204G}{GW191204\_17}{GW191216G}{GW191216}{GW200129D}{GW200129}{GW200225B}{GW200225}{GW200316I}{GW200316}{GW190412A}{GW190412}{GW190412A}{GW190412}{GW151226A}{GW151226}{GW170608A}{GW170608}}}

      \DeclareRobustCommand{\chipinfinityonlyprecavgminus}[1]{\IfEqCase{#1}{{GW200322G}{0.32}{GW200316I}{0.20}{GW200311L}{0.35}{GW200308G}{0.30}{GW200306A}{0.31}{GW200302A}{0.28}{GW200225B}{0.38}{GW200224H}{0.36}{GW200220H}{0.37}{GW200220E}{0.38}{GW200219D}{0.35}{GW200216G}{0.35}{GW200210B}{0.12}{GW200209E}{0.37}{GW200208K}{0.29}{GW200208G}{0.29}{GW200202F}{0.22}{GW200129D}{0.38}{GW200128C}{0.40}{GW200115A}{0.16}{GW200112H}{0.30}{200105F}{0.07}{GW191230H}{0.38}{GW191222A}{0.32}{GW191219E}{0.07}{GW191216G}{0.15}{GW191215G}{0.38}{GW191204G}{0.27}{GW191204A}{0.38}{GW191129G}{0.19}{GW191127B}{0.41}{GW191126C}{0.26}{GW191113B}{0.16}{GW191109A}{0.38}{GW191105C}{0.24}{GW191103A}{0.27}}}
\DeclareRobustCommand{\chipinfinityonlyprecavgmed}[1]{\IfEqCase{#1}{{GW200322G}{0.42}{GW200316I}{0.29}{GW200311L}{0.45}{GW200308G}{0.41}{GW200306A}{0.43}{GW200302A}{0.37}{GW200225B}{0.53}{GW200224H}{0.49}{GW200220H}{0.50}{GW200220E}{0.51}{GW200219D}{0.47}{GW200216G}{0.45}{GW200210B}{0.15}{GW200209E}{0.51}{GW200208K}{0.41}{GW200208G}{0.39}{GW200202F}{0.28}{GW200129D}{0.52}{GW200128C}{0.56}{GW200115A}{0.20}{GW200112H}{0.40}{200105F}{0.09}{GW191230H}{0.51}{GW191222A}{0.41}{GW191219E}{0.09}{GW191216G}{0.23}{GW191215G}{0.51}{GW191204G}{0.40}{GW191204A}{0.49}{GW191129G}{0.26}{GW191127B}{0.52}{GW191126C}{0.39}{GW191113B}{0.20}{GW191109A}{0.63}{GW191105C}{0.30}{GW191103A}{0.40}}}
\DeclareRobustCommand{\chipinfinityonlyprecavgplus}[1]{\IfEqCase{#1}{{GW200322G}{0.41}{GW200316I}{0.38}{GW200311L}{0.39}{GW200308G}{0.42}{GW200306A}{0.39}{GW200302A}{0.45}{GW200225B}{0.35}{GW200224H}{0.37}{GW200220H}{0.38}{GW200220E}{0.37}{GW200219D}{0.40}{GW200216G}{0.42}{GW200210B}{0.22}{GW200209E}{0.39}{GW200208K}{0.38}{GW200208G}{0.42}{GW200202F}{0.40}{GW200129D}{0.41}{GW200128C}{0.34}{GW200115A}{0.34}{GW200112H}{0.38}{200105F}{0.17}{GW191230H}{0.38}{GW191222A}{0.42}{GW191219E}{0.07}{GW191216G}{0.35}{GW191215G}{0.37}{GW191204G}{0.39}{GW191204A}{0.40}{GW191129G}{0.36}{GW191127B}{0.40}{GW191126C}{0.40}{GW191113B}{0.54}{GW191109A}{0.28}{GW191105C}{0.45}{GW191103A}{0.41}}}
\DeclareRobustCommand{\chipinfinityonlyprecavgtenthpercentile}[1]{\IfEqCase{#1}{{GW200322G}{0.15}{GW200316I}{0.12}{GW200311L}{0.16}{GW200308G}{0.15}{GW200306A}{0.17}{GW200302A}{0.12}{GW200225B}{0.22}{GW200224H}{0.19}{GW200220H}{0.18}{GW200220E}{0.19}{GW200219D}{0.18}{GW200216G}{0.15}{GW200210B}{0.05}{GW200209E}{0.20}{GW200208K}{0.16}{GW200208G}{0.14}{GW200202F}{0.09}{GW200129D}{0.20}{GW200128C}{0.23}{GW200115A}{0.06}{GW200112H}{0.14}{200105F}{0.03}{GW191230H}{0.20}{GW191222A}{0.14}{GW191219E}{0.03}{GW191216G}{0.10}{GW191215G}{0.19}{GW191204G}{0.17}{GW191204A}{0.18}{GW191129G}{0.10}{GW191127B}{0.17}{GW191126C}{0.17}{GW191113B}{0.06}{GW191109A}{0.33}{GW191105C}{0.09}{GW191103A}{0.18}}}
\DeclareRobustCommand{\chipinfinityonlyprecavgnintiethpercentile}[1]{\IfEqCase{#1}{{GW200322G}{0.75}{GW200316I}{0.58}{GW200311L}{0.77}{GW200308G}{0.74}{GW200306A}{0.75}{GW200302A}{0.73}{GW200225B}{0.82}{GW200224H}{0.79}{GW200220H}{0.82}{GW200220E}{0.81}{GW200219D}{0.80}{GW200216G}{0.80}{GW200210B}{0.32}{GW200209E}{0.84}{GW200208K}{0.72}{GW200208G}{0.72}{GW200202F}{0.59}{GW200129D}{0.89}{GW200128C}{0.84}{GW200115A}{0.46}{GW200112H}{0.70}{200105F}{0.19}{GW191230H}{0.84}{GW191222A}{0.75}{GW191219E}{0.14}{GW191216G}{0.48}{GW191215G}{0.82}{GW191204G}{0.70}{GW191204A}{0.83}{GW191129G}{0.54}{GW191127B}{0.88}{GW191126C}{0.70}{GW191113B}{0.63}{GW191109A}{0.87}{GW191105C}{0.65}{GW191103A}{0.72}}}
\DeclareRobustCommand{\luminositydistanceminus}[1]{\IfEqCase{#1}{{GW200322G}{2.2}{GW200316I}{0.44}{GW200311L}{0.40}{GW200308G}{4.4}{GW200306A}{1.1}{GW200302A}{0.70}{GW200225B}{0.53}{GW200224H}{0.65}{GW200220H}{2.2}{GW200220E}{3.1}{GW200219D}{1.5}{GW200216G}{2.0}{GW200210B}{0.34}{GW200209E}{1.8}{GW200208K}{2.0}{GW200208G}{0.85}{GW200202F}{0.16}{GW200129D}{0.37}{GW200128C}{1.8}{GW200115A}{0.10}{GW200112H}{0.46}{200105F}{0.11}{GW191230H}{1.9}{GW191222A}{1.7}{GW191219E}{0.16}{GW191216G}{0.13}{GW191215G}{0.86}{GW191204G}{0.26}{GW191204A}{1.1}{GW191129G}{0.33}{GW191127B}{1.9}{GW191126C}{0.74}{GW191113B}{0.62}{GW191109A}{0.65}{GW191105C}{0.48}{GW191103A}{0.47}{GW190412A}{0.17}{GW151226A}{0.20}{GW170608A}{0.13}}}
\DeclareRobustCommand{\luminositydistancemed}[1]{\IfEqCase{#1}{{GW200322G}{3.5}{GW200316I}{1.12}{GW200311L}{1.17}{GW200308G}{7.1}{GW200306A}{2.1}{GW200302A}{1.48}{GW200225B}{1.15}{GW200224H}{1.71}{GW200220H}{4.0}{GW200220E}{6.0}{GW200219D}{3.4}{GW200216G}{3.8}{GW200210B}{0.94}{GW200209E}{3.4}{GW200208K}{4.1}{GW200208G}{2.23}{GW200202F}{0.41}{GW200129D}{0.89}{GW200128C}{3.4}{GW200115A}{0.29}{GW200112H}{1.25}{200105F}{0.27}{GW191230H}{4.3}{GW191222A}{3.0}{GW191219E}{0.55}{GW191216G}{0.34}{GW191215G}{1.93}{GW191204G}{0.64}{GW191204A}{1.9}{GW191129G}{0.79}{GW191127B}{3.4}{GW191126C}{1.62}{GW191113B}{1.37}{GW191109A}{1.29}{GW191105C}{1.15}{GW191103A}{0.99}{GW190412A}{0.74}{GW151226A}{0.46}{GW170608A}{0.34}}}
\DeclareRobustCommand{\luminositydistanceplus}[1]{\IfEqCase{#1}{{GW200322G}{12.5}{GW200316I}{0.48}{GW200311L}{0.28}{GW200308G}{13.9}{GW200306A}{1.7}{GW200302A}{1.02}{GW200225B}{0.51}{GW200224H}{0.50}{GW200220H}{2.8}{GW200220E}{4.8}{GW200219D}{1.7}{GW200216G}{3.0}{GW200210B}{0.43}{GW200209E}{1.9}{GW200208K}{4.4}{GW200208G}{1.02}{GW200202F}{0.15}{GW200129D}{0.26}{GW200128C}{2.1}{GW200115A}{0.15}{GW200112H}{0.43}{200105F}{0.12}{GW191230H}{2.1}{GW191222A}{1.7}{GW191219E}{0.24}{GW191216G}{0.12}{GW191215G}{0.89}{GW191204G}{0.20}{GW191204A}{1.7}{GW191129G}{0.26}{GW191127B}{3.1}{GW191126C}{0.74}{GW191113B}{1.15}{GW191109A}{1.13}{GW191105C}{0.43}{GW191103A}{0.50}{GW190412A}{0.14}{GW151226A}{0.16}{GW170608A}{0.12}}}
\DeclareRobustCommand{\luminositydistancetenthpercentile}[1]{\IfEqCase{#1}{{GW200322G}{1.7}{GW200316I}{0.76}{GW200311L}{0.87}{GW200308G}{3.5}{GW200306A}{1.2}{GW200302A}{0.91}{GW200225B}{0.74}{GW200224H}{1.19}{GW200220H}{2.2}{GW200220E}{3.5}{GW200219D}{2.2}{GW200216G}{2.2}{GW200210B}{0.67}{GW200209E}{2.0}{GW200208K}{2.4}{GW200208G}{1.54}{GW200202F}{0.28}{GW200129D}{0.60}{GW200128C}{1.9}{GW200115A}{0.21}{GW200112H}{0.89}{200105F}{0.17}{GW191230H}{2.7}{GW191222A}{1.6}{GW191219E}{0.42}{GW191216G}{0.23}{GW191215G}{1.22}{GW191204G}{0.43}{GW191204A}{1.0}{GW191129G}{0.52}{GW191127B}{1.8}{GW191126C}{1.01}{GW191113B}{0.85}{GW191109A}{0.76}{GW191105C}{0.76}{GW191103A}{0.60}}}
\DeclareRobustCommand{\luminositydistancenintiethpercentile}[1]{\IfEqCase{#1}{{GW200322G}{11.6}{GW200316I}{1.49}{GW200311L}{1.39}{GW200308G}{17.2}{GW200306A}{3.4}{GW200302A}{2.26}{GW200225B}{1.55}{GW200224H}{2.10}{GW200220H}{6.2}{GW200220E}{9.6}{GW200219D}{4.7}{GW200216G}{6.0}{GW200210B}{1.26}{GW200209E}{4.9}{GW200208K}{7.2}{GW200208G}{2.99}{GW200202F}{0.53}{GW200129D}{1.11}{GW200128C}{5.0}{GW200115A}{0.40}{GW200112H}{1.60}{200105F}{0.36}{GW191230H}{5.9}{GW191222A}{4.3}{GW191219E}{0.73}{GW191216G}{0.44}{GW191215G}{2.64}{GW191204G}{0.81}{GW191204A}{3.2}{GW191129G}{1.00}{GW191127B}{5.7}{GW191126C}{2.19}{GW191113B}{2.21}{GW191109A}{2.13}{GW191105C}{1.49}{GW191103A}{1.38}}}
\DeclareRobustCommand{\massonesourceminus}[1]{\IfEqCase{#1}{{GW200322G}{22}{GW200316I}{2.9}{GW200311L}{3.8}{GW200308G}{29}{GW200306A}{7.7}{GW200302A}{8.5}{GW200225B}{3.0}{GW200224H}{4.5}{GW200220H}{8.6}{GW200220E}{23}{GW200219D}{6.9}{GW200216G}{13}{GW200210B}{4.6}{GW200209E}{6.8}{GW200208K}{30}{GW200208G}{6.2}{GW200202F}{1.4}{GW200129D}{3.1}{GW200128C}{8.1}{GW200115A}{2.5}{GW200112H}{4.5}{200105F}{1.7}{GW191230H}{9.6}{GW191222A}{8.0}{GW191219E}{2.8}{GW191216G}{2.2}{GW191215G}{4.1}{GW191204G}{1.7}{GW191204A}{5.9}{GW191129G}{2.1}{GW191127B}{20}{GW191126C}{2.2}{GW191113B}{14}{GW191109A}{11}{GW191105C}{1.6}{GW191103A}{2.2}{GW190412A}{5.1}{GW151226A}{3.6}{GW170608A}{1.4}}}
\DeclareRobustCommand{\massonesourcemed}[1]{\IfEqCase{#1}{{GW200322G}{38}{GW200316I}{13.1}{GW200311L}{34.2}{GW200308G}{60}{GW200306A}{28.3}{GW200302A}{37.8}{GW200225B}{19.3}{GW200224H}{40.0}{GW200220H}{38.9}{GW200220E}{87}{GW200219D}{37.5}{GW200216G}{51}{GW200210B}{24.1}{GW200209E}{35.6}{GW200208K}{51}{GW200208G}{37.7}{GW200202F}{10.1}{GW200129D}{34.5}{GW200128C}{42.2}{GW200115A}{5.9}{GW200112H}{35.6}{200105F}{9.1}{GW191230H}{49.4}{GW191222A}{45.1}{GW191219E}{31.1}{GW191216G}{12.1}{GW191215G}{24.9}{GW191204G}{11.7}{GW191204A}{27.3}{GW191129G}{10.7}{GW191127B}{53}{GW191126C}{12.1}{GW191113B}{29}{GW191109A}{65}{GW191105C}{10.7}{GW191103A}{11.8}{GW190412A}{30.1}{GW151226A}{14.2}{GW170608A}{10.6}}}
\DeclareRobustCommand{\massonesourceplus}[1]{\IfEqCase{#1}{{GW200322G}{130}{GW200316I}{10.2}{GW200311L}{6.4}{GW200308G}{166}{GW200306A}{17.1}{GW200302A}{8.7}{GW200225B}{5.0}{GW200224H}{6.7}{GW200220H}{14.1}{GW200220E}{40}{GW200219D}{10.1}{GW200216G}{22}{GW200210B}{7.5}{GW200209E}{10.5}{GW200208K}{103}{GW200208G}{9.3}{GW200202F}{3.5}{GW200129D}{9.9}{GW200128C}{11.6}{GW200115A}{2.0}{GW200112H}{6.7}{200105F}{1.7}{GW191230H}{14.0}{GW191222A}{10.9}{GW191219E}{2.2}{GW191216G}{4.6}{GW191215G}{7.1}{GW191204G}{3.3}{GW191204A}{10.8}{GW191129G}{4.1}{GW191127B}{47}{GW191126C}{5.5}{GW191113B}{12}{GW191109A}{11}{GW191105C}{3.7}{GW191103A}{6.2}{GW190412A}{4.7}{GW151226A}{11.1}{GW170608A}{4.0}}}
\DeclareRobustCommand{\massonesourcetenthpercentile}[1]{\IfEqCase{#1}{{GW200322G}{18}{GW200316I}{10.6}{GW200311L}{31.1}{GW200308G}{33}{GW200306A}{21.8}{GW200302A}{31.0}{GW200225B}{16.8}{GW200224H}{36.3}{GW200220H}{31.7}{GW200220E}{68}{GW200219D}{31.8}{GW200216G}{40}{GW200210B}{20.7}{GW200209E}{30.1}{GW200208K}{23}{GW200208G}{32.7}{GW200202F}{8.9}{GW200129D}{31.9}{GW200128C}{35.7}{GW200115A}{3.7}{GW200112H}{31.9}{200105F}{8.1}{GW191230H}{41.6}{GW191222A}{38.5}{GW191219E}{29.1}{GW191216G}{10.1}{GW191215G}{21.5}{GW191204G}{10.2}{GW191204A}{22.4}{GW191129G}{8.9}{GW191127B}{36}{GW191126C}{10.2}{GW191113B}{17}{GW191109A}{57}{GW191105C}{9.3}{GW191103A}{9.9}}}
\DeclareRobustCommand{\massonesourcenintiethpercentile}[1]{\IfEqCase{#1}{{GW200322G}{102}{GW200316I}{20.0}{GW200311L}{38.9}{GW200308G}{164}{GW200306A}{39.7}{GW200302A}{44.5}{GW200225B}{23.0}{GW200224H}{45.0}{GW200220H}{49.0}{GW200220E}{113}{GW200219D}{45.1}{GW200216G}{68}{GW200210B}{29.7}{GW200209E}{43.4}{GW200208K}{97}{GW200208G}{44.8}{GW200202F}{12.7}{GW200129D}{42.2}{GW200128C}{50.9}{GW200115A}{7.4}{GW200112H}{40.8}{200105F}{10.0}{GW191230H}{60.0}{GW191222A}{53.2}{GW191219E}{32.9}{GW191216G}{15.2}{GW191215G}{30.1}{GW191204G}{14.1}{GW191204A}{34.8}{GW191129G}{13.8}{GW191127B}{88}{GW191126C}{16.0}{GW191113B}{36}{GW191109A}{73}{GW191105C}{13.3}{GW191103A}{16.1}}}
\DeclareRobustCommand{\masstwosourceminus}[1]{\IfEqCase{#1}{{GW200322G}{6.0}{GW200316I}{2.9}{GW200311L}{5.9}{GW200308G}{13}{GW200306A}{6.4}{GW200302A}{5.7}{GW200225B}{3.5}{GW200224H}{7.2}{GW200220H}{9.0}{GW200220E}{25}{GW200219D}{8.4}{GW200216G}{16}{GW200210B}{0.42}{GW200209E}{7.8}{GW200208K}{5.5}{GW200208G}{7.3}{GW200202F}{1.7}{GW200129D}{9.3}{GW200128C}{9.2}{GW200115A}{0.28}{GW200112H}{5.9}{200105F}{0.24}{GW191230H}{12}{GW191222A}{10.5}{GW191219E}{0.06}{GW191216G}{1.9}{GW191215G}{4.1}{GW191204G}{1.7}{GW191204A}{6.0}{GW191129G}{1.7}{GW191127B}{14}{GW191126C}{2.4}{GW191113B}{1.3}{GW191109A}{13}{GW191105C}{1.9}{GW191103A}{2.4}{GW190412A}{0.9}{GW151226A}{2.8}{GW170608A}{1.9}}}
\DeclareRobustCommand{\masstwosourcemed}[1]{\IfEqCase{#1}{{GW200322G}{11.3}{GW200316I}{7.8}{GW200311L}{27.7}{GW200308G}{24}{GW200306A}{14.8}{GW200302A}{20.0}{GW200225B}{14.0}{GW200224H}{32.7}{GW200220H}{27.9}{GW200220E}{61}{GW200219D}{27.9}{GW200216G}{30}{GW200210B}{2.83}{GW200209E}{27.1}{GW200208K}{12.3}{GW200208G}{27.4}{GW200202F}{7.3}{GW200129D}{29.0}{GW200128C}{32.6}{GW200115A}{1.44}{GW200112H}{28.3}{200105F}{1.91}{GW191230H}{37}{GW191222A}{34.7}{GW191219E}{1.17}{GW191216G}{7.7}{GW191215G}{18.1}{GW191204G}{8.4}{GW191204A}{19.2}{GW191129G}{6.7}{GW191127B}{24}{GW191126C}{8.3}{GW191113B}{5.9}{GW191109A}{47}{GW191105C}{7.7}{GW191103A}{7.9}{GW190412A}{8.3}{GW151226A}{7.5}{GW170608A}{7.8}}}
\DeclareRobustCommand{\masstwosourceplus}[1]{\IfEqCase{#1}{{GW200322G}{24.3}{GW200316I}{2.0}{GW200311L}{4.1}{GW200308G}{36}{GW200306A}{6.5}{GW200302A}{8.1}{GW200225B}{2.8}{GW200224H}{4.8}{GW200220H}{9.2}{GW200220E}{26}{GW200219D}{7.4}{GW200216G}{14}{GW200210B}{0.47}{GW200209E}{7.8}{GW200208K}{9.2}{GW200208G}{6.3}{GW200202F}{1.1}{GW200129D}{3.3}{GW200128C}{9.5}{GW200115A}{0.85}{GW200112H}{4.4}{200105F}{0.33}{GW191230H}{11}{GW191222A}{9.3}{GW191219E}{0.07}{GW191216G}{1.6}{GW191215G}{3.8}{GW191204G}{1.3}{GW191204A}{5.5}{GW191129G}{1.5}{GW191127B}{17}{GW191126C}{1.9}{GW191113B}{4.4}{GW191109A}{15}{GW191105C}{1.4}{GW191103A}{1.7}{GW190412A}{1.6}{GW151226A}{2.4}{GW170608A}{1.2}}}
\DeclareRobustCommand{\masstwosourcetenthpercentile}[1]{\IfEqCase{#1}{{GW200322G}{6.4}{GW200316I}{5.6}{GW200311L}{23.2}{GW200308G}{12}{GW200306A}{9.6}{GW200302A}{15.3}{GW200225B}{11.2}{GW200224H}{27.3}{GW200220H}{20.8}{GW200220E}{41}{GW200219D}{21.6}{GW200216G}{16}{GW200210B}{2.49}{GW200209E}{21.4}{GW200208K}{7.9}{GW200208G}{21.6}{GW200202F}{6.0}{GW200129D}{21.1}{GW200128C}{25.4}{GW200115A}{1.21}{GW200112H}{23.7}{200105F}{1.77}{GW191230H}{28}{GW191222A}{26.6}{GW191219E}{1.12}{GW191216G}{6.3}{GW191215G}{14.9}{GW191204G}{7.0}{GW191204A}{14.6}{GW191129G}{5.4}{GW191127B}{12}{GW191126C}{6.4}{GW191113B}{4.9}{GW191109A}{36}{GW191105C}{6.2}{GW191103A}{6.0}}}
\DeclareRobustCommand{\masstwosourcenintiethpercentile}[1]{\IfEqCase{#1}{{GW200322G}{30.4}{GW200316I}{9.4}{GW200311L}{31.0}{GW200308G}{49}{GW200306A}{20.0}{GW200302A}{26.3}{GW200225B}{16.3}{GW200224H}{36.5}{GW200220H}{34.9}{GW200220E}{81}{GW200219D}{33.8}{GW200216G}{41}{GW200210B}{3.16}{GW200209E}{33.2}{GW200208K}{19.2}{GW200208G}{32.4}{GW200202F}{8.3}{GW200129D}{31.7}{GW200128C}{40.0}{GW200115A}{2.12}{GW200112H}{31.8}{200105F}{2.08}{GW191230H}{45}{GW191222A}{42.1}{GW191219E}{1.22}{GW191216G}{9.1}{GW191215G}{21.1}{GW191204G}{9.5}{GW191204A}{23.7}{GW191129G}{8.0}{GW191127B}{37}{GW191126C}{9.8}{GW191113B}{9.1}{GW191109A}{58}{GW191105C}{8.8}{GW191103A}{9.3}}}
\DeclareRobustCommand{\chieffinfinityonlyprecavgminus}[1]{\IfEqCase{#1}{{GW200322G}{0.58}{GW200316I}{0.10}{GW200311L}{0.20}{GW200308G}{0.49}{GW200306A}{0.46}{GW200302A}{0.26}{GW200225B}{0.28}{GW200224H}{0.16}{GW200220H}{0.33}{GW200220E}{0.38}{GW200219D}{0.29}{GW200216G}{0.36}{GW200210B}{0.21}{GW200209E}{0.30}{GW200208K}{0.46}{GW200208G}{0.27}{GW200202F}{0.06}{GW200129D}{0.16}{GW200128C}{0.25}{GW200115A}{0.42}{GW200112H}{0.15}{200105F}{0.18}{GW191230H}{0.31}{GW191222A}{0.25}{GW191219E}{0.09}{GW191216G}{0.06}{GW191215G}{0.21}{GW191204G}{0.05}{GW191204A}{0.26}{GW191129G}{0.08}{GW191127B}{0.36}{GW191126C}{0.11}{GW191113B}{0.29}{GW191109A}{0.31}{GW191105C}{0.09}{GW191103A}{0.10}}}
\DeclareRobustCommand{\chieffinfinityonlyprecavgmed}[1]{\IfEqCase{#1}{{GW200322G}{0.27}{GW200316I}{0.13}{GW200311L}{-0.02}{GW200308G}{0.16}{GW200306A}{0.32}{GW200302A}{0.01}{GW200225B}{-0.12}{GW200224H}{0.10}{GW200220H}{-0.07}{GW200220E}{0.06}{GW200219D}{-0.08}{GW200216G}{0.10}{GW200210B}{0.02}{GW200209E}{-0.12}{GW200208K}{0.45}{GW200208G}{-0.07}{GW200202F}{0.04}{GW200129D}{0.11}{GW200128C}{0.12}{GW200115A}{-0.15}{GW200112H}{0.06}{200105F}{0.00}{GW191230H}{-0.05}{GW191222A}{-0.04}{GW191219E}{0.00}{GW191216G}{0.11}{GW191215G}{-0.04}{GW191204G}{0.16}{GW191204A}{0.05}{GW191129G}{0.06}{GW191127B}{0.18}{GW191126C}{0.21}{GW191113B}{0.00}{GW191109A}{-0.29}{GW191105C}{-0.02}{GW191103A}{0.21}}}
\DeclareRobustCommand{\chieffinfinityonlyprecavgplus}[1]{\IfEqCase{#1}{{GW200322G}{0.54}{GW200316I}{0.27}{GW200311L}{0.16}{GW200308G}{0.58}{GW200306A}{0.28}{GW200302A}{0.25}{GW200225B}{0.17}{GW200224H}{0.15}{GW200220H}{0.27}{GW200220E}{0.40}{GW200219D}{0.23}{GW200216G}{0.34}{GW200210B}{0.22}{GW200209E}{0.24}{GW200208K}{0.42}{GW200208G}{0.21}{GW200202F}{0.13}{GW200129D}{0.11}{GW200128C}{0.24}{GW200115A}{0.23}{GW200112H}{0.15}{200105F}{0.13}{GW191230H}{0.26}{GW191222A}{0.20}{GW191219E}{0.07}{GW191216G}{0.13}{GW191215G}{0.17}{GW191204G}{0.08}{GW191204A}{0.25}{GW191129G}{0.16}{GW191127B}{0.34}{GW191126C}{0.15}{GW191113B}{0.37}{GW191109A}{0.42}{GW191105C}{0.13}{GW191103A}{0.16}}}
\DeclareRobustCommand{\chieffinfinityonlyprecavgtenthpercentile}[1]{\IfEqCase{#1}{{GW200322G}{-0.17}{GW200316I}{0.04}{GW200311L}{-0.17}{GW200308G}{-0.22}{GW200306A}{-0.06}{GW200302A}{-0.17}{GW200225B}{-0.34}{GW200224H}{-0.02}{GW200220H}{-0.33}{GW200220E}{-0.22}{GW200219D}{-0.30}{GW200216G}{-0.17}{GW200210B}{-0.13}{GW200209E}{-0.35}{GW200208K}{0.08}{GW200208G}{-0.28}{GW200202F}{-0.01}{GW200129D}{0.00}{GW200128C}{-0.07}{GW200115A}{-0.51}{GW200112H}{-0.05}{200105F}{-0.10}{GW191230H}{-0.28}{GW191222A}{-0.23}{GW191219E}{-0.06}{GW191216G}{0.06}{GW191215G}{-0.20}{GW191204G}{0.12}{GW191204A}{-0.14}{GW191129G}{0.00}{GW191127B}{-0.10}{GW191126C}{0.12}{GW191113B}{-0.20}{GW191109A}{-0.54}{GW191105C}{-0.09}{GW191103A}{0.13}}}
\DeclareRobustCommand{\chieffinfinityonlyprecavgnintiethpercentile}[1]{\IfEqCase{#1}{{GW200322G}{0.75}{GW200316I}{0.32}{GW200311L}{0.10}{GW200308G}{0.68}{GW200306A}{0.55}{GW200302A}{0.20}{GW200225B}{0.02}{GW200224H}{0.22}{GW200220H}{0.13}{GW200220E}{0.37}{GW200219D}{0.10}{GW200216G}{0.36}{GW200210B}{0.20}{GW200209E}{0.07}{GW200208K}{0.81}{GW200208G}{0.09}{GW200202F}{0.13}{GW200129D}{0.20}{GW200128C}{0.31}{GW200115A}{0.04}{GW200112H}{0.18}{200105F}{0.08}{GW191230H}{0.15}{GW191222A}{0.11}{GW191219E}{0.05}{GW191216G}{0.20}{GW191215G}{0.09}{GW191204G}{0.22}{GW191204A}{0.24}{GW191129G}{0.19}{GW191127B}{0.45}{GW191126C}{0.32}{GW191113B}{0.27}{GW191109A}{0.00}{GW191105C}{0.07}{GW191103A}{0.33}}}
\DeclareRobustCommand{\chipminus}[1]{\IfEqCase{#1}{{GW200322G}{0.31}{GW200316I}{0.20}{GW200311L}{0.35}{GW200308G}{0.30}{GW200306A}{0.31}{GW200302A}{0.28}{GW200225B}{0.38}{GW200224H}{0.35}{GW200220H}{0.37}{GW200220E}{0.37}{GW200219D}{0.35}{GW200216G}{0.35}{GW200210B}{0.12}{GW200209E}{0.37}{GW200208K}{0.29}{GW200208G}{0.29}{GW200202F}{0.22}{GW200129D}{0.37}{GW200128C}{0.40}{GW200115A}{0.16}{GW200112H}{0.30}{200105F}{0.07}{GW191230H}{0.39}{GW191222A}{0.32}{GW191219E}{0.07}{GW191216G}{0.15}{GW191215G}{0.38}{GW191204G}{0.27}{GW191204A}{0.38}{GW191129G}{0.19}{GW191127B}{0.41}{GW191126C}{0.26}{GW191113B}{0.16}{GW191109A}{0.37}{GW191105C}{0.24}{GW191103A}{0.26}}}
\DeclareRobustCommand{\chipmed}[1]{\IfEqCase{#1}{{GW200322G}{0.42}{GW200316I}{0.29}{GW200311L}{0.45}{GW200308G}{0.41}{GW200306A}{0.43}{GW200302A}{0.37}{GW200225B}{0.53}{GW200224H}{0.48}{GW200220H}{0.49}{GW200220E}{0.50}{GW200219D}{0.48}{GW200216G}{0.45}{GW200210B}{0.15}{GW200209E}{0.51}{GW200208K}{0.41}{GW200208G}{0.39}{GW200202F}{0.28}{GW200129D}{0.51}{GW200128C}{0.57}{GW200115A}{0.20}{GW200112H}{0.39}{200105F}{0.09}{GW191230H}{0.52}{GW191222A}{0.41}{GW191219E}{0.09}{GW191216G}{0.23}{GW191215G}{0.50}{GW191204G}{0.40}{GW191204A}{0.50}{GW191129G}{0.26}{GW191127B}{0.52}{GW191126C}{0.39}{GW191113B}{0.20}{GW191109A}{0.63}{GW191105C}{0.30}{GW191103A}{0.40}}}
\DeclareRobustCommand{\chipplus}[1]{\IfEqCase{#1}{{GW200322G}{0.42}{GW200316I}{0.38}{GW200311L}{0.40}{GW200308G}{0.42}{GW200306A}{0.39}{GW200302A}{0.45}{GW200225B}{0.34}{GW200224H}{0.37}{GW200220H}{0.39}{GW200220E}{0.37}{GW200219D}{0.40}{GW200216G}{0.42}{GW200210B}{0.22}{GW200209E}{0.39}{GW200208K}{0.38}{GW200208G}{0.41}{GW200202F}{0.40}{GW200129D}{0.42}{GW200128C}{0.34}{GW200115A}{0.34}{GW200112H}{0.39}{200105F}{0.17}{GW191230H}{0.38}{GW191222A}{0.41}{GW191219E}{0.07}{GW191216G}{0.35}{GW191215G}{0.37}{GW191204G}{0.38}{GW191204A}{0.40}{GW191129G}{0.36}{GW191127B}{0.41}{GW191126C}{0.40}{GW191113B}{0.54}{GW191109A}{0.29}{GW191105C}{0.45}{GW191103A}{0.41}}}
\DeclareRobustCommand{\chiptenthpercentile}[1]{\IfEqCase{#1}{{GW200322G}{0.16}{GW200316I}{0.12}{GW200311L}{0.16}{GW200308G}{0.15}{GW200306A}{0.17}{GW200302A}{0.12}{GW200225B}{0.22}{GW200224H}{0.19}{GW200220H}{0.18}{GW200220E}{0.19}{GW200219D}{0.18}{GW200216G}{0.15}{GW200210B}{0.05}{GW200209E}{0.20}{GW200208K}{0.16}{GW200208G}{0.14}{GW200202F}{0.09}{GW200129D}{0.20}{GW200128C}{0.24}{GW200115A}{0.06}{GW200112H}{0.13}{200105F}{0.03}{GW191230H}{0.19}{GW191222A}{0.15}{GW191219E}{0.03}{GW191216G}{0.10}{GW191215G}{0.19}{GW191204G}{0.17}{GW191204A}{0.18}{GW191129G}{0.10}{GW191127B}{0.17}{GW191126C}{0.17}{GW191113B}{0.06}{GW191109A}{0.33}{GW191105C}{0.09}{GW191103A}{0.18}}}
\DeclareRobustCommand{\chipnintiethpercentile}[1]{\IfEqCase{#1}{{GW200322G}{0.75}{GW200316I}{0.58}{GW200311L}{0.77}{GW200308G}{0.74}{GW200306A}{0.74}{GW200302A}{0.73}{GW200225B}{0.82}{GW200224H}{0.78}{GW200220H}{0.82}{GW200220E}{0.82}{GW200219D}{0.80}{GW200216G}{0.80}{GW200210B}{0.32}{GW200209E}{0.84}{GW200208K}{0.71}{GW200208G}{0.72}{GW200202F}{0.59}{GW200129D}{0.89}{GW200128C}{0.85}{GW200115A}{0.46}{GW200112H}{0.70}{200105F}{0.19}{GW191230H}{0.84}{GW191222A}{0.75}{GW191219E}{0.14}{GW191216G}{0.48}{GW191215G}{0.82}{GW191204G}{0.70}{GW191204A}{0.83}{GW191129G}{0.54}{GW191127B}{0.88}{GW191126C}{0.70}{GW191113B}{0.63}{GW191109A}{0.87}{GW191105C}{0.65}{GW191103A}{0.72}}}
\DeclareRobustCommand{\chieffminus}[1]{\IfEqCase{#1}{{GW200322G}{0.58}{GW200316I}{0.10}{GW200311L}{0.20}{GW200308G}{0.49}{GW200306A}{0.46}{GW200302A}{0.26}{GW200225B}{0.28}{GW200224H}{0.16}{GW200220H}{0.33}{GW200220E}{0.38}{GW200219D}{0.29}{GW200216G}{0.36}{GW200210B}{0.21}{GW200209E}{0.30}{GW200208K}{0.46}{GW200208G}{0.27}{GW200202F}{0.06}{GW200129D}{0.16}{GW200128C}{0.25}{GW200115A}{0.42}{GW200112H}{0.15}{200105F}{0.18}{GW191230H}{0.31}{GW191222A}{0.25}{GW191219E}{0.09}{GW191216G}{0.06}{GW191215G}{0.21}{GW191204G}{0.05}{GW191204A}{0.26}{GW191129G}{0.08}{GW191127B}{0.36}{GW191126C}{0.11}{GW191113B}{0.29}{GW191109A}{0.31}{GW191105C}{0.09}{GW191103A}{0.10}{GW190412A}{0.11}{GW151226A}{0.08}{GW170608A}{0.05}}}
\DeclareRobustCommand{\chieffmed}[1]{\IfEqCase{#1}{{GW200322G}{0.27}{GW200316I}{0.13}{GW200311L}{-0.02}{GW200308G}{0.16}{GW200306A}{0.32}{GW200302A}{0.01}{GW200225B}{-0.12}{GW200224H}{0.10}{GW200220H}{-0.07}{GW200220E}{0.06}{GW200219D}{-0.08}{GW200216G}{0.10}{GW200210B}{0.02}{GW200209E}{-0.12}{GW200208K}{0.45}{GW200208G}{-0.07}{GW200202F}{0.04}{GW200129D}{0.11}{GW200128C}{0.12}{GW200115A}{-0.15}{GW200112H}{0.06}{200105F}{0.00}{GW191230H}{-0.05}{GW191222A}{-0.04}{GW191219E}{0.00}{GW191216G}{0.11}{GW191215G}{-0.04}{GW191204G}{0.16}{GW191204A}{0.05}{GW191129G}{0.06}{GW191127B}{0.18}{GW191126C}{0.21}{GW191113B}{0.00}{GW191109A}{-0.29}{GW191105C}{-0.02}{GW191103A}{0.21}{GW190412A}{0.25}{GW151226A}{0.20}{GW170608A}{0.05}}}
\DeclareRobustCommand{\chieffplus}[1]{\IfEqCase{#1}{{GW200322G}{0.54}{GW200316I}{0.27}{GW200311L}{0.16}{GW200308G}{0.58}{GW200306A}{0.28}{GW200302A}{0.25}{GW200225B}{0.17}{GW200224H}{0.15}{GW200220H}{0.27}{GW200220E}{0.40}{GW200219D}{0.23}{GW200216G}{0.34}{GW200210B}{0.22}{GW200209E}{0.24}{GW200208K}{0.42}{GW200208G}{0.21}{GW200202F}{0.13}{GW200129D}{0.11}{GW200128C}{0.24}{GW200115A}{0.23}{GW200112H}{0.15}{200105F}{0.13}{GW191230H}{0.26}{GW191222A}{0.20}{GW191219E}{0.07}{GW191216G}{0.13}{GW191215G}{0.17}{GW191204G}{0.08}{GW191204A}{0.25}{GW191129G}{0.16}{GW191127B}{0.34}{GW191126C}{0.15}{GW191113B}{0.37}{GW191109A}{0.42}{GW191105C}{0.13}{GW191103A}{0.16}{GW190412A}{0.08}{GW151226A}{0.23}{GW170608A}{0.13}}}
\DeclareRobustCommand{\chiefftenthpercentile}[1]{\IfEqCase{#1}{{GW200322G}{-0.17}{GW200316I}{0.04}{GW200311L}{-0.17}{GW200308G}{-0.22}{GW200306A}{-0.06}{GW200302A}{-0.17}{GW200225B}{-0.34}{GW200224H}{-0.02}{GW200220H}{-0.33}{GW200220E}{-0.22}{GW200219D}{-0.30}{GW200216G}{-0.17}{GW200210B}{-0.13}{GW200209E}{-0.35}{GW200208K}{0.08}{GW200208G}{-0.28}{GW200202F}{-0.01}{GW200129D}{0.00}{GW200128C}{-0.07}{GW200115A}{-0.51}{GW200112H}{-0.05}{200105F}{-0.10}{GW191230H}{-0.28}{GW191222A}{-0.23}{GW191219E}{-0.06}{GW191216G}{0.06}{GW191215G}{-0.20}{GW191204G}{0.12}{GW191204A}{-0.14}{GW191129G}{0.00}{GW191127B}{-0.10}{GW191126C}{0.12}{GW191113B}{-0.20}{GW191109A}{-0.54}{GW191105C}{-0.09}{GW191103A}{0.13}}}
\DeclareRobustCommand{\chieffnintiethpercentile}[1]{\IfEqCase{#1}{{GW200322G}{0.75}{GW200316I}{0.32}{GW200311L}{0.10}{GW200308G}{0.68}{GW200306A}{0.55}{GW200302A}{0.20}{GW200225B}{0.02}{GW200224H}{0.22}{GW200220H}{0.13}{GW200220E}{0.37}{GW200219D}{0.10}{GW200216G}{0.36}{GW200210B}{0.20}{GW200209E}{0.07}{GW200208K}{0.81}{GW200208G}{0.09}{GW200202F}{0.13}{GW200129D}{0.20}{GW200128C}{0.31}{GW200115A}{0.04}{GW200112H}{0.18}{200105F}{0.08}{GW191230H}{0.15}{GW191222A}{0.11}{GW191219E}{0.05}{GW191216G}{0.20}{GW191215G}{0.09}{GW191204G}{0.22}{GW191204A}{0.24}{GW191129G}{0.19}{GW191127B}{0.45}{GW191126C}{0.32}{GW191113B}{0.27}{GW191109A}{0.00}{GW191105C}{0.07}{GW191103A}{0.33}}}
\DeclareRobustCommand{\chirpmasssourceminus}[1]{\IfEqCase{#1}{{GW200322G}{4.0}{GW200316I}{0.55}{GW200311L}{2.0}{GW200308G}{18}{GW200306A}{3.0}{GW200302A}{3.0}{GW200225B}{1.4}{GW200224H}{2.7}{GW200220H}{5.1}{GW200220E}{15}{GW200219D}{3.8}{GW200216G}{8.5}{GW200210B}{0.40}{GW200209E}{4.2}{GW200208K}{5.2}{GW200208G}{3.1}{GW200202F}{0.20}{GW200129D}{2.3}{GW200128C}{5.5}{GW200115A}{0.07}{GW200112H}{2.1}{200105F}{0.08}{GW191230H}{5.6}{GW191222A}{5.0}{GW191219E}{0.17}{GW191216G}{0.19}{GW191215G}{1.7}{GW191204G}{0.28}{GW191204A}{3.2}{GW191129G}{0.28}{GW191127B}{9.1}{GW191126C}{0.71}{GW191113B}{1.0}{GW191109A}{7.5}{GW191105C}{0.45}{GW191103A}{0.57}{GW190412A}{0.3}{GW151226A}{0.3}{GW170608A}{0.2}}}
\DeclareRobustCommand{\chirpmasssourcemed}[1]{\IfEqCase{#1}{{GW200322G}{15.0}{GW200316I}{8.75}{GW200311L}{26.6}{GW200308G}{34}{GW200306A}{17.5}{GW200302A}{23.4}{GW200225B}{14.2}{GW200224H}{31.1}{GW200220H}{28.2}{GW200220E}{62}{GW200219D}{27.6}{GW200216G}{32.9}{GW200210B}{6.56}{GW200209E}{26.7}{GW200208K}{19.8}{GW200208G}{27.7}{GW200202F}{7.49}{GW200129D}{27.2}{GW200128C}{32.0}{GW200115A}{2.43}{GW200112H}{27.4}{200105F}{3.42}{GW191230H}{36.5}{GW191222A}{33.8}{GW191219E}{4.31}{GW191216G}{8.33}{GW191215G}{18.4}{GW191204G}{8.56}{GW191204A}{19.8}{GW191129G}{7.31}{GW191127B}{29.9}{GW191126C}{8.65}{GW191113B}{10.7}{GW191109A}{47.5}{GW191105C}{7.82}{GW191103A}{8.34}{GW190412A}{13.3}{GW151226A}{8.9}{GW170608A}{7.9}}}
\DeclareRobustCommand{\chirpmasssourceplus}[1]{\IfEqCase{#1}{{GW200322G}{29.5}{GW200316I}{0.62}{GW200311L}{2.4}{GW200308G}{44}{GW200306A}{3.5}{GW200302A}{4.7}{GW200225B}{1.5}{GW200224H}{3.3}{GW200220H}{7.3}{GW200220E}{23}{GW200219D}{5.6}{GW200216G}{9.3}{GW200210B}{0.38}{GW200209E}{6.0}{GW200208K}{10.5}{GW200208G}{3.7}{GW200202F}{0.24}{GW200129D}{2.1}{GW200128C}{7.5}{GW200115A}{0.05}{GW200112H}{2.6}{200105F}{0.08}{GW191230H}{8.2}{GW191222A}{7.1}{GW191219E}{0.12}{GW191216G}{0.22}{GW191215G}{2.2}{GW191204G}{0.41}{GW191204A}{3.6}{GW191129G}{0.43}{GW191127B}{11.7}{GW191126C}{0.95}{GW191113B}{1.1}{GW191109A}{9.6}{GW191105C}{0.61}{GW191103A}{0.66}{GW190412A}{0.4}{GW151226A}{0.3}{GW170608A}{0.2}}}
\DeclareRobustCommand{\chirpmasssourcetenthpercentile}[1]{\IfEqCase{#1}{{GW200322G}{11.6}{GW200316I}{8.31}{GW200311L}{25.0}{GW200308G}{18}{GW200306A}{15.0}{GW200302A}{21.0}{GW200225B}{13.1}{GW200224H}{29.1}{GW200220H}{24.1}{GW200220E}{50}{GW200219D}{24.6}{GW200216G}{26.2}{GW200210B}{6.25}{GW200209E}{23.3}{GW200208K}{15.4}{GW200208G}{25.3}{GW200202F}{7.33}{GW200129D}{25.5}{GW200128C}{27.5}{GW200115A}{2.38}{GW200112H}{25.7}{200105F}{3.36}{GW191230H}{31.9}{GW191222A}{29.7}{GW191219E}{4.19}{GW191216G}{8.18}{GW191215G}{17.0}{GW191204G}{8.34}{GW191204A}{17.2}{GW191129G}{7.08}{GW191127B}{22.8}{GW191126C}{8.09}{GW191113B}{9.9}{GW191109A}{41.5}{GW191105C}{7.46}{GW191103A}{7.88}}}
\DeclareRobustCommand{\chirpmasssourcenintiethpercentile}[1]{\IfEqCase{#1}{{GW200322G}{36.7}{GW200316I}{9.26}{GW200311L}{28.4}{GW200308G}{64}{GW200306A}{20.2}{GW200302A}{26.8}{GW200225B}{15.3}{GW200224H}{33.6}{GW200220H}{33.7}{GW200220E}{79}{GW200219D}{32.0}{GW200216G}{40.0}{GW200210B}{6.85}{GW200209E}{31.3}{GW200208K}{28.3}{GW200208G}{30.5}{GW200202F}{7.68}{GW200129D}{28.8}{GW200128C}{37.5}{GW200115A}{2.47}{GW200112H}{29.3}{200105F}{3.49}{GW191230H}{42.7}{GW191222A}{39.4}{GW191219E}{4.41}{GW191216G}{8.51}{GW191215G}{20.1}{GW191204G}{8.89}{GW191204A}{22.7}{GW191129G}{7.66}{GW191127B}{38.9}{GW191126C}{9.40}{GW191113B}{11.5}{GW191109A}{54.5}{GW191105C}{8.30}{GW191103A}{8.88}}}
\newcommand{\chipuncert}[1]{\ensuremath{ \chipmed{#1}_{-\chipminus{#1}}^{+\chipplus{#1}}  } } 
\newcommand{\spinonexuncert}[1]{\ensuremath{ \spinonexmed{#1}_{-\spinonexminus{#1}}^{+\spinonexplus{#1}}  } } 
\newcommand{\masstwosourceuncert}[1]{\ensuremath{ \masstwosourcemed{#1}_{-\masstwosourceminus{#1}}^{+\masstwosourceplus{#1}}  } } 
\newcommand{\spintwozuncert}[1]{\ensuremath{ \spintwozmed{#1}_{-\spintwozminus{#1}}^{+\spintwozplus{#1}}  } } 
\newcommand{\massonedetuncert}[1]{\ensuremath{ \massonedetmed{#1}_{-\massonedetminus{#1}}^{+\massonedetplus{#1}}  } } 
\newcommand{\finalmasssourceuncert}[1]{\ensuremath{ \finalmasssourcemed{#1}_{-\finalmasssourceminus{#1}}^{+\finalmasssourceplus{#1}}  } } 
\newcommand{\spinonezuncert}[1]{\ensuremath{ \spinonezmed{#1}_{-\spinonezminus{#1}}^{+\spinonezplus{#1}}  } } 
\newcommand{\loglikelihooduncert}[1]{\ensuremath{ \loglikelihoodmed{#1}_{-\loglikelihoodminus{#1}}^{+\loglikelihoodplus{#1}}  } } 
\newcommand{\phitwouncert}[1]{\ensuremath{ \phitwomed{#1}_{-\phitwominus{#1}}^{+\phitwoplus{#1}}  } } 
\newcommand{\spinoneuncert}[1]{\ensuremath{ \spinonemed{#1}_{-\spinoneminus{#1}}^{+\spinoneplus{#1}}  } } 
\newcommand{\tilttwouncert}[1]{\ensuremath{ \tilttwomed{#1}_{-\tilttwominus{#1}}^{+\tilttwoplus{#1}}  } } 
\newcommand{\iotauncert}[1]{\ensuremath{ \iotamed{#1}_{-\iotaminus{#1}}^{+\iotaplus{#1}}  } } 
\newcommand{\psiuncert}[1]{\ensuremath{ \psimed{#1}_{-\psiminus{#1}}^{+\psiplus{#1}}  } } 
\newcommand{\chirpmassdetuncert}[1]{\ensuremath{ \chirpmassdetmed{#1}_{-\chirpmassdetminus{#1}}^{+\chirpmassdetplus{#1}}  } } 
\newcommand{\massonesourceuncert}[1]{\ensuremath{ \massonesourcemed{#1}_{-\massonesourceminus{#1}}^{+\massonesourceplus{#1}}  } } 
\newcommand{\spintwoyuncert}[1]{\ensuremath{ \spintwoymed{#1}_{-\spintwoyminus{#1}}^{+\spintwoyplus{#1}}  } } 
\newcommand{\finalspinuncert}[1]{\ensuremath{ \finalspinmed{#1}_{-\finalspinminus{#1}}^{+\finalspinplus{#1}}  } } 
\newcommand{\radiatedenergyuncert}[1]{\ensuremath{ \radiatedenergymed{#1}_{-\radiatedenergyminus{#1}}^{+\radiatedenergyplus{#1}}  } } 
\newcommand{\geocenttimeuncert}[1]{\ensuremath{ \geocenttimemed{#1}_{-\geocenttimeminus{#1}}^{+\geocenttimeplus{#1}}  } } 
\newcommand{\spintwoxuncert}[1]{\ensuremath{ \spintwoxmed{#1}_{-\spintwoxminus{#1}}^{+\spintwoxplus{#1}}  } } 
\newcommand{\comovingdistuncert}[1]{\ensuremath{ \comovingdistmed{#1}_{-\comovingdistminus{#1}}^{+\comovingdistplus{#1}}  } } 
\newcommand{\luminositydistanceuncert}[1]{\ensuremath{ \luminositydistancemed{#1}_{-\luminositydistanceminus{#1}}^{+\luminositydistanceplus{#1}}  } } 
\newcommand{\decuncert}[1]{\ensuremath{ \decmed{#1}_{-\decminus{#1}}^{+\decplus{#1}}  } } 
\newcommand{\spintwouncert}[1]{\ensuremath{ \spintwomed{#1}_{-\spintwominus{#1}}^{+\spintwoplus{#1}}  } } 
\newcommand{\tiltoneuncert}[1]{\ensuremath{ \tiltonemed{#1}_{-\tiltoneminus{#1}}^{+\tiltoneplus{#1}}  } } 
\newcommand{\phijluncert}[1]{\ensuremath{ \phijlmed{#1}_{-\phijlminus{#1}}^{+\phijlplus{#1}}  } } 
\newcommand{\chirpmasssourceuncert}[1]{\ensuremath{ \chirpmasssourcemed{#1}_{-\chirpmasssourceminus{#1}}^{+\chirpmasssourceplus{#1}}  } } 
\newcommand{\costhetajnuncert}[1]{\ensuremath{ \costhetajnmed{#1}_{-\costhetajnminus{#1}}^{+\costhetajnplus{#1}}  } } 
\newcommand{\spinoneyuncert}[1]{\ensuremath{ \spinoneymed{#1}_{-\spinoneyminus{#1}}^{+\spinoneyplus{#1}}  } } 
\newcommand{\symmetricmassratiouncert}[1]{\ensuremath{ \symmetricmassratiomed{#1}_{-\symmetricmassratiominus{#1}}^{+\symmetricmassratioplus{#1}}  } } 
\newcommand{\costilttwouncert}[1]{\ensuremath{ \costilttwomed{#1}_{-\costilttwominus{#1}}^{+\costilttwoplus{#1}}  } } 
\newcommand{\rauncert}[1]{\ensuremath{ \ramed{#1}_{-\raminus{#1}}^{+\raplus{#1}}  } } 
\newcommand{\redshiftuncert}[1]{\ensuremath{ \redshiftmed{#1}_{-\redshiftminus{#1}}^{+\redshiftplus{#1}}  } } 
\newcommand{\totalmassdetuncert}[1]{\ensuremath{ \totalmassdetmed{#1}_{-\totalmassdetminus{#1}}^{+\totalmassdetplus{#1}}  } } 
\newcommand{\masstwodetuncert}[1]{\ensuremath{ \masstwodetmed{#1}_{-\masstwodetminus{#1}}^{+\masstwodetplus{#1}}  } } 
\newcommand{\chipinfinityonlyprecavguncert}[1]{\ensuremath{ \chipinfinityonlyprecavgmed{#1}_{-\chipinfinityonlyprecavgminus{#1}}^{+\chipinfinityonlyprecavgplus{#1}}  } } 
\newcommand{\thetajnuncert}[1]{\ensuremath{ \thetajnmed{#1}_{-\thetajnminus{#1}}^{+\thetajnplus{#1}}  } } 
\newcommand{\totalmasssourceuncert}[1]{\ensuremath{ \totalmasssourcemed{#1}_{-\totalmasssourceminus{#1}}^{+\totalmasssourceplus{#1}}  } } 
\newcommand{\cosiotauncert}[1]{\ensuremath{ \cosiotamed{#1}_{-\cosiotaminus{#1}}^{+\cosiotaplus{#1}}  } } 
\newcommand{\massratiouncert}[1]{\ensuremath{ \massratiomed{#1}_{-\massratiominus{#1}}^{+\massratioplus{#1}}  } } 
\newcommand{\chieffinfinityonlyprecavguncert}[1]{\ensuremath{ \chieffinfinityonlyprecavgmed{#1}_{-\chieffinfinityonlyprecavgminus{#1}}^{+\chieffinfinityonlyprecavgplus{#1}}  } } 
\newcommand{\costiltoneuncert}[1]{\ensuremath{ \costiltonemed{#1}_{-\costiltoneminus{#1}}^{+\costiltoneplus{#1}}  } } 
\newcommand{\finalmassdetuncert}[1]{\ensuremath{ \finalmassdetmed{#1}_{-\finalmassdetminus{#1}}^{+\finalmassdetplus{#1}}  } } 
\newcommand{\phionetwouncert}[1]{\ensuremath{ \phionetwomed{#1}_{-\phionetwominus{#1}}^{+\phionetwoplus{#1}}  } } 
\newcommand{\phaseuncert}[1]{\ensuremath{ \phasemed{#1}_{-\phaseminus{#1}}^{+\phaseplus{#1}}  } } 
\newcommand{\chieffuncert}[1]{\ensuremath{ \chieffmed{#1}_{-\chieffminus{#1}}^{+\chieffplus{#1}}  } } 
\newcommand{\phioneuncert}[1]{\ensuremath{ \phionemed{#1}_{-\phioneminus{#1}}^{+\phioneplus{#1}}  } } 
      \DeclareRobustCommand{\networkmatchedfiltersnrIMRPminus}[1]{\IfEqCase{#1}{{GW200322G}{3.0}{GW200316I}{0.7}{GW200311L}{0.2}{GW200308G}{2.9}{GW200306A}{0.6}{GW200302A}{0.4}{GW200225B}{0.4}{GW200224H}{0.2}{GW200220H}{0.5}{GW200220E}{0.7}{GW200219D}{0.5}{GW200216G}{0.5}{GW200210B}{0.7}{GW200209E}{0.5}{GW200208K}{1.2}{GW200208G}{0.4}{GW200202F}{0.4}{GW200129D}{0.2}{GW200128C}{0.4}{GW200115A}{0.5}{GW200112H}{0.2}{GW191230H}{0.4}{GW191222A}{0.3}{GW191219E}{0.8}{GW191216G}{0.2}{GW191215G}{0.4}{GW191204G}{0.3}{GW191204A}{0.6}{GW191129G}{0.3}{GW191127B}{0.6}{GW191126C}{0.5}{GW191113B}{1.1}{GW191109A}{0.5}{GW191105C}{0.5}{GW191103A}{0.5}{200105F}{0.4}{GW190412A}{0.4}{GW151226A}{0.4}{GW170608A}{0.3}}}
\DeclareRobustCommand{\networkmatchedfiltersnrIMRPmed}[1]{\IfEqCase{#1}{{GW200322G}{4.5}{GW200316I}{10.3}{GW200311L}{17.8}{GW200308G}{4.7}{GW200306A}{7.8}{GW200302A}{10.8}{GW200225B}{12.5}{GW200224H}{20.0}{GW200220H}{8.5}{GW200220E}{7.2}{GW200219D}{10.7}{GW200216G}{8.1}{GW200210B}{8.4}{GW200209E}{9.6}{GW200208K}{7.4}{GW200208G}{10.8}{GW200202F}{10.8}{GW200129D}{26.8}{GW200128C}{10.6}{GW200115A}{11.3}{GW200112H}{19.8}{GW191230H}{10.4}{GW191222A}{12.5}{GW191219E}{9.1}{GW191216G}{18.6}{GW191215G}{11.2}{GW191204G}{17.4}{GW191204A}{8.9}{GW191129G}{13.1}{GW191127B}{9.2}{GW191126C}{8.3}{GW191113B}{7.9}{GW191109A}{17.3}{GW191105C}{9.7}{GW191103A}{8.9}{200105F}{13.7}{GW190412A}{19.0}{GW151226A}{12.7}{GW170608A}{15.3}}}
\DeclareRobustCommand{\networkmatchedfiltersnrIMRPplus}[1]{\IfEqCase{#1}{{GW200322G}{2.7}{GW200316I}{0.4}{GW200311L}{0.2}{GW200308G}{2.5}{GW200306A}{0.4}{GW200302A}{0.3}{GW200225B}{0.3}{GW200224H}{0.2}{GW200220H}{0.3}{GW200220E}{0.4}{GW200219D}{0.3}{GW200216G}{0.4}{GW200210B}{0.5}{GW200209E}{0.4}{GW200208K}{1.4}{GW200208G}{0.3}{GW200202F}{0.2}{GW200129D}{0.2}{GW200128C}{0.3}{GW200115A}{0.3}{GW200112H}{0.1}{GW191230H}{0.3}{GW191222A}{0.2}{GW191219E}{0.5}{GW191216G}{0.2}{GW191215G}{0.3}{GW191204G}{0.2}{GW191204A}{0.4}{GW191129G}{0.2}{GW191127B}{0.7}{GW191126C}{0.2}{GW191113B}{0.5}{GW191109A}{0.5}{GW191105C}{0.3}{GW191103A}{0.3}{200105F}{0.2}{GW190412A}{0.2}{GW151226A}{0.3}{GW170608A}{0.2}}}
\DeclareRobustCommand{\networkmatchedfiltersnrIMRPtenthpercentile}[1]{\IfEqCase{#1}{{GW200322G}{1.9}{GW200316I}{9.8}{GW200311L}{17.7}{GW200308G}{2.3}{GW200306A}{7.3}{GW200302A}{10.5}{GW200225B}{12.2}{GW200224H}{19.8}{GW200220H}{8.1}{GW200220E}{6.7}{GW200219D}{10.3}{GW200216G}{7.7}{GW200210B}{7.9}{GW200209E}{9.2}{GW200208K}{6.6}{GW200208G}{10.5}{GW200202F}{10.5}{GW200129D}{26.6}{GW200128C}{10.3}{GW200115A}{10.9}{GW200112H}{19.6}{GW191230H}{10.1}{GW191222A}{12.3}{GW191219E}{8.5}{GW191216G}{18.4}{GW191215G}{10.9}{GW191204G}{17.2}{GW191204A}{8.4}{GW191129G}{12.9}{GW191127B}{8.7}{GW191126C}{7.9}{GW191113B}{7.1}{GW191109A}{16.9}{GW191105C}{9.3}{GW191103A}{8.5}{200105F}{13.4}}}
\DeclareRobustCommand{\networkmatchedfiltersnrIMRPnintiethpercentile}[1]{\IfEqCase{#1}{{GW200322G}{6.7}{GW200316I}{10.6}{GW200311L}{18.0}{GW200308G}{7.1}{GW200306A}{8.1}{GW200302A}{11.1}{GW200225B}{12.8}{GW200224H}{20.1}{GW200220H}{8.7}{GW200220E}{7.6}{GW200219D}{10.9}{GW200216G}{8.4}{GW200210B}{8.8}{GW200209E}{9.9}{GW200208K}{8.4}{GW200208G}{11.0}{GW200202F}{11.0}{GW200129D}{27.0}{GW200128C}{10.9}{GW200115A}{11.5}{GW200112H}{19.9}{GW191230H}{10.7}{GW191222A}{12.7}{GW191219E}{9.6}{GW191216G}{18.7}{GW191215G}{11.5}{GW191204G}{17.6}{GW191204A}{9.2}{GW191129G}{13.3}{GW191127B}{9.7}{GW191126C}{8.5}{GW191113B}{8.3}{GW191109A}{17.7}{GW191105C}{9.9}{GW191103A}{9.1}{200105F}{13.9}}}
\newcommand{\radiatedenergyIMRPuncert}[1]{\ensuremath{ \radiatedenergyIMRPmed{#1}_{-\radiatedenergyIMRPminus{#1}}^{+\radiatedenergyIMRPplus{#1}}  } } 
\newcommand{\spinonezIMRPuncert}[1]{\ensuremath{ \spinonezIMRPmed{#1}_{-\spinonezIMRPminus{#1}}^{+\spinonezIMRPplus{#1}}  } } 
\newcommand{\masstwosourceIMRPuncert}[1]{\ensuremath{ \masstwosourceIMRPmed{#1}_{-\masstwosourceIMRPminus{#1}}^{+\masstwosourceIMRPplus{#1}}  } } 
\newcommand{\spintwoyIMRPuncert}[1]{\ensuremath{ \spintwoyIMRPmed{#1}_{-\spintwoyIMRPminus{#1}}^{+\spintwoyIMRPplus{#1}}  } } 
\newcommand{\decIMRPuncert}[1]{\ensuremath{ \decIMRPmed{#1}_{-\decIMRPminus{#1}}^{+\decIMRPplus{#1}}  } } 
\newcommand{\spinoneyIMRPuncert}[1]{\ensuremath{ \spinoneyIMRPmed{#1}_{-\spinoneyIMRPminus{#1}}^{+\spinoneyIMRPplus{#1}}  } } 
\newcommand{\thetajnIMRPuncert}[1]{\ensuremath{ \thetajnIMRPmed{#1}_{-\thetajnIMRPminus{#1}}^{+\thetajnIMRPplus{#1}}  } } 
\newcommand{\phitwoIMRPuncert}[1]{\ensuremath{ \phitwoIMRPmed{#1}_{-\phitwoIMRPminus{#1}}^{+\phitwoIMRPplus{#1}}  } } 
\newcommand{\costilttwoIMRPuncert}[1]{\ensuremath{ \costilttwoIMRPmed{#1}_{-\costilttwoIMRPminus{#1}}^{+\costilttwoIMRPplus{#1}}  } } 
\newcommand{\chieffinfinityonlyprecavgIMRPuncert}[1]{\ensuremath{ \chieffinfinityonlyprecavgIMRPmed{#1}_{-\chieffinfinityonlyprecavgIMRPminus{#1}}^{+\chieffinfinityonlyprecavgIMRPplus{#1}}  } } 
\newcommand{\chieffIMRPuncert}[1]{\ensuremath{ \chieffIMRPmed{#1}_{-\chieffIMRPminus{#1}}^{+\chieffIMRPplus{#1}}  } } 
\newcommand{\raIMRPuncert}[1]{\ensuremath{ \raIMRPmed{#1}_{-\raIMRPminus{#1}}^{+\raIMRPplus{#1}}  } } 
\newcommand{\networkoptimalsnrIMRPuncert}[1]{\ensuremath{ \networkoptimalsnrIMRPmed{#1}_{-\networkoptimalsnrIMRPminus{#1}}^{+\networkoptimalsnrIMRPplus{#1}}  } } 
\newcommand{\totalmasssourceIMRPuncert}[1]{\ensuremath{ \totalmasssourceIMRPmed{#1}_{-\totalmasssourceIMRPminus{#1}}^{+\totalmasssourceIMRPplus{#1}}  } } 
\newcommand{\psiIMRPuncert}[1]{\ensuremath{ \psiIMRPmed{#1}_{-\psiIMRPminus{#1}}^{+\psiIMRPplus{#1}}  } } 
\newcommand{\chirpmasssourceIMRPuncert}[1]{\ensuremath{ \chirpmasssourceIMRPmed{#1}_{-\chirpmasssourceIMRPminus{#1}}^{+\chirpmasssourceIMRPplus{#1}}  } } 
\newcommand{\massonedetIMRPuncert}[1]{\ensuremath{ \massonedetIMRPmed{#1}_{-\massonedetIMRPminus{#1}}^{+\massonedetIMRPplus{#1}}  } } 
\newcommand{\phionetwoIMRPuncert}[1]{\ensuremath{ \phionetwoIMRPmed{#1}_{-\phionetwoIMRPminus{#1}}^{+\phionetwoIMRPplus{#1}}  } } 
\newcommand{\finalspinIMRPuncert}[1]{\ensuremath{ \finalspinIMRPmed{#1}_{-\finalspinIMRPminus{#1}}^{+\finalspinIMRPplus{#1}}  } } 
\newcommand{\symmetricmassratioIMRPuncert}[1]{\ensuremath{ \symmetricmassratioIMRPmed{#1}_{-\symmetricmassratioIMRPminus{#1}}^{+\symmetricmassratioIMRPplus{#1}}  } } 
\newcommand{\finalmassdetnonevolvedIMRPuncert}[1]{\ensuremath{ \finalmassdetnonevolvedIMRPmed{#1}_{-\finalmassdetnonevolvedIMRPminus{#1}}^{+\finalmassdetnonevolvedIMRPplus{#1}}  } } 
\newcommand{\tiltoneIMRPuncert}[1]{\ensuremath{ \tiltoneIMRPmed{#1}_{-\tiltoneIMRPminus{#1}}^{+\tiltoneIMRPplus{#1}}  } } 
\newcommand{\spintwozIMRPuncert}[1]{\ensuremath{ \spintwozIMRPmed{#1}_{-\spintwozIMRPminus{#1}}^{+\spintwozIMRPplus{#1}}  } } 
\newcommand{\geocenttimeIMRPuncert}[1]{\ensuremath{ \geocenttimeIMRPmed{#1}_{-\geocenttimeIMRPminus{#1}}^{+\geocenttimeIMRPplus{#1}}  } } 
\newcommand{\phaseIMRPuncert}[1]{\ensuremath{ \phaseIMRPmed{#1}_{-\phaseIMRPminus{#1}}^{+\phaseIMRPplus{#1}}  } } 
\newcommand{\spinonexIMRPuncert}[1]{\ensuremath{ \spinonexIMRPmed{#1}_{-\spinonexIMRPminus{#1}}^{+\spinonexIMRPplus{#1}}  } } 
\newcommand{\luminositydistanceIMRPuncert}[1]{\ensuremath{ \luminositydistanceIMRPmed{#1}_{-\luminositydistanceIMRPminus{#1}}^{+\luminositydistanceIMRPplus{#1}}  } } 
\newcommand{\spinoneIMRPuncert}[1]{\ensuremath{ \spinoneIMRPmed{#1}_{-\spinoneIMRPminus{#1}}^{+\spinoneIMRPplus{#1}}  } } 
\newcommand{\redshiftIMRPuncert}[1]{\ensuremath{ \redshiftIMRPmed{#1}_{-\redshiftIMRPminus{#1}}^{+\redshiftIMRPplus{#1}}  } } 
\newcommand{\iotaIMRPuncert}[1]{\ensuremath{ \iotaIMRPmed{#1}_{-\iotaIMRPminus{#1}}^{+\iotaIMRPplus{#1}}  } } 
\newcommand{\cosiotaIMRPuncert}[1]{\ensuremath{ \cosiotaIMRPmed{#1}_{-\cosiotaIMRPminus{#1}}^{+\cosiotaIMRPplus{#1}}  } } 
\newcommand{\finalmasssourceIMRPuncert}[1]{\ensuremath{ \finalmasssourceIMRPmed{#1}_{-\finalmasssourceIMRPminus{#1}}^{+\finalmasssourceIMRPplus{#1}}  } } 
\newcommand{\logpriorIMRPuncert}[1]{\ensuremath{ \logpriorIMRPmed{#1}_{-\logpriorIMRPminus{#1}}^{+\logpriorIMRPplus{#1}}  } } 
\newcommand{\chipIMRPuncert}[1]{\ensuremath{ \chipIMRPmed{#1}_{-\chipIMRPminus{#1}}^{+\chipIMRPplus{#1}}  } } 
\newcommand{\networkmatchedfiltersnrIMRPuncert}[1]{\ensuremath{ \networkmatchedfiltersnrIMRPmed{#1}_{-\networkmatchedfiltersnrIMRPminus{#1}}^{+\networkmatchedfiltersnrIMRPplus{#1}}  } } 
\newcommand{\chirpmassdetIMRPuncert}[1]{\ensuremath{ \chirpmassdetIMRPmed{#1}_{-\chirpmassdetIMRPminus{#1}}^{+\chirpmassdetIMRPplus{#1}}  } } 
\newcommand{\massonesourceIMRPuncert}[1]{\ensuremath{ \massonesourceIMRPmed{#1}_{-\massonesourceIMRPminus{#1}}^{+\massonesourceIMRPplus{#1}}  } } 
\newcommand{\massratioIMRPuncert}[1]{\ensuremath{ \massratioIMRPmed{#1}_{-\massratioIMRPminus{#1}}^{+\massratioIMRPplus{#1}}  } } 
\newcommand{\costiltoneIMRPuncert}[1]{\ensuremath{ \costiltoneIMRPmed{#1}_{-\costiltoneIMRPminus{#1}}^{+\costiltoneIMRPplus{#1}}  } } 
\newcommand{\tilttwoIMRPuncert}[1]{\ensuremath{ \tilttwoIMRPmed{#1}_{-\tilttwoIMRPminus{#1}}^{+\tilttwoIMRPplus{#1}}  } } 
\newcommand{\loglikelihoodIMRPuncert}[1]{\ensuremath{ \loglikelihoodIMRPmed{#1}_{-\loglikelihoodIMRPminus{#1}}^{+\loglikelihoodIMRPplus{#1}}  } } 
\newcommand{\chipinfinityonlyprecavgIMRPuncert}[1]{\ensuremath{ \chipinfinityonlyprecavgIMRPmed{#1}_{-\chipinfinityonlyprecavgIMRPminus{#1}}^{+\chipinfinityonlyprecavgIMRPplus{#1}}  } } 
\newcommand{\costhetajnIMRPuncert}[1]{\ensuremath{ \costhetajnIMRPmed{#1}_{-\costhetajnIMRPminus{#1}}^{+\costhetajnIMRPplus{#1}}  } } 
\newcommand{\phijlIMRPuncert}[1]{\ensuremath{ \phijlIMRPmed{#1}_{-\phijlIMRPminus{#1}}^{+\phijlIMRPplus{#1}}  } } 
\newcommand{\finalmassdetIMRPuncert}[1]{\ensuremath{ \finalmassdetIMRPmed{#1}_{-\finalmassdetIMRPminus{#1}}^{+\finalmassdetIMRPplus{#1}}  } } 
\newcommand{\totalmassdetIMRPuncert}[1]{\ensuremath{ \totalmassdetIMRPmed{#1}_{-\totalmassdetIMRPminus{#1}}^{+\totalmassdetIMRPplus{#1}}  } } 
\newcommand{\comovingdistIMRPuncert}[1]{\ensuremath{ \comovingdistIMRPmed{#1}_{-\comovingdistIMRPminus{#1}}^{+\comovingdistIMRPplus{#1}}  } } 
\newcommand{\spintwoIMRPuncert}[1]{\ensuremath{ \spintwoIMRPmed{#1}_{-\spintwoIMRPminus{#1}}^{+\spintwoIMRPplus{#1}}  } } 
\newcommand{\phioneIMRPuncert}[1]{\ensuremath{ \phioneIMRPmed{#1}_{-\phioneIMRPminus{#1}}^{+\phioneIMRPplus{#1}}  } } 
\newcommand{\spintwoxIMRPuncert}[1]{\ensuremath{ \spintwoxIMRPmed{#1}_{-\spintwoxIMRPminus{#1}}^{+\spintwoxIMRPplus{#1}}  } } 
\newcommand{\finalspinnonevolvedIMRPuncert}[1]{\ensuremath{ \finalspinnonevolvedIMRPmed{#1}_{-\finalspinnonevolvedIMRPminus{#1}}^{+\finalspinnonevolvedIMRPplus{#1}}  } } 
\newcommand{\masstwodetIMRPuncert}[1]{\ensuremath{ \masstwodetIMRPmed{#1}_{-\masstwodetIMRPminus{#1}}^{+\masstwodetIMRPplus{#1}}  } } 

      \chapter{Testing for Black hole mimickers in GW data}
\label{chap:siqm}

\section{Introduction}
\label{sec:siqm-intro}

With dozens of gravitational wave events detected by the LIGO-Virgo detectors, and a majority of those being consistent with binary black hole mergers, one naturally asks the fundamental question of whether other binaries, mimicking binary black hole systems, are contaminating the population. We discussed in Sec.~\ref{subsec:intro-siqm} that the nature of the compact objects in a compact binary merger is imprinted on the gravitational wave signal in the form of spin-induced quadrupole moment parameter ($\kappa$), and testing the deviation ($\delta\kappa$) on this parameter can give us an idea about the nature of the source. In this chapter, we present the results for spin-induced quadrupole moment (SIQM) test using different waveforms for various GW signals from simulated black hole binaries and real events from the gravitational wave catalogues. We start by outlining the definition of the deviation parameter in Sec.~\ref{sec:siqm-setup} along with the details of the waveforms used in the analysis. First, we report on the results (Sec.~\ref{sec:siqm-gwtc-3-results}) which were obtained by the LVK collaboration from SIQM analysis of events in the 3\textsuperscript{rd} Gravitational wave transient catalog (GWTC-3) and were included in \cite{LIGOScientific:2021sio}. These were analysed using a dominant mode, single spin-precessing waveform model (\texttt{IMRPhenomPv2}). Here, we extend the analysis for the waveform models with double spin-precession and higher modes (Secs.~\ref{sec:siqm-sec-simulation} and \ref{sec:siqm-real-event}).

\begin{table}[t]
\def\arraystretch{1.3}
\begin{tabular}{|p{0.25\linewidth}|p{0.2\linewidth}|p{0.45\linewidth}|}
\hline
\multicolumn{1}{|c|}{\textbf{Waveform name}} & \multicolumn{1}{c|}{\textbf{Modes ($\bm{\ell, |m|}$)}} & \multicolumn{1}{c|}{\textbf{Features}} \\ \hline
\texttt{IMRPhenomPv2} & (2,2) & Single spin precession \citep{Hannam:2013oca} \\ \hline
\texttt{IMRPhenomXP} & (2,2) & Double spin-precession \citep{Pratten:2020ceb} \\ \hline
\texttt{IMRPhenomXHM} & (2,2), (2,1), (3,3), (3,2), (4,4) & Higher modes, aligned spin \citep{Garcia-Quiros:2020qpx} \\ \hline
\texttt{IMRPhenomXPHM} & (2,2), (2,1), (3,3), (3,2), (4,4) & Higher modes, double-spin precession \citep{Pratten:2020ceb} \\ \hline
\end{tabular}
\caption[List of waveform models used in SIQM chapter]{List of waveform models used in this chapter. The columns denote the name of the waveform approximant coded up in \texttt{LALSuite}, the spherical harmonic modes (in the co-precessing frame for waveforms with spin-precession), and main features of the waveforms along with relevant publications.}
\end{table}

The effects of spin-precession and higher-order modes have been extensively studied in the context of testing general relativity (TGR) using binary BHs \citep[see for instance][]{Bustillo:2016gid, Krishnendu:2021cyi, Krishnendu:2021fga, Puecher:2022sfm, Mehta:2022pcn, Islam:2021pbd, Breschi:2019wki}. 
By injecting the most up-to-date phenomenological waveform models with full spin-precession (\texttt{IMRPhenomXP}) and higher modes (\texttt{IMRPhenomXHM}, \texttt{IMRPhenomXPHM} \citep{Pratten:2020ceb, Pratten:2020fqn, Garcia-Quiros:2020qpx}) for binary black hole signals of varying masses and spins, we investigate the effects of spin-precession and higher modes on $\delta\kappa$ measurements. Specifically, our injections include binaries with mass ratios ($q=m_1/m_2$, where $m_1$ and $m_2$ are the detector-frame component masses and $m_1 > m_2$) in the range $q\in[1,5]$, and in-plane \& out of plane spin effects for a fixed mass binary ($M = 30 M_\odot$). Section \ref{subsec:siqm-XP-Pv2-comp} outlines the differences in the bounds of $\delta\kappa$ between \texttt{IMRPhenomXP} (double spin-precessing) and \texttt{IMRPhenomPv2} (single spin-precessing waveform model) in three scenarios: varying mass ratio, varying effective aligned spin parameter $\chi_\text{eff}$ (Eq.~\ref{eq:chieff}), and varying effective spin-precession parameter, $\chi_\text{p}$ (Eq.~\ref{eq:chip}). Next, we employ the higher mode waveform models \texttt{IMRPhenomXHM} and \texttt{IMRPhenomXPHM} to study the effect of HMs on $\delta\kappa$ in Sec. \ref{subsec:siqm-effect_of_hm}. 
%We observe that for higher mass ratios, the higher mode waveform models perform better compared to the dominant mode model \texttt{IMRPhenomXP}. 
Finally, following the SIQM analyses on GWTC-2 \citep{LIGOScientific:2020tif} and GWTC-3 \citep{LIGOScientific:2021sio} events, which were performed using \texttt{IMRPhenomPv2}, we measure $\delta\kappa$ of the binary systems using  \texttt{IMRPhenomXP} and \texttt{IMRPhenomXPHM} instead (Sec.~\ref{sec:siqm-real-event}). We find that together with the effect of spin-precession, the inclusion of higher modes plays a critical role when analysing binaries with mass-asymmetries similar to that in the event GW190412 \citep{LIGOScientific:2020stg} (mass ratio $q\approx3.7$). In fact, for GW190412, the bounds on $\delta\kappa$ obtained with \texttt{IMRPhenomXPHM} are constrained enough to rule out the boson star binaries, subject to the assumptions in the current work. %\ckm{Since multiple waveforms are being used, it might be a good idea to summarize them in a table here with the effects. Only the waveforms that have been used in this chapter.}

\section{Analysis Setup}
\label{sec:siqm-setup}

For a binary system composed of two BHs with $\kappa_i$, following \cite{Krishnendu:2019tjp}, we define $\kappa_i=1+\delta\kappa_i$, where $i=1,2$, and $\delta\kappa_i=0$ gives the BH limit. With the current detector sensitivities and dominant mode waveform models, the simultaneous measurements of both $\delta\kappa_i$ lead to uninformative results \cite{Krishnendu:2019tjp}. Thus, we stick to the proposal of~\cite{Krishnendu:2019tjp}, where a symmetric combination of $\delta\kappa_i$ is measured keeping the anti-symmetric combination ($\delta\kappa_a$) to zero. Henceforth, we call this symmetric combination SIQM deviation parameter and use the definition $\delta\kappa_s =(\delta\kappa_1 + \delta\kappa_2)/2$.

GW data analysis has routinely employed phenomenological waveform models with varying properties. For instance, the SIQM analysis in the past was carried out using the precessing dominant mode phenomenological waveform model {\tt IMRPhenomPv2}~\citep{Khan:2019kot}. 
With recent developments in phenomenological modelling, it is now possible to describe quasi-circular binaries involving generic spin components. Specifically, the waveform family {\tt IMRPhenomXP}, developed in \cite{Pratten:2020fqn}, is the current state-of-the-art phenomenological model where double spin-precession effects are introduced. 
Further, the {\tt IMRPhenomXPHM} model (which is also double spin-precessing) includes higher order modes $(\ell, |m|)= (3, 3), (4, 4), (2, 1), (3, 2)$ in the co-precessing frame, in addition to the dominant modes $(\ell, |m|)= (2, 2)$~\citep{Garcia-Quiros:2020qpx, Pratten:2020ceb}. 
We modify the inspiral phase coefficients of these waveform models at 2~PN and 3~PN orders by introducing explicit dependence on SIQM parameters, and study the measurement probabilities for simulated GW signals from BBHs and the detected GW events by performing parameter estimation using Bayesian inference (introduced in Sec.~\ref{subsec:intro-bayesian}). While there have been Fisher Matrix analyses \citep[such as][]{Krishnendu:2017shb, Krishnendu:2018nqa} estimating bounds on SIQM parameter(s) for aligned-spin systems, Fisher matrix formalism may not yield very good estimates for spin-precessing systems due to higher dimensionality of the parameter space (6 spin vector components compared to only 2 for aligned-spin case), and may require very high SNRs. Hence we use Bayesian inference for the studies performed here. The explicit expressions for modified phase are given in Appendix~\ref{chap:appn:siqm}.
Adding to the generic binary black hole parameter set, $\delta\kappa_s$ is included as a free parameter to be constrained from the data. We can then extract the $\delta\kappa_s$ posterior by marginalizing over all other parameters $\theta_{\rm BBH}$ from the multi-dimensional posterior samples as 
\begin{equation}
     p(\delta\kappa_s|d) = \int p(\theta | d) \text{d}\theta_\text{BBH}.
\end{equation}
We use {\tt LALSimulation} \citep{lalsuite} for generating all waveforms, and \emph{nested} sampling algorithm (described in Sec.\ref{subsec:intro-samplers}) implemented through {\tt dynesty} sampler in {\tt bilby} and {\tt bilby\_pipe} for parameter estimation in Secs.~\ref{sec:siqm-sec-simulation} and \ref{sec:siqm-real-event}. For analysing events from GWTC-3 using \texttt{IMRPhenomPv2} (Sec.~\ref{sec:siqm-gwtc-3-results}), \textit{nested} sampling algorithm implemented in \texttt{LALInference} is used.

\begin{PE_table}
    \begin{table}[t]
    %\begin{ruledtabular}
\begin{tabular}{l c c c c c c c}
Candidate & $\underset{\displaystyle (M_\odot)}{\mathcal{M}}$ & $\underset{\displaystyle (M_\odot)}{m_1}$ & $\underset{\displaystyle (M_\odot)}{m_2}$ & $\chi_\mathrm{{eff}}$ & $\underset{\displaystyle ({\rm Gpc})}{D_\mathrm{L}}$ & $\mathrm{SNR}$\\ \hline
\makebox[5pt][l]{\fboxsep0pt\colorbox{lightgray}{\mystrut\hspace*{1.0\linewidth}}}\!\!
\MINIMALNAME{GW151226A} & $\chirpmasssourceuncert{GW151226A}$ & $\massonesourceuncert{GW151226A}$ & $\masstwosourceuncert{GW151226A}$ & $\chieffuncert{GW151226A}$ & $\luminositydistanceuncert{GW151226A}$ & $\networkmatchedfiltersnrIMRPuncert{GW151226A}$\\
\makebox[5pt][l]{\fboxsep0pt\colorbox{white}{\mystrut\hspace*{1.0\linewidth}}}\!\!
\MINIMALNAME{GW170608A} & $\chirpmasssourceuncert{GW170608A}$ & $\massonesourceuncert{GW170608A}$ & $\masstwosourceuncert{GW170608A}$ & $\chieffuncert{GW170608A}$ & $\luminositydistanceuncert{GW170608A}$ & $\networkmatchedfiltersnrIMRPuncert{GW170608A}$\\
\makebox[5pt][l]{\fboxsep0pt\colorbox{lightgray}{\mystrut\hspace*{1.0\linewidth}}}\!\!
\MINIMALNAME{GW190412A} & $\chirpmasssourceuncert{GW190412A}$ & $\massonesourceuncert{GW190412A}$ & $\masstwosourceuncert{GW190412A}$ & $\chieffuncert{GW190412A}$ & $\luminositydistanceuncert{GW190412A}$ & $\networkmatchedfiltersnrIMRPuncert{GW190412A}$\\
\makebox[5pt][l]{\fboxsep0pt\colorbox{white}{\mystrut\hspace*{1.0\linewidth}}}\!\!
\MINIMALNAME{GW191129G} & $\chirpmasssourceuncert{GW191129G}$ & $\massonesourceuncert{GW191129G}$ & $\masstwosourceuncert{GW191129G}$ & $\chieffinfinityonlyprecavguncert{GW191129G}$ & $\luminositydistanceuncert{GW191129G}$ & $\networkmatchedfiltersnrIMRPuncert{GW191129G}$\\
\makebox[5pt][l]{\fboxsep0pt\colorbox{lightgray}{\mystrut\hspace*{1.0\linewidth}}}\!\!
\MINIMALNAME{GW191204G} & $\chirpmasssourceuncert{GW191204G}$ & $\massonesourceuncert{GW191204G}$ & $\masstwosourceuncert{GW191204G}$ & $\chieffinfinityonlyprecavguncert{GW191204G}$ & $\luminositydistanceuncert{GW191204G}$ & $\networkmatchedfiltersnrIMRPuncert{GW191204G}$\\
\makebox[5pt][l]{\fboxsep0pt\colorbox{white}{\mystrut\hspace*{1.0\linewidth}}}\!\!
\MINIMALNAME{GW191216G} & $\chirpmasssourceuncert{GW191216G}$ & $\massonesourceuncert{GW191216G}$ & $\masstwosourceuncert{GW191216G}$ & $\chieffinfinityonlyprecavguncert{GW191216G}$ & $\luminositydistanceuncert{GW191216G}$ & $\networkmatchedfiltersnrIMRPuncert{GW191216G}$\\
\makebox[5pt][l]{\fboxsep0pt\colorbox{lightgray}{\mystrut\hspace*{1.0\linewidth}}}\!\!
\MINIMALNAME{GW200129D} & $\chirpmasssourceuncert{GW200129D}$ & $\massonesourceuncert{GW200129D}$ & $\masstwosourceuncert{GW200129D}$ & $\chieffinfinityonlyprecavguncert{GW200129D}$ & $\luminositydistanceuncert{GW200129D}$ & $\networkmatchedfiltersnrIMRPuncert{GW200129D}$\\
\makebox[5pt][l]{\fboxsep0pt\colorbox{white}{\mystrut\hspace*{1.0\linewidth}}}\!\!
\MINIMALNAME{GW200225B} & $\chirpmasssourceuncert{GW200225B}$ & $\massonesourceuncert{GW200225B}$ & $\masstwosourceuncert{GW200225B}$ & $\chieffinfinityonlyprecavguncert{GW200225B}$ & $\luminositydistanceuncert{GW200225B}$ & $\networkmatchedfiltersnrIMRPuncert{GW200225B}$\\
\makebox[5pt][l]{\fboxsep0pt\colorbox{lightgray}{\mystrut\hspace*{1.0\linewidth}}}\!\!
\MINIMALNAME{GW200316I} & $\chirpmasssourceuncert{GW200316I}$ & $\massonesourceuncert{GW200316I}$ & $\masstwosourceuncert{GW200316I}$ & $\chieffinfinityonlyprecavguncert{GW200316I}$ & $\luminositydistanceuncert{GW200316I}$ & $\networkmatchedfiltersnrIMRPuncert{GW200316I}$\\
\end{tabular}%\end{ruledtabular}
    \caption[List of events analysed for SIQM test]{Median and $90\%$ symmetric credible intervals for selected source parameters, for GW events analysed in this chapter.  
    The columns show source chirp mass $\Mc$, component masses $m_i$, effective inspiral spin $\chieff$, luminosity distance $\DL$, and the network matched-filter SNR. All quoted results are calculated using BBH waveform models. The first two rows show events from first and second observing runs and the values of the source parameters are taken from Table VII of \cite{LIGOScientific:2021usb}. Next is GW190412, which is taken from GWTC-2 \citep[Table VI of][]{LIGOScientific:2020ibl}. The rest of the events are from GWTC-3 \citep[Table IV of][]{LIGOScientific:2021djp}.
    \label{table:siqm-events}
    }
    \end{table}
\end{PE_table}

%%%%%%%%%%%%%%%%%%%%%%%%%%%%%%%%%%%%%%%%%%

\section{Results from GWTC-3}
\label{sec:siqm-gwtc-3-results}

\subsubsection{Declaration}

\textit{We declare that the work included in this section is part of the LVK Collaboration paper \cite{LIGOScientific:2021sio} and has been included in this thesis after obtaining permission from the concerned authorities. The author's contribution to the above-mentioned paper was the Spin-Induced Quadrupole Moment section of the paper, in particular the data analysis of various GWTC-3 events which passed the threshold for SIQM test of GR. Only that part of the paper has been included in this thesis. Various funding agencies that support the work carried out by the LVK Collaboration have been duly acknowledged in the Acknowledgements section of this thesis.}

In this section, we include the results from the SIQM analysis of events from GWTC-3 catalog which were included in the GWTC-3 Test of GR paper \citep{LIGOScientific:2021sio}. Events from the second half of the third observing run (O3b) with False Alarm Rate ($FAR$) $< 10^{-3}~\text{yr}^{-1}$ that were confidently observed in two or more detectors as determined by any search pipeline used in the catalog of O3b events \citep{LIGOScientific:2021djp} were analysed. In addition to the $FAR$ and detector criteria, two additional conditions were considered. First, inspiral-dominated events having an inspiral network $SNR \geq 6$ were selected. Since the test relies on at least one of the binary's components having non-zero spin, events whose effective inspiral spin parameter measurements included zero at the 68\% credible level were dropped. The waveform used for this analysis was \texttt{IMRPhenomPv2}.

\begin{figure}[t!]
    \centering
    \includegraphics[width=0.95\linewidth]{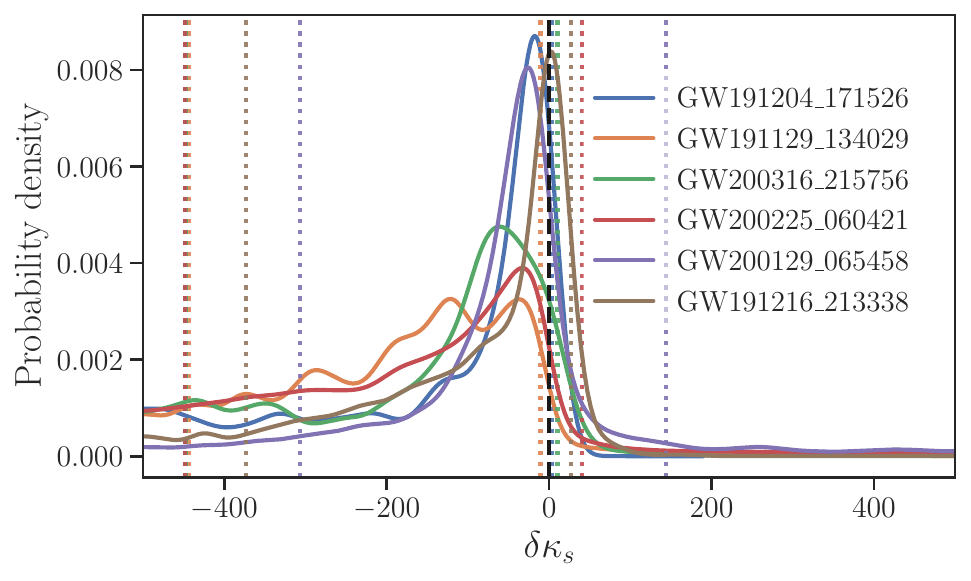}
    \caption[The posterior probability distribution on the SIQM deviation parameter, $\delta\kappa_s$, from the O3b events listed Table \ref{table:siqm-events}]{\takenfrom{\cite{LIGOScientific:2021sio}} The posterior probability distribution on the SIQM deviation parameter, $\delta\kappa_s$, from the O3b events listed Table \ref{table:siqm-events}. The black dashed vertical line indicates the BBH value ($\delta\kappa_s = 0$). The coloured vertical lines show the 90\% symmetric bounds on $\delta\kappa_s$ calculated from the individual events assuming a uniform prior ranging between $[-500, 500]$ on $\delta\kappa_s$.}
    \label{fig:siqm-gwtc3}
\end{figure}

Figure \ref{fig:siqm-gwtc3} shows the posterior distributions on $\delta\kappa_s$ for the O3b events listed in Table \ref{table:siqm-events} assuming a uniform prior on $\delta\kappa_s$ ranging between $[-500, 500]$. Individual events constrain positive values of $\delta\kappa_s$ more strongly than negative ones. This is
primarily because of how these parameters correlate with $\chi_\text{eff}$ \citep{Krishnendu:2019tjp}. As most of the events included here have small but positive $\chi_\text{eff}$, the combined posterior and the 90\% bounds are expected to
show this feature. The combined
symmetric 90\% bound on $\delta\kappa_s$ considering GWTC-3 events was estimated to be $\delta\kappa_s = -16.0^{+13.6}_{-16.7}$ and, conditional on positive values, $\delta\kappa_s < 6.66$ from the joint likelihood analysis. The Bayes factor in support of Kerr nature of the compact binary system was also calculated against $\delta\kappa_s\neq 0$ hypothesis defined as (see Eq.~\ref{eq:bayes_factor} for detailed discussion on Bayes factor):
\begin{equation}
    \mathcal{B}_{\text{Kerr}/\delta\kappa_s\neq 0} = \frac{\mathcal{Z}(d|\mathcal{H}_\text{Kerr})}{\mathcal{Z}(d|\mathcal{H}_{\delta\kappa_s \neq 0})},
\end{equation}
where $\mathcal{H_\text{Kerr}}$ represents parameter estimation using binary black hole waveform model ($\delta\kappa_s = 0$), whereas $\mathcal{H_{\delta\kappa_s \neq 0}}$ represents the analysis with waveform model where SIQM deviation parameter ($\delta\kappa_s$) is included and sampled over. The combined log Bayes factor of $\log_{10}\mathcal{B}_{\text{Kerr}/{\delta\kappa_s \neq 0}} = 0.9$ was obtained supporting the BBH hypothesis over the hypothesis of all events being non-BBH. 
%A more generic approach has recently been proposed \citep{Saleem:2021vph} that uses a hierarchical mixture-likelihood formalism to estimate the fraction of events in the population that deviated from BBH nature. With the increased number of detections in the future, it would be more natural to employ generic approaches that consider the population to be comprised of BBH and non-BBH subpopulations.

%%%%%%%%%%%%%%%%%%%%%%%%%%%%%%%%%%%%%%%%%%

\section{Results from simulated signals}
\label{sec:siqm-sec-simulation}

The analysis of events from GWTC-3 was done using a single spin-precessing, dominant mode waveform, \texttt{IMRPhenomPv2}, and must be revised by employing waveforms which include an improved prescription for spin-precession and higher modes so as to be able to find better constraints on $\delta\kappa_s$. This motivates the investigations presented here that assess the importance of using a waveform model with double spin-precession and higher modes for analysing binary black hole signals with varying properties on the SIQM test. Specifically, we look into different binaries of varying mass-asymmetries and in-plane \& out-of-plane spin parameters while fixing all other parameters to look into the effect of mass and spin variations on the $\delta\kappa_s$ measurements. We consider four waveform models, {\tt IMRPhenomXPHM} (waveform with two-spin effects and higher modes), {\tt IMRPhenomXP} (waveform with two-spin effects and dominant mode),\footnote{ The solutions employed in \texttt{IMRPhenomX} family are precession-averaged.} {\tt IMRPhenomXHM} (waveform with no spin-precession effects but higher modes), and {\tt IMRPhenomPv2} (waveform model with single-spin precession approximation and dominant mode).

\begin{table}[t]
\centering
\def\arraystretch{1.3}
\begin{tabular}{|c|c|c|}
\hline
\textbf{Parameter} & \textbf{Prior} & \textbf{Range} \\ \hline
$q_\text{inv}$ & Uniform & $0.125 \text{ - } 1$ \\ \hline
$d_L$ & \begin{tabular}[c]{@{}c@{}}Uniform \\ Source Frame\end{tabular} & $100 \text{ - } 1000$ Mpc \\ \hline
$\cos{\theta_\text{jn}}$ & Uniform & $-1 \text{ - } 1$ \\ \hline
$t_c$ & Uniform & $t_\text{gps} \pm 2$~s \\ \hline
$\phi_c$ & Uniform & $0 \text{ - } 2\pi$ \\ \hline
$a_1$, $a_2$ & Uniform & $0 \text{ - } 0.99$ \\ \hline
$\theta_1$, $\theta_2$$^*$ & Uniform sine & $0 \text{ - } \pi$ \\ \hline
$\phi_{12}$ & Uniform & $0 \text{ - } 2\pi$ \\ \hline
$\phi_{\text{JL}}$ & Uniform & $0 \text{ - } 2\pi$ \\ \hline
$\alpha$ & Uniform & $0 \text{ - } 2\pi$ \\ \hline
$\delta$ & Uniform cos & $-\pi/2 \text{ - } \pi/2$ \\ \hline
$\psi$ & Uniform & $0 \text{ - } \pi$ \\ \hline
$\delta\kappa_s$ & Uniform & $-500 \text{ - } 500$ \\ \hline
\end{tabular}
\caption[Priors for parameters used in precessing spin recoveries]{Priors for parameters used in precessing spin recoveries. We use the subscript "inv" in $q_\text{inv}$ to indicate that this is the inverse of the \textit{mass ratio} we have used throughout the chapter; \texttt{bilby} uses the variable \texttt{mass\_ratio} for this. $^*$ - Denoted by the spin angle variables \texttt{tilt\_1} and \texttt{tilt\_2} in \texttt{bilby}.}
\label{table:siqm-priors}
\end{table}

\begin{figure}[t!]
    \centering
    \includegraphics[trim=0 0 0 0, clip, width=0.85\linewidth]{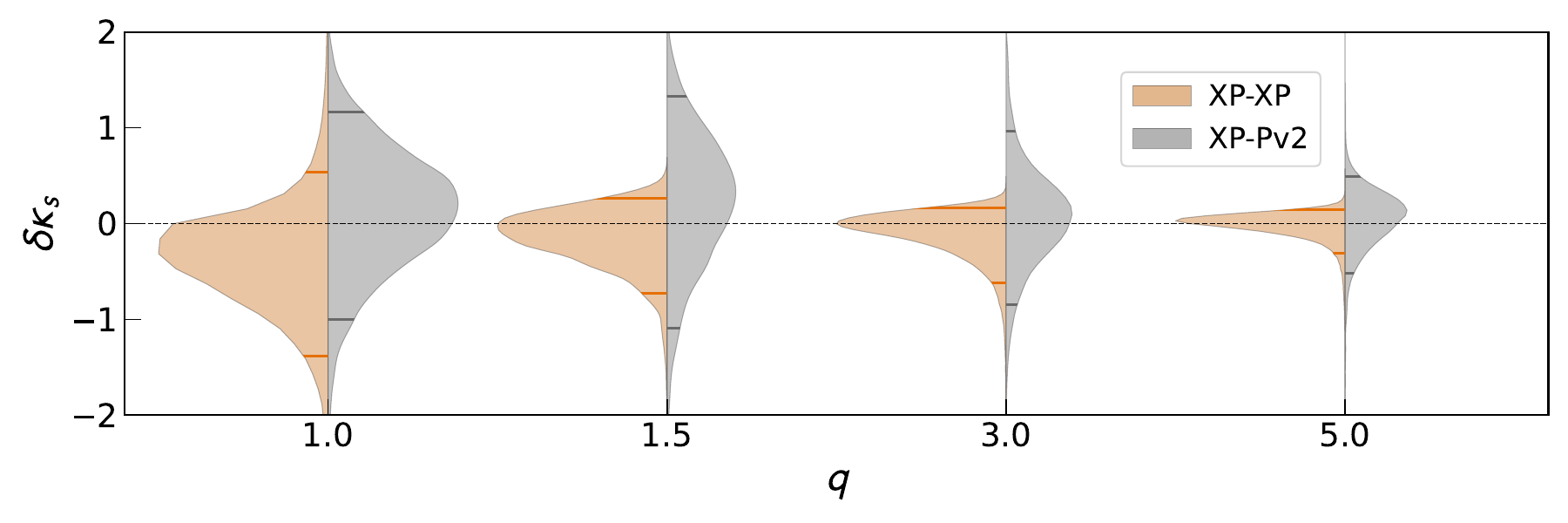}\hfill
    \includegraphics[trim=0 0 0 0, clip, width=0.49\linewidth]{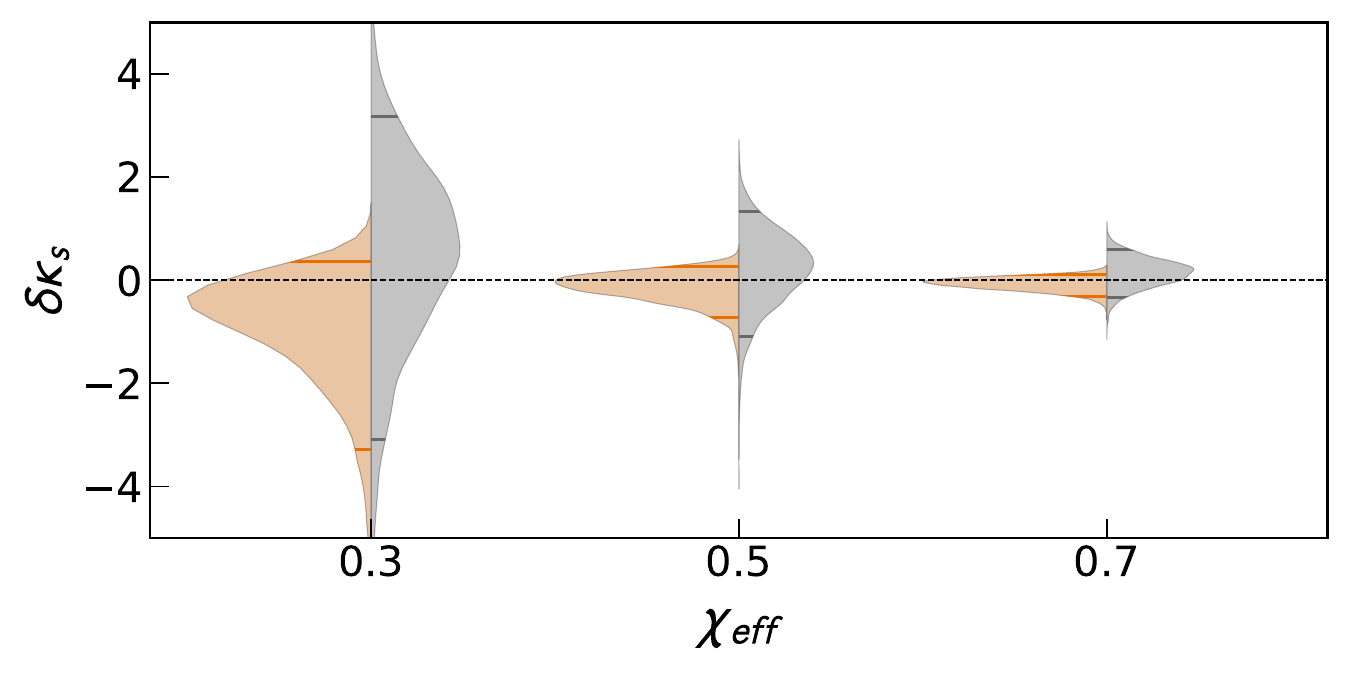}
    \includegraphics[trim=0 0 0 0, clip, width=0.5\linewidth]{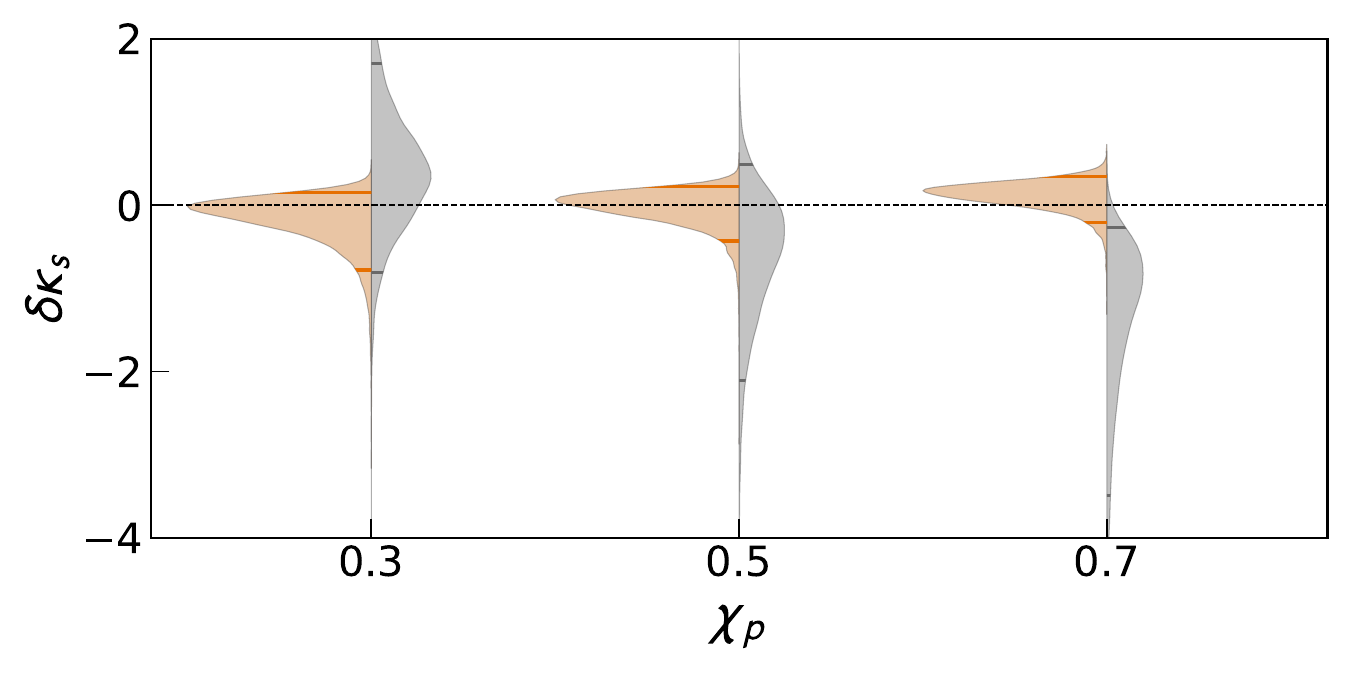}
    \caption[Violin plots showing posterior distributions for the SIQM-deviation parameter for various injection studies using \texttt{IMRPhenomXP}]{The violin plots show posterior distributions for the SIQM-deviation parameter for various injection studies. Top: Four different mass ratio cases are chosen, $q=(1, 1.5, 3, 5)$; the spin magnitudes and related angles are fixed to the values included in Table \ref{table:siqm-fixed_spin_angles}. Bottom Left: Three different values for the effective spin parameter are chosen, $\chi_\text{eff}$ = $0.3, 0.5, 0.7$ (see Table \ref{table:siqm-fixed_chip_q} for complete information); mass ratio is taken to be $q=1.5$ and $\rm{\chi_\mathrm{p}=0.3}$. Bottom right: Three different values for spin-precession parameter are chosen, $\chi_\mathrm{p}$ = 0.3, 0.5, 0.7 (see Table \ref{table:siqm-fixed_chieff_q} for information on other spin parameters); mass ratio is taken to be $q=3$ and $\chi_\text{eff}=0.5$. The total mass is fixed to $M=30~M_\odot$, and the network SNR is 40 for all cases. The injections are performed using the fully spin-precessing dominant mode waveform (\texttt{IMRPhenomXP}) and recovered with the same (orange) as well as with single spin-precessing dominant mode waveform \texttt{IMRPhenomPv2} (grey). The horizontal black-dashed lines denote the injected value, and the coloured lines inside the violins indicate the 90\% credible intervals for the respective posterior distributions. The legend follows the pattern "injected waveform - recovery waveform".}
    \label{fig:siqm-violins}
\end{figure}

For injections, we fix the total mass of the binary to be $30~M_{\odot}$ and vary mass ratios and spins. The distance is scaled in each case such that we get a fixed network SNR of 40 in a three-detector network consisting of two advanced LIGO and one advanced Virgo detectors, with advanced sensitivities~\citep{H1L1V1Dcc}. We have chosen the angles $\alpha$, $\delta$, $\psi$, and $\phi_c$ as 0, and $t_\text{gps}=1126259462$~s for injections. While a different choice of these angles and $t_\text{gps}$ may lead to a different SNR value, since we are scaling the distance in each case to fix the network SNR, choosing a different value for these angles should not affect the results. All the injections considered in this section are zero-noise, i.e. we do not introduce any noise in the injected signals and the data on which parameter estimation is performed is generated directly from the waveform models. All of our injections represent BBH mergers ($\delta\kappa_s = 0$). The parameter space taken for precessing-spin analysis is $\{\mathcal{M}$, $q_\text{inv}$, $a_1$, $a_2$, $\theta_1$, $\theta_2$, $\phi_{12}$, $\phi_{\mathrm{JL}}$, $d_L$, $\cos{\theta_{\mathrm{jn}}}$, $t_c$, $\phi_c$, $\alpha$, $\delta$, $\psi$, $\delta\kappa_s\}$. The priors for chirp mass have been modified according to the injection value such that the lower and upper limits are $\sim 0.5$ and $\sim 3$ times the injected value, respectively. Note that this is a rather conservative choice compared to the standard priors used while performing parameter estimation with \texttt{bilby} \citep[see for instance, Table B1 of]{Romero-Shaw:2020owr}. Moreover, it can be observed that while our injected values for chirp mass lie in the range $\sim 9 \text{-} 13 \Msun$ (depending on the choice of injected mass ratio keeping total mass constant), the 90\% error bounds on the chirp mass never exceed $0.2 \Msun$ in width around the median value (see for instance the chirp mass posterior in Figs.~\ref{fig:siqm-corner_fixed_spin_angles_q_1}-\ref{fig:siqm-corner_vary_chieff_0.7}). Hence, we can safely assert that the prior limits on chirp mass are wide enough not to affect the final posterior distribution. The prior constraints on component masses are also modified accordingly, which means that the sampler rejects the samples drawn outside the range of these constraints during the sampling process. This is done to speed up the parameter estimation process and to avoid samples with very low or very high masses. The priors for the rest of the parameters are given in Table \ref{table:siqm-priors}. For aligned spin recovery, instead of the six spin parameters ($a_1, a_2, \theta_1, \theta_2, \phi_{12}, \phi_{\mathrm{JL}}$), we use the parameters $\chi_1$ and $\chi_2$ using the \texttt{AlignedSpin} prior mentioned in \texttt{bilby} which puts a uniform range of $[0, 0.99]$ on the spin magnitudes. For real event analyses with \texttt{IMRPhenomXP} and \texttt{IMRPhenomXPHM}, we have used the same priors as were used for the respective analyses of the events in GWTC-2~\citep{LIGOScientific:2020tif} and GWTC-3~\citep{LIGOScientific:2021sio} TGR papers.
%\clearpage

\subsection{Comparison of ${\delta\kappa_s}$ estimates: {\tt IMRPhenomXP} and {\tt IMRPhenomPv2}}
\label{subsec:siqm-XP-Pv2-comp}

The effect of $\rm{\chi_{eff}}$ on the posteriors of $\delta\kappa_s$ is well established and was explored in \cite{Krishnendu:2019tjp}, albeit using \texttt{IMRPhenomPv2}. Furthermore, while the study in~\cite{Krishnendu:2019tjp} was performed for aligned-spin systems, here we choose systems with precessing spins and hence also explore the effect of varying $\chi_\mathrm{p}$ on $\delta\kappa_s$. In this section, we wish to compare the bounds on $\delta\kappa_s$ obtained from \texttt{IMRPhenomXP} and \texttt{IMRPhenomPv2}. We consider three cases: 
\begin{itemize}
    \item Fixed spins (magnitudes and angles) with varying mass ratio ($q$). 
    \item Fixed mass ratio \& spin-precession parameter ($\chi_\mathrm{p}$), varying effective spin parameter ($\rm{\chi_\text{eff}}$). 
    \item Different spin-precession values, keeping the mass ratio and aligned-spin components fixed.
\end{itemize} 

For all cases, we inject binary black hole ($\delta\kappa_s=0$) signals with precessing spins using \texttt{IMRPhenomXP} model and recover these with \texttt{IMRPhenomXP} and \texttt{IMRPhenomPv2}. We present the results in the form of violin plots in Fig.
\ref{fig:siqm-violins}. These are 1D kernel density estimate (KDE) plots for the posterior samples, and the posterior distribution is plotted vertically such that the parameter being plotted appears on the vertical axis. The posteriors are plotted for two cases simultaneously, denoted in different colours (like orange and grey in Fig.~\ref{fig:siqm-violins}), such that the height of the posteriors (or width of the violin) is "normalised by area" for each pair, i.e. each violin in a pair has the same area. The horizontal axis simply acts as a separator for various violin pairs, denoting different cases, to be depicted in the same plot.

\subsubsection{Effect of mass ratio variation on $\delta\kappa_s$}
\label{subsec:siqm-q_variation}

\begin{table}[t]
\centering
\def\arraystretch{1.7}
\begin{tabular}{|c|c|c|c|c|c|c|c|c|}
\hline
\textbf{$q=\frac{m_1}{m_2}$} & \textbf{$\rm{\chi_{1x}}$} & \textbf{$\rm{\chi_{1y}}$} & \textbf{$\rm{\chi_{1z}}$} & \textbf{$\rm{\chi_{2x}}$} & \textbf{$\rm{\chi_{2y}}$} & \textbf{$\rm{\chi_{2z}}$} & $\chi_\text{eff}$ & \textbf{$\chi_\mathrm{p}$} \\ \hline
\textbf{1} & 0.0992 & 0.1008 & 0.6 & 0.3343 & 0.3397 & 0.35 & 0.48 & 0.48 \\ \hline
\textbf{1.5} & 0.1013 & 0.0987 & 0.6 & 0.3414 & 0.3326 & 0.35 & 0.5 & 0.3 \\ \hline
\textbf{3} & 0.1015 & 0.0984 & 0.6 & 0.3422 & 0.3318 & 0.35 & 0.54 & 0.14 \\ \hline
\textbf{5} & 0.0997 & 0.1003 & 0.6 & 0.3359 & 0.3381 & 0.35 & 0.56 & 0.14 \\ \hline
\end{tabular}
\caption[Values of spin parameters for mass ratio variation study.]{Values of dimensionless spin components ($\rm{\chi_{1x}...\chi_{2z}}$), effective spin parameter ($\chi_\text{eff}$), and precessing-spin parameter ($\chi_\mathrm{p}$) for different mass ratios when the dimensionless spin magnitude and spin angles have been fixed as follows: $a_1=0.6164, ~a_2=0.5913, ~\theta_\text{JN}=0.4606, ~\phi_{\text{JL}}=3.7926, ~\theta_1=0.2315, ~\theta_2=0.9374, ~\phi_{12}=0.0$. This has been used for studying the effect of higher modes (Sec.~\ref{subsec:siqm-effect_of_hm}) and the effect of mass ratio variation (Sec.~\ref{subsec:siqm-q_variation}).}
\label{table:siqm-fixed_spin_angles}
\end{table}

To see the effect of mass ratio on $\delta\kappa_s$ measurements, we inject binaries with fixed spin magnitudes and angles, and vary the mass ratio as $q=1,1.5,3,5$. The spin parameter values are listed in Table \ref{table:siqm-fixed_spin_angles}. The injection signals are created using \texttt{IMRPhenomXP} and are analysed with both {\tt IMRPhenomPv2} and {\tt IMRPhenomXP}.

For all the cases, we observe that {\tt IMRPhenomXP} outperforms {\tt IMRPhenomPv2}, especially for binaries with higher mass asymmetry. Moreover, we observe from the top panel of Fig.~\ref{fig:siqm-violins}, that an increase in mass ratio results in better constraints on $\delta\kappa_s$. This is consistent with the findings of  \cite{Krishnendu:2018nqa} where the dependence of mass ratio on the errors of SIQM parameter ($\Delta\kappa_s$) are discussed in detail [see Fig.~2 and the discussion around Eq.~(4.2)-(4.4) there]. Moreover, the values of $\chi_\text{eff}$ also increase gradually as we go from equal mass case to unequal mass while keeping spin magnitudes and angles fixed. Both of these effects result in the improvement of $\delta\kappa_s$ bounds with increasing mass ratio. We show the corresponding corner plots in Figs. \ref{fig:siqm-corner_fixed_spin_angles_q_1}-\ref{fig:siqm-corner_fixed_spin_angles_q_5}.

\begin{figure}[p!]
    \centering
    \includegraphics[trim=40 40 70 60, clip, width=\linewidth]{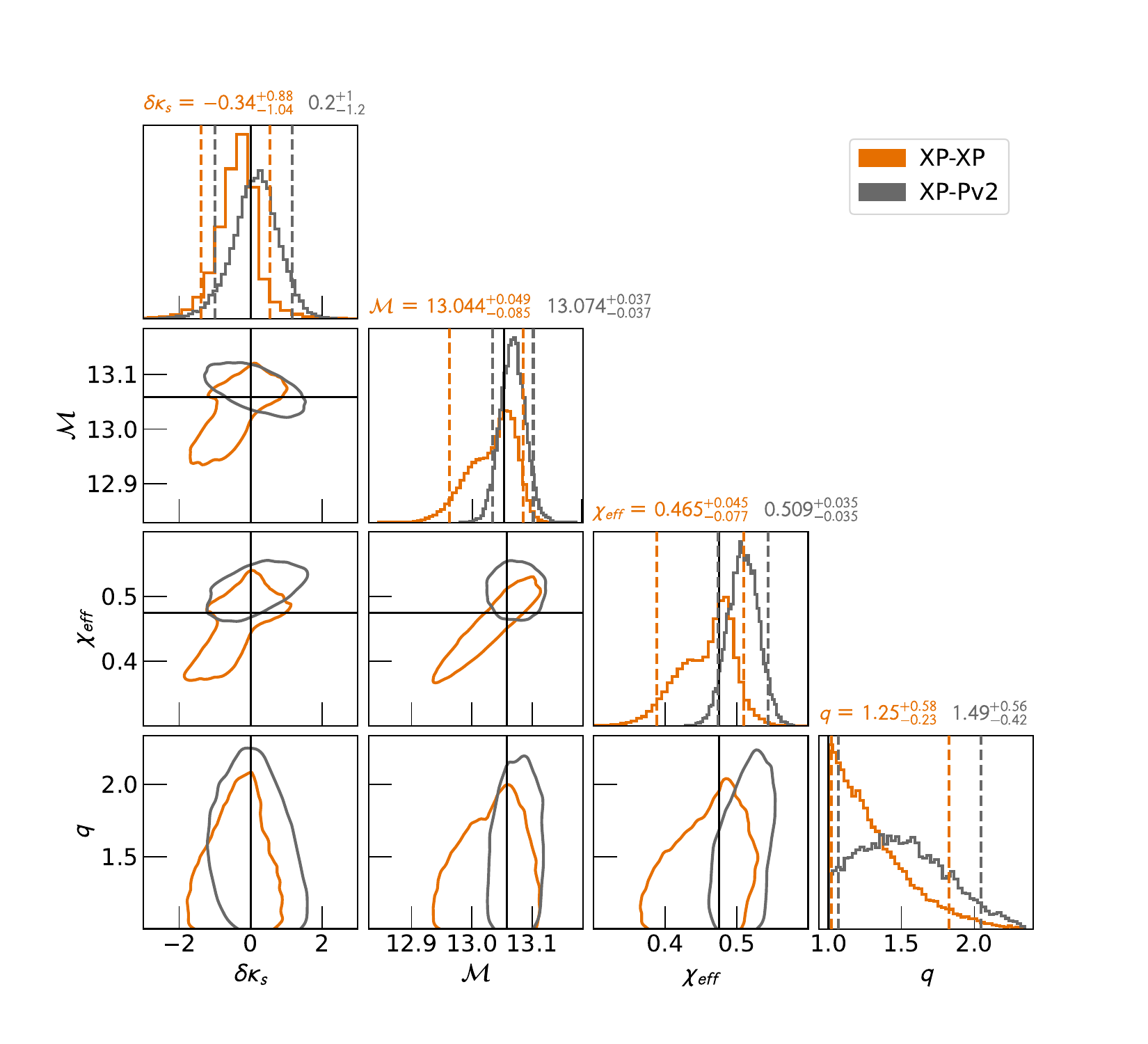}
    \caption[Corner plot for fixed spin magnitudes and angles given in Table \ref{table:siqm-fixed_spin_angles} for $q=1$.]{Corner plot for fixed spin magnitudes and angles given in Table \ref{table:siqm-fixed_spin_angles} for $q=1$. The plots show 1D and 2D posteriors for the SIQM deviation parameter ($\delta\kappa_s$), chirp mass ($\mathcal{M}$), effective spin ($\chi_\text{eff}$), and mass ratio ($q$). Injections are performed using the fully spin-precessing dominant mode waveform (\texttt{IMRPhenomXP}) and recovered with the same (orange) as well as with single spin-precessing dominant mode waveform \texttt{IMRPhenomPv2} (grey). The histograms shown on the diagonal of the plots are 1D marginalized posteriors for the respective parameters with vertical dashed lines denoting $90\%$ credible intervals and black lines indicating the injected value of the parameters. The contours in the 2D plots are also drawn for 90\% credible interval. The titles on the 1D marginalized posteriors for respective parameters and recoveries indicate 50\% quantiles with error bounds at 5\% and 95\% quantiles.}
    \label{fig:siqm-corner_fixed_spin_angles_q_1}
\end{figure}

\begin{figure}[p!]
    \centering
    \includegraphics[trim=40 40 70 60, clip, width=\linewidth]{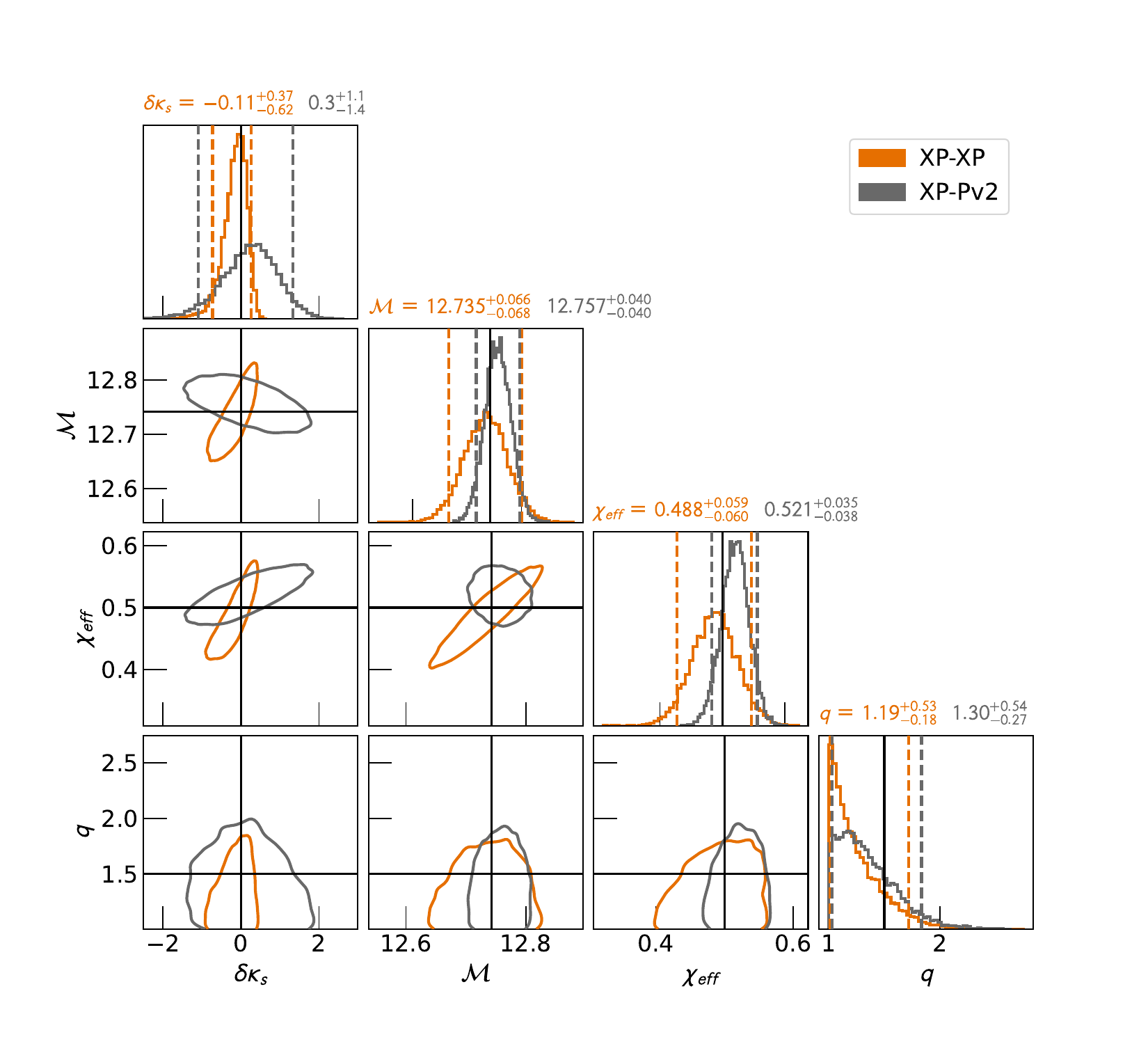}
    \caption{Same as Fig.~\ref{fig:siqm-corner_fixed_spin_angles_q_1} but for $q=1.5$.}
    \label{fig:siqm-corner_fixed_spin_angles_q_1.5}
\end{figure}

\begin{figure}[p!]
    \centering
    \includegraphics[trim=40 40 70 60, clip, width=\linewidth]{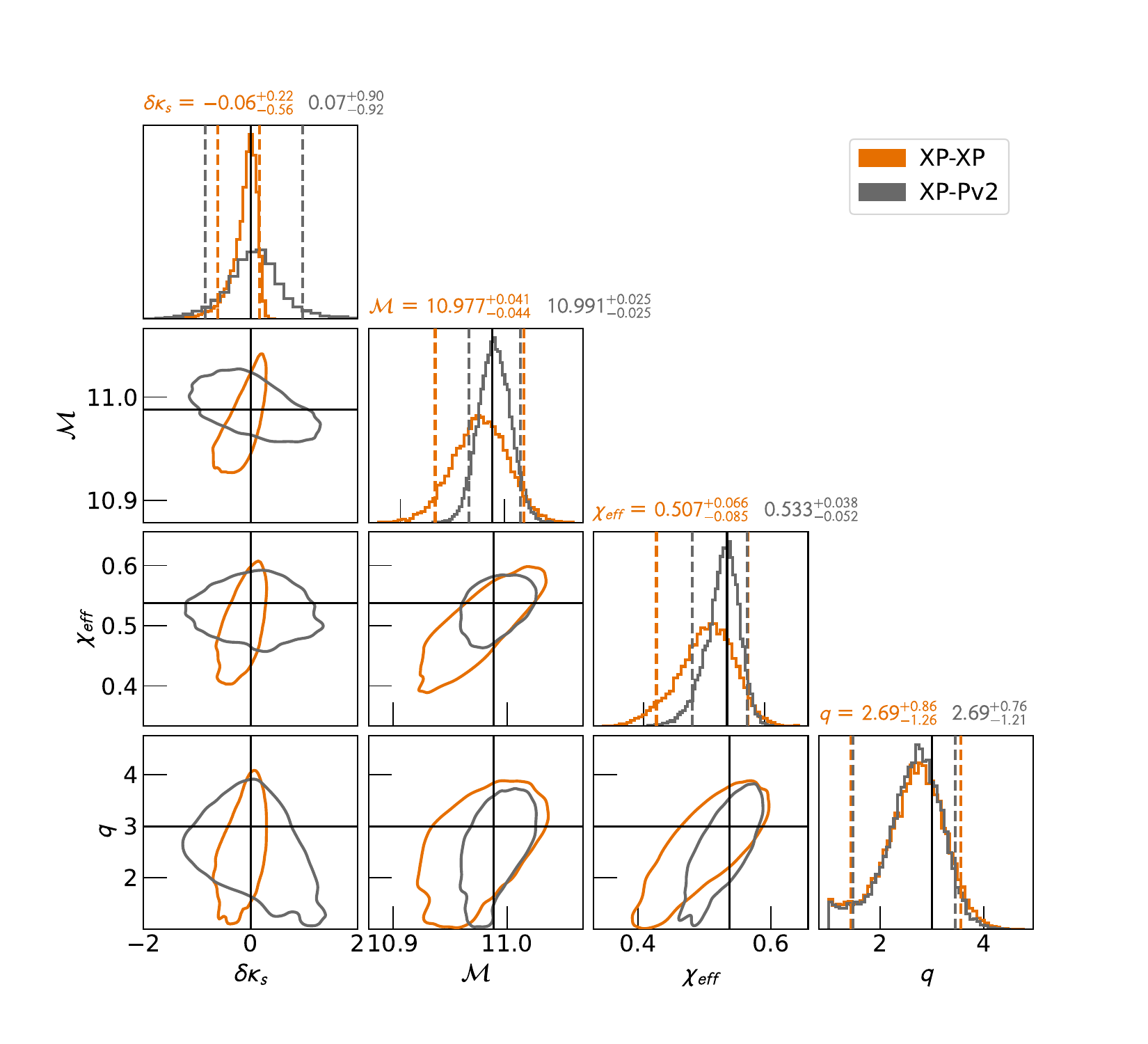}
    \caption{Same as Fig.~\ref{fig:siqm-corner_fixed_spin_angles_q_1} but for $q=3$.}
    \label{fig:siqm-corner_fixed_spin_angles_q_3}
\end{figure}

\begin{figure}[p!]
    \centering
    \includegraphics[trim=40 40 70 60, clip, width=\linewidth]{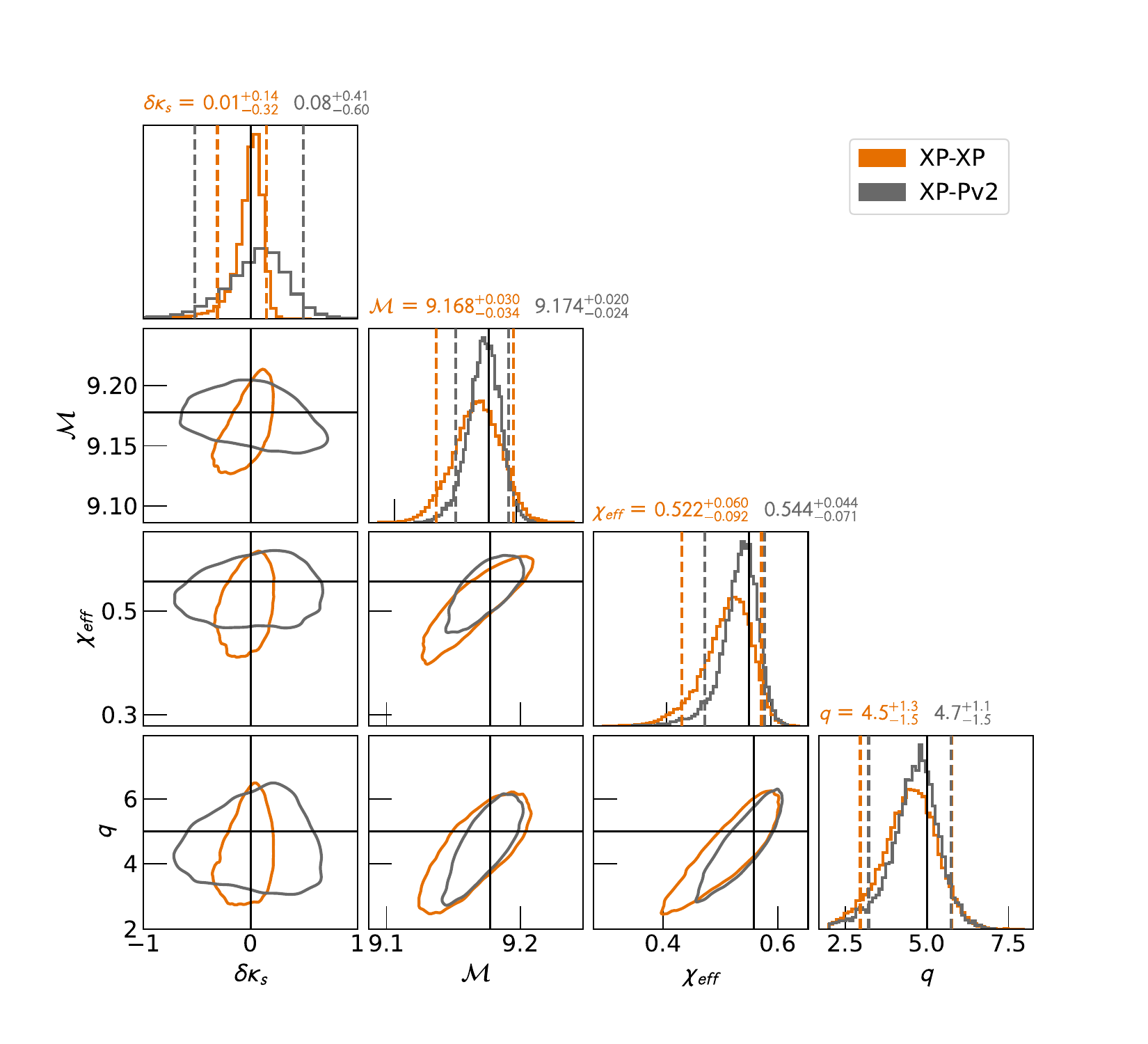}
    \caption{Same as Fig.~\ref{fig:siqm-corner_fixed_spin_angles_q_1} but for $q=5$.}
    \label{fig:siqm-corner_fixed_spin_angles_q_5}
\end{figure}
\clearpage

%\vspace{3em}

\begin{table}[t]
\centering
\def\arraystretch{1.5}
\begin{tabular}{|c|c|c|c|c|c|c|}
\hline
\textbf{$\chi_\mathrm{eff}$} & \textbf{$\rm{\chi_{1x}}$} & \textbf{$\rm{\chi_{2x}}$} & \textbf{$\rm{\chi_{1y}}$} & \textbf{$\rm{\chi_{2y}}$} & \textbf{$\rm{\chi_{1z}}$} & \textbf{$\rm{\chi_{2z}}$} \\ \hline
\textbf{0.3} & \multirow{3}{*}{0.1013} & \multirow{3}{*}{0.3414} & \multirow{3}{*}{0.0987} & \multirow{3}{*}{0.3326} & 0.4 & 0.15 \\ \cline{1-1} \cline{6-7} 
\textbf{0.5} &  &  &  &  & 0.6 & 0.35 \\ \cline{1-1} \cline{6-7} 
\textbf{0.7} &  &  &  &  & 0.8 & 0.55 \\ \hline
\end{tabular}
\caption[Values of spin parameters for $\chi_\text{eff}$ variation study.]{Here we fix the mass ratio $q=1.5$, inclination angle $\iota=0.5162$ rad, and $\rm{x-}$ and $\rm{y-}$ components of dimensionless spin vectors while varying the $\rm{z-}$ components in order to obtain different values of $\chi_\text{eff}$. The $\chi_\mathrm{p}$ value remains constant at 0.3. This has been used to study the effect of $\chi_\text{eff}$ on $\delta\kappa_s$ (Sec.~\ref{subsec:siqm-chi_eff_variation}).}
\label{table:siqm-fixed_chip_q}
\end{table}

\subsubsection{Effect of $\chi_\mathrm{eff}$ variation on $\delta\kappa_s$}
\label{subsec:siqm-chi_eff_variation}

Here we choose a nearly equal-mass binary, with mass ratio $q=1.5$ and $\rm{\chi_\mathrm{p}=0.3}$. Since we are not exploring the effect of HMs in this injection set, a nearly equal-mass system serves the purpose well. We vary the $\chi_\mathrm{eff}$ parameter as $\rm{\chi_\mathrm{eff}=0.3, 0.5, 0.7}$. This is done by fixing the $\rm{x}$- and $\rm{y}$- components of the two spin vectors and varying the $\rm{z}$- components $\chi_\text{1z}$ and $\chi_\text{2z}$ to obtain three values of $\chi_\text{eff}$ as $0.3$, $0.5$, and $0.7$ (see Table \ref{table:siqm-fixed_chip_q}).

As observed in ~\cite{Krishnendu:2019tjp}, the estimates on $\delta\kappa_s$ improve as we choose large positive $\chi_\mathrm{eff}$ values. Also, for all values of $\chi_\mathrm{eff}$, the bounds obtained using \texttt{IMRPhenomXP} are better than \texttt{IMRPhenomPv2}. These improvements can be explained by looking at correlations between $\chi_\text{eff}$ and $\delta\kappa_s$ shown in Figs. \ref{fig:siqm-corner_vary_chieff_0.3}-\ref{fig:siqm-corner_vary_chieff_0.7}.

%\vspace{3em}

\begin{figure}[t!]
    \centering
    \includegraphics[trim=10 10 0 30, clip, width=\linewidth]{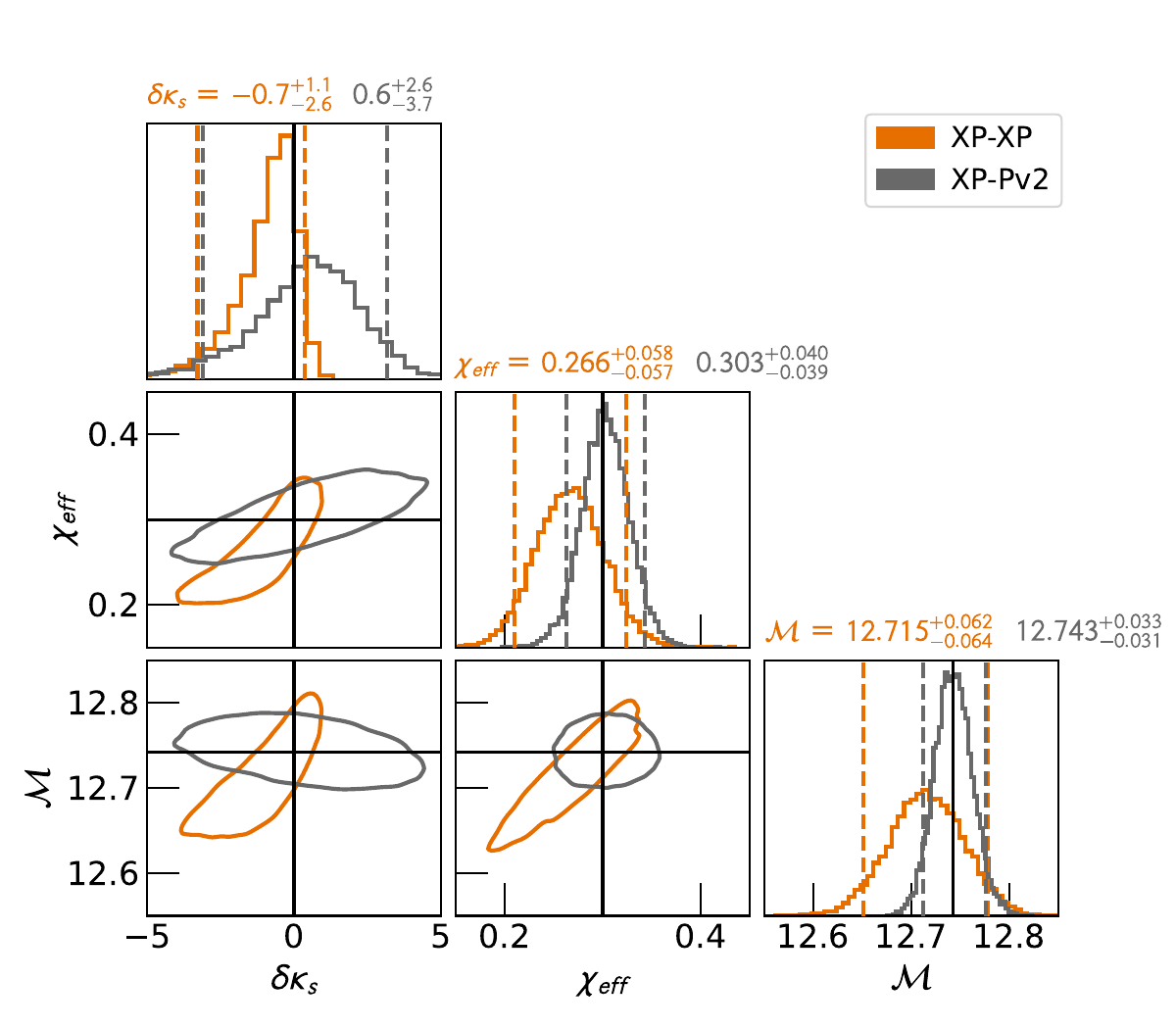}
    \caption[Corner plot for $q=1.5$, $\chi_p=0.3$, and $\chi_\text{eff}=0.3$.]{Corner plot for $q=1.5$, $\chi_p=0.3$, and $\chi_\text{eff}=0.3$ with the other spin values given in Table \ref{table:siqm-fixed_chip_q}. The plots show 1D and 2D posteriors for the SIQM deviation parameter ($\delta\kappa_s$), effective spin ($\chi_\text{eff}$), and chirp mass ($\mathcal{M}$). Injections are performed using the fully spin-precessing dominant mode waveform (\texttt{IMRPhenomXP}) and recovered with the same (orange) as well as with single spin-precessing dominant mode waveform \texttt{IMRPhenomPv2} (grey). The histograms shown on the diagonal of the plots are 1D marginalized posteriors for the respective parameters with vertical dashed lines denoting $90\%$ credible intervals and black lines indicating the injected value of the parameters. The contours in the 2D plots are also drawn for 90\% credible interval. The titles on the 1D marginalized posteriors for respective parameters and recoveries indicate 50\% quantiles with error bounds at 5\% and 95\% quantiles.}
    \label{fig:siqm-corner_vary_chieff_0.3}
\end{figure}

\begin{figure}[t!]
    \centering
    \includegraphics[trim=10 10 0 30, clip, width=\linewidth]{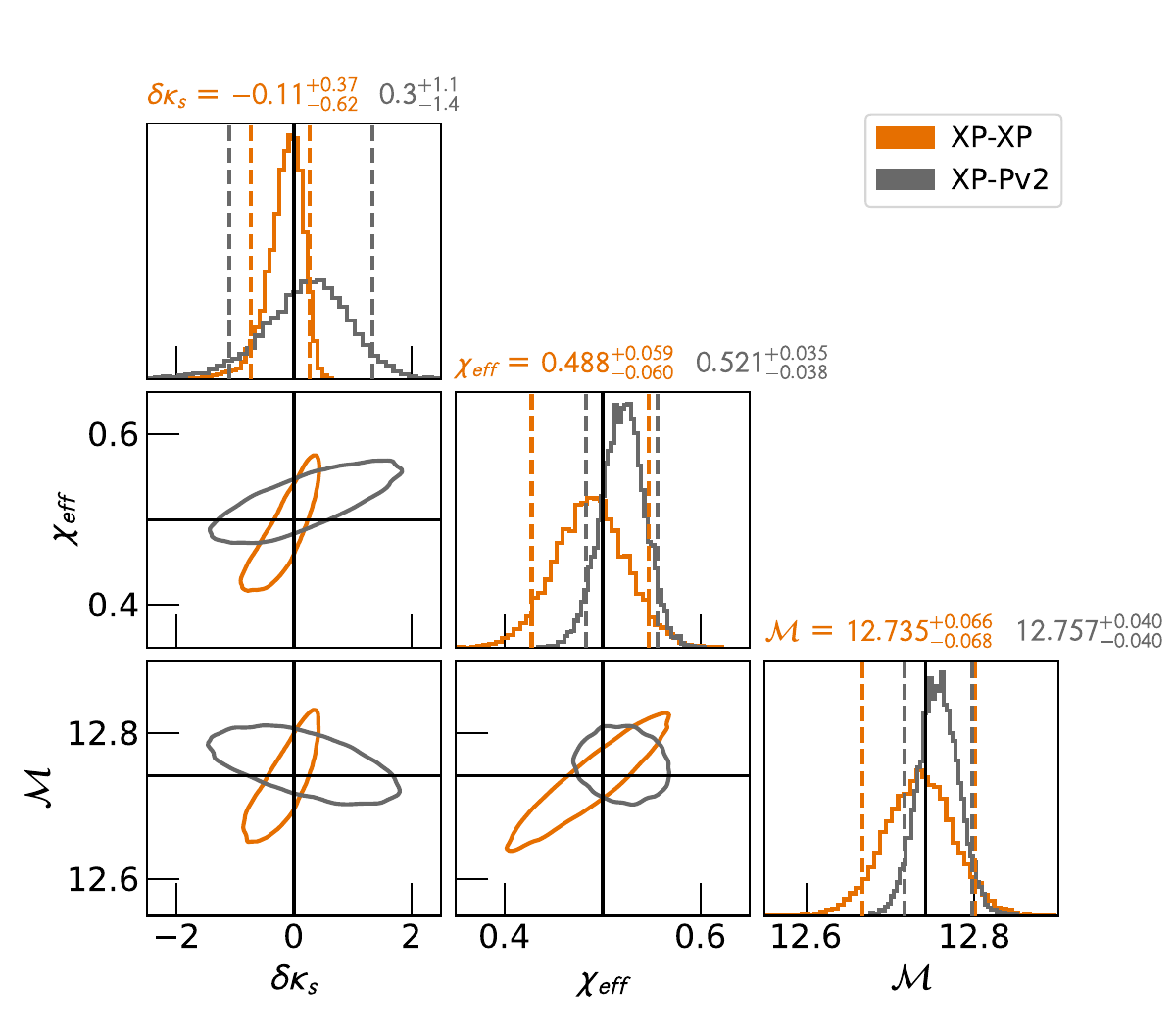}
    \caption{Same as Fig.~\ref{fig:siqm-corner_vary_chieff_0.3} but for $\chi_\text{eff}=0.5$}
    \label{fig:siqm-corner_vary_chieff_0.5}
\end{figure}

\begin{figure}[t!]
    \centering
    \includegraphics[trim=10 10 0 30, clip, width=\linewidth]{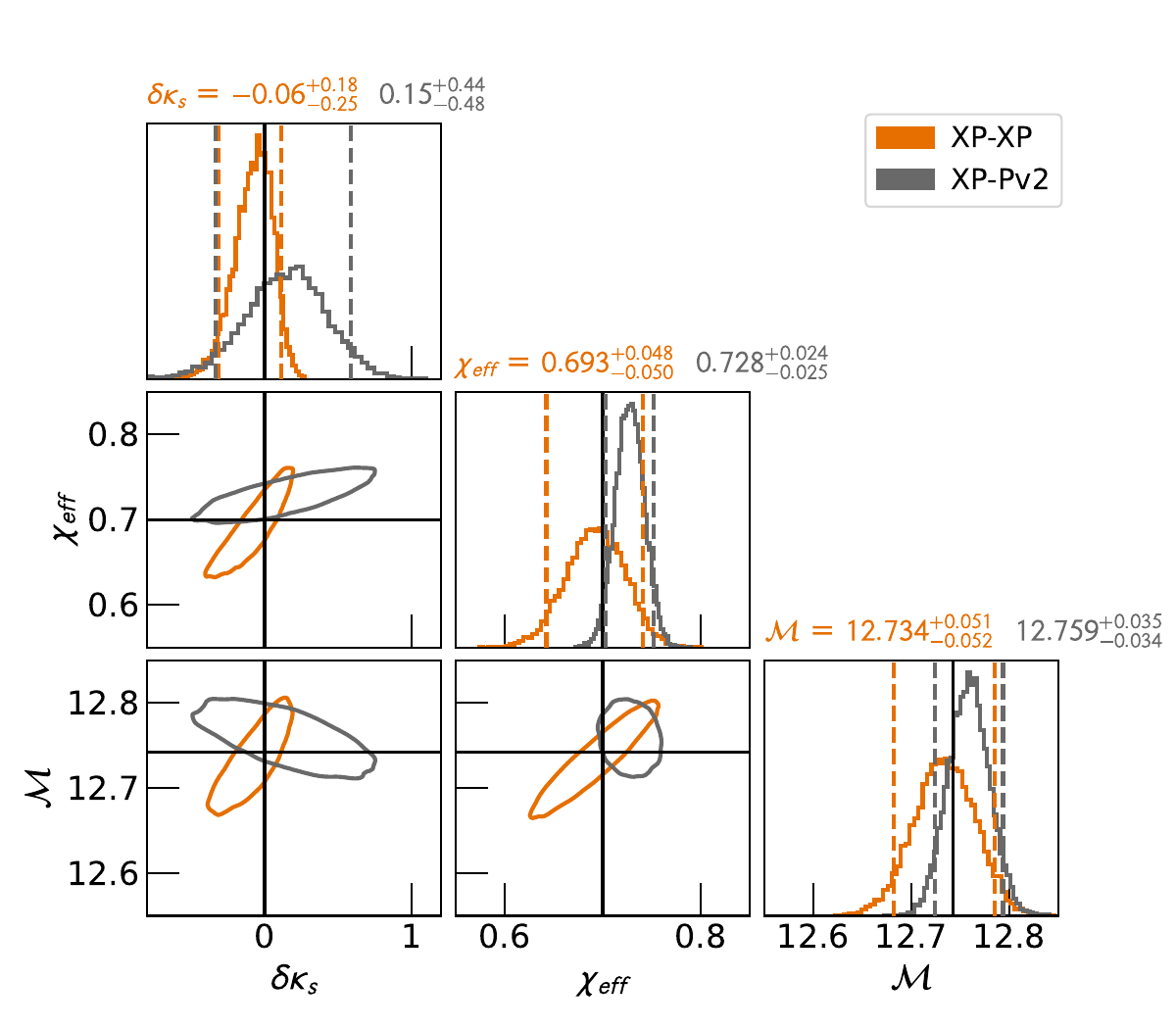}
    \caption{Same as Fig.~\ref{fig:siqm-corner_vary_chieff_0.3} but for $\chi_\text{eff}=0.7$}
    \label{fig:siqm-corner_vary_chieff_0.7}
\end{figure}
\clearpage

\subsubsection{Effect of $\chi_\mathrm{p}$ variation on $\delta\kappa_s$}
\label{subsec:siqm-chi_p_variation}

Here we choose a mass ratio of $q=3$ and a moderate value of $\chi_\text{eff}=0.5$. Compared to the previous section, a slightly larger mass ratio is chosen here to avoid the uninformative inference on the analyses due to unconstrained spin-precession effects for near-equal mass binaries. Keeping the $\rm{z}$- component of spin vectors same, we vary the $\rm{x}$- and $\rm{y}$- components to obtain three distinct values of $\chi_\mathrm{p}$ as 0.3, 0.5, and 0.7 (see Table \ref{table:siqm-fixed_chieff_q}). 

As the values of $\chi_\mathrm{p}$ increase, the bounds on $\delta\kappa_s$ with \texttt{IMRPhenomXP} become tighter, enhancing the differences between \texttt{IMRPhenomXP} and \texttt{IMRPhenomPv2} waveforms. \texttt{IMRPhenomPv2} bounds shift away from the injected value as we move from low to high $\chi_\mathrm{p}$ values, excluding 0 from the 90\% credible interval for $\chi_\mathrm{p}=0.7$. Additionally, they become increasingly worse (the posteriors become broader) compared to \texttt{IMRPhenomXP} as we go to higher values of $\chi_\mathrm{p}$. We suspect that the double spin-precessing model \texttt{IMRPhenomXP} is helping to break certain degeneracies between the SIQM parameter and the spins leading to a more symmetric estimate of $\delta\kappa_s$ for all the $\chi_\mathrm{p}$ values compared to the \texttt{IMRPhenomPv2} waveform model.

%\vspace{3em}

\begin{table}[t!]
\centering
\def\arraystretch{1.5}
\begin{tabular}{|c|c|c|c|c|c|c|}
\hline
\textbf{$\chi_\mathrm{p}$} & \textbf{$\rm{\chi_{1x}}$} & \textbf{$\rm{\chi_{2x}}$} & \textbf{$\rm{\chi_{1y}}$} & \textbf{$\rm{\chi_{2y}}$} & \textbf{$\rm{\chi_{1z}}$} & \textbf{$\rm{\chi_{2z}}$} \\ \hline
\textbf{0.3} & 0.2792 & 0.1 & 0.11 & 0.1 & \multirow{3}{*}{0.56} & \multirow{3}{*}{0.32} \\ \cline{1-5}
\textbf{0.5} & 0.4 & 0.2 & 0.3 & 0.2 &  &  \\ \cline{1-5}
\textbf{0.7} & 0.5524 & 0.3 & 0.43 & 0.3 &  &  \\ \hline
\end{tabular}
\caption[Values of spin parameters for $\chi_\text{p}$ variation study.]{Here we fix the mass ratio $q=3$, inclination angle $\iota=0.5149$ rad, and $\rm{z-}$ components of the dimensionless spin vectors while varying the $\rm{x-}$ and $\rm{y-}$ components in order to obtain different values of $\chi_\mathrm{p}$. The value of $\chi_\text{eff}$ remains constant at 0.5. This has been used to study the effect of spin-precession on $\delta\kappa_s$ (Sec.~\ref{subsec:siqm-chi_p_variation}).}
\label{table:siqm-fixed_chieff_q}
\end{table}

\subsection{Effect of higher modes  and possible systematic biases}
\label{subsec:siqm-effect_of_hm}

\begin{figure}[t!]
    \centering
    \includegraphics[trim=10 10 10 10, clip, width=0.49\linewidth]{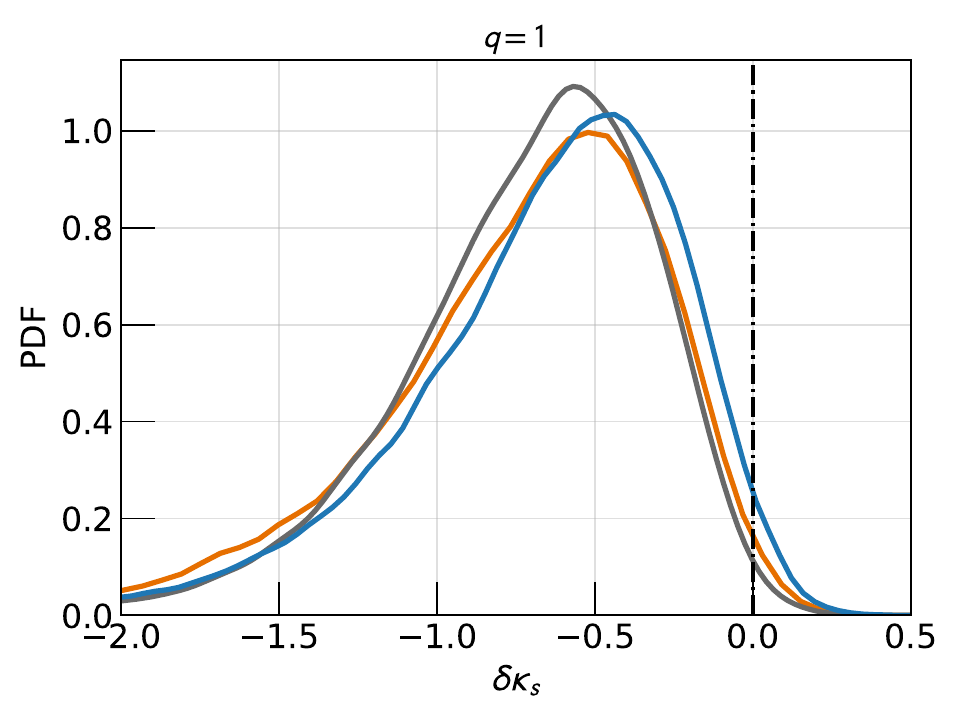}
    \includegraphics[trim=10 10 10 10, clip, width=0.49\linewidth]{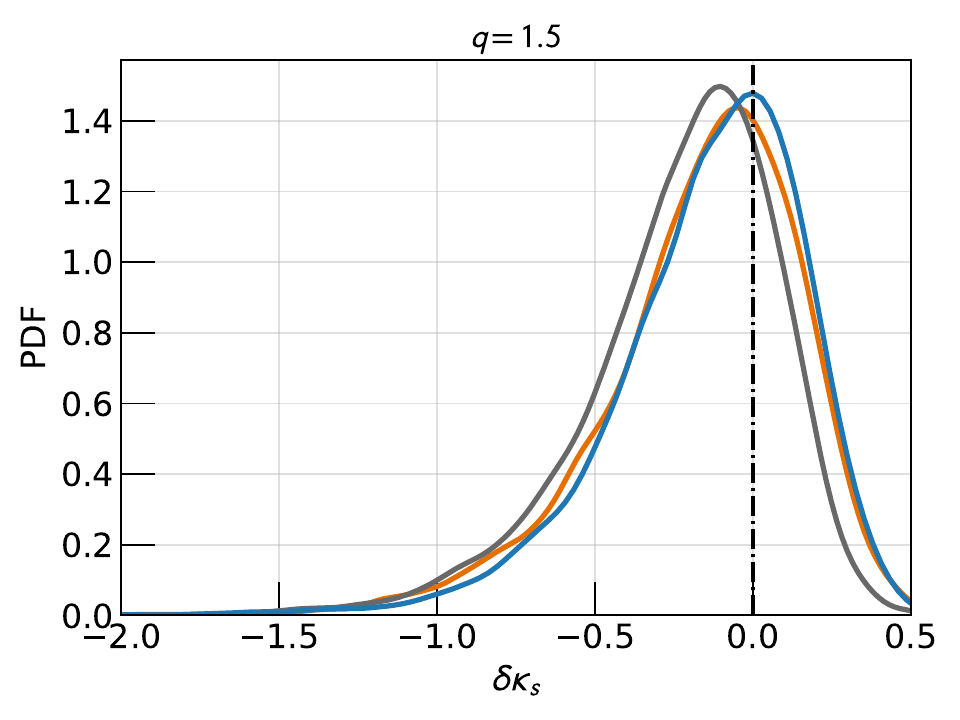}
    \includegraphics[trim=10 10 10 10, clip, width=0.49\linewidth]{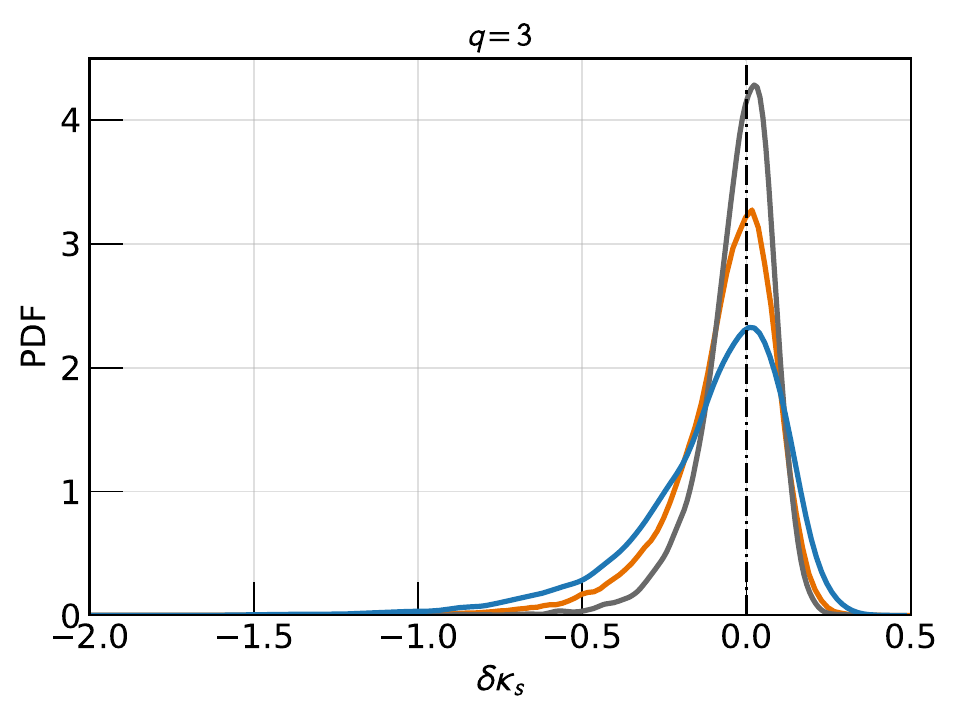}
    \includegraphics[trim=10 10 10 10, clip, width=0.49\linewidth]{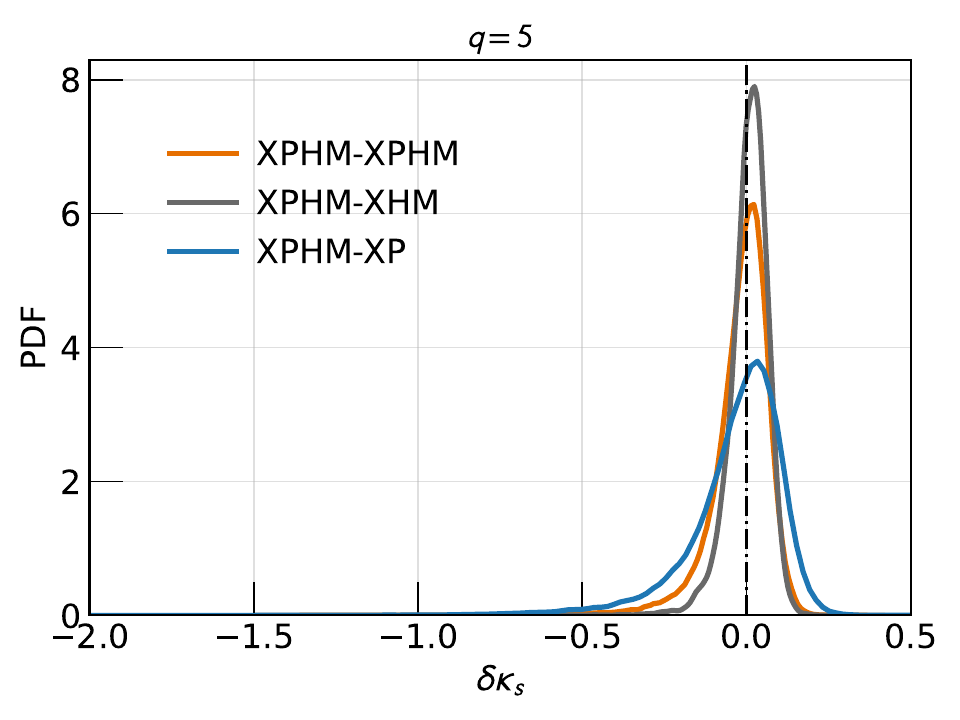}
    \caption[Posterior distributions of the SIQM deviation parameter for higher mode, precessing-spin injections]{Posterior distributions of the SIQM deviation parameter for higher mode, precessing-spin injections with mass ratios $q=1, 1.5, 3, 5$. The total mass is fixed to $M=30~M_{\odot}$, and network SNR is 40 for all cases. The spin magnitudes and angles are fixed (see Table \ref{table:siqm-fixed_spin_angles}). We use {\tt{IMRPhenomXPHM}} for injections and recover using the same (orange), {\tt{IMRPhenomXHM}} (grey), and {\tt{IMRPhenomXP}} (blue) to study the effect of higher modes on $\delta\kappa_s$ measurements. The vertical black-dashed lines denote the injected value. The legend follows the pattern "injected waveform - recovery waveform".}
    \label{fig:siqm-hm-inj-post}
\end{figure}

In the previous sections we compared the two waveforms with single and double spin-precession effects but using the dominant modes only. Here, we are interested in extending the analysis to higher modes. In the discussion in Sec.~\ref{sec:intro-hm}, we noted that the relative contribution from higher modes increases as the binary systems become more asymmetric in masses. Thus, we simulate GW signals from BBHs with a total mass of $30~M_{\odot}$ and vary the mass ratios as $q=1, 1.5, 3, 5$. We fix the spin magnitudes ($a_1=0.6164$, $a_2=0.5913$), spin angles ($\phi_{\text{JL}}=3.7926$, $\theta_1=0.2315$, $\theta_2=0.9374$, $\phi_{12}=0.0$), and inclination angle ($\theta_\text{JN}=0.4606$), taking a different mass ratio in each case, leading to different values for the dimensionless spin components ($\rm{\chi_{1x}, \chi_{1y}, \chi_{1z}, \chi_{2x}, \chi_{2y}, \chi_{2z}}$) and hence different values of $\chi_\text{eff}$ and $\chi_\mathrm{p}$ as listed in Table \ref{table:siqm-fixed_spin_angles}.
We generate simulations assuming {\tt IMRPhenomXPHM} model and analyse them using the same (orange), {\tt{IMRPhenomXHM}} (grey), and {\tt{IMRPhenomXP}} (blue) shown in Fig.~\ref{fig:siqm-hm-inj-post}. The aim is to demonstrate the importance of using a waveform model with higher modes in measuring $\delta\kappa_s$ and to examine the possible biases that could arise by neglecting them. As expected, we see the significance of higher modes when we go to higher mass ratio binaries. The posteriors are tightly constrained to the true value ($\delta\kappa_s=0$) as the binary becomes asymmetric, and the $90\%$ bound on $\delta\kappa_s$ improves from 1.3 to 0.18 (nearly seven times) when moving from near-equal mass binary ($q=1.5$) to the most asymmetric binary ($q=5$) when using the higher mode waveform \texttt{IMRPhenomXHM}. For comparison, the improvement is only three times when we use the dominant mode waveform \texttt{IMRPhenomXP}. For both $q=3$ and $q=5$ cases, {\tt IMRPhenomXHM} constraints are better than the other models. This is because the relative contribution from the higher modes increases for higher mass ratios, and hence \texttt{IMRPhenomXHM} performs better than \texttt{IMRPhenomXP}. While \texttt{IMRPhenomXPHM} is also a higher mode waveform, the binary's spin parameters (six in total) increase the dimensionality of the parameter space. This is not the case with \texttt{IMRPhenomXHM} as it is an aligned-spin waveform model and has only two spin parameters. Since the system is only mildly precessing ($\rm{\chi_\mathrm{p} \sim 0.14}$), the additional spin parameters do not contribute much towards improving the bounds on $\delta\kappa_s$, and the four extra parameters in \texttt{IMRPhenomXPHM} add a disadvantage compared to recovery with \texttt{IMRPhenomXHM}. Additionally, we observe that the time of coalescence ($t_c$) posteriors show a bias when the injected signal includes the effect of spin-precession and the recovery waveforms do not. The bias increases with an increase in the mass ratio of the system. While right ascension and declination do not improve while using \texttt{IMRPhenomXPHM} waveform, the polarization angle posterior is drastically improved with the use of higher modes in the waveform. Thus, a higher mode waveform plays a significant role in improving the sky localization of the binary system.

%While all three waveforms are unable to recover the injected $\delta\kappa_s=0$ value for $q=1$ case, 
We believe the non-recovery of the injected $\delta\kappa_s=0$ value for $q=1$ case 
%that this 
is because of the prior railing effect which arises when the injection is exactly at $q=1$. Since nearly the entire posterior volume lies outside the injected value of the mass ratio (see Fig.~\ref{fig:siqm-corner_q1_xphm-xphm}), this causes a bias in the recovery of the chirp mass parameter. Given that both $\chi_\text{eff}$ and $\delta\kappa_s$ are correlated with chirp mass, it causes a bias in both of these parameters, and the injected value of $\delta\kappa_s$ is not recovered. These correlations can be seen in the corner plot (Fig.~\ref{fig:siqm-corner_q1_xphm-xphm}) where we see biases in the values of $\mathcal{M}$ and $\chi_\text{eff}$, resulting in a biased $\delta\kappa_s$ posterior. In fact, by fixing $q=1$ for recovery waveforms, we have verified that the \texttt{IMRPhenomXPHM} injections are recovered with both \texttt{IMRPhenomXP} and \texttt{IMRPhenomXPHM} templates and thus confirming the claim about prior railing issue for $q=1$ case.

\begin{figure}[t!]
    \centering
    \includegraphics[trim=60 60 90 80, clip, width=0.93\linewidth]{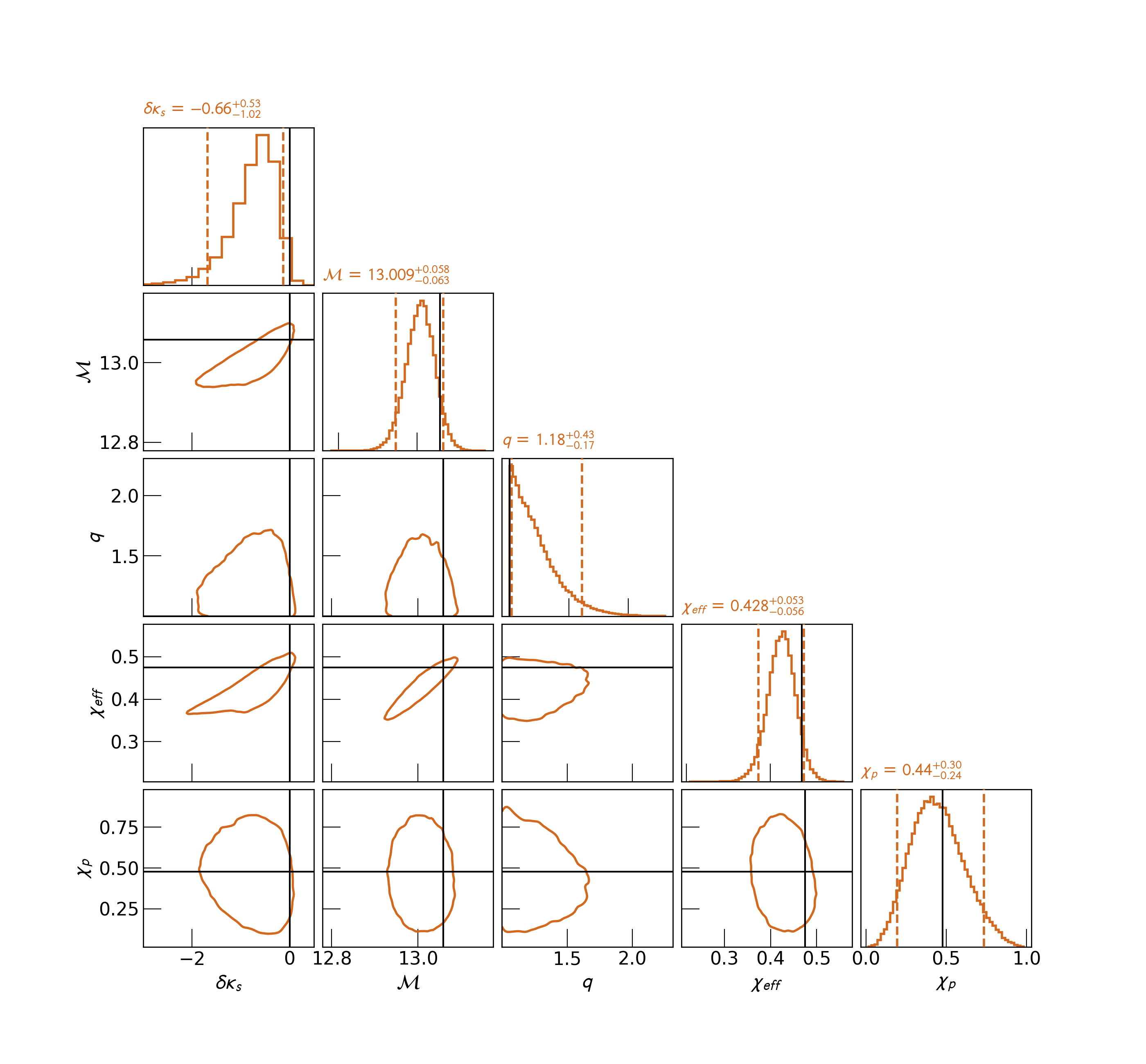}
    \caption[Corner plot for $q=1$ showing posterior on $\delta\kappa_s$, $\mathcal{M}$, $q$, $\chi_\text{eff}$, and $\chi_\mathrm{p}$]{Corner plot for the equal-mass case ($q=1$) showing posterior on the SIQM deviation parameter ($\delta\kappa_s$), chirp mass ($\mathcal{M}$), mass ratio ($q$), the effective spin parameter ($\chi_\text{eff}$), and the spin-precession parameter ($\chi_\mathrm{p}$). We have used a fully spin-precessing waveform model including HMs (\texttt{IMRPhenomXPHM}) for both injection and recovery. The histograms shown on the diagonal of the plot are 1D marginalized posteriors for the respective parameters with vertical dashed lines denoting $90\%$ credible intervals. The contours in the 2D plots are also drawn for 90\% credible interval. The black lines denote the injected value of the parameters, and the titles on the 1D marginalized posteriors for respective parameters indicate 50\% quantiles with error bounds at 5\% and 95\% quantiles.}
    \label{fig:siqm-corner_q1_xphm-xphm}
\end{figure}
\clearpage

\section{Revisiting real events with models including the effect of spin-precession and higher modes}
\label{sec:siqm-real-event}

In this section, we return to the re-analysis of selected LVK events, with the aim of comparing the performance of different waveform models in constraining  $\delta\kappa_s$ from the GW transient catalogues. % We choose the events with {\red best bounds} \ckm{quantify} \divya{Quantifying will lead us to trouble because we didn't choose ALL the events with best bounds. The runs for a couple of them failed and didn't go in the paper. So I'm removing this sentence entirely.} on $\delta\kappa_s$ from GWTC-1, 2, and 3 to demonstrate the effect of waveform models on constraining the $\delta\kappa_s$ parameter in Fig.~\ref{fig:siqm-posteriors_realevents}. 
We show the results for different waveforms used for each event in Fig.~\ref{fig:siqm-posteriors_realevents}. The {\tt IMRPhenomPv2} results~\citep{LIGOScientific:2020tif, LIGOScientific:2021sio} are shown with blue dot-dashed lines along with the {\tt IMRPhemomXPHM} (orange) and {\tt IMRPhemomXP} (grey) results.  

For events with nearly equal mass ($q \approx 1.4$ - 1.7), GW151226 and GW170608, even if we use the more informed models {\tt IMRPhemomXPHM} and {\tt IMRPhemomXP}, the bounds do not alter considerably compared to {\tt IMRPhenomPv2}. %\ckm{Is this an aligned spin run?} \divya{No, it is a precession spin run but with Pv2} 
On the other hand, the posteriors show a considerable difference from the {\tt IMRPhenomPv2} counterpart for GW190412. Note that GW190412 is the first asymmetric BBH event ($q \approx 3.75$) with an indication for moderate spins and higher modes~\citep{LIGOScientific:2020stg}. Hence, we expect the most noticeable effect on the bounds of $\delta\kappa_s$ for this event. The requirement of using waveform models with higher modes is also evident from the {\tt IMRPhemomXPHM} and  {\tt IMRPhemomXP} comparison for GW190412 (bottom-left panel of Fig.~\ref{fig:siqm-posteriors_realevents}). This can also be seen in the corner plot shown in Fig.~\ref{fig:siqm-corner_190412} where the bounds on mass ratio and $\chi_\text{eff}$ are considerably different for \texttt{IMRPhenomXPHM} compared to the dominant mode waveform models. As discussed in the previous sections, bounds on $\delta\kappa_s$ strongly depend on these parameters, and hence, for an event with non-negligible higher mode content, we observe that \texttt{IMRPhenomXPHM} performs much better. In fact, the bounds for GW190412 obtained using \texttt{IMRPhenomXPHM} exclude boson star binaries as the source, subject to the assumptions made in our study (such as neglect of the tidal corrections, assumption that $\delta\kappa_a=0$, and that spin-induced effects are accounted for only in the inspiral part of the waveform).

\begin{figure*}[t!]
    \centering
    \includegraphics[trim=10 10 10 10, clip, width=0.49\linewidth]{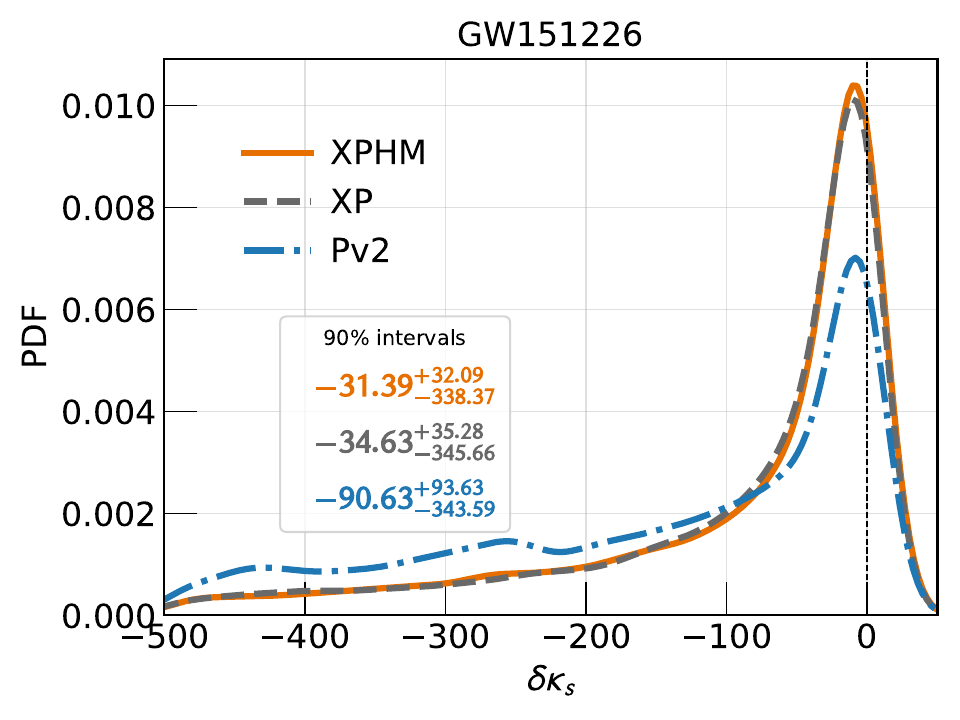}    \includegraphics[trim=10 10 10 10, clip, width=0.49\linewidth]{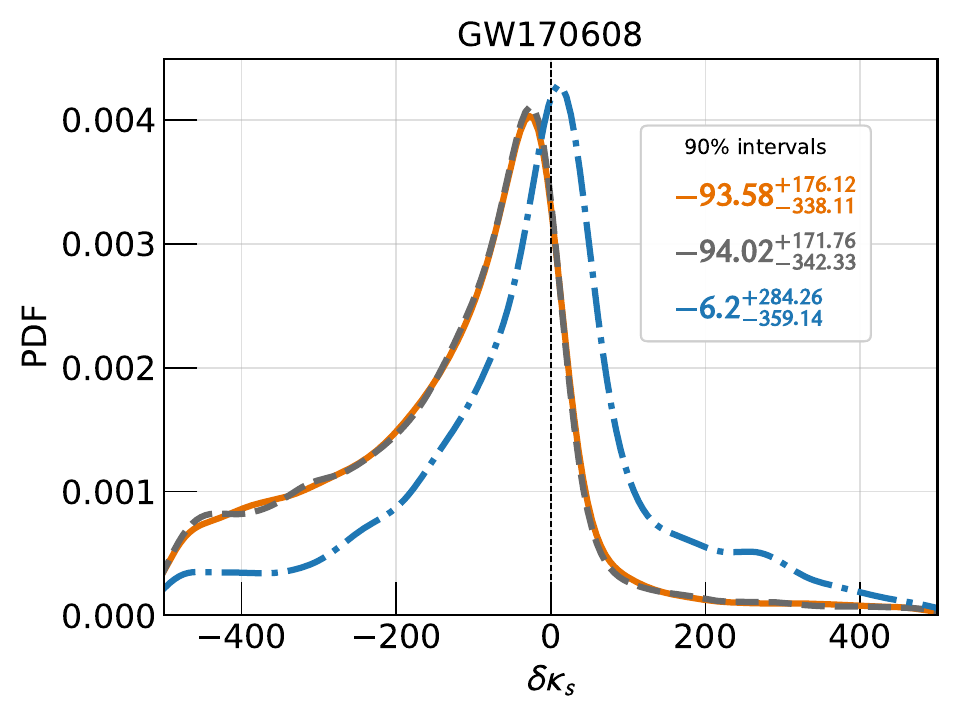}
    \includegraphics[trim=10 10 10 10, clip, width=0.49\linewidth]{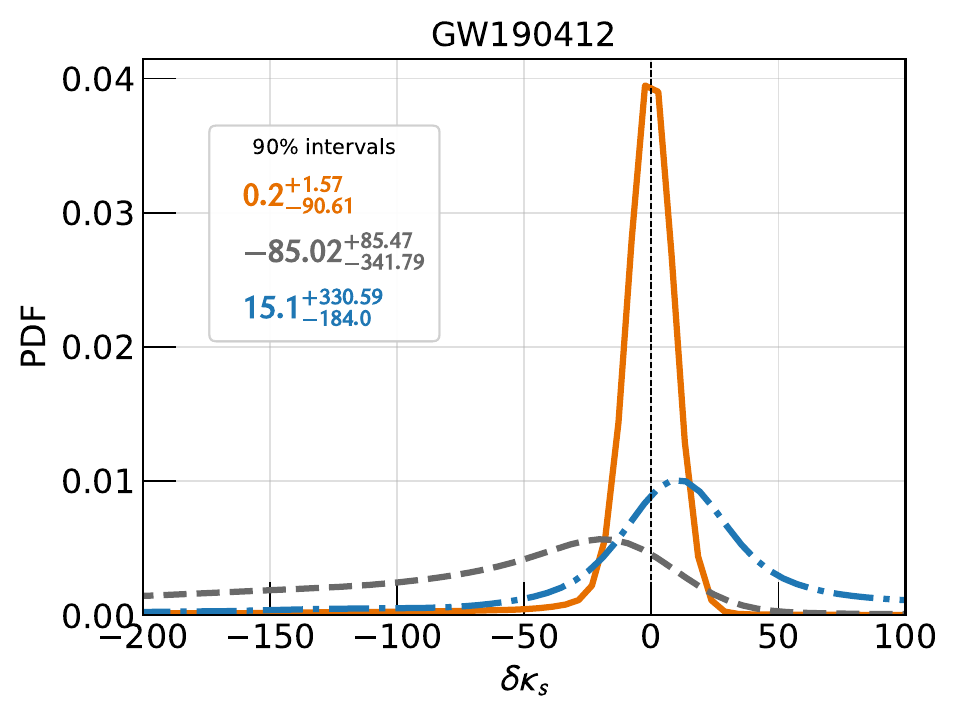}
    \includegraphics[trim=10 10 10 10, clip, width=0.49\linewidth]{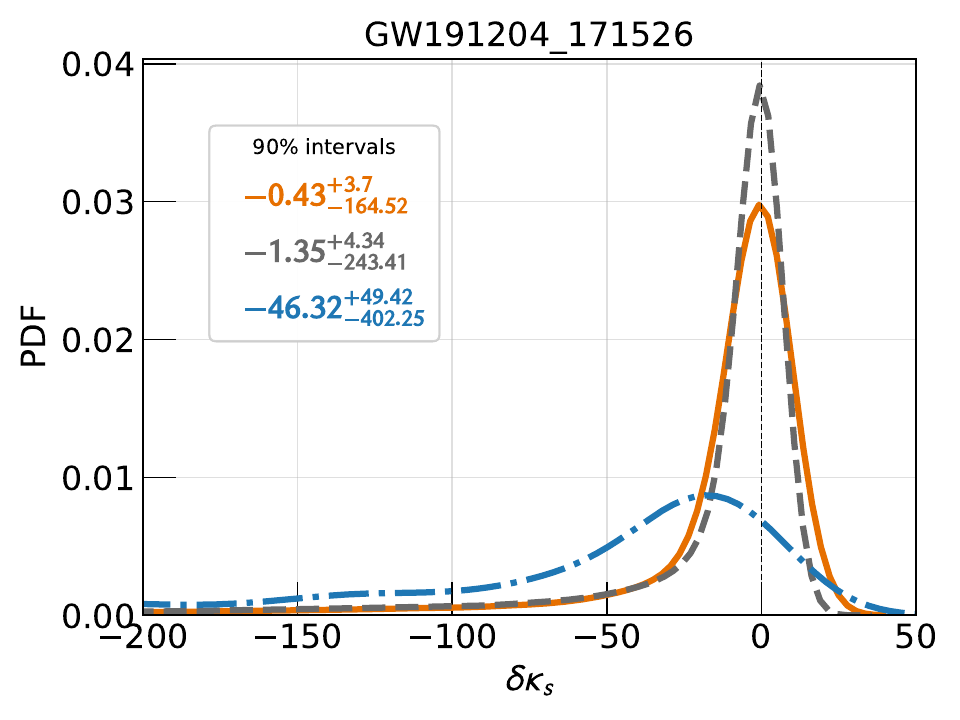}
    \caption[Posterior distributions on $\delta\kappa_s$ for observed GW events]{Posterior distributions on SIQM-deviation parameter for observed GW events. The curves labeled with {\tt{IMRPhenomPv2}} (blue) correspond to the previous results \citep{LIGOScientific:2020tif, LIGOScientific:2021sio} using {\tt{IMRPhenomPv2}} waveform model. These are being compared with the new models \texttt{IMRPhenomXP} (grey) and \texttt{IMRPhenomXPHM} (orange). The vertical dotted line indicates the BBH limit $\delta\kappa_s=0$ and the numbers written inside the plots denote the 50\% quantiles with error bounds at 5\% and 95\% quantiles for different waveform model recoveries in respective colours.}
\label{fig:siqm-posteriors_realevents}
\end{figure*}

\begin{figure*}[t!]
    \centering
    \includegraphics[trim=10 10 0 20, clip, width=\linewidth]{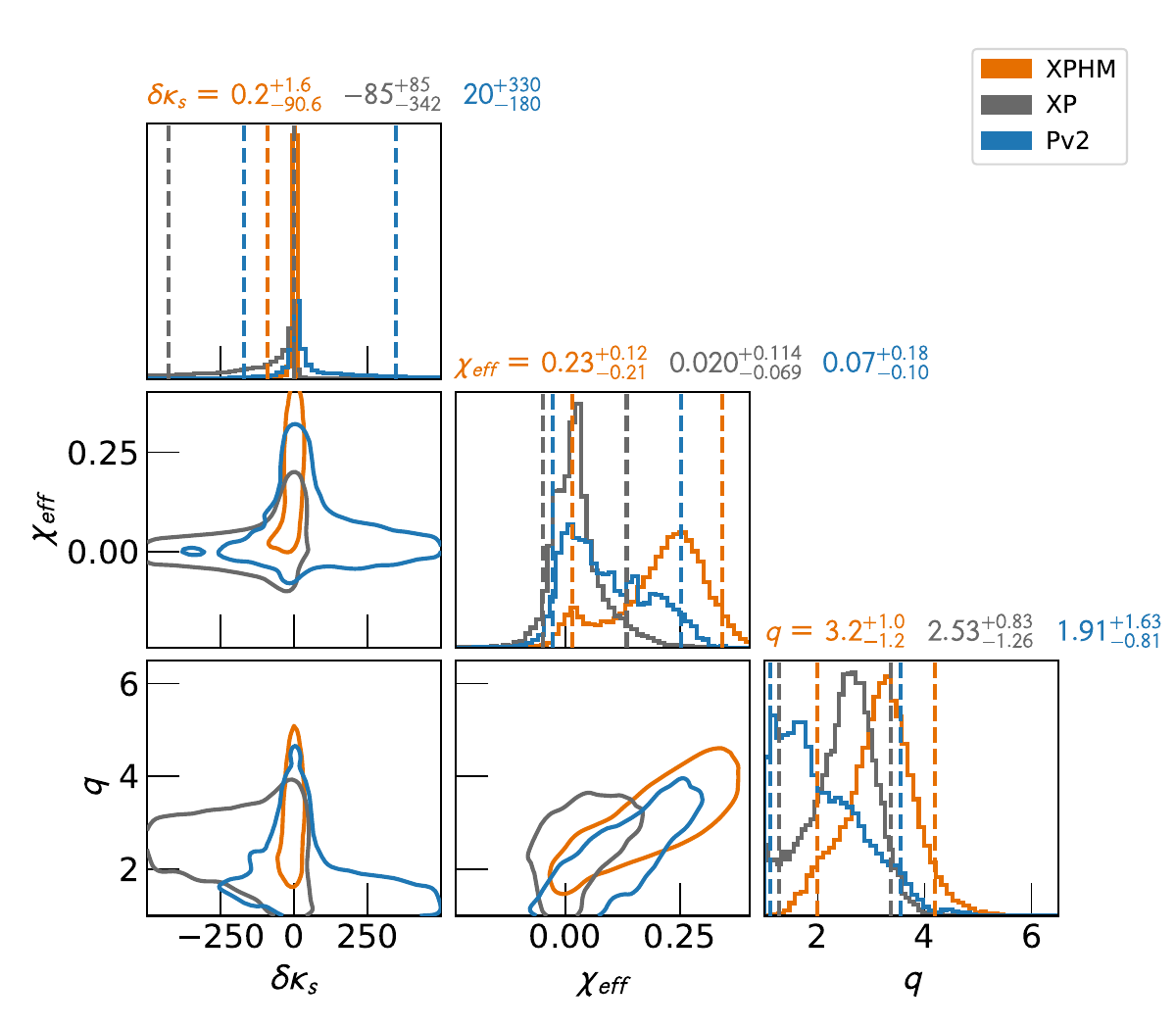}
    \caption[Corner plot for GW190412 showing posteriors on $\delta\kappa_s$, $\chi_\text{eff}$, and$q$]{Corner plot for GW190412 showing posteriors on the SIQM deviation parameter ($\delta\kappa_s$), effective spin ($\chi_\text{eff}$), and mass ratio ($q=m_1/m_2$). We show on the same plot results from single spin-precessing dominant mode waveform model \texttt{IMRPhenomPv2} (blue), fully spin-precessing dominant mode waveform model \texttt{IMRPhenomXP} (grey), and fully spin-precessing HM waveform model \texttt{IMRPhenomXPHM} (orange). The histograms shown on the diagonal of the plot are 1D marginalized posteriors for the respective parameters with vertical dashed lines denoting $90\%$ credible intervals. The contours in the 2D plots are also drawn for 90\% credible interval. The titles on the 1D marginalized posteriors for respective parameters and recoveries indicate 50\% quantiles with error bounds at 5\% and 95\% quantiles.}
    \label{fig:siqm-corner_190412}
\end{figure*}
\clearpage

\section{Summary}
\label{sec:siqm-summary}

Spin-induced multipole moment-based tests were routinely employed to determine the nature of compact binary signal during the first three observing runs of the advanced LIGO and advanced Virgo detectors~\citep{LIGOScientific:2020tif, LIGOScientific:2021sio}. In this study, we extend the applicability of the test to binaries with double spin-precession effects not considered in previous versions of the test, and discuss the possible improvements in the measurement of the SIQM deviation parameter using a more informed waveform model containing two spin-precession effects and higher modes. 
Starting with a simulation study, we demonstrate the applicability of the SIQM test on binaries with large spin-precession and moderate mass asymmetries. We find considerable differences in the bounds of $\delta\kappa_s$ obtained using \texttt{IMRPhenomPv2} compared to \texttt{IMRPhenomXP} for systems with high spin values. We also report on the improvements and biases observed in the $\delta\kappa_s$ bounds with the choice of different mass ratios and compare them between \texttt{IMRPhenomXP} and \texttt{IMRPhenomPv2} waveform models. Further, by injecting higher mode spin-precessing signals, we find that higher mode waveforms are essential when analysing GW signals with high mass asymmetries. Finally, we re-analyse selected events from GWTC 1-3 with the most up-to-date waveform models including double spin-precession effects and higher modes. 
Our findings show that {\tt IMRPhemomXPHM} may be preferred for analysing GW events such as GW190412~\citep{LIGOScientific:2020stg}, where there is evidence for mass asymmetry and non-negligible spin effects.% While the current paper studies waveform systematics on the SIQM tests for various spin and mass ratio configurations by injecting BBH waveforms consistent with GR, a detailed follow-up study with non-GR injections may be carried out in a future work. 
      \chapter{Detectability of non-quadrupole modes in future detectors}
\label{chap-hm}

\section{Introduction}
\label{sec:hm-intro}

We discussed in Section \ref{sec:intro-hm}, how the gravitational wave strain can be expanded into multipoles and how these multipoles play a role in extending the GW spectrum to lower frequencies. In this chapter, we explore the detection of these non-quadrupole modes/higher-order modes (HMs) in detail, especially in the context of future GW detectors. While there was a hint of higher mode presence in the data for the event GW170729\,\citep{Chatziioannou:2019dsz}, clear evidence of a higher order mode was found during the analysis of two events namely, GW190412~\citep{LIGOScientific:2020stg} and GW190814~\citep{LIGOScientific:2020zkf}, both quite asymmetric in component masses. Further, for about six events (all observed during the first part of the third observing run of the LIGO-Virgo network), the inclusion of higher modes in waveform models was found to improve the parameter estimation accuracies~\citep{Abbott:2020tfl, LIGOScientific:2020ibl}, hinting at their presence. As ground-based detectors continue to improve their sensitivities over the next few years, they will detect more massive and more distant BBHs, should they exist \citep{Borhanian:2022czq}. The increased mass reach is mostly due to the improved lower cutoff frequency of these detectors, which may be as low as a few Hz. Note that including higher modes also improves the detector's mass reach, as discussed earlier. The increased distance reach is due to the improved sensitivity at different frequency bands. Going by the present estimates \citep{Baibhav:2019gxm},  these observations would unravel more asymmetric binary systems, many of which may not be face-on. This should facilitate the detection of several higher modes by the next-generation detectors. As these higher modes would very likely bring in improvements to the parameter estimation in various contexts, a study of their detectability is a very important first step towards understanding the impact they will have on GW science. This forms the context of this chapter, where we quantify the detectability of higher modes using a network of future ground-based gravitational wave detectors.

%\divya{Need to finalize the next 2 paras.}

%Our study on a 3G detector network, using quasicircular, nonprecessing higher mode waveforms of Ref. [41], finds that about 33\% of the population will detect the subleading mode, $\ell=3, |m|=3$ (or simply the 33 mode) and $\sim 28$\% of the population will detect the $\ell=4, |m|=4$ (or 44 mode) mode, in addition to the dominant 22 mode. Further, for about 10\% of the population, it is possible to detect five leading spherical harmonic modes (i.e. 22, 21, 33, 32, and 44). These should have a profound impact on the planned astrophysics, cosmology, and fundamental physics using these detectors.

%Section\,\ref{sec:hm-method} includes details of the waveform employed and our choice of detector network(s) used in the analysis. We start Sec.~\ref{sec:hm-fiducial} by discussing the detection criteria (used throughout the paper) followed by results of a study concerning the detectability of higher order modes in the mass ratio ($q$) and inclination angle ($\iota$) plane. Additionally, the detection of higher modes in selected GWTC-2 events, assuming a 3G detector was operational during the O3a run of LIGO and Virgo, is explored. Section\,\ref{sec:hm-pop} presents the results of a full population study (based on the observed BBH population reported in \cite{LIGOScientific:2020ibl}) using a 3G detector network, along with a comparison study with a network of 2G detectors and their future upgrades.

\section{Detection criteria and detector networks}
\label{sec:hm-det-crit-netw}

For our study, we choose to work with an inspiral-merger-ringdown (IMR) waveform model of \texttt{Phenom} family for BBHs in quasi-circular orbits including the effect of higher order modes and non-precessing spins (coded up in \texttt{LALSuite} with the name \texttt{IMRPhenomHM}) \citep{London:2017bcn}. In addition to the dominant 22 mode, this model also contains the higher modes $(l,|m|) = (2,1)$, $(3,3)$, $(3,2)$, $(4,4)$, and $(4,3)$ and is valid for mass ratios up to $18$, and component dimensionless spin magnitudes up to $0.85$ \citep[up to 0.98 for the equal mass case. See][for details]{London:2017bcn}.

\subsection{Detection criteria}
\label{subsec:hm-det-criteria}

A robust method to quantify confident detection of a weak gravitational wave signal in noisy detector data involves computing the signal-to-noise ratio (SNR) discussed in Sec.~\ref{sec:intro-searches-matched-filtering}. Following the definition of optimal SNR [given by Eq.~\eqref{eq:opt-snr}], we can quantify the power in higher-order modes by defining the optimal SNR tied to individual modes. We define 
\begin{equation}
\rho^2_{\ell m} = (h_{\ell m}|h_{\ell m})=4\bigintssss_{0}^{\infty}\frac{|\tilde{h}_{\ell m}(f)|^2}{S_h(f)}df\,,
\label{eq:opt-snr-hm}
\end{equation}
where $\tilde{h}_{\ell m}(f)$, analogous to the GW strain in the frequency domain, represents strain for any $(\ell, \pm m)$ mode pair and can be expressed as a linear combination of associated polarizations, $\tilde{h}_+^{\ell m}(f)$ and $\tilde{h}_\times^{\ell m}(f)$, as 
\begin{equation}
\tilde{h}_{\ell m}(f) = F_+(\theta, \phi, \psi)\,\tilde{h}^{\ell m}_+(f)+F_\times(\theta, \phi, \psi) \,\tilde{h}^{\ell m}_\times(f)
\label{eq:hlmf}
\end{equation}
where the antenna pattern functions $F_+(\theta, \phi, \psi)$ and $F_{\times}(\theta, \phi, \psi)$ are functions of two angles ($\theta$, $\phi$) giving binary's location in the sky and the polarization angle ($\psi$). The two polarizations associated with each $(\ell, \pm m)$ mode pair [$\tilde{h}_+^{\ell m}(f)$, $\tilde{h}_\times^{\ell m}(f)$] can suitably be expressed using a basis of spin-weighted spherical harmonics of weight $-2$ in the frequency domain as \citep[see Appendix C of][for details and derivation]{Mehta:2017jpq}
\begin{subequations}
\label{eq:mode-pol}
\begin{align}
\tilde{h}_+^{\ell m}(f) & = \bigg[(-)^\ell \frac{d_2^{\ell, -m}(\iota)}{d_2^{\ell m}(\iota)} + 1\bigg] Y_{-2}^{\ell m}(\iota, \varphi_0) \tilde{h}^{\rm R}_{\ell m}(f)
\\
\tilde{h}_\times^{\ell m}(f) & = -{\rm i} \bigg[(-)^\ell \frac{d_2^{\ell, -m}(\iota)}{d_2^{\ell m}(\iota)} - 1\bigg] Y_{-2}^{\ell m}(\iota, \varphi_0) \tilde{h}^{\rm R}_{\ell m}(f)
\end{align}
\end{subequations}
where $\tilde{h}^{\rm R}_{\ell m}(f)$ represents the Fourier transform of real part of $h_{\ell m}(t)$ appearing in Eq.~\eqref{eq:hlm-sph-time-domain}, $d^{\ell, m}_2(\iota)$ are the Wigner $d$ functions, and $Y^{\ell, m}_{-2}(\iota, \varphi_0)$ are spin-weighted spherical harmonics of weight $-2$ \citep[see for example,][]{Wiaux:2005fm}.\footnote{In writing Eq.\,\eqref{eq:mode-pol} we have set the spherical angles appearing in Eq.\,\eqref{eq:hlm-sph-time-domain} as, $(\Theta, \Phi)\equiv(\iota, \varphi_0)$ , where $\iota$ is binary's inclination angle and $\varphi_0$ is a reference phase.} Note that $\tilde{h}^{\rm R}_{\ell m}(f)$ can be expressed in terms of an amplitude and a phase associated with each mode as
\begin{equation}
\label{eq:hlm-R}
    \tilde{h}^{\rm R}_{\ell m}(f) = A_{\ell m}(f)\,e^{i\varphi_{\ell m}(f)}\,,
\end{equation}
where $A_{\ell m}(f)$ and $\varphi_{\ell m}(f)$, for instance, can be obtained by performing fits to a set of target waveforms chosen appropriately \citep[see, for instance,][]{Mehta:2017jpq}. Details and the explicit expressions for the amplitude and the phase models used in this work can be found in Eqs.\,(4)-(9) of \cite{London:2017bcn}.

This definition of mode SNR [given by Eq.~\eqref{eq:opt-snr-hm}] closely follows the one in \cite{Mills:2020thr} which was used for quantifying the SNR of the 33 mode for GW190814~\citep{LIGOScientific:2020zkf} and GW190412~\citep{LIGOScientific:2020stg}. For our purposes, we choose to work with a threshold of $3$ on SNR for individual higher modes ($\rho_{\ell m}$) defined above, and of 10 for the dominant 22 mode. This would mean that a confident detection of a source in the dominant 22 mode requires the corresponding SNR to be above 10, and that of a higher mode requires the corresponding SNR to be above 3. 
The choice of the higher mode SNR threshold of $3$ is motivated by the measures adopted in \cite{LIGOScientific:2020stg} that discusses the detection of the 33 mode in data for the event GW190412.

It is important to note that the observed gravitational waveform is a superposition of different spherical harmonic modes [as in Eq.~\eqref{eq:hlm-sph-time-domain}], and hence the total SNR would contain contributions from the {\it interference} terms between different harmonics~\citep{Arun:2007qv, VanDenBroeck:2006ar}. They are likely to contribute negligibly to the total SNR, compared to the dominant contributions [given by Eq.~\eqref{eq:opt-snr-hm}] as shown in \cite{Mills:2020thr} in the context of aLIGO detectors. Regardless of the magnitude of the interference terms, Eq.~\eqref{eq:opt-snr-hm} may be seen as a definition of SNR in different modes. 

For all practical purposes, we can choose to work with a low- and high-frequency cutoff ($f_{\rm low}, f_{\rm cut}$) and re-express the optimal SNR for each mode as
\begin{equation}
\rho^2_{\ell m} =4\bigintssss_{f_{\rm low}}^{f_{\rm cut}}\frac{|\tilde{h}_{\ell m}(f)|^2}{S_h(f)}df\,.
\label{eq:opt-snr-hm-fcut}
\end{equation}
We choose a universal lower frequency cutoff of 5Hz following \cite{Chamberlain:2017fjl}. The high-frequency cutoff ($f_{\rm cut}$), though, is decided by the mass of the binary and chosen automatically by the waveform module with a high enough value so as not to lose any signal power \citep{London:2017bcn}. The investigations presented in Secs.~\ref{sec:hm-gwtc2} and \ref{sec:hm-q-iota} are in the context of a single 3G configuration using Cosmic Explorer (CE), whereas we use different detector networks in our population study described in Sec.~\ref{sec:hm-pop}. 

It is worth mentioning that the inclusion of higher harmonics in search pipelines may come at the cost of a higher false alarm rate because of the ability of the template to fit the noise \citep{Pekowsky:2012sr, Brown:2012nn, Bustillo:2015qty}. Studies such as \cite{Capano:2013raa} deal with this by splitting the template bank such that HM templates are only used for a part of the parameter space. Vetoes based on modified false alarm probability are beyond the scope of this work and we use only SNR as a detection criteria in this study.

%%%%%%%%%%%%%%%%%%%%%%%%%%%%%%%%%%%%%%%%%%%%%

\section{Detecting higher modes of GWTC-2 events using 3G detectors}
\label{sec:hm-gwtc2}

In this section, we investigate the detectability of higher modes for a few selected GWTC-2 catalog events, assuming the sensitivity of 3G detectors. This helps us assess the improved detection rates that can be expected due to the use of an advanced detector configuration over what we already have from the present detectors. For this, we have chosen a few representative events from the GWTC-2 catalog \citep{LSC:GWTC-GWOSC} with high detection significance, either because higher modes have already been detected for them by the LIGO-Virgo observations (GW190814, GW190412), or because the inclusion of higher modes in the waveform significantly improved the parameter estimation of the events~\citep{LIGOScientific:2020ibl}.

\begin{table}[t!]
\def\arraystretch{1.3}
\centering
\begin{tabular}{|c|c|c|c|c|c|c|c|c|}
\hline
\multirow{2}{*}{\textbf{Event}} &
  \multirow{2}{*}{\textbf{M}} &
  \multirow{2}{*}{\textbf{q}} &
  \multirow{2}{*}{\textbf{$\chi_{\text{eff}}$}} &
  \multicolumn{5}{c|}{\textbf{SNR in mode}} \\ \cline{5-9} 
                 &        &      &      & \textbf{22} & \textbf{33} & \textbf{44} & \textbf{21} & \textbf{32} \\ \hline
GW190412         & 42.6  & 3.2 & 0.2 & 649      & 81        & 17       & 14        &    3.9    \\ \hline
GW190519\_153544 & 159.5 & 1.6 & 0.4 & 685      & 79       & 48       & 19       & 14       \\ \hline
GW190521         & 279.8 & 1.4 & 0.1 & 424      & 22       & 19       & 7.1        & 7.5        \\ \hline
GW190602\_175927 & 173.8 & 1.4  & 0.1 & 330      & 15       & 9.4        &  4.3      &    4.7   \\ \hline
GW190630\_185205 & 69.9   & 1.5 & 0.1  & 708      & 31       & 14       & 7.9        & 5.8        \\ \hline
GW190706\_222641 & 183.5 & 1.7 & 0.3 & 223      & 18       & 7.6        &  3.4      & 3.3       \\ \hline
GW190814         & 27.2  & 9.0 & 0    & 982      & 172      & 33       & 32       &   4.4      \\ \hline
GW190828\_065509 & 44.4  & 2.4 & 0.1 & 418      & 34       & 7.2        & 6.7        &  3.0      \\ \hline
\end{tabular}
\caption[Detectability of higher modes in GWTC-2 events using a 3G detector]{Detectability of higher modes in GWTC-2 events using a 3G detector. We have sampled the parameter values from the posteriors of these events\citep{LIGOScientific:2020stg, Abbott:2020tfl, LIGOScientific:2020zkf, LIGOScientific:2020ibl, LSC:GWTC-GWOSC}, and have quoted the median values of SNRs obtained from 10,000 posterior samples. The total mass values quoted above are the detector frame masses. While in GWTC-2, only GW190412 and GW190814 showed detection of the 33 mode, it can be seen here that many more events would have shown detectability of HMs in a 3G detector configuration.}
\label{table: GWTC-2 events}
\end{table}

In this analysis, we consider the SNRs corresponding to a single CE detector placed at the location L (see Table \ref{table:det-loc}). For each event, to compute the distribution of SNRs for different higher-order modes, we take 10,000 random posterior samples from the corresponding dataset available for that event \citep[complete datasets can be found at][]{LSC:GWTC-GWOSC}. We then take the median value of SNR from this distribution of 10,000 points and quote this value for each mode in Table \ref{table: GWTC-2 events}.

As expected, there is a significant improvement in the detection rate of higher modes compared to the aLIGO and Virgo with their current sensitivities. The single 3G detector shows promise of detecting 33, 44, 21, and 32 modes for all of the above-mentioned events. It is noteworthy that GW190814 shows the highest SNR values for the higher modes, as well as the highest relative SNR for 33 mode. This can clearly be explained by the higher mass ratio value ($q \sim 9$) of this event. The 33 mode network SNR reported by LIGO-Virgo for this event was $\sim$6.6, whereas we can see that for a single CE detector, this number becomes $\sim$170. Similarly, the relative contribution of higher modes is considerable (by a factor larger than 20) for GW190412. Such high SNR can be attributed to the event's mass ratio ($q \sim 3.2$) and nonzero effective spin. To summarize, several GWTC-2 events would have led to reliable detection of all four leading modes of gravitational waveforms with the sensitivities of the proposed 3G detectors.

%%%%%%%%%%%%%%%%%%%%%%%%%%%%%%%%%%%%%%%%%%%%%

\section{Higher modes in different regions of the parameter space}
\label{sec:hm-q-iota}

We discussed at the beginning of this chapter (Sec.~\ref{sec:hm-intro}) that HMs not only become relevant when binary has mass asymmetry ($q>1$) and is not optimally inclined, ($\iota\neq0$) but might also be detected frequently by the next-generation detectors such as those in the 3G era. Hence, quantifying the detectability of higher modes in the ($q-\iota$) plane is very important for the physics associated with compact binary mergers. This section explores the parameter space in the ($q-\iota$) plane which will be accessible through higher-order modes in the 3G era. The results of this investigation are summarized in Figs.~\ref{fig:hm-q-iota} and \ref{fig:hm-q-iota-low-spins}.

\begin{figure}[p!]
\centering
\includegraphics[trim=40 10 40 10, clip, width=0.49\linewidth]{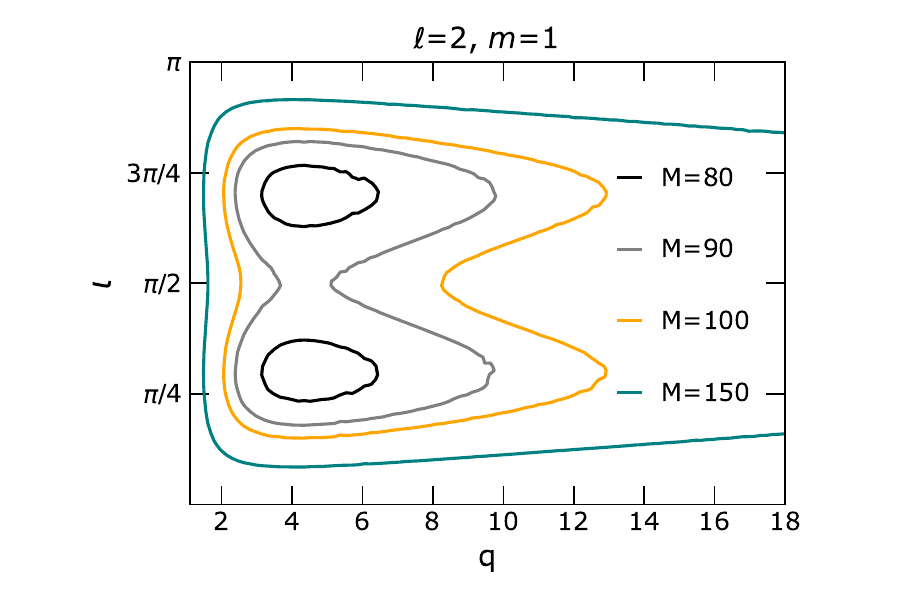}
\includegraphics[trim=40 10 40 10, clip, width=0.49\linewidth]{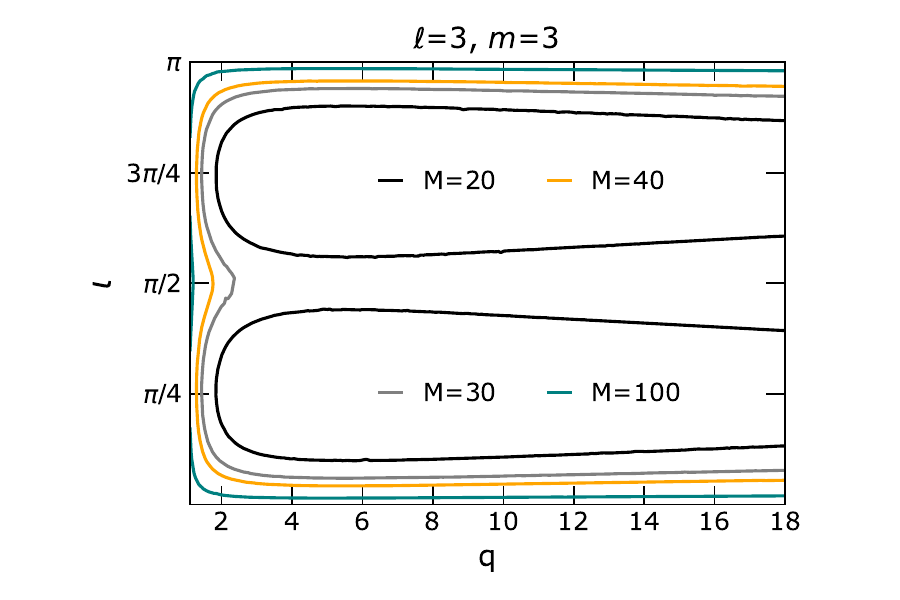}
\includegraphics[trim=40 10 40 10, clip, width=0.49\linewidth]{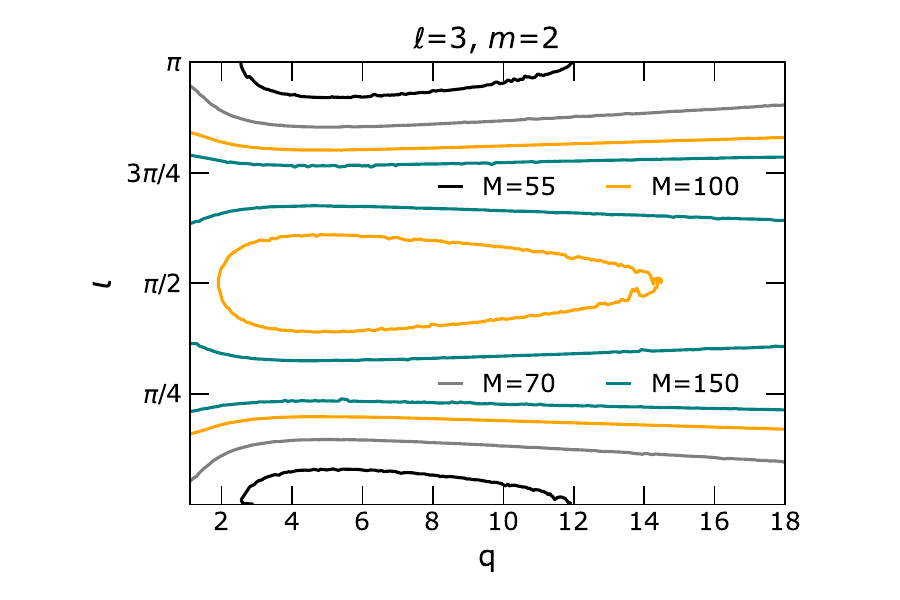}
\includegraphics[trim=40 10 40 10, clip, width=0.49\linewidth]{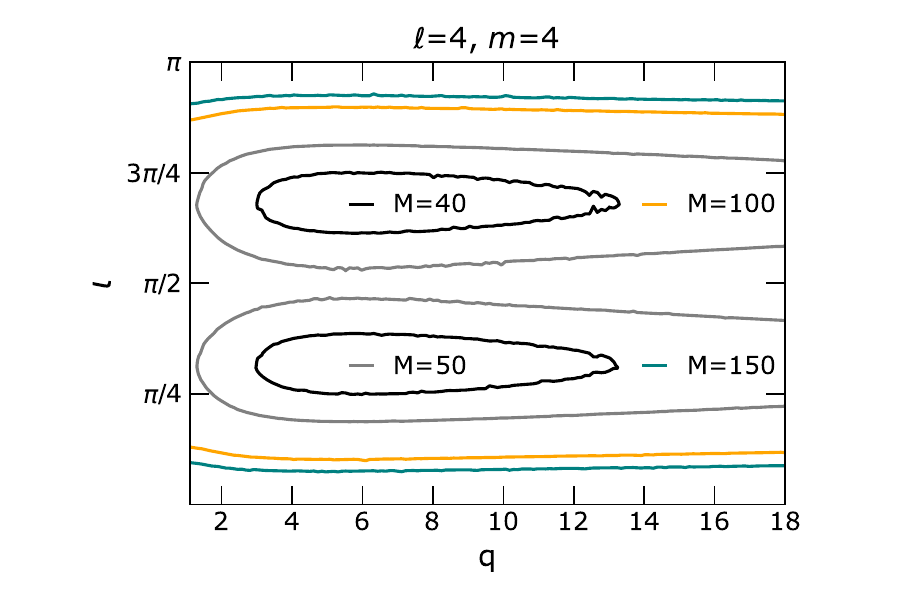}
\includegraphics[trim=40 10 40 10, clip, width=0.49\linewidth]{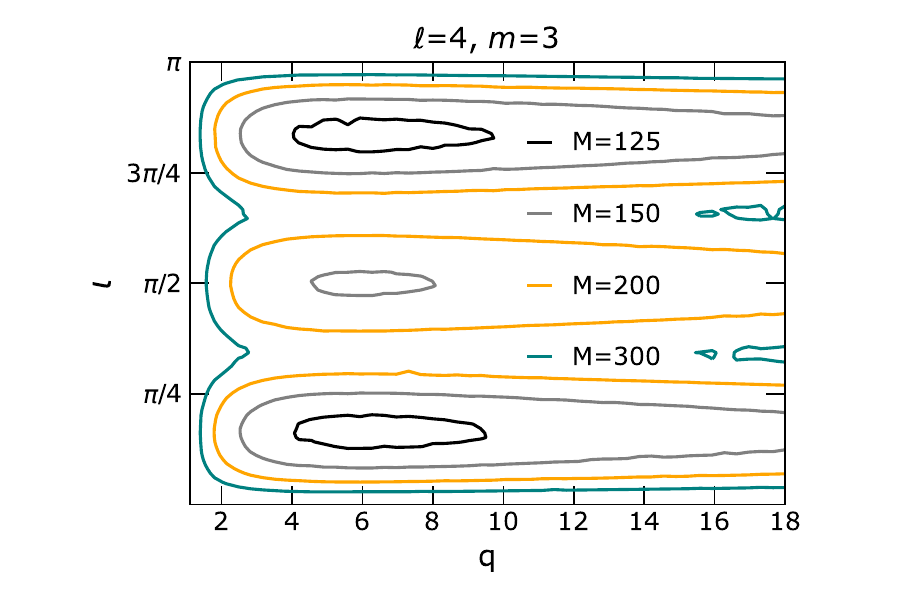}
\caption[Fixed SNR (=3) contours for various higher modes, corresponding to different total mass systems in the ($q-\iota$) plane]{Figure shows fixed SNR (=3) contours for various higher modes, corresponding to different total mass systems in the ($q-\iota$) plane. Each plot corresponds to a particular mode (see panel title). In a particular plot, the contours correspond to different values of total mass. All the systems have been taken at a fixed distance of 3\,Gpc, and with spin values $\chi_\text{1z}=0.9$ and $\chi_\text{2z}=0.8$.}
\label{fig:hm-q-iota}
\end{figure}

The analysis here considers the SNR corresponding to a single detector: CE placed at the location L (refer Table \ref{table:det-loc}). Binaries, which act as representative systems, have a fixed value of dimensionless spin components as $\chi_{\rm 1z}$=0.9 and $\chi_{\rm 2z}$=0.8 and are assumed to be kept at a distance of 3\,Gpc, with sky location and polarization angles as $\theta=30^0$, $\phi=45^0$, and $\psi=60^0$, a choice that, although arbitrary, has no impact on the conclusions, since the change of angles will only result in an overall change of SNR. Here, we have drawn contours of fixed SNR for each mode; hence, the trends and shape of the contours should remain unchanged.

Figure \ref{fig:hm-q-iota} shows the fixed SNR contours for various higher-order modes in the ($q-\iota$) plane. Each contour corresponds to a fixed (single-detector) SNR of 3, and the region inside each contour has an SNR higher than 3. Contours corresponding to different choices of total mass are displayed in the plot. The contours provide the regions in the ($q-\iota$) plane where detection of different subdominant modes will be plausible. In other words, sources that lie inside the contours will be detectable, whereas those that lie outside will not. We observe that while the detection of 33 mode is possible for masses as low as $20M_{\odot}$, 44 mode can only be detected in heavier systems as displayed by contours in the middle right panel of Fig.~\ref{fig:hm-q-iota}. This is unsurprising since the 44 mode (compared to the 33 mode) is more sensitive to high frequencies. As heavier systems merge at lower frequencies, they bring the higher mode content to the sweet spot of the detector band, allowing the accumulation of SNR. Note also, that for a given binary, the 44 mode amplitudes are relatively lower than the 33 amplitude and more or less increase linearly with its total mass. This naturally affects the power in a given mode and can explain the non-detection of 44 mode in lighter systems. Similar arguments (based on the frequency sweep of each mode in the detector's band and their relative amplitude) can be outlined to explain the trends seen in Fig.~\ref{fig:hm-q-iota} with respect to the minimum mass for which a certain mode is detected.

The trends in $q$ and $\iota$ are distinct for each mode. For a particular total mass value, as $q$ increases, the contours become narrower until they close at a point. Any binary with a mass ratio value higher than this point will not be detectable. The maximum mass ratio for detectable binary is very different for each mode. For a total mass of 100 M$_\odot$, the mass ratio reach of 33, 44, and 32 modes is well beyond 18, whereas for 21 mode it is only up to 13, and for 43 mode, the binary is not even detectable. This is also because the SNR keeps reducing, so the detectability of HMs also reduces. But, as we have discussed earlier, the relative contribution of HMs increases as we go to higher values of $q$.

\begin{figure}[p!]
\centering
\includegraphics[trim=20 10 20 10, clip, width=0.49\linewidth]{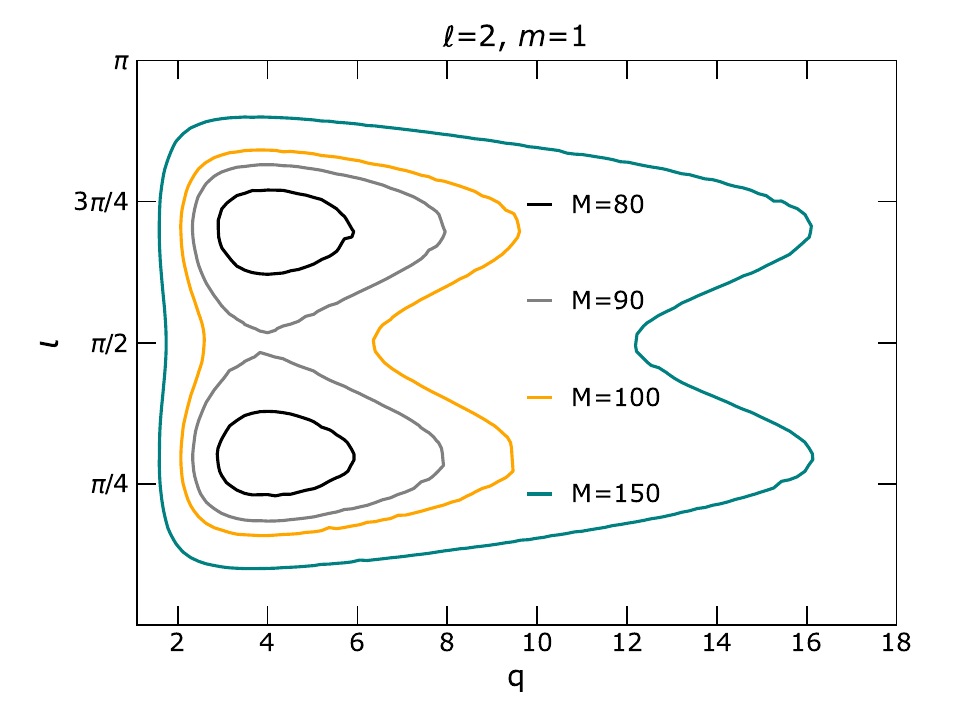}
\includegraphics[trim=20 10 20 10, clip, width=0.49\linewidth]{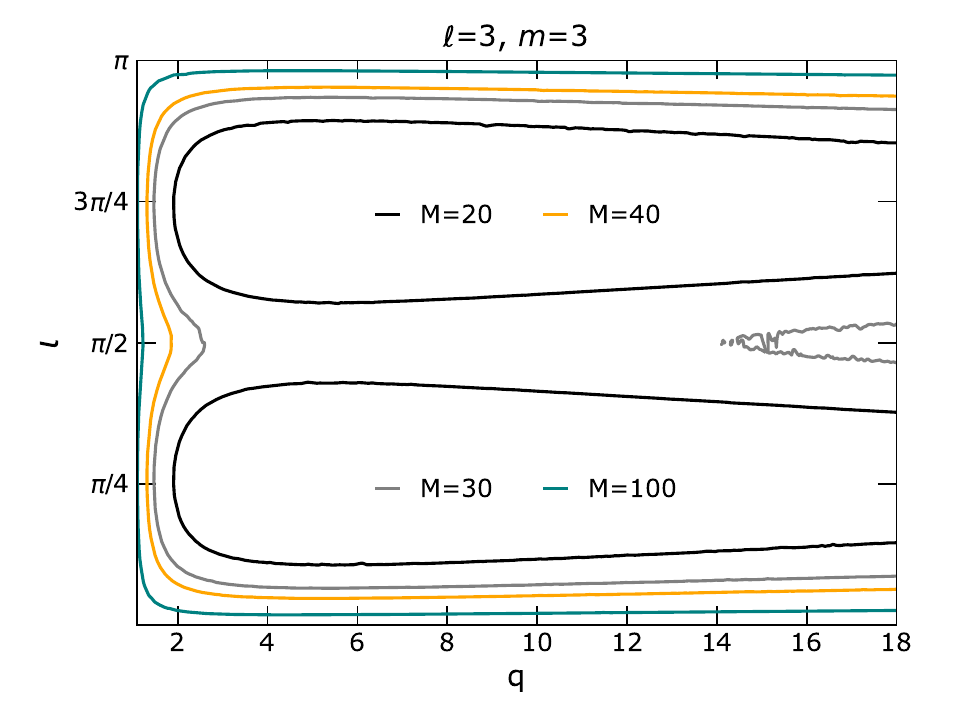}
\includegraphics[trim=20 10 20 10, clip, width=0.49\linewidth]{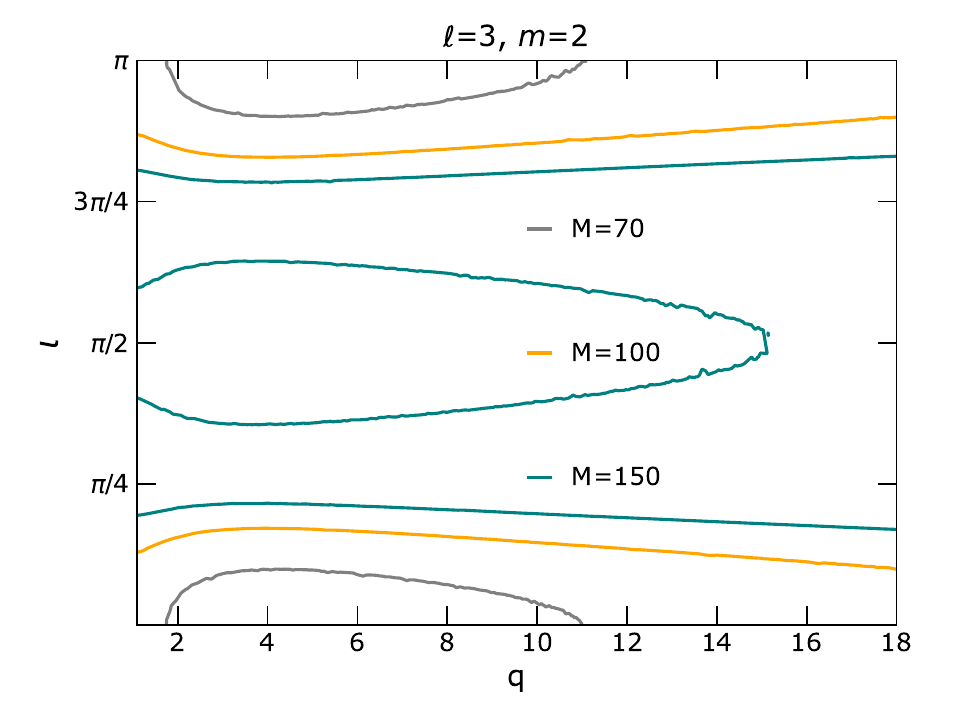}
\includegraphics[trim=20 10 20 10, clip, width=0.49\linewidth]{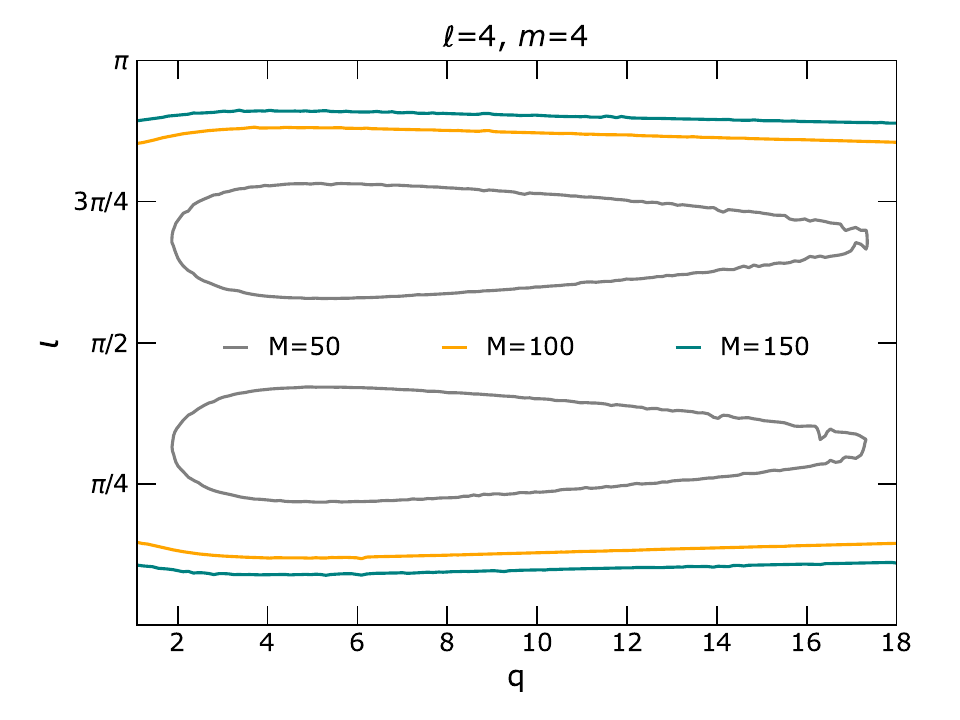}
\includegraphics[trim=20 10 20 10, clip, width=0.49\linewidth]{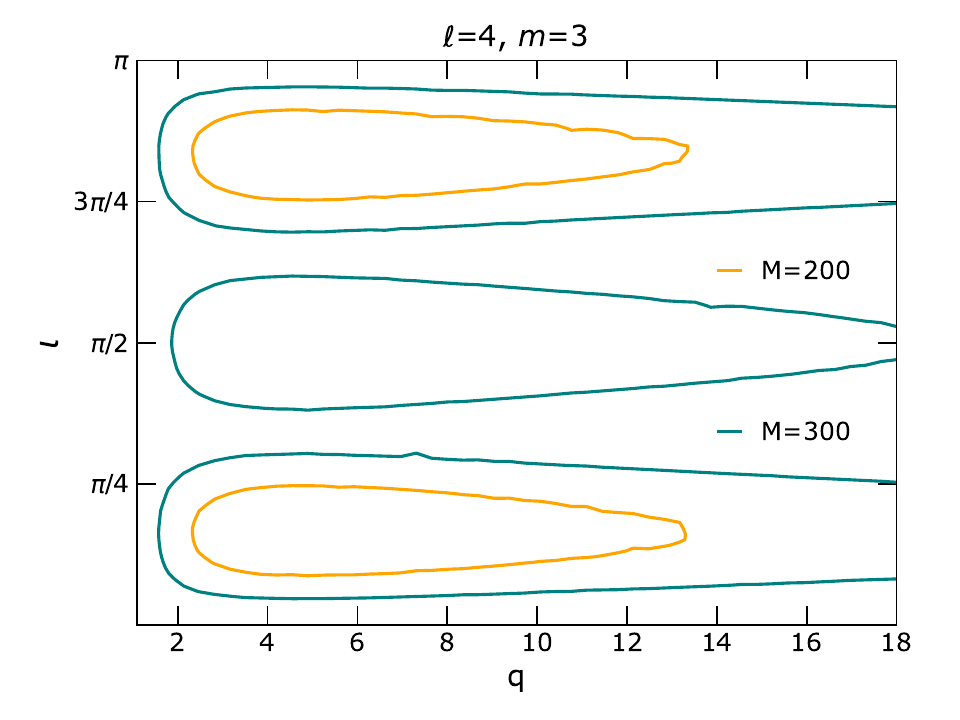}
\caption{Same as Fig.~\ref{fig:hm-q-iota} but for low spin values $\chi_\text{1z}=0.3$ and $\chi_\text{2z}=0.2$.}
\label{fig:hm-q-iota-low-spins}
\end{figure}

We can see these distinct (and somewhat complementary) trends in $\iota$ as well. Again, for a total mass of 100 M$_\odot$, and a fixed mass ratio (say $q=10$), 33 mode covers almost the entire $\iota$ range, whereas for 44 mode it is somewhat restricted. It is interesting to note that for 21 and 32 modes, the $\iota$ coverage is almost complementary, with 21 covering (a little more than) the range between ($\pi/4$, $3\pi/4$) and 32 covering the rest. It can also be seen that as the total mass increases, the contours include a larger region of the parameter space. The bi- and tri-modality of these contours reflects the symmetries these modes possess with respect to change in $\iota$. 

While Fig.~\ref{fig:hm-q-iota} shows the contours for high spin case, the trends remain the same for low spins too. Going from low spins ($\chi_{1z}=0.3$, $\chi_{2z}=0.2$) to high spins ($\chi_{1z}=0.9$, $\chi_{2z}=0.8$) the SNR increases very slightly in 33 mode (less than 15\% increase), moderately in 44 and 32 modes (nearly 30\% increase), and visibly in 21 and 43 modes (nearly 50\% increase). This, however, does not change the overall shape of the contour (Fig.~\ref{fig:hm-q-iota-low-spins}). Due to the increase in SNR when going from lower to higher spins, some of the mass contours in the respective mode plots are absent in Fig.~\ref{fig:hm-q-iota-low-spins} compared to Fig.~\ref{fig:hm-q-iota}. For instance, we see that $M=55$ is absent for 32 mode in Fig.~\ref{fig:hm-q-iota-low-spins}.

%%%%%%%%%%%%%%%%%%%%%%%%%%%%%%%%%%%%%%%%%%%%%%%%

\section{Population Study}
\label{sec:hm-pop}

Our knowledge of the BBH population in the universe has evolved from the first observation run through the third observation run of the LIGO/Virgo detectors \citep{LIGOScientific:2020kqk, LIGOScientific:2018jsj, KAGRA:2021duu}. Here, we employ the population models of \cite{LIGOScientific:2020kqk} to synthesize a BBH population and assess the detectability of various subdominant modes by using the method introduced earlier. 

\subsection{Population models}
\label{subsec:hm-pop-models}

We consider two different mass distribution models, \textsc{Power Law + Peak (PL+P)} and \textsc{Broken Power Law (BPL)} outlined in \cite{LIGOScientific:2020kqk}. The primary mass distribution for the PL+P model is given by
\begin{subequations}
\begin{align}
p(m_1) &= \big[(1-\lambda_{\rm peak})\mathcal{B}(m_1) + \lambda_{\rm peak}G(m_1)\big]S(m_1), \\
\mathcal{B}(m) &= \mathcal{C}m^{-\alpha},\ \ m < m_\text{max},\text{\ \ \ \ and}\\
G(m) &= \frac{1}{\sqrt{2\pi}\sigma_m}\big[e^{-(m-\mu_m)^2/2\sigma_m^2}\big].
\end{align}
\end{subequations}
Here $\mathcal{C}$ is a normalization constant, and $S(m_1)$ is the smoothing function given by Eq.~(B6) of \cite{LIGOScientific:2020kqk}. The mass ratio for both mass distribution models (PL+P and BPL) is given by a power law that also includes the smoothing term and is given as
\begin{equation}
    p(q) = q^\beta S(m_1 q).
\end{equation}
For the BPL model, the primary mass ($m_1$) is distributed as follows
\begin{equation}
p(m_1) \propto 
\begin{cases}
m_1^{-\alpha_1}S(m_1), & m_1 < m_{\rm break}\\
m_1^{-\alpha_2}S(m_1), & m_1 > m_{\rm break}\\
0, & \text{otherwise,}
\end{cases}
\end{equation}
where $m_{\rm break} = m_{\rm min} + b(m_{\rm max} - m_{\rm min})$.
The values of hyper-parameters used in the above-mentioned models are given in Table \ref{table: pop models}. We have distributed the primary mass ($m_1$) in the limit of [5, 100]$M_\odot$ and the mass ratio ($q=m_1/m_2$) in the range of [1, 18], which is the maximum $q$ up to which the waveform used here is calibrated \citep{London:2017bcn}. 

\begin{table}[t!]
\def\arraystretch{1.3}
\centering
\begin{tabular}{|c|c|c|c|}
\hline
\multicolumn{2}{|c|}{\textbf{\textsc{Power Law + Peak}}} & \multicolumn{2}{c|}{\textbf{\textsc{Broken Power Law}}} \\ \hline
Parameter        & Value & Parameter  & Value \\ \hline
$\alpha$         & 2.63  & $\alpha_1$ & 1.58  \\ \hline
$\mu_m$          & 33.07 & $\alpha_2$ & 5.59  \\ \hline
$\sigma_m$       & 5.69  & b          & 0.43  \\ \hline
$\delta_m$       & 4.82  & $\delta_m$ & 4.83  \\ \hline
$\beta$          & 1.26  & $\beta$    & 1.4   \\ \hline
$m_\text{min}$        & 4.59  & $m_\text{min}$  & 3.96  \\ \hline
$m_\text{max}$        & 86.22 & $m_\text{max}$  & 87.14 \\ \hline
$\lambda_\text{peak}$ & 0.10  &            &       \\ \hline
\end{tabular}
\caption{Values of model parameters for mass models used in the population study.}
\label{table: pop models}
\end{table}

To distribute these sources to redshifts accessible to a 3G detector network, we have employed the Madau-Dickinson-Belczynski-Ng model for field BBH merger rate of \cite{Ng:2020qpk}. The volumetric merger rate reads
\begin{equation}
    \dot{n}_{\rm F}(z) \propto \frac{(1+z)^{ \alpha_{\rm F}}}{1+\big[ (1+z)/C_{\rm F} \big]^{\beta_{\rm F}}},
\end{equation}
with $\alpha_{\rm F} = 2.57, \beta_{\rm F} = 5.83, C_{\rm F} = 3.36$. Further details can be found in Appendix~B of \cite{Ng:2020qpk}. Using this model for the merger rates, the redshift is then distributed as follows:
\begin{equation}
    p(z) \propto \frac{4\pi\dot{n}_\text{F}(z)}{1+z} \left(\frac{dV_{\rm c}}{dz}\right),
\end{equation}
where $V_{\rm c}$ represents the comoving volume. We have used the recently developed \texttt{python} package \texttt{gwbench}~\citep{Borhanian:2020ypi} for the distribution of redshift. The range for $z$ has been taken as [0, 10]. The redshift distribution of population of sources, along with the resultant (total) mass distribution from the PL+P and BPL mass models, is shown in Fig.~\ref{fig:hm-mass-models}. 

\begin{figure}[p!]
\centering
\includegraphics[trim=10 10 10 10, clip, width=0.64\linewidth]{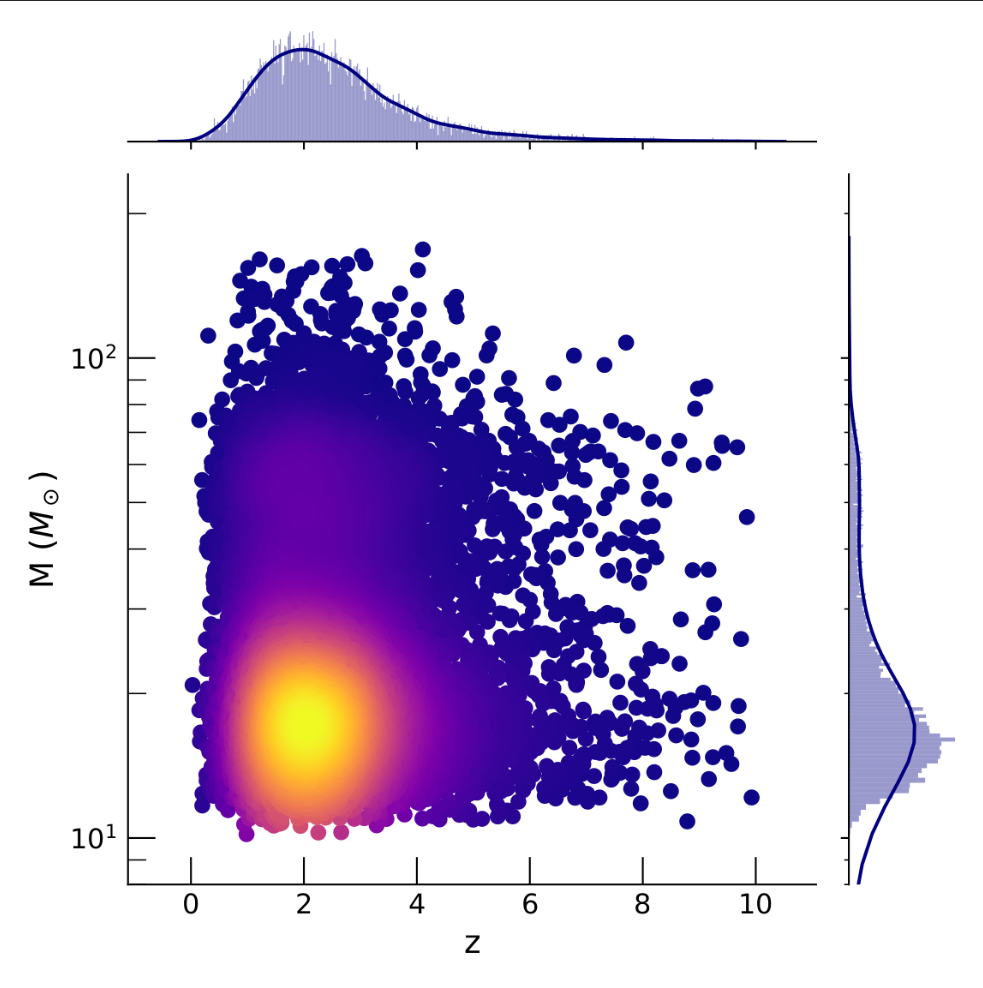}
\includegraphics[trim=10 10 10 10, clip, width=0.64\linewidth]{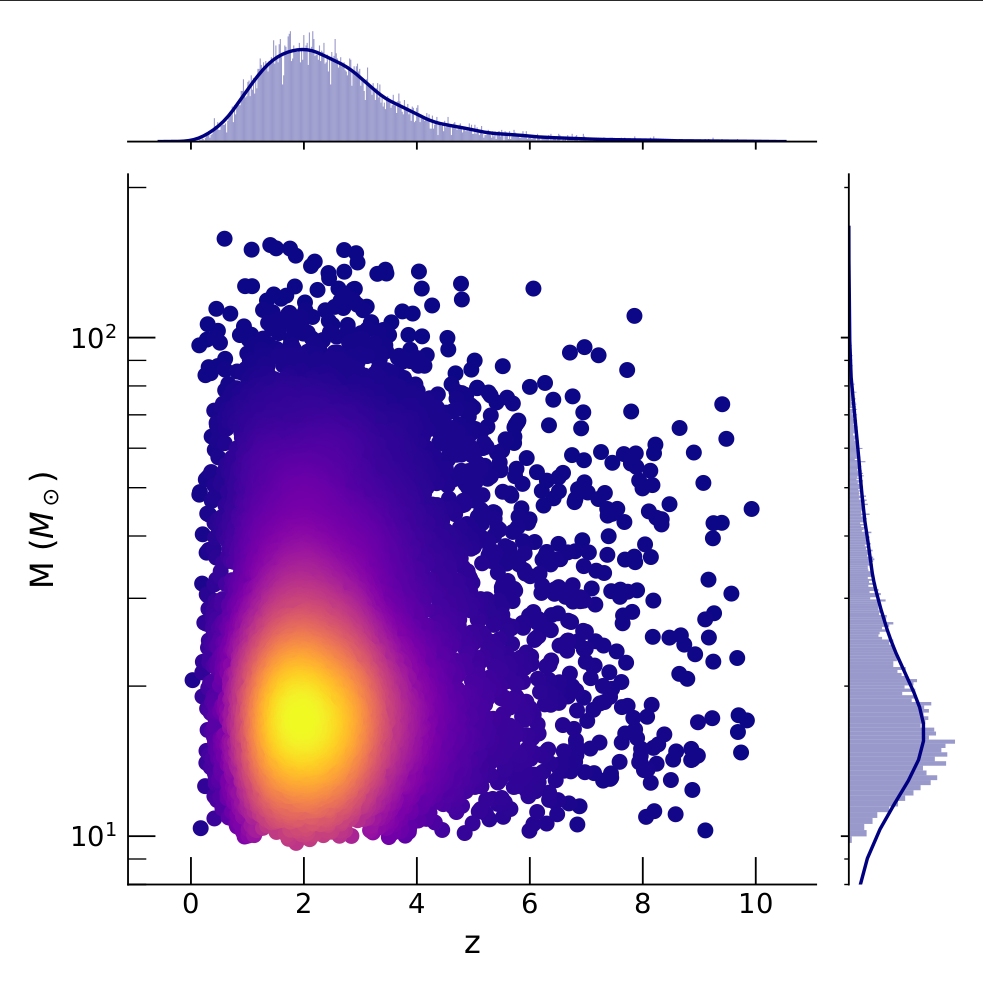}
\caption[2D distribution of injected values of total mass ($M$) and redshift($z$)]{2D distribution of injected values of total mass ($M$) and redshift($z$). Top: Mass model is \textsc{Power Law + Peak}, Bottom: Mass model is \textsc{Broken Power Law}. Redshift has been distributed according to the MDBN merger rate and taken to be the same for both mass distributions. The colours (in the scatter plot) from yellow to blue denote the number density of samples going from most to least dense regions.}
\label{fig:hm-mass-models}
\end{figure}

The spins are distributed using the \textsc{Default Model} of \cite{LIGOScientific:2020kqk} for dimensionless spin magnitude ($\chi_{1,2}$) given by
\begin{equation}
p(\chi) = \text{Beta}(\alpha_\chi, \beta_\chi).
\label{eq: spin mag}
\end{equation}
The values of $\alpha_\chi$ and $\beta_\chi$ are computed using the mean ($\mu_\chi$) and variance ($\sigma^2_\chi$) for the Beta distribution. We found these values to be $\alpha_\chi=$ 6.3788 and $\beta_\chi=$ 2.2412 and used them in constructing the distribution for dimensionless spin magnitudes ($\chi_{1,2}$). Further the cosine of the tilt angle, defined as $z_\chi = \cos(\theta_{1,2})$, is distributed as $p(z_\chi)$ \citep[see Sec.~D1 of][for related details]{LIGOScientific:2020kqk}. The distribution $\chi_{\rm 1z}$ and $\chi_{\rm 2z}$ reads
\begin{equation}
p(\chi_{\rm 1z,2z}) = p(\chi_{\rm 1z,2z})\ p(z_\chi).
\label{eq:spin-dist}
\end{equation}
We have taken the same distribution for $\chi_{\rm 1z}$ and $\chi_{\rm 2z}$, in the range [-1,1].\footnote{Note that the waveform we use (\texttt{IMRPhenomHM}) is calibrated up to spin magnitude of 0.85 (0.98 for equal mass systems); however, we have extended this up to 1 to include the complete range of spin magnitudes.} We simulate 10,000 sources following the above-mentioned prescription. We vary all the nine parameters \{$d_L$, $\iota$, $M$, $q$, $\theta, \phi$, $\psi$, $\chi_{\rm 1z}$, $\chi_{\rm 2z}$\}, following the population models. The ranges for total mass, mass ratio, spins, and redshift have been mentioned with their respective population models above. The cosines of the angles $\iota$ and $\theta$ have been varied uniformly between [-1, 1], and $\phi$ and $\psi$ are uniform between [$0, 2\pi$]. It is worth mentioning that due to the low mass of the secondary, which falls in the NS-BH mass gap, GW190814 is an outlier. Thus, the O3a population models do not include it while calculating the hyper-parameters \citep{LIGOScientific:2020kqk}.

\subsection{Detector networks}
\label{subsec:hm-det-network}

\begin{table}[t!]
\def\arraystretch{1.3}
\begin{tabular}{|c|c|c|c|c|c|}
\hline
\textbf{Label} & \textbf{Location} & \textbf{Latitude} & \textbf{Longitude} & \textbf{Orientation} & \textbf{Type(s)} \\ \hline
L & \begin{tabular}[c]{@{}c@{}}Louisiana, \\ USA\end{tabular} & 0.53 & -1.58 & -1.26 & CE/A+/Voyager \\ \hline
H & \begin{tabular}[c]{@{}c@{}}Washington, \\ USA\end{tabular} & 0.81 & -2.08 & -2.51 & CE/A+/Voyager \\ \hline
V & Cascina, Italy & 0.76 & 0.18 & 2.8 & CE/A+/Voyager \\ \hline
A & \begin{tabular}[c]{@{}c@{}}New South \\ Wales, Australia\end{tabular} & -0.59 & 2.53 & 0.78 & CE \\ \hline
E & Cascina, Italy & 0.76 & 0.18 & 2.8 & ET \\ \hline
\end{tabular}
\caption[The detector locations used in HM detectability study]{The detector locations which have been used in this study~\citep{Borhanian:2020ypi}. All angle values are in radians. Some of these sites have not been finalized yet and have been planned/proposed for future detectors.}
\label{table:det-loc}
\end{table}

We consider the detection of higher modes with network(s) consisting of two kinds of 3G detectors: CE and ET. We also compare the detection of higher modes in the 3G network with that in the upgraded 2G networks with LIGO A+ configuration \citep{McClelland:T1500290-v3, Shoemaker:2019bqt} and LIGO Voyager \citep{McClelland:T1500290-v3}. Both LIGO A+ and LIGO Voyager are expected to have an overall improved sensitivity compared to current generation detectors.

A three-detector network of 3G detectors is used in the analyses presented here. Our primary 3G network consists of a detector with CE configuration in the US (LIGO-Livingston site), ET in Europe (at the Virgo site), and another CE detector in Australia. We refer to this network as the LAE network of 3G detectors. Additionally, three different 3-detector networks have been used to study the detectability of higher-order modes as detectors evolve through LIGO A+, LIGO Voyager, and CE configurations. The sensitivity curves and locations of these detectors are shown in Fig.~\ref{fig:intro-PSDs} and Table \ref{table:det-loc}, respectively. For each detector, we put the lower frequency bound ($f_{\text{low}}$) as 5\,Hz following \cite{Chamberlain:2017fjl}.\footnote{Note that in \cite{Chamberlain:2017fjl}, authors have used a low-frequency cutoff of 1 Hz for ET configuration; however, we work with a universal low-frequency cutoff of 5 Hz for all detector configurations (LIGO A+, LIGO Voyager, CE, or ET) in this work.} While signals in 3G detector networks are expected to last much longer than the current ground-based detectors, in our study we have not included the effect of rotation of the Earth on the observed signals. This is because majority of the systems taken in our study lie in the mass range of [10, 30] $M_\odot$ and the signals will last only a few minutes. Thus we posit that the conclusions drawn in our study are not affected by ignoring this effect.

\subsection{SNR distribution of higher modes}
\label{sec:hm-pop hist}

Figure \ref{fig:hm-SNR histograms} shows the cumulative histograms for the SNRs of various modes in the LAE network (two CE detectors at L \& A, and one ET detector at E) of 3G detectors (see Table \ref{table:det-loc} for details). The numbers in brackets are the median values of SNR for each mode. For a particular value of SNR (on the $x$-axis), the $y$-axis corresponds to the fraction of the population having SNR up to that value. The dashed lines correspond to the population distributed according to the BPL mass model, while the solid lines correspond to the PL+P distribution. Since the results from the two models are very close, we only quote numbers corresponding to the PL+P model. We find that 99.8\% of the simulated population following the PL+P model has SNR $>$10 in 22 mode. Note that the detection fractions we quote in the subsequent sections are obtained by putting a minimum cutoff of SNR $>10$ for the 22 mode, and SNR $>3$ for all the other modes.

\begin{figure}[t!]
\centering
\includegraphics[trim=30 0 30 30, clip, width=0.7\linewidth]{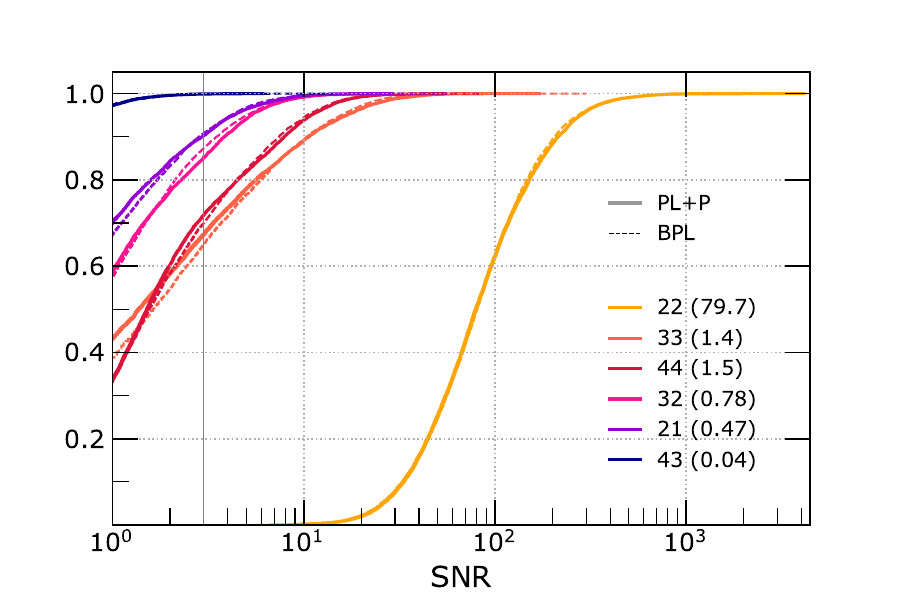}
\caption[Cumulative SNR histograms for six leading modes, with the two mass distribution models (PL+P and BPL) coupled with the MDBN model for redshift evolution]{Cumulative SNR histograms for six leading modes, with the two mass distribution models (PL+P and BPL) coupled with the MDBN model for redshift evolution. The plot shows trends for each mode for SNRs $>1$. The solid and dashed coloured lines correspond to the population drawn from the PL+P and BPL models, respectively. The values in the brackets are the median values of SNR for each mode. The detector network used here consists of a detector with CE configuration in the U.S. (at the LIGO-Livingston site), a detector with ET design in Europe (at the Virgo site), and another CE detector in Australia, and referred to as 3G LAE network in this chapter.} 
\label{fig:hm-SNR histograms}
\end{figure}

We find that the 33 mode is detectable in nearly 33\% of the sources, and 44 mode is detected in $\sim$28\% of the population. 32 and 21 modes are detectable in nearly 15\% and 10\% sources, respectively, while the detected fraction is only $\sim$0.1\% for the 43 mode. This demonstrates that for a population of sources similar to those detected in GWTC-2, with the increased reach of 3G detectors to higher redshifts, we will detect most higher modes in many sources.

\subsubsection{Comparison between various generations of detector networks}
\label{subsec:hm-2-2.5-3G hist}

\begin{figure}[t!]
\centering
\includegraphics[trim=30 0 30 30, clip, width=0.7\linewidth]{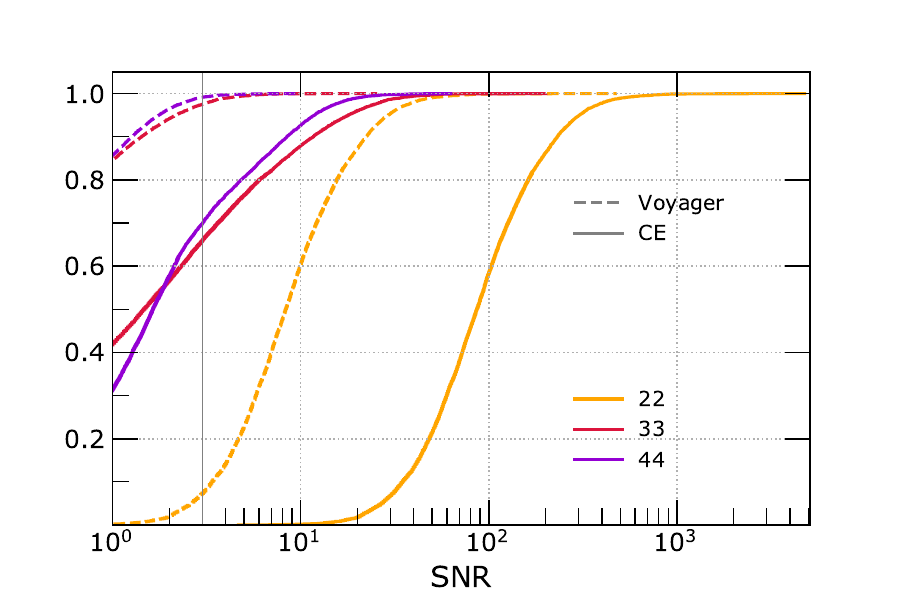}
\caption[Cumulative histograms of SNR for various higher modes as a comparison between various generations of detectors]{Cumulative histograms of SNR for various higher modes as a comparison between various generations of detectors. The solid, coloured lines denote the 3G detector network with CE, while the dashed lines denote the detector network using LIGO Voyager. Both the detectors have been placed at the locations of the LHV network. We have taken $q_{max}=18$. 
}
\label{fig:hm-SNR 2.5-3g}
\end{figure}

In this section, we investigate how the detectability of higher modes changes for a population across various generations of detector networks. We consider three kinds of detector networks for this study: LIGO-A+, LIGO-Voyager, and CE detector networks. To avoid detector location bias, we have chosen the same locations, Livingston and Hanford in the U.S. and Cascina in Italy (LHV), for all three networks. For constructing this population, we have considered the \textsc{Power Law + Peak} mass model. We have shown the results for CE and Voyager networks in Fig.~\ref{fig:hm-SNR 2.5-3g}. A quick look at Fig.~\ref{fig:hm-SNR 2.5-3g} reveals that the LIGO Voyager detector network can barely detect the two additional modes besides the (dominant) 22 mode, whereas 3G detectors have a good detection percentage for 33 and 44 modes. As expected, the improvement with 3G detectors is highly significant compared to the upgraded 2G configurations. Note that here, we quote the detection percentages of higher modes, over and above the detection of 22 mode. 
Again, as seen in the previous section, nearly 100\% sources are detectable in 22 mode in the detector network formed by the CE detectors at locations of current LIGO and Virgo detector sites. 
Out of these, nearly 35\% and 30\% of the sources show a detection in 33 and 44 modes, respectively. This detection percentage drastically decreases for Voyager and A+. For Voyager, only 40\% of the simulated population is detectable in 22 mode, out of which only $\sim$6\% and $\sim$2\% show a detection in 33 and 44 modes, respectively. For A+, only about 7\% of the systems show a detection in 22 mode, out of which $\sim$1\% show 33 mode, and $\sim$0.5\% show a detection in 44 mode. Considering the detection rate numbers from \cite{Baibhav:2019gxm}, these percentages can still result in the detection of a considerable number of HMs in the detected population. This shows the tremendous potential of a network of 3G detectors, and how the number of higher mode detections will significantly rise compared to the currently operating 2G detectors. This, in turn, will also have a profound impact on the overall parameter estimation capabilities of the 3G detectors, and hence influence the astrophysics and fundamental physics in the 3G era in a big way.

\section{Summary} 
\label{sec:hm-concl}

We have investigated the detectability of higher modes of gravitational radiation in mass ratio and inclination angle space (Sec. \ref{sec:hm-q-iota}). The effect of increasing total mass (in discrete values) was also noted in Sec.~\ref{sec:hm-q-iota}. We find that various modes activate different regions of the ($q$-$\iota$) plane, and show various symmetries in $\iota$, leading to bi- and tri-modality in the contours. We also explored the detectability of higher modes for a few events from the GWTC-2 catalog, assuming they were detected during the 3G era (Sec. \ref{sec:hm-gwtc2}). For the GWTC-2 events, we observe a massive improvement of SNR (by a factor larger than 20 times) for the events reported to show detection in 33 mode by LIGO/Virgo observations (GW190814, GW190412). In the 3G era, events like these promise to show detection in other higher harmonics as well. Additionally, we also see detectable SNRs in higher modes for other GWTC-2 events, which emphasizes that the number of events that permit the measurement of higher modes will also increase in the 3G era.

We also performed a population study for 10,000 sources, which will be detected by the 3G network, and quote the fraction of the population which will show the presence of higher modes (Sec.~\ref{sec:hm-pop hist}). It is found that nearly 33\% and 28\% of the sources will have detectable SNRs in 33 and 44 modes, respectively, and other modes will also be detectable in a small percentage of the population. Additionally, we compared this fraction with the fraction of higher modes detectable in the upgraded 2G gravitational wave detector networks, such as LIGO A+ and LIGO Voyager, using the PL+P mass model. We conclude that this fraction significantly increases from less than 6\% to nearly 35\% for the 33 mode, with the 3G network.

All of the above-mentioned investigations were performed using a spinning inspiral-merger-ringdown higher mode waveform of the \texttt{Phenom} family (\texttt{IMRPhenomHM}), and using a different waveform may alter the numbers only slightly.
      \chapter{Conclusions and future directions}
\label{chap:concl}

In this \textbf{thesis} we have explored the effects of orbital eccentricity, spin-precession, and higher modes on data analysis of gravitational signals from binary black hole systems. We have studied the interplay of these effects from various angles performing a variety of injection studies. We started with investigating the loss of signals when searches do not include eccentric waveform models for a population which contained eccentric signals, by calculating the signal recovery fraction. We found that high mass ratio binaries with non-negligible values of eccentricities show the maximum loss of signal recovery fraction. This was followed by a detailed parameter estimation study of simulated eccentric signals, which were analysed with eccentric and quasi-circular waveforms in various spin configurations. We observed that quasi-circular waveforms in any spin-configuration were incapable of correctly characterizing the chirp mass of the source. This also shed some light on the interplay of orbital eccentricity with spin-precession and we concluded that these effects do not mimic each other for signals with long inspirals and moderate eccentricities. We also performed a few injections using eccentric higher mode signals and showed that quasi-circular higher mode waveforms also bias the recovered parameters, same as was observed using only dominant mode waveforms. 

Next, the effect of spin-precession on spin-induced quadrupole moment parameter, which presents as a null test for binary black hole nature of the GW signal source, was explored. For various low and high precessing spin systems, comparisons were made between the analyses with two different waveforms, highlighting the difference between them and recommending the use of waveforms with double spin-precession for the test when dealing with signals which have non-negligible spin precession. Here too, the interplay of spin-precession was studied along with higher modes and it was observed that for real events which have non-negligible spins and mass asymmetry, use of higher mode waveforms with double-spin precession would lead to better results.

Finally, we studied the detectability of higher modes in the next-generation ground-based GW detectors such as Cosmic Explorer and Einstein telescope. We showed that the improvement in sensitivities, especially in the lower range of these detector frequencies, will lead to a considerable signal-to-noise ratio in the higher modes in addition of the dominant mode. We also quantified the regions in mass ratio and inclination angle space to highlight how various modes get excited in different portions of the parameter space and sometimes complement each other. For the 3G detector networks, the population of binary black holes showed detection of several higher-order modes, which was not the case for the upgraded 2G network of detectors.

All of the studies performed above lead to various implications about the formation channels of the black hole binary systems. Orbital eccentricity and spin-precession are important markers to indicate that the binary may not have formed in isolation. Studying these effects in details can help us categorize the populations better and estimate the merger rates. Injection studies which highlight the biases when using different waveform models, both for null tests of general relativity and for parameter estimation of binary black hole systems, help keep the scientific community informed of the limitations that the present waveform models possess, and caution the community against various inferences drawn from the results. The detection and inclusion of higher modes in waveform models helps break degeneracies between various parameters that aid in their more accurate and precise estimation. This in turn has larger applications in cosmology such as the estimation of Hubble constant. 

While at present there are no IMR models that include all three physical effects (orbital eccentricity, spin-precession, and higher order modes), several groups are actively working on their development. Once these models become available, studies included in Chapter \ref{chap:ecc} of this \textbf{thesis} can be extended further. While the analysis presented in Chapter \ref{chap:siqm} studies waveform systematics on the SIQM tests for various spin and mass ratio configurations by injecting BBH waveforms consistent with GR, a detailed follow-up study with non-GR injections may be carried out in a future work.
While in Chapter \ref{chap-hm} we have shown the detectability of various modes for spinning systems, a more detailed study can be done by including the effect of spin-precession along with higher modes. Further, the effect of higher modes in the error analysis of various parameters in 3G detector era can be explored.
%Also, an independent study can be conducted where the injected signals are not from binary black hole mergers but instead have some other values for the SIQM deviation parameter. These can then be used to probe the effect of higher modes, orbital eccentricity, and spin-precession on the null tests of general relativity. %A more detailed study can be done for the detectability of higher modes by including the effect of precession along with the higher modes. While we have touched upon the detectability of higher modes in third generation detectors, the parameter estimation studies mentioned above can also be extended for 3G network. Finally, we can injected signals other than binary black hole merger and see the effect of higher modes, orbital eccentricity, and spin-precession on the null tests of general relativity.
      
      \appendix
      \chapter{Spin-induced quadrupole moment}
\label{chap:appn:siqm}

The definition of a Kerr black hole (BH) is closely associated with the "no-hair" conjecture, which states that all the characteristics of a Kerr black hole are entirely determined by its mass and spin. Measuring the spin-induced quadrupole moments (SIQM) of a compact binary's constituents can test for the black hole nature of the binary's compact objects since these values are unique for Kerr BHs due to the no-hair conjecture. For an isolated Kerr BH, the quadrupole moment scalar is given by $Q \propto \chi^2 m^3$ \citep{Poisson:1995ef}, where $m$ is the mass and the spin magnitude $\chi$ is defined as $\chi = \frac{|\Vec{S}|}{m^2}$ with $\Vec{S}$ as the spin angular momentum of the BH. For a non-black hole compact object, this can be generalized to $Q = -\kappa m^3 \chi^2 $, where $\kappa = 1$ represents the black hole limit. For a binary system, one can define the symmetric and anti-symmetric combinations of $\kappa$ as:
\begin{align}
    \kappa_s &= \frac{\kappa_1 + \kappa_2}{2} \\
    \kappa_a &= \frac{\kappa_1 - \kappa_2}{2}.
\end{align}
Now, in order to pose this as a null test of binary black hole nature, one can introduce a deviation parameter $\delta\kappa$ such that $\kappa_i = 1 + \delta\kappa_i$. In this parameterization, $\delta\kappa_i = 0$ represents the black hole limit. Again, we can define the symmetric and anti-symmetric combinations as:
\begin{align}
    \delta\kappa_s &= \frac{\delta\kappa_1 + \delta\kappa_2}{2} \\
    \delta\kappa_a &= \frac{\delta\kappa_1 - \delta\kappa_2}{2}.
\end{align}
Here, choosing $\delta\kappa_a = 0$, a measurement of $\delta\kappa_s = 0$ recovers the \textit{binary} black hole limit.

%%%%%%%%%%%%%%%%%%%%%%%%%%%%%%%%%%%%%%%%%%%%

Here, we explicitly write the phasing coefficients which have been changed to include the SIQM deviation parameter ($\delta\kappa_i$). These have been derived from \cite{Krishnendu:2017shb}\footnote{See the supplemental material of \cite{Krishnendu:2017shb} for complete waveform expressions.}. Schematically, the expression for phase in Stationary Phase Approximation (SPA) \citep{Arun:2008kb} can be written as:
\begin{equation}
 \Psi_\text{SPA}(f) = 2\pi f t_c - \phi_c + \bigg\{\frac{3}{128\eta \varv^5}[\psi_\text{NS} + \psi_\text{spin}]\bigg\}.
\end{equation}
Here, $\varv = (\pi M f)^{1/3}$, where the total mass $M$ is taken in natural units, and $\phi_c$ denotes the phase at the time of coalescence $t_c$. Now, we can separate the spinning part of phase $\psi_\text{spin}$ as:
\begin{equation}
    \psi_\text{spin} \equiv \psi_\text{SO} + \psi_\text{SS} + \psi_\text{SSS} = \varv^3[\mathcal{P}_3 + \mathcal{P}_4 \varv + \mathcal{P}_5 \varv^2 + \mathcal{P}_6 \varv^3 + ...].
\end{equation}
Expressions for the coefficients $\mathcal{P}_n$ can be found in \cite{Arun:2008kb, Vallisneri:2007ev}, where the explicit dependence on $\kappa_s$ and $\kappa_a$ is omitted by setting them to their respective values for Kerr BBHs. Here, we list the expressions which contain the symmetric and anti-symmetric forms of SIQM deviation parameters ($\delta\kappa_s$ and $\delta\kappa_a$) that have been included in the waveform used for SIQM studies in this \textbf{thesis}.
\begin{subequations}
\begin{eqnarray} 
\label{eq:P4} 
{\mathcal P}_4 &=&
-\frac{5}{8}(\boldsymbol{\chi}_\mathrm{s}\cdot\hat{\boldsymbol{L}}_\mathrm{N})^2\,
\Bigl[1+156 \,\eta +80 \,\nu  \,(1 + \delta\kappa_a) + 80 (1-2 \,\eta ) (1+\delta\kappa_s)\Bigr] \nonumber\\
&+&(\boldsymbol{\chi}_\mathrm{a}\cdot\hat{\boldsymbol{L}}_\mathrm{N})^2
\left[-\frac{5}{8}-50 \,\nu  \,(1+\delta\kappa_a) - 50 (1+\delta\kappa_s)+100 \,\eta
\left(2+\delta\kappa_s\right)\right] \nonumber\\
&-&\frac{5}{4}(\boldsymbol{\chi}_\mathrm{a}\cdot\hat{\boldsymbol{L}}_\mathrm{N})(\boldsymbol{\chi}_\mathrm{s}\cdot\hat{\boldsymbol{L}}_\mathrm{N})
\Bigl[\nu +80\,(1-2 \,\eta )\,(1+\delta\kappa_a)+80 \,\nu  \,(1+\delta\kappa_s)\Bigr]\,,\\
%%%%%%%%%
{\mathcal P}_6 &=& \pi\,\Biggl[\frac{2270}{3}\,\nu
\,\boldsymbol{\chi}_\mathrm{a}\cdot\hat{\boldsymbol{L}}_\mathrm{N}
+\left(\frac{2270}{3}-520\,\eta\right)
\boldsymbol{\chi}_\mathrm{s}\cdot\hat{\boldsymbol{L}}_\mathrm{N}\Biggr] + (\boldsymbol{\chi}_\mathrm{s}\cdot\hat{\boldsymbol{L}}_\mathrm{N})^2
\left[-\frac{1344475}{2016} \right. \nonumber\\
&+&
\left. \frac{829705}{504}\,\eta+\frac{3415}{9}\,\eta ^2
+\nu \left(\frac{26015}{28}-\frac{1495}{6}\,\eta\right) (1+\delta\kappa_a) + \left(\frac{26015}{28}-\frac{44255}{21} \,\eta \right. \right. \nonumber\\
&-&
\left. \left. 240 \,\eta ^2\right) (1+\delta\kappa_s)\right]
+ \boldsymbol{\chi}_\mathrm{a}\cdot\hat{\boldsymbol{L}}_\mathrm{N})^2
\left[-\frac{1344475}{2016}+\frac{267815}{252} \,\eta - 240 \,\eta ^2 \right. \nonumber\\
&+&
\left. \nu \left(\frac{26015}{28} - \frac{1495}{6} \,\eta \right) (1+\delta\kappa_a) + \left(\frac{26015}{28} -
\frac{44255}{21} \,\eta -240 \,\eta ^2\right) (1+\delta\kappa_s)\right] \nonumber\\
&+& (\boldsymbol{\chi}_\mathrm{a}\cdot\hat{\boldsymbol{L}}_\mathrm{N})(\boldsymbol{\chi}_\mathrm{s}\cdot\hat{\boldsymbol{L}}_\mathrm{N})
\left[\left(\frac{26015}{14}-\frac{88510}{21} \,\eta -480 \,\eta ^2\right)
(1+\delta\kappa_a) + \nu\, \biggl[-\frac{1344475}{1008} \right. \nonumber\\
&+&
\left. \frac{745}{18} \,\eta
+\left(\frac{26015}{14} - \frac{1495
}{3} \,\eta\right) (1+\delta\kappa_s) \biggr]\right]\,, \label{eq:P6}
\end{eqnarray} 
\end{subequations}
where $\nu$ is the difference mass ratio given by $|m_1 - m_2|/(m_1 + m_2)$. Since $\mathcal{P}_3$ and $\mathcal{P}_5$ only contain spin-orbit terms, they do not exhibit a dependence on $\kappa$ parameters and hence have not been included here.

These changes have been included in the \texttt{LALSimulation} module of the \texttt{LALSuite} package and can be called by various \texttt{Phenom} waveforms such as \texttt{IMRPhenomD}, \texttt{IMRPhenomPv2}, \texttt{IMRPhenomPv3}, \texttt{IMRPhenomHM}, \texttt{IMRPhenomPv3HM}, and \texttt{PhenomX} family of waveforms such as \texttt{IMRPhenomXAS}, \texttt{IMRPhenomXHM}, \texttt{IMRPhenomXP}, \texttt{IMRPhenomXPHM}. We have used these waveforms for the injection studies and real event analyses presented in Chapter \ref{chap:siqm} of this \textbf{thesis}.
  \end{MainMatter}
  \begin{BackMatter}
      \printBibliographyReferences{master.bib}
      %\input{bibliography}
      %\printCV
      %\printCommittee
  \end{BackMatter}
\end{document}